\documentclass[fleqn,usenatbib]{mnras}
\usepackage{newtxtext,newtxmath}
\usepackage[T1]{fontenc}
\usepackage{graphicx}	
\usepackage{amsmath}	
\edef\next{\toks0=%
   {\catcode`\noexpand\@=\the\catcode`\@\toks0{\the\toks0}}%
}
\next
\catcode`\@11
\def\next#1#2#3{\expandafter \def \csname bb@#1\endcsname##1%
  {#2\csname bb@#3##1\endcsname}}
\next{00}00 \next{01}01 \next{02}10 \next{03}11
\next{04}20 \next{05}21 \next{06}30 \next{07}31
\next{08}40 \next{09}41 \next{10}50 \next{11}51
\next{12}60 \next{13}61 \next{14}70 \next{15}71
\next{16}80 \next{17}81 \next{18}90 \next{19}91
\expandafter \def \csname bb@0+\endcsname {+0}
\expandafter \def \csname bb@1+\endcsname {+1}
\def\bb@endbinary#1+{\fi\fi}
\expandafter \def \csname bb@0-\endcsname {0+-\bb@dobinary}
\expandafter\def\csname bb@0m\endcsname#1+{#1+0}
\expandafter\def\csname bb@1m\endcsname#1+{#1+1}
\def\bb@dobinary#1#2{\if#10\if m\string#2\else\bb@endbinary\fi\fi
 \expandafter\bb@dobinary\number\csname bb@0#1\endcsname#2}
\def\nbinary#1#2{\expandafter\bb@dobinary\number\number#2%
 \romannumeral\number\number#1 000+}

\def \next #1#2{\expandafter \def
 \csname bb@h\number +#1\endcsname ##1+{\bb@dohex ##1+#2}%
}
\next   {0}0 \next   {1}1 \next  {10}2 \next  {11}3
\next {100}4 \next {101}5 \next {110}6 \next {111}7
\next{1000}8 \next{1001}9 \next{1010}A \next{1011}B
\next{1100}C \next{1101}D \next{1110}E \next{1111}F
\def\bb@dohex #1{\csname bb@x#1\endcsname}
\def\bb@x\endcsname#1{ \bb@xm{m\endcsname}}
\def\bb@xm #1\endcsname #2#3+{#2#3%
 \csname bb@h\number+\endcsname
 #1\endcsname m#3+}
\def\bb@nbinbased #1#2#3{\expandafter \bb@dobinary \number#1%
 \expandafter \bb@dohex
 \romannumeral \number\number #2 000\expandafter\endcsname
 \romannumeral \number\number #3 000+}
\def\nbinbased #1#2#3{\expandafter\bb@nbinbased
 \expandafter {\number#3}{#2}{#1}}

\the\toks0
\usepackage{pdfpages}
\usepackage{epsfig}  
\usepackage{ulem}
\usepackage{longtable, lscape}
\usepackage{tablefootnote}
\usepackage{subfloat}
\usepackage{mathrsfs}
\usepackage{verbatim}
\usepackage{aas_macros}  
\usepackage{longtable}
\usepackage{ltablex}
\usepackage{booktabs}
\usepackage[flushleft]{threeparttable}
\usepackage[percent]{overpic}
\title[RRd Stars Observed by {\it K2}]{Double-Mode RR\,Lyrae Stars Observed by {\it K2}: Analysis of \\ High-Precision {\it Kepler} Photometry }

\author[James M.\,Nemec et al.]{James M.\,Nemec$^{1}$\thanks{E-mail: jmn@isr.bc.ca}, Amanda\,F.\,Linnell Nemec$^{1}$,    Pawel\,Moskalik$^{2}$, L\'aszl\'o\,Moln\'ar$^{3,4,5}$, 
\newauthor Emese\,Plachy$^{3,4,5}$, R\'obert\,Szab\'o$^{3,4,5}$ and Katrien\,Kolenberg$^{6,7}$
\\
$^{1}$International Statistics \& Research Corporation, Brentwood Bay, British Columbia, Canada\\
$^{2}$Nicolaus Copernicus Astronomical Center, Warszawa, Poland\\
$^{3}$Konkoly Observatory, HUN-REN CSFK, Konkoly-Thege Mikl\'os \'ut 15-17, H-1121, Budapest, Hungary\\
$^{4}$CSFK, MTA Centre of Excellence, Budapest, Konkoly-Thege Mikl\'os \'ut 15-17, H-1121, Budapest,  Hungary\\
$^{5}$ELTE E\"otv\"os Lor\'and University, Institute of Physics \& Astronomy, 1117, P\'azm\'any P\'eters\'et\'any 1/A, Budapest, Hungary\\
$^{6}$Institute of Astronomy, KU Leuven, Heverlee, Belgium\\
$^{7}$Physics Department, University of Antwerp, Antwerpen, Belgium   
}

\date{Accepted February 6, 2024;  Received February 6, 2024;  in original form December 5, 2023}

\pubyear{2024}

\begin{document}
\label{firstpage}
\pagerange{\pageref{firstpage}--\pageref{lastpage}}
\maketitle

\begin{abstract} 

The results of a  Fourier analysis of high-precision {\it Kepler} photometry of
75 double-mode RR~Lyrae (RRd) stars observed during NASA's {\it K2}
Mission (2014-18) are presented.  Seventy-two of the stars are
`classical' RRd (cRRd) stars lying along a well-defined curve in the
Petersen diagram and showing no evidence of Blazhko modulations.  
The remaining three stars are `anomalous' RRd (aRRd) stars that lie well below the
cRRd curve in the Petersen diagram.  These stars have
larger fundamental-mode amplitudes than first-overtone amplitudes and
exhibit Blazhko variations.  
Period-amplitude relations for the individual
pulsation components of the cRRd stars are examined, as well
as correlations involving Fourier phase-difference and amplitude-ratio
parameters that characterize the light curves for the two radial modes.
A simple statistical model relating the fundamental ($P_0$) and first-overtone ($P_1$) periods to [Fe/H] provides
insight into the functional form of the Petersen diagram.  A
calibration equation for estimating [Fe/H]$_{\rm phot}$ abundances of
`classical' RRd stars is derived by inverting the model and using 211 field and 57 globular cluster cRRd
stars with spectroscopic metallicities to estimate the model coefficients.  
The equation is used to
obtain [Fe/H]$_{\rm phot}$ for the full sample of 72 {\it K2} cRRd
stars and for 2130 cRRd stars observed by the ESA {\it Gaia} Mission.
Of the 49 {\it K2} cRRd stars that are in the {\it Gaia} DR3 catalogue
only five were found to be correctly classified, the remainder having
been misclassified `RRc' or `RRab'. 

\end{abstract}


\section{INTRODUCTION}


\begin{figure} 
\begin{center}
\begin{overpic}[width=8.4cm]{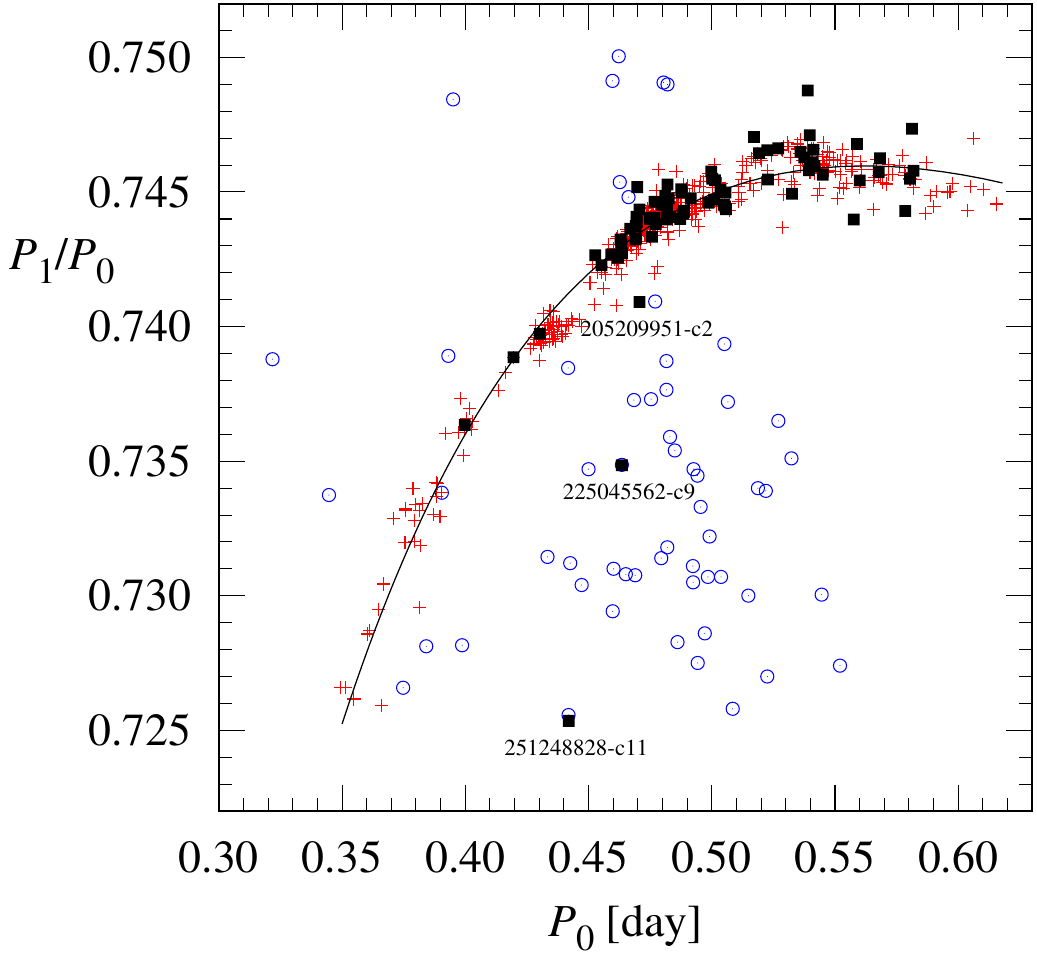} \end{overpic} 
\end{center}

\caption{Petersen diagram for the 72 `classical' and three `anomalous' RRd
stars observed by {\it K2} (solid black squares).  Also plotted are 458
Galactic Disk and Bulge  cRRd stars (red plus signs) observed by \texttt{OGLE}
(Soszy\'nski et al. 2019) and 54 aRRd stars (blue circles).  The aRRd stars are
in the globular clusters M3 (4 stars; Jurcsik {\it et al.} 2015) and NGC\,6362
(2 stars; Smolec et al. 2017a), in the Magellanic Clouds (20 in the LMC,
two in the SMC; Soszy\'nski et al. 2016b), and in the Galactic Bulge (28 stars;
Soszy\'nski et al. 2019).  The equation of the fitted  {\it K2}+\texttt{OGLE} curve is given in $\S$4.4.2.} 

\label{Fig1} 
\end{figure}

Double-mode RR Lyrae (RRd) stars are old low-mass stars burning helium in their
cores and pulsating simultaneously in the fundamental and first-overtone {\it
radial} modes.  As the two modes go in and out of phase the observed 
amplitude of the pulsation varies from cycle to cycle.   In
Hertzsprung-Russell diagrams RRd stars are found in the instability strip
between the cooler RRab stars and the hotter RRc stars, and have effective
temperatures 6200$<$$T_{\rm eff}$$<$7000\,K,  luminosities
20$<$$L/L_{\sun}$$<$60, and masses 0.55$<$$M/M_{\sun}$$<$0.90 (Christy
1966; Cox {\it et al.} 1980, 1983; Simon \& Cox 1991; 
Kov\'acs \& Karamiqucham 2021; Netzel \& Smolec 2022).

The first RR~Lyrae star in which double-mode pulsation was observed was the
high galactic latitude star AQ~Leo (Jerzykiewicz \& Wenzel 1977; see also
Gruberbauer {\it et al.} 2007).  The  subsequent discovery of dozens of RRd
stars in several globular clusters and dwarf galaxies  led to a much-improved
understanding of their properties.  For instance,  most RRd stars are now known to
lie along a well-defined  curve in a Petersen (1973) diagram (see {\bf
Figure\,1}), with the radial first-overtone pulsation mode usually dominating
over the fundamental mode (see {\bf Figure\,2}).  Such stars will
hereafter be referred to as `classical' RRd (cRRd) stars.  The RRd stars that
lie off the `Petersen curve', usually below but sometimes above, are commonly
referred to as `anomalous' RRd (aRRd) stars (Soszy\'nski et al. 2016b).    The
aRRd stars tend
to have larger fundamental than first-overtone amplitudes (blue open circles in
Fig.2) and often exhibit Blazhko variations.  Both cRRd and aRRd stars exhibit
the same approximately-linear relationship between the two periods, with the
aRRd stars showing more scatter than the cRRd stars (see Fig.15a of Nemec \&
Moskalik 2021, hereafter NM21).

It is now well-established from observations and pulsation models that the
locations of cRRd stars in Petersen diagrams correlate with metal abundance and
mass:  the shorter the period the smaller the period ratio, the greater the
metal abundance, and the smaller the mass.   The period-metallicity correlation
was first established when it was observed that the RRd stars in the most
metal-poor globular clusters (i.e., Oosterhoff type II GCs), such as M15
(Sandage, Katem \& Sandage 1981; Cox, Hodson \& Clancy 1983; Nemec 1985a) and
M68 (Clement et al.  1993), have fundamental-mode periods $P_0$ greater than
0.51\,day, while those in intermediate metallicity (Oo\,I) GCs, such as IC4499
(Clement {\it et al.} 1986; Walker \& Nemec 1996) and M3 (Nemec \& Clement
1989; Jurcsik et al.  2014, 2015), have shorter periods, 0.45$<$$P_0$$<$0.51
day.  Many RRd stars have also been found in all the nearby dwarf galaxies (see
Clementini et al. 2023 for references) and are being used to study the
metallicity variations in those systems (see Braga et al. 2022) and their
relationship to the history of our Galaxy.  The discovery of large numbers of
RRd stars in the Magellanic Clouds by the \texttt{MACHO} (Alcock et al.  1997,
2000, 2004) and \texttt{OGLE} (Soszy\'nski et al. 2009, 2010, 2016a,b) surveys,
and in the Bulge and Disk of our Galaxy (Soszy\'nski {\it et al.} 2010, 2011,
2017a,b, 2019), extended the period-metallicity trend to shorter periods,
revealing that most cRRd stars with periods $P_0$ between 0.42 and 0.45\,day
are located in a prominent clump of stars (Soszy\'nski et al., 2014b; Kunder et al. 2019),
and that the most metal-rich RRd stars (found mainly in the Galactic Bulge)
have the shortest periods, with periods as short as $P_0$=0.35\,d,  period
ratios as small as $P_1$/$P_0$=0.725, and metallicities as rich as
[Fe/H]$\sim -0.35$\,dex (see Soszy\'nski {\it et al.} 2011, 2014b).

Early theoretical models by Cox et al. (1980, 1983), which used Los Alamos
opacities, hypothesized that the radial pulsation periods and their ratios are
determined mainly by the mass $M$ and metal abundance [Fe/H], and to a lesser
degree by luminosity $L$ and effective temperature $T_{\rm eff}$.  Popielski,
Dziembowski \& Cassisi (2000, fig.2) used stellar evolution and pulsation
models with the newer opacities of Iglesias \& Rogers (1991, 1996; see also
Simon 1982 and Seaton 1994) to illustrate the impact on location in the
Petersen diagram of varying $M$, $L$, $T_{\rm eff}$ and [Fe/H].  Theoretical
curves of constant mass and constant metallicity derived from such models are
commonly overlaid onto the Petersen diagram as a means of inferring [Fe/H] and
mass (see fig.4 of Simon \& Cox 1991, fig.1 of Bono et al. 1996, fig.1 of
Alcock et al. 1997;  fig.3 of Kov\'acs 2001; fig.2 of Soszy\'nski et al.  2014a;
fig.7 of Coppola et al. 2015;  fig.4 of Braga et al. 2022). 

High-precision  surveys from space,  such as the MOST, CoRoT, {\it Kepler/K2},
{\it Gaia} and TESS missions (see Moln\'ar et al. 2022 for recent references),
have led to the recognition that, in addition to radial pulsations, most, if
not all, RRd stars exhibit low-amplitude {\it non-radial} pulsations (see
Moskalik et al. 2018a,b). Such pulsations are present in all of the
well-studied {\it K2} RRd stars, the  most prominent having a period $P_{\rm
nr} \sim 0.61P_1$, and will be discussed in detail elsewhere (Moskalik et al., in
preparation).  Other discoveries from space surveys include `period doubling',
{\it i.e.}, intermittent amplitude alternation (Kolenberg et al. 2010;  Szab\'o
{\it et al.} 2010),   `peculiar' RRd (pRRd) stars which have unusually low
period ratios (Prudil et al. 2017; Nemec \& Moskalik 2021), and various other
types of Blazhko and multimode amplitude and phase modulations (Gruberbauer
{\it et al.} 2007; Chadid 2010; Benko et al.  2010, 2014; Poretti et al.  2010;
Nemec et al. 2011, 2013;  Jursik et al.  2015; Kurtz {\it et al.} 2016; Smolec
{\it et al.} 2015a,b, 2016, 2017a,b; Netzel \& Smolec 2022;  Netzel, Moln\'ar
\& Joyce 2023).  

\begin{figure} 
\begin{center}
\begin{overpic}[width=8.4cm] {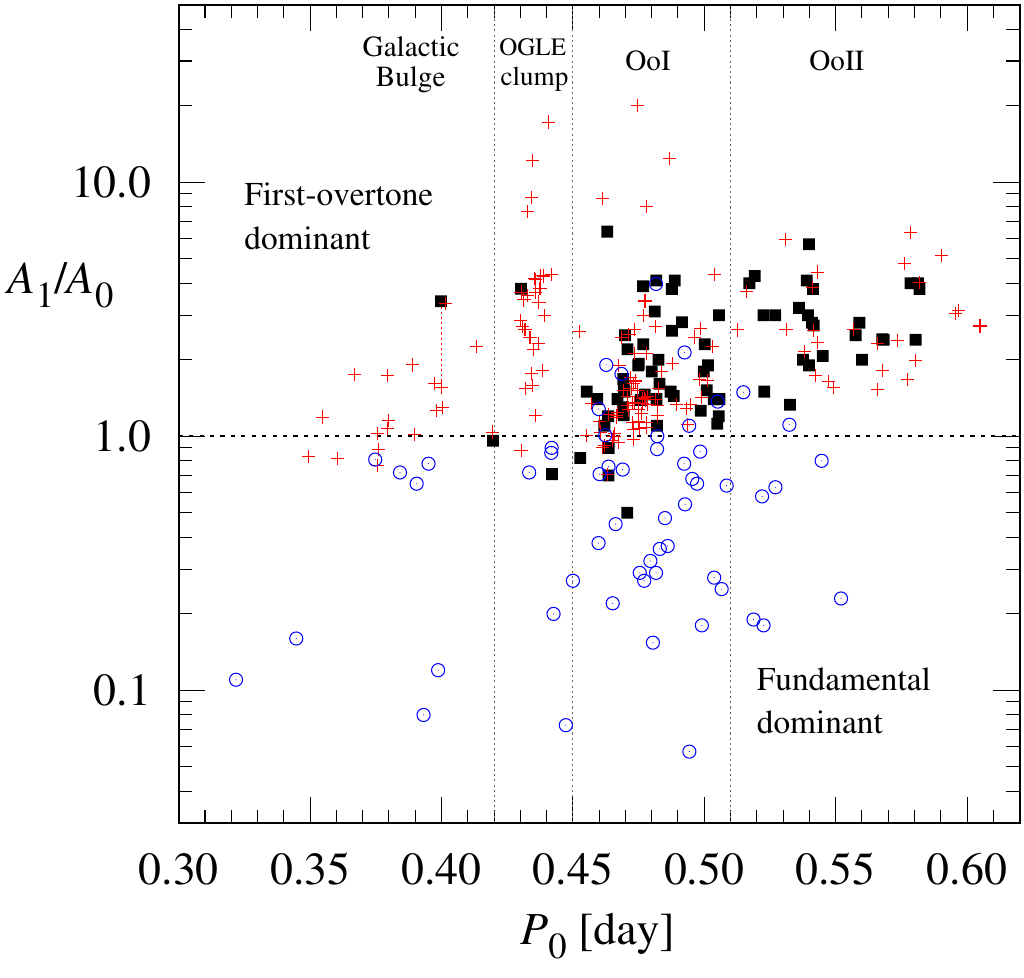}     \end{overpic} 
\end{center}

\caption{Amplitude-ratio diagram for the RRd stars observed by {\it K2} (same
stars and same symbols as Fig.\,1), where the fundamental and first-overtone
amplitudes for the {\it K2} stars are Fourier first-term {\it Kp}-amplitudes
and those for the \texttt{OGLE} stars are trough-to-peak (min-max) $I$-amplitudes.  Three of the six
{\it K2} stars with $A_1$$<$$A_0$ are `anomalous' RRd stars lying off the
Petersen curve while the other three  are `classical' RRd stars.  Graphs
of $A_1/A_0$ vs period for RRd stars in the Large Magellanic Cloud previously
were plotted by Alcock et al. (1997) and Soszy\'nski et al. (2009).} 

\label{Fig2} 
\end{figure}

This paper presents the results of a detailed analysis of the {\it radial}
pulsation properties of 75 RRd stars distributed around the Ecliptic Plane.
The stars were observed during NASA's {\it K2} Mission (Howell {\it et al.} 2014;
Moln\'ar {\it et al.} 2015).  Empirical relationships among the periods and
amplitudes of the pulsations and the Fourier parameters are
investigated.  Such studies are important for identifying significant trends
and correlations, which in turn, are important for the development and
validation of theoretical models.  For example, the Petersen diagram has, as
discussed above, played a key role in understanding double-mode pulsations.
Empirical studies also have many practical applications, including [Fe/H] and mass
estimation.

\section{K2 PHOTOMETRY}

\begin{table}
\centering
\caption{{\it K2} Campaigns, observation dates and time intervals, and number
of RRd stars  in each field.  In column (4) the numbers in parentheses
are numbers of re-observed stars (see Tables~2-3).}
\label{tab:one}
\begin{tabular}{llrc} 
\hline
\multicolumn{1}{c}{K2\,C} & \multicolumn{1}{c}{Observation} & \multicolumn{1}{c}{ Duration\,[days]} & \multicolumn{1}{c}{N(RRd)}   \\
	\multicolumn{1}{c}{ }   & \multicolumn{1}{c}{Dates}  &  \multicolumn{1}{c}{(BJD-2450000)}  &   \\               
\multicolumn{1}{c}{(1)} &  \multicolumn{1}{c}{(2)} & \multicolumn{1}{c}{(3)} & \multicolumn{1}{c}{(4)}  \\             
\hline 
\\
E2  & 2014\,Feb4-Feb13  &  8.9\,(6693.1-6702.0) &  2 \\ 
C0  & 2014\,Mar12-May27 & 76.7\,(6728.5-6805.2) &  0 \\
C1  & 2014\,May30-Aug20 & 80.0\,(6810.3-6890.3) &  1 \\
C2  & 2014\,Aug23-Nov10 & 77.7\,(6894.3-6972.0) &  1 \\
C3  & 2014\,Nov15-Jan23 & 69.2\,(6977.1-7046.3) &  0 \\
C4  & 2015\,Feb8-Apr20  & 69.1\,(7061.8-7130.9) &  4 \\
C5  & 2015\,Apr27-Jul10 & 74.6\,(7139.6-7214.2) &  3 \\
C6  & 2015\,Jul13-Sep30 & 78.9\,(7217.5-7296.4) &  4 \\       
C7  & 2015\,Oct4-Dec26  & 82.6\,(7300.3-7382.9) &  6 \\
C8  & 2016\,Jan3-Mar23  & 78.7\,(7392.1-7470.8) &  4 \\   
C9  & 2016\,Apr22-Jul02 & 71.3\,(7501.1-7572.4) &  3 \\   
C10 & 2016\,Jul6-Sep20  & 69.1\,(7582.6-7651.7) &  6 \\   
C11 & 2016\,Sep24-Dec08 & 74.2\,(7656.3-7730.5) & 10 \\   
C12 & 2016\,Dec15-Mar04 & 78.9\,(7738.4-7817.3) &  3 \\   
C13 & 2017\,Mar8-May27  & 80.6\,(7820.6-7901.1) &  1 \\   
C14 & 2017\,May31-Aug19 & 79.7\,(7905.7-7985.4) & 12 \\   
C15 & 2017\,Aug23-Nov20 & 88.0\,(7989.4-8077.4) &  2 \\   
C16 & 2017\,Dec7-Feb25  & 79.5\,(8095.5-8175.0) &  1(+1) \\   
C17 & 2018\,Mar2-May08  & 67.1\,(8179.6-8246.6) & 12(+2) \\   
C18 & 2018\,May10-Jul02 & 50.8\,(8251.6-8302.4) &  0(+2) \\   
C19 & 2018\,Aug30-Sep26 & 26.4\,(8361.1-8387.5) &  0(+2) \\ [0.2cm]  
\hline 
\end{tabular}
\end{table}

\begin{table*}
\fontsize{6}{7.2}\selectfont  
\centering

\caption{Coordinates and cross-identifications for 75 RRd stars observed during
NASA's {\it K2} Mission, ordered by {\it K2} campaign number and then Ecliptic
Plane Input Catalog (EPIC) number.   For the three aRRd stars the EPIC numbers
are given in {\it italics} (column 1), and the campaign numbers for stars observed
at both short cadence (1-min) and long cadence (30-min) have been
\underline{underlined}. The very precise Right Ascension and Declination
(J2000) values are from {\it Gaia} DR3 if available, otherwise from {\it Gaia} DR2 (see Table 8 for {\it Gaia} Identification numbers).  
Additional cross-identifications are given in the `Notes on
Individual Stars' (Appendix A).}  

\label{tab:two}
\begin{tabular}{lllcccccl}
\hline
\multicolumn{1}{c}{ Star } &  \multicolumn{1}{c}{ RA } & \multicolumn{1}{c}{DEC}   & UCAC & 2MASS  &   SDSS  & CSS/MLS/SSS  & CSS Id No.  \\ 
\multicolumn{1}{c}{(1)} & \multicolumn{1}{c}{(2)} &\multicolumn{1}{c}{(3)} &  (4)  &  (5)  &  \multicolumn{1}{c}{(6)} &   \multicolumn{1}{c}{(7)} &  \multicolumn{1}{c}{(8)} \\
\hline  
\\
60018653\,(E2)                      & 359.1797269 & +03.022410116 &  \dots     &   \dots          & J235643.13+030120.2  & J235643.2+030119  & 1104128004573    \\   
60018662\,(E2)                      & 355.5305737 & +00.547990965 &  \dots     & 23420733+0032527 & J234207.33+003252.8  & J234207.3+003252  & 1101127011298    \\  [0.1cm]   

201585823\,(C\underline{1})         & 176.83406987 & +01.8239430144 & 460-049450 & 11472018+0149261 & J114720.17+014926.2  & J114720.1+014926  & 1101063037246  \\ [0.1cm]   

{\it 205209951}\,(C\underline{2})   & 239.15812379 & $-$18.84713525 & 356-076676 & 15563794--1850494& J155637.94--185049.5 & J155638.0--185049 & 1018082033899  \\  [0.1cm]  

210600482\,(C4)                     & 054.206234255 & +16.972638035 &  \dots     & 03364949+1658215 & J033649.49+165821.41 & J033649.5+165821  & 1118019001177  \\   
210831816\,(C4)                     & 058.05126063  & +20.353092411 & 552-007885 & 03521230+2021111 &     \dots            & J035212.2+202111  & 1121020015334  \\   
210933539\,(C\underline{4})         & 052.991116616 & +21.920979811 & 560-007163 & 03315785+2155158 &     \dots            & J033157.8+215515  & 1121018050281  \\   
211072039\,(C4)                     & 062.041413241 & +24.126412764 &  \dots     & 04080993+2407349 &     \dots            & J040809.9+240735  & 1123021048042  \\ [0.1cm]   

211694449\,(C5,\underline{18})      & 126.202990    & +15.841135    &  \dots     & 08244871+1550280 & J082448.72+155028.2  & J082448.7+155028  & 1115044045716  \\   
211888680\,(C5,16)                  & 135.59286728  & +18.561096716 &  \dots     &    \dots         & J090222.28+183339.9  & J090222.2+183339  & 1118047037354  \\  
211898723\,(C\underline{5},\underline{18}) 
			            & 126.10948466  & +18.70793751  &  \dots     & 08242627+1842287 & J082426.27+184228.6  & J082426.3+184228  & 1118044044198  \\ [0.1cm]  

212335848\,(C6)                     & 201.22962833 & $-$16.50728188 &  \dots     & 13245510--1630262&     \dots            & J132455.0--163026 & 1015070008472  \\ 
212449019\,(C6)                     & 197.45153792 & $-$13.82287849 & 381-064183 & 13094837--1349224&     \dots            & J130948.3--134920 & 2012179019185  \\ 
212455160\,(C6,\underline{17})      & 204.4464111  & $-$13.6915226  &  \dots     & 13374714--1341294&     \dots            & J133747.0--134128 & 2012186019145  \\ 
212547473\,(C\underline{6},\underline{17}) 
			            & 204.20923696 & $-$11.72830043 & 392-057021 & 13365022--1143418&     \dots            & J133650.2--114341 & 1012072054664  \\  [0.1cm]  

213514736\,(C7)                     & 282.6946106 & $-$29.1564026 &  \dots     & 18504670--2909230&    \dots             &     \dots         &   \dots   \\ 
214147122\,(C7)                     & 282.77500119& $-$27.25085983 & 314-224182 & 18510600--2715029&    \dots             &     \dots         &   \dots   \\ 
229228175\,(C7)                     & 281.93738788 & $-$28.3133953   &  \dots     &     \dots        &    \dots             &     \dots         &   \dots      \\ 
229228184\,(C7)                     & 282.02556684 & $-$29.21242517  &  \dots     &     \dots        &    \dots             &     \dots         &   \dots      \\  
229228194\,(C7)                     & 282.13188061 & $-$29.00425155  &  \dots     &     \dots        &    \dots             &     \dots         &   \dots      \\ 
229228220\,(C7)                     & 282.42495918 & $-$25.94913418  &  \dots     &     \dots        &    \dots             &     \dots         &   \dots      \\ [0.1cm] 

220254937\,(C8)                     &022.74793882  & +01.6735085972 &  \dots     &  \dots           & J013059.49+014024.6  & J013059.5+014024  & 1101009031457 \\ 
220604574\,(C8)                     &018.29136574  & +09.127136574  &  \dots     & 01130993+0907380 & J011309.92+090737.6  & J011309.9+090738  & 1109007012508 \\ 
220636134\,(C\underline{8})         &019.791889692 & +09.8452556167 &  \dots     &  \dots           & J011910.05+095042.9  & J011910.0+095043  & 1109007026639 \\ 
229228811\,(C8)                     &016.932617903 & +05.9752227748 &  \dots     &  \dots           & J010743.83+055830.7  & J010743.7+055831  & 1107006006816 \\ [0.1cm] 

223051735\,(C9)                     & 271.284699 & $-$26.588769 &  318-136990 &18050845--2635196&   \dots              &     \dots   &    \dots  \\ 
224366356\,(C9)                     & 275.447105 & $-$24.373911 &  329-142168 &18214730--2422262&   \dots              &     \dots   &    \dots  \\ 
{\it 225045562}\,(C9)               & 266.691452 & $-$23.213318 &  334-120104 &17464595--2312479&   \dots              &     \dots   &    \dots  \\ [0.1cm] 

201152424\,(C10)                    & 182.52057895 & $-$05.238513136 &  \dots      &12100492--0514186&    \dots             & J121004.8--051418 & 1004065008955 \\ 
201440678\,(C10)                    & 181.21246874 & $-$00.351834551 &  \dots      &12045100--0021065& J120450.99--002106.6 & J120451.0--002106 & 1001065053812 \\
201519136\,(C10)                    & 185.41194533 &  +00.8164293962 &  \dots      &12213890+0048599 & J122138.86+004859.1  & J122138.8+004859  & 1101066018344 \\
228800773\,(C10)                    & 191.092577   & $-$06.750381    &  \dots      &   \dots         & J124422.21--064501.3 & J124422.2--064501 & 1007068037006 \\ 
228952519\,(C10)                    & 192.12454944 & $-$02.346772672 &  \dots      &   \dots         &    \dots             & J124829.9--022049 & 1001069010092 \\ 
248369176\,(C10)                    & 185.55124222 &  +00.0526260871 &  \dots      &    \dots        & J122212.29+000309.4  & J122212.3+000309  & 1101066001707 \\ [0.1cm]  

225326517\,(C11)                    & 263.67547509 & $-$22.72459028 &  \dots      &17344233--2243295&    \dots         &  \dots    &  \dots \\
225456697\,(C11)                    & 263.731894   & $-$22.48951    & 338-101883  &17345565--2229222&   \dots          &  \dots    &  \dots \\
235631055\,(C11)                    & 259.58450133 & $-$29.83886112 &  \dots      &17182028--2950198&   \dots          &  \dots    &  \dots \\
235794591\,(C11)                    & 261.90005336 & $-$29.33767559 &  \dots      &17273602--2920156&     \dots        &  \dots    &  \dots \\
236212613\,(C11)                    & 258.64080782 & $-$28.00833788 &  \dots      &17143376--2800301&     \dots        &  \dots    &  \dots \\
251248825\,(C11)                    & 261.74790938 & $-$28.45405684 &  \dots      &     \dots       &     \dots        &  \dots    &  \dots \\
251248826\,(C11)                    & 262.16480265 & $-$28.49276597 &  \dots      &     \dots       &     \dots        &  \dots    &  \dots \\
251248827\,(C11)                    & 262.4679167  & $-$28.4059167  &  \dots      &     \dots       &     \dots        &  \dots    &  \dots \\
{\it 251248828}\,(C11)              & 264.6770417  & $-$23.3172778  &  \dots      &     \dots       &     \dots        &  \dots    &  \dots \\
251248830\,(C11)                    & 266.0330417  & $-$25.9605278  &  \dots      &     \dots       &     \dots        &  \dots    &  \dots \\ [0.1cm]  

245974758\,(C\underline{12},\underline{19})
				    & 349.306822   & $-$10.148914 & \dots & 23171363--1008560& J231713.66--100855.9  & J231713.6--100855 & 1009124026999      \\
246058914\,(C\underline{12},\underline{19})
				    & 351.42089314 &  $-$08.049997582 & \dots & 23254104--0803002& J232541.00--080259.8  & J232540.9--080259 & 1007124007872  \\
251456808\,(C12)                    & 350.4464167  &  +00.2357222     & \dots &    \dots         & J232147.14+001408.6   & J232147.2+001408  & 1101125004451  \\ [0.1cm] 

247334376\,(C13)                    & 078.638079082 & +20.852896452   &   \dots     & 05143315+2051098 &   \dots              &  J051433.1+205110 & 1121027045355 \\ [0.1cm] 

201749391\,(C\underline{14}) & 165.264565   & +04.472659     & 473-046469 & 11010348+0428214 & J110103.49+042821.6  & J110103.5+042821 & 1104059033335 \\
248426222\,(C14)             & 160.48146706 & +00.5889994022 &  \dots     & 10415554+0035202 & J104155.55+003520.4  & J104155.6+003520 & 1101058012922 \\
248509474\,(C\underline{14}) & 158.3716221  & +03.4408391135 &   \dots    & 10332919+0326271 & J103329.19+032627.0  & J103329.2+032627 & 1104057013286 \\
248514834\,(C\underline{14}) & 157.53471727 & +03.6022996457 & 469-043448 & 10300833+0336084 & J103008.33+033608.2  & J103008.3+033608 & 1104057016830 \\
248653210\,(C14)             & 158.52769285 & +07.2022393032 &   \dots    &     \dots        & J103406.64+071208.0  & J103406.6+071207 & 1107056030993 \\
248653582\,(C\underline{14}) & 160.49248756 & +07.2116772807 & 487-051343 & 10415821+0712422 & J104158.19+071242.0  & J104158.2+071242 & 1107057030584 \\
248667792\,(C14)             & 157.14515297 & +07.5659142694 &   \dots    & 10283482+0733567 & J102834.83+073357.3  & J102834.8+073357 & 1107056038736 \\
248730795\,(C14)             & 160.16983198 & +09.0358273696 & 496-056499 & 10404076+0902089 & J104040.76+090209.0  &   \dots          & \dots         \\
248731983\,(C\underline{14}) & 160.77562409 & +09.0612043096 & 496-056556 & 10430615+0903405 & J104306.15+090340.3  &   \dots          & \dots         \\
248827979\,(C14)             & 155.557981   & +11.355388     &   \dots    & 10221391+1121193 & J102213.92+112119.4  &   \dots          & \dots         \\
248845745\,(C14)             & 160.43736748 & +11.796589092  &   \dots    &   \dots          & J104144.96+114747.7  &   \dots          & \dots         \\
248871792\,(C\underline{14}) & 162.66135792 & +12.457650633  & 513-052553 & 10503874+1227276 & J105038.73+122727.6  &   \dots          & \dots         \\ [0.1cm] 

249790928\,(C15)             & 234.3329736  & $-$18.01575394 & 360-074594 &15371992--1800567 &       \dots          &J153719.9--180056 & 1018081063248   \\  
250056977\,(C\underline{15}) & 234.77375229 & $-$14.80469274 & \dots      &15390570--1448168 & J153905.70--144816.8 &J153905.8--144816 & 1015081070790   \\  [0.1cm]  

211665293\,(C16)             & 136.01875492 & +15.444972769  & 528-049596 & 09040449+1526417 & J090404.49+152642.0  & J090404.5+152642 & 1115047028617 \\ [0.1cm] 

212467099\,(C17)                & 199.90526047 & $-$13.43885865  &  \dots     & 13193728--1326195&     \dots            & J131937.3--132619 & 1012070013928  \\ 
212498188\,(C17)                & 204.4953766  & $-$12.7721872   &  \dots     & 13375888--1246198&     \dots            & J133758.9--124619 & 1012072031579  \\ 
212615778\,(C17)                & 198.42722343 & $-$10.2322055   &   \dots    & 13134252--1013557&     \dots            & J131342.5--101356 & 1009071023139  \\ 
212819285\,(C17)                & 207.0258789  & $-$04.9317350   &   \dots    &    \dots         & J134806.20--045554.2 & J134806.2--045554 & 1004074016553  \\ 
251521080\,(C17)                & 196.93288681 & $-$03.280491255 &   \dots    &    \dots         & J130743.89--031649.7 & J130743.8--031641 & 1004071051158  \\ 
251629085\,(C17)                & 199.52759172 & $-$00.550094147 & 448-055695 & 13180662-0033002 & J131806.62--003300.2 & J131806.6--003300 & 1001071051356  \\ 
251809772\,(C17)                & 204.6996200  & $-$08.5891100   &   \dots    &    \dots         &      \dots           & J133847.9--083520 & 2008188002549  \\ 
251809814\,(C17)                & 201.8208117  & $-$05.6359357   &   \dots    &    \dots         &      \dots           & J132716.9--053809 & 1004072000498  \\ 
251809825\,(C17)                & 204.43143696 & $-$07.755244459 &   \dots    &    \dots         &      \dots           & J133743.5--074518 & 1007073015747  \\ 
251809832\,(C17)                & 204.83307773 & $-$13.67106856  &   \dots    &    \dots         &      \dots           & J133920.0--134016 & 1012072010870  \\ 
251809860\,(C17)                & 201.4751985  & $-$01.0399776   &   \dots    &    \dots         & J132554.04--010224.0 & J132553.9--010222 & 1001072041910  \\ 
251809870\,(C17)                & 207.1282046  & $-$12.2034152   &   \dots    &    \dots         &      \dots           & J134830.9--121210 & 1012073044969  \\ [0.1cm] 
\hline
\end{tabular}
\end{table*}

The data that were analyzed are the high-precision photometric measurements
made with the CCD cameras onboard NASA's {\it Kepler} space telescope during
the {\it K2} Mission (Howell {\it et al.} 2014).      In total more than 3000
RR\,Lyrae stars were observed.  The stars were proposed
for observation by the `RR~Lyrae and Cepheid Working Group' of the {\it Kepler}
Asteroseismic Science Consortium (KASC), the same team that worked on the
RR~Lyrae stars observed in the original {\it Kepler}-field (Szab\'o {\it et
al.} 2010, 2017; Kolenberg {\it et al.} 2010; Nemec {\it et al.} 2011, 2013;
Moskalik {\it et al.} 2015).  For a discussion of the RR~Lyrae  selection
process see Plachy {\it et al.} (2016).   The wide bandpass of the {\it Kepler}
filter (420-900\,nm), the milli-magnitude (mmag) precision of the photometry,
the long ($\sim$67-88\,d) time-baseline of the continuous observations, and the short
integration times (30-minute for all the stars and 1-min for 17 stars) combine to make the {\it K2}
photometry a unique and excellent dataset for studying RRd stars.  The {\it K2}
observations began in February 2014 with the successful nine-day `Two-wheel
Concept Engineering Test' (see Moln\'ar {\it et al.} 2015), and ended in
September 2018 when the telescope ran out of hydrazine fuel.  

A preliminary screening of the $\sim$3000 RR~Lyrae stars observed by {\it K2}
was conducted to identify RRd stars from  including those previously not known to be RRd
stars.  Non-parametric methods (Lomb 1976; Scargle 1982; Stellingwerf
1978,2011; Zechmeister \& Kurster 2009; VanderPlas 2018), specifically the
methods available in the \texttt{vartools} package (Hartman \& Bakos 2016),
were used for this purpose.  This initial search provided frequency estimates
for all the RR~Lyrae stars.  When dealing with so many stars
this procedure had the advantage of being fully automatic. After the initial
screening,  improved pulsation periods were obtained for all candidate RRd
stars using Fourier methods.  The \texttt{Period04} package (Lenz \& Breger
2005) was employed for this purpose.  In addition to identifying the main
frequencies this program can be used to identify combination and alias
frequencies, as well as low-amplitude non-radial frequencies.  

A total of 75 RRd stars (excluding the four `peculiar' RRd stars discussed by
Nemec \& Moskalik 2021) were identified from among the initial list of
$\sim$3000 stars.  {\bf Table~1} summarizes the number of RRd stars observed
during each campaign.  Coordinates (RA,DEC) for the stars and
cross-identifications are given in {\bf Table 2}.  The source catalogues are:
the U.S.  Naval Observatory Astrograph Catalog (DR4), \texttt{UCAC}; the
Two-Micron All Sky Survey catalog, \texttt{2MASS}; the Sloan Digital Sky Survey
(DR14), \texttt{SDSS}; and the Catalina Surveys, \texttt{CSS/MLS/SSS} (see
Drake {\it et al.} 2009, 2017).  Star names from the \texttt{LINEAR} and
\texttt{OGLE} surveys, and from the Cseresnjes (2001) study of the RR~Lyrae
stars in the direction of the Sagittarius dwarf galaxy, are given in {\bf
Appendix\,A} (Notes on Individual Stars).   Cross-identifications with the {\it
{\it Gaia}} catalogue (DR2, DR3) are discussed in $\S$4.

\subsection{{\it K2} long-cadence photometry}

All 75 RRd stars were observed at an interval close to 29.4 minutes, {\it
i.e.}, long cadence (LC).  With 49 observations per day and typical radial
pulsation periods of 0.54\,d (fundamental) and 0.40\,d (first-overtone) the
number of LC brightness measurements typically amounted to 26 fundamental and
20 first-overtone measurements per pulsation cycle.  Over an 80-day campaign
$\sim$3918 LC brightness measurements were made per star, covering 148
(fundamental) and 200 (first-overtone) pulsation cycles.  The photometric
measurements provided by the NASA-Ames {\it K2}  `Pre-Search Data Conditioning'
(\texttt{PDCsap})  pipeline were usually  analyzed first.  In all cases the
photometry was found to exhibit some type of slow trend and to have a small
number of outliers.  After removal of obvious outliers (i.e., $>$5$\sigma$) the
low-level trends were removed by fitting a polynomial to the flux data.  


All of the LC data up to and including the Campaign\,6 stars as well as stars
observed during Campaigns 8, 10 and 12-14 also were pre-processed ({\it i.e.},
outliers removed, detrended, etc.) using the Extended Aperture Photometry
(\texttt{EAP}) pipeline (see Plachy {\it et al.} 2019).  An important feature
of this procedure was that new apertures were created for every star using the
\texttt{PyKE} software (Still \& Barclay 2012).  The
photometric apertures contained most of the star movement within the target
pixel masks but were not so large as to be contaminated by light from nearby
stars.   For all the photometry except that for Campaign\,2 the points with
\texttt{SAP\_QUALITY} flags larger than 0 were removed (for Campaign\,2  the
large number of points with a 16384 flag were retained -- this flag dominated
the second half of Campaign\,2 but does not affect most targets -- see the
Campaign\,2 Data Release Notes for details.)  An automated Fourier analysis
script  was run on the light curves to construct an initial fit, which was then
subtracted from the data and all three-sigma (or more) outliers were removed.
The script was then rerun  on the cleaned data and this second iteration was
used to derive preliminary pulsation periods and pulsation modes.  Although
this method produced slightly increased noise from background pixels, in many
cases it preserved the pulsation amplitudes much better than either the
\texttt{SAP} or \texttt{PDCsap} fluxes.  For the later Campaigns (7-18) the
detrending and pre-processing was done by hand using the \texttt{Pyke}
software.

\subsection{{\it K2} short-cadence photometry}


%

In addition to the long-cadence observations, 17 of the RRd stars were also
observed at a sampling rate of (29.4/30=) 0.98 min, {\it i.e.}, short
cadence (SC).   These data permit the detection of high frequencies, up to
the SC Nyquist frequency of 734.7 d$^{-1}$ (or equivalently, periods as short as
two minutes).   For frequencies lower than the LC Nyquist frequency the
frequencies detectable are similar for the SC and LC data.  Where the two data
sets differ is in the background noise levels of the amplitude spectra:  the SC
data tend to have higher signal-to-noise ratio peaks than the LC data, and
hence are more suitable for the detection of low-amplitude frequencies (such as
non-radial modes). In Table\,2 (and elsewhere throughout this paper) the EPIC
numbers of stars with SC data have been underlined.  During Campaign\,14 six of
the 15 stars were observed at SC; the other eight stars with SC data were
observed during seven other campaigns.  To take advantage of the more frequent
sampling offered by the SC data, brightness measurements were also made using
the \texttt{PyKE} software.  Apertures larger than the \texttt{PDCSAP}
apertures were defined for each star (as for the \texttt{EAP} pipeline) and
aperture photometry was performed.  To minimize outliers all data points with
\texttt{SAPQUALITY} flags greater than zero were excluded from the analysis.
EPIC\,205209951 is the only star with SC data that is not a classical RRd star;
it lies below the cRRd curve in the Petersen diagram, exhibits distinct Blazhko
modulations (see Plachy {\it et al.} 2017), and is an aRRd star similar to
those found in Messier\,3 and in the Magellanic Clouds.

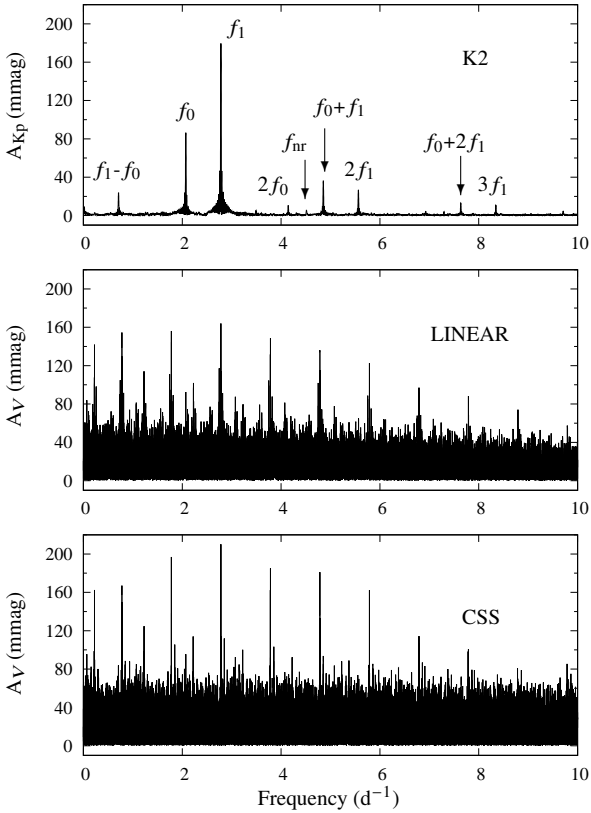
\begin{figure} \begin{center}
\begin{overpic}[width=7.5cm]{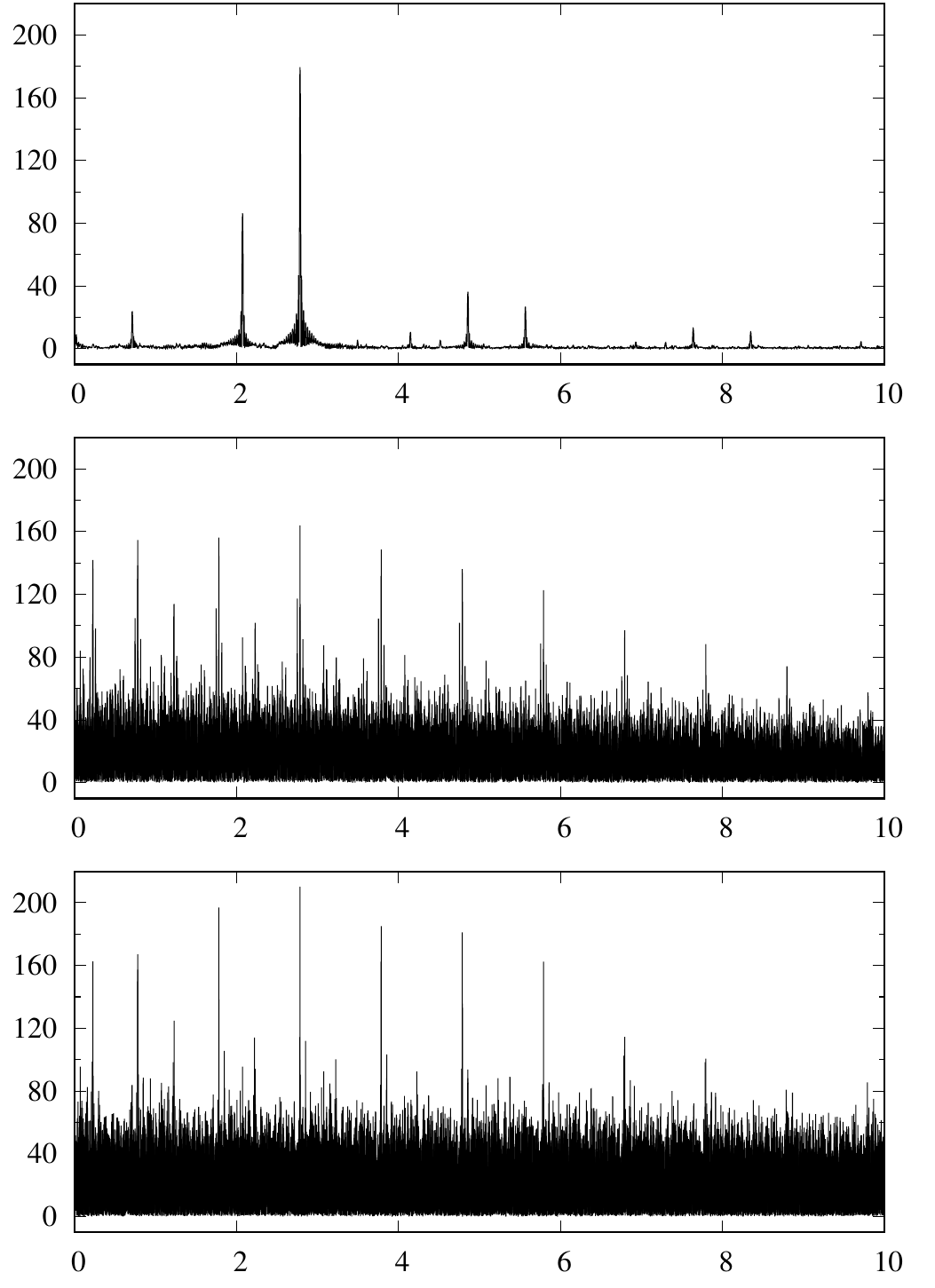} 
	\put(28,-2){Frequency (d$^{-1}$)}  
	\put(54,92){K2} 
	\put(50,57){LINEAR}  
	\put(54,21){CSS}  
	\put(17.5,85.5){$f_0$} 
	\put(24,96){$f_1$}
        \put(31,81.5){$f_{\rm nr}$}   \put(34,80){\vector(0,-1){5}}    
	\put(56,76){3$f_1$}
	\put(49,81.5){$f_0$+2$f_1$}   \put(53.8,80.5){\vector(0,-1){5}}   
	\put(39,78){2$f_1$}
	\put(28,76){2$f_0$}
	\put(7,78){$f_1$-$f_0$}
	\put(35,86){$f_0$+$f_1$}   \put(36.5,84){\vector(0,-3){5.5}}    
        \put(-4,12){\rotatebox{90}{A$_V$\,(mmag)}}    
        \put(-4,46){\rotatebox{90}{A$_V$\,(mmag)}}   
        \put(-4,80){\rotatebox{90}{A$_{\rm Kp}$\,(mmag)}}  \end{overpic}
\end{center} 

\caption{Fourier amplitude spectra for EPIC\,201585823 (C1):  (Top) the
{\it K2} photometry (3671 brightness measurements made over 80 days);  (Middle) the
LINEAR Survey data (519 observations made over 5.5 years with no filter and
transformed to the $V$-passband);  (Bottom) the Catalina Sky Survey data
(354\,CSS and 243\,MLS photometric observations over 8.2 years).  }

\label{Fig3} 
\end{figure}

\section{ANALYSIS OF K2 PHOTOMETRY}

Fourier amplitude spectra  (\texttt{Period04}) for EPIC\,201585823, which is a
classical RRd star observed during Campaign~1, are shown in {\bf Figure\,3}.
This star has SC data and was studied in detail by Kurtz et al. (2016).   The
figure illustrates the considerably better quality of the {\it K2} photometry
(top panel) compared with earlier ground-based observations made by the
\texttt{LINEAR} Survey (middle panel) and by the Catalina Sky Survey (bottom
panel).  All three data sets show peaks at the dominant first-overtone ($f_1$)
and fundamental-mode ($f_0$) frequencies.  For the K2 data the two radial
pulsation components have Fourier 1st-term amplitudes $A_1$(Kp)$\sim$180 mmag
for the first-overtone and $A_0$(Kp)$\sim$90 mmag for the fundamental mode (see
Table 3).  Only the {\it K2} spectrum is of sufficient quality to detect low
amplitude peaks due to combinations of these two frequencies and the even lower
amplitude peak of the non-radial pulsation.  The non-radial frequency (labelled
$f_{\rm nr}$)  at 4.5154($\pm$1)\,d$^{-1}$ with amplitude 5.24$\pm$0.18 mmag is
barely visible at the scale shown.  At higher resolution the background level
of the {\it K2} data in the vicinity of this non-radial peak is seen to be
$\sim$0.2 mmag and thus the signal-to-noise ratio of the peak is high,
$\sim$30, despite its low amplitude.  The numbers given here are consistent
with those found by Kurtz et al.  (2016).  

Detrended photometry and fitted light curves for the `classical' RRd stars
observed during Campaigns 1-6 of the {\it K2} Mission are plotted in {\bf
Figure 4}.  For each star only the first five days of the available data are
shown.  Panel labels give the EPIC number and  the {\it K2} campaign number in
which the star was observed.  The smooth fitted curves (green lines) illustrate
the quality of the \texttt{Period04} analysis for each star.  Typically more
than 25 frequencies were included in the least squares fits with the brightest
stars usually revealing the greatest number of significant frequencies.   In
addition to the independent radial frequencies, many of the detected
periodicities are harmonics that describe the non-sinusoidal nature of the
light curves of the individual modes or observational aliases (for example, due
to the thruster firings every six hours).   In addition, combination
frequencies involving the two radial modes are commonly detected.  In general
the fits are exceptionally good with root-mean-squared errors ranging from 5 to
100 mmag depending on the quality of the photometry.  

Light curves for the RRd stars observed during Campaigns\,7-18 are given in
{\bf Appendix B}.   The graphs
illustrate the similarities and differences seen in the light curves for the
various stars and campaigns.  These include photometric variations from star to
star and from one campaign to the next, gaps in the data, systemic brightness
differences, cycle-to-cycle differences that could not previously be seen in
earlier ground-based observations, etc.

\begin{figure} \begin{center}
\begin{overpic}[width=7.9cm]{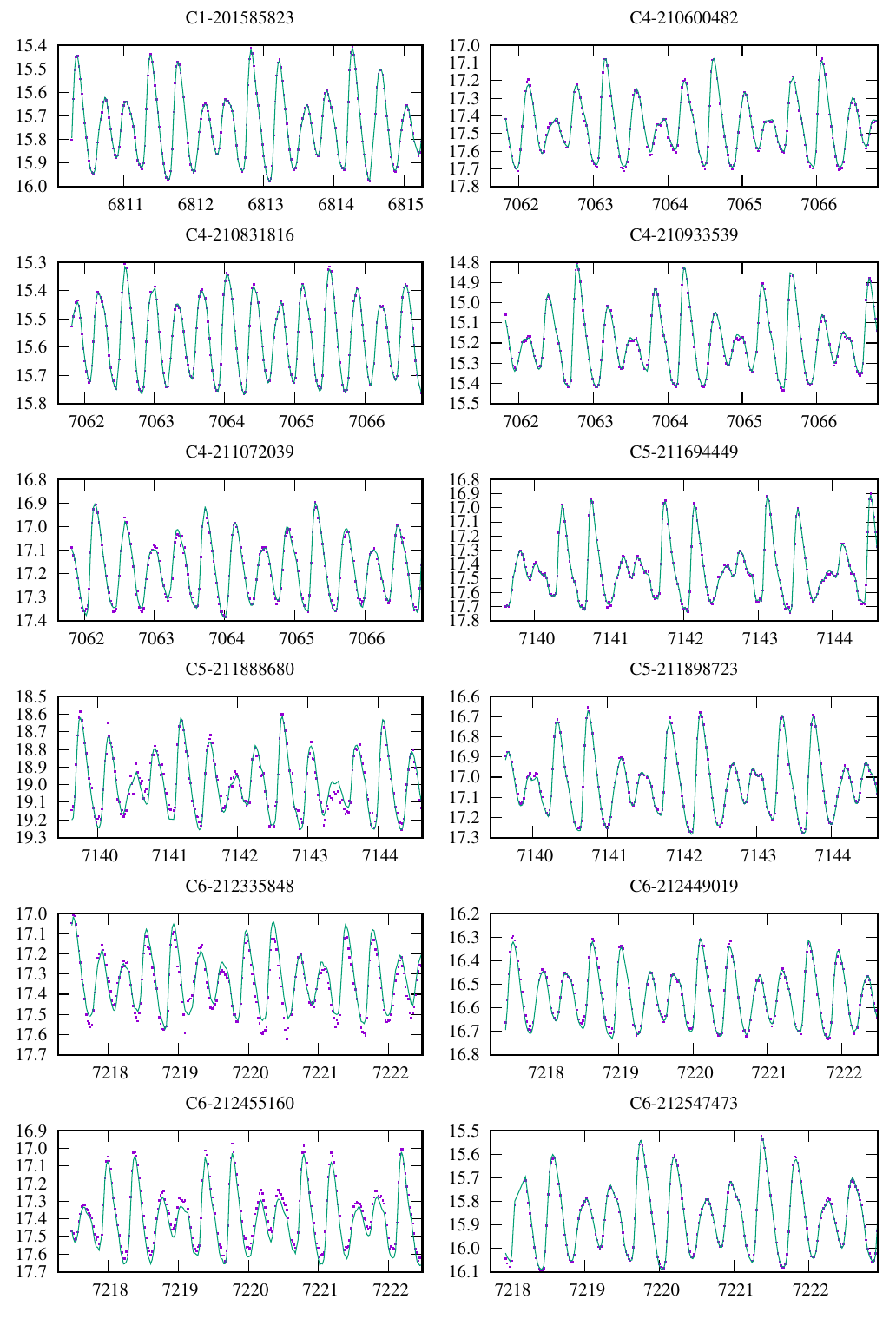} \put(26,-2){BJD - 2450000}    \end{overpic}
\end{center} 

\caption{Observed {\it Kepler} Kp photometry (purple squares) and fitted light curves (\texttt{Period04}, green
lines) for the 12 `classical' RRd stars observed during \underline{Campaigns 1-6} of the {\it K2} Mission.  The
observed Kp magnitudes  were derived using the long-cadence fluxes
(e$^{-}$/s, \texttt{EAP} pipeline) transformed to magnitudes using Kp\,$=
25.3-2.5\log F$.  Only the first five days of data are shown.    } 

\label{Fig4} 	
\end{figure}

\begin{table*}

{\fontsize{6}{7.2}\selectfont  
\centering

\caption{Mean magnitudes, pulsation periods and  amplitudes for the 72 cRRd and
three aRRd stars observed during the {\it K2} Mission (ordered by increasing
pulsation period).  Column (1) contains EPIC identification numbers and, in
parentheses, the {\it K2} campaign(s) in which the star was observed (an
\underline{underlined} campaign number indicates that short-cadence photometry
was analyzed).  The mean magnitudes are through the $V$ filter
(column 2), the {\it Gaia} G-filter (column 3), and the {\it Kepler} filter, where
the {\it Kp} magnitudes in column (4) are from the \texttt{MAST} catalogue
(Huber et al. 2016), and those in column (5) are from the present analysis.
Columns (6)-(9) contain first-overtone and fundamental-mode periods,
respectively, and Fourier first-term amplitudes.  For the stars observed during
Campaigns\,9 and 11 the adopted periods are the very precise values from the
\texttt{OGLE-IV} (2010-17) survey, and the associated amplitudes are from the
high-precision {\it K2} photometry (this paper).  Columns (10) and (11) give
ratios of the periods and amplitudes.  Empty rows in the table separate stars
with multiple observations (LC, SC, multiple campaigns) from
those with LC observations only. }  

\label{tab:three}
\begin{tabular}{llccllrlrll}
\hline

\multicolumn{1}{c}{ EPIC  } &\multicolumn{1}{c}{$<$V$>$} & \multicolumn{1}{c}{$<$G$>$} & \multicolumn{1}{c}{ $<$Kp$>$} & \multicolumn{1}{c}{ $<$Kp$>$} & \multicolumn{1}{c}{ $P_1$} & \multicolumn{1}{c}{$A_1$(Kp)}   & \multicolumn{1}{c}{$P_0$} & \multicolumn{1}{c} {$A_0$(Kp)}  & \multicolumn{1}{c}{$P_1 / P_0$} &  \multicolumn{1}{c}{$A_1 / A_0$}  \\
\multicolumn{1}{c}{No.}&[mag]&\multicolumn{1}{c}{\it Gaia}&\multicolumn{1}{c}{\texttt{MAST}}&\multicolumn{1}{c}{\it K2}&\multicolumn{1}{c}{[day]}&\multicolumn{1}{c}{[mmag]} &\multicolumn{1}{c}{[day]} &\multicolumn{1}{c}{[mmag]} && \\ 
\multicolumn{1}{c}{(1)} & \multicolumn{1}{c}{(2)} &\multicolumn{1}{c}{(3)}  &\multicolumn{1}{c}{(4)} & \multicolumn{1}{c}{(5)} &
\multicolumn{1}{c}{(6)}  & \multicolumn{1}{c}{(7) } & \multicolumn{1}{c}{(8)} & \multicolumn{1}{c}{(9)}  & \multicolumn{1}{c}{(10)}  & \multicolumn{1}{c}{(11)} \\

\hline  
\\
\multicolumn{11}{c}{{\bf (a) `Classical' RRd (cRRd) stars  } } \\ [0.1cm]

225456697\,(C11) &16.84&\dots&[14.50] & 16.85  & 0.2943527($\pm$1)  &  97.5$\pm$0.4    & 0.3998380($\pm$4) & 53.4$\pm$0.4  & 0.73618($\pm$1)  & 1.82$\pm$0.02 \\

251248830\,(C11) &19.9:&\dots& 19.86  & 20.29  & 0.310017($\pm$8)   & 225.7$\pm$3.7    & 0.419592($\pm$14) & 235.4$\pm$3.7 & 0.73886($\pm$4)  & 0.96$\pm$0.03 \\
\hline
\end{tabular}
}
\end{table*}

\begin{table*}
{\fontsize{6}{7.2}\selectfont  
\centering
\contcaption{  }
\label{tab:three}
\begin{tabular}{llccllrlrll}
\hline

\multicolumn{1}{c}{ EPIC } & \multicolumn{1}{c}{$<$V$>$} & \multicolumn{1}{c}{$<$G$>$} & \multicolumn{1}{c}{ $<$Kp$>$ } & \multicolumn{1}{c}{ $<$Kp$>$ } & \multicolumn{1}{c}{ $P_1$} & \multicolumn{1}{c}{$A_1$(Kp)} & \multicolumn{1}{c}{$P_0$} & \multicolumn{1}{c} {$A_0$(Kp)} & \multicolumn{1}{c}{$P_1 / P_0$} &  \multicolumn{1}{c}{$A_1 / A_0$} \\
\multicolumn{1}{c}{No. }  &[mag]  & \multicolumn{1}{c}{\it Gaia} &  \multicolumn{1}{c}{\texttt{MAST}} & \multicolumn{1}{c}{\it K2} &  \multicolumn{1}{c}{[day]} & \multicolumn{1}{c}{[mmag]} & \multicolumn{1}{c}{[day]} &\multicolumn{1}{c}{[mmag]} &  &  \\  
\multicolumn{1}{c}{(1)} & \multicolumn{1}{c}{(2)} &\multicolumn{1}{c}{(3)}  &\multicolumn{1}{c}{(4)} & \multicolumn{1}{c}{(5)} &
\multicolumn{1}{c}{(6)}  & \multicolumn{1}{c}{(7) } & \multicolumn{1}{c}{(8)} & \multicolumn{1}{c}{(9)}  &  \multicolumn{1}{c}{(10)}  &  \multicolumn{1}{c}{(11)}   \\

\hline  
\\

248730795\,(C14)&16.1&16.03  & 16.10  & 16.11  & 0.390117($\pm$1)  & 185.2$\pm$0.2  & 0.522545($\pm$6)  &  62.6$\pm$0.2 & 0.74657($\pm$1) & 2.96$\pm$0.02  \\
\hline
\end{tabular}
}
\end{table*}

\begin{table*}
\fontsize{7}{8.4}\selectfont
\centering

\caption{Fourier phase-difference parameters, $\phi_{\rm21}^s$ and
$\phi_{\rm31}^s$, and  amplitude-ratio parameters, $R_{\rm 21}$ and $R_{\rm 31}$,
for the  {\bf radial first-overtone} pulsations of the 72 classical double-mode
RR~Lyrae (cRRd) stars observed during the {\it K2} mission.  All are derived from Fourier decomposition of the {\it K2}
photometry and thus are on the {\it Kp} photometric system.  Note that the phase-difference parameters are sine values (which are related
to cosine values according to  $\phi_{21}^c$ = $\phi_{21}^s+\pi$/2 and
$\phi_{31}^c$ = $\phi_{31}^s-\pi$).    }

\label{tab:three}
\begin{tabular}{lclcclll}
\hline
\multicolumn{1}{c}{ EPIC  } & \multicolumn{1}{c}{K2} & \multicolumn{1}{c}{ $P_1$} & \multicolumn{1}{c}{$\phi_{\rm21,1}^s$(Kp)}   & \multicolumn{1}{c}{$\phi_{\rm31,1}^s$(Kp)} & \multicolumn{1}{c} {$R_{\rm 21,1}$(Kp) }  & \multicolumn{1}{c}{$R_{\rm 31,1}$(Kp) }   \\
\multicolumn{1}{c}{ no.   } & \multicolumn{1}{c}{ Campaign } & \multicolumn{1}{c}{[day]}   & \multicolumn{1}{c}{[radians]} & \multicolumn{1}{c} {[radians]}  & \multicolumn{1}{c}{ }   \\	
\multicolumn{1}{c}{(1)} & \multicolumn{1}{c}{(2)} &\multicolumn{1}{c}{(3)}  &\multicolumn{1}{c}{(4)} & \multicolumn{1}{c}{(5)} & \multicolumn{1}{c}{(6)}   & \multicolumn{1}{c}{(7)} \\
\hline  
\\	    
060018653 &E2      & 0.4023084($\pm$4)  & 3.35 $\pm$ 0.08 & 6.33 $\pm$ 0.07 & 0.177 $\pm$ 0.014 & 0.080 $\pm$ 0.005   \\
060018662 &E2      & 0.4175081($\pm$3)  & 3.27 $\pm$ 0.08 & 6.20 $\pm$ 0.08 & 0.230 $\pm$ 0.017 & 0.083 $\pm$ 0.005   \\ [0.00cm]

201585823 &C1      & 0.3594190($\pm$2)  & 3.26 $\pm$ 0.01 & 6.32 $\pm$ 0.01 & 0.153 $\pm$ 0.001 & 0.065 $\pm$ 0.001   \\ [0.00cm]

210600482 &C4      & 0.362471($\pm$1)   & 3.38 $\pm$ 0.01 & 6.39 $\pm$ 0.02 & 0.157 $\pm$ 0.001 & 0.058 $\pm$ 0.001   \\
210831816 &C4      & 0.3638034($\pm$9)  & 3.32 $\pm$ 0.01 & 6.65 $\pm$ 0.02 & 0.124 $\pm$ 0.001 & 0.062 $\pm$ 0.001   \\
210933539 &C4      & 0.35861619($\pm$2) & 3.31 $\pm$ 0.01 & 6.19 $\pm$ 0.01 & 0.171 $\pm$ 0.001 & 0.060 $\pm$ 0.001   \\
211072039 &C4      & 0.393491($\pm$2)   & 3.28 $\pm$ 0.01 & 6.25 $\pm$ 0.03 & 0.181 $\pm$ 0.002 & 0.075 $\pm$ 0.002   \\ [0.00cm]

211694449 &C5,18   & 0.34436220($\pm$4) & 3.31 $\pm$ 0.01 & 5.63 $\pm$ 0.04 & 0.193 $\pm$ 0.002 & 0.035 $\pm$ 0.001  \\ 

211888680&C5,16    & 0.3593838($\pm$1)  & 3.31 $\pm$ 0.01 & 6.38 $\pm$ 0.03 & 0.155 $\pm$ 0.002 & 0.059 $\pm$ 0.002   \\ 

211898723 &C5,18   & 0.37651088($\pm$4) & 3.45 $\pm$ 0.01 & 6.40 $\pm$ 0.02 & 0.162 $\pm$ 0.001 & 0.060 $\pm$ 0.001   \\ [0.00cm] 

212335848 &C6      & 0.355068($\pm$2)   & 3.23 $\pm$ 0.01 & 6.37 $\pm$ 0.03 & 0.145 $\pm$ 0.002 & 0.063 $\pm$ 0.002   \\
212449019 &C6      & 0.363388($\pm$1)   & 3.21 $\pm$ 0.01 & 6.33 $\pm$ 0.03 & 0.144 $\pm$ 0.002 & 0.064 $\pm$ 0.002   \\
212455160 &C6,17   & 0.349321($\pm$2)   & 3.29 $\pm$ 0.02 & 6.27 $\pm$ 0.05 & 0.157 $\pm$ 0.003 & 0.056 $\pm$ 0.003   \\
212547473 &C6      & 0.406430($\pm$1)   & 3.41 $\pm$ 0.01 & 6.39 $\pm$ 0.02 & 0.171 $\pm$ 0.001 & 0.068 $\pm$ 0.001   \\ [0.00cm]

213514736 &C7      & 0.375213($\pm$4)   & 3.35 $\pm$ 0.02 & 6.13 $\pm$ 0.07 & 0.199 $\pm$ 0.004 & 0.060 $\pm$ 0.004   \\
214147122 &C7      & 0.403654($\pm$3)   & 3.36 $\pm$ 0.01 & 6.41 $\pm$ 0.03 & 0.167 $\pm$ 0.002 & 0.071 $\pm$ 0.002   \\
229228175 &C7      & 0.349452($\pm$11)  & 3.35 $\pm$ 0.07 & 6.77 $\pm$ 0.14 & 0.135 $\pm$ 0.010 & 0.068 $\pm$ 0.010   \\
229228184 &C7      & 0.341053($\pm$15)  & 3.46 $\pm$ 0.12 & 6.38 $\pm$ 0.23 & 0.130 $\pm$ 0.015 & 0.066 $\pm$ 0.015   \\
229228194 &C7      & 0.38971($\pm$2)    & 3.70 $\pm$ 0.06 & 6.67 $\pm$ 0.16 & 0.217 $\pm$ 0.004 & 0.077 $\pm$ 0.004   \\
229228220 &C7      & 0.363387($\pm$6)   & 3.21 $\pm$ 0.06 & 6.36 $\pm$ 0.11 & 0.127 $\pm$ 0.007 & 0.072 $\pm$ 0.007   \\ [0.00cm] 

220254937 &C8      & 0.400127($\pm$3)   & 3.34 $\pm$ 0.02 & 6.33 $\pm$ 0.05 & 0.179 $\pm$ 0.003 & 0.077 $\pm$ 0.003   \\
220604574 &C8      & 0.354733($\pm$1)   & 3.28 $\pm$ 0.01 & 6.59 $\pm$ 0.02 & 0.131 $\pm$ 0.001 & 0.069 $\pm$ 0.001   \\
220636134 &C8      & 0.3737393($\pm$7)  & 3.29 $\pm$ 0.01 & 6.24 $\pm$ 0.03 & 0.175 $\pm$ 0.002 & 0.068 $\pm$ 0.002   \\
229228811 &C8      & 0.372911($\pm$2)   & 3.37 $\pm$ 0.01 & 6.53 $\pm$ 0.03 & 0.145 $\pm$ 0.002 & 0.065 $\pm$ 0.002   \\ [0.00cm]

224366356 &C9      & 0.3430486($\pm$4)  & 3.13 $\pm$ 0.07 & 5.71 $\pm$ 0.23 & 0.180 $\pm$ 0.012 & 0.053 $\pm$ 0.012   \\
223051735 &C9      & 0.4233213($\pm$2)  & 3.44 $\pm$ 0.07 & 6.21 $\pm$ 0.16 & 0.199 $\pm$ 0.014 & 0.087 $\pm$ 0.014   \\ [0.00cm] 

201152424 &C10     & 0.358500($\pm$2)   & 3.41 $\pm$ 0.02 & 6.82 $\pm$ 0.03 & 0.122 $\pm$ 0.002 & 0.068 $\pm$ 0.002   \\
201440678 &C10     & 0.376676($\pm$2)   & 3.39 $\pm$ 0.02 & 6.62 $\pm$ 0.03 & 0.132 $\pm$ 0.002 & 0.066 $\pm$ 0.002   \\
201519136 &C10     & 0.344312($\pm$2)   & 3.36 $\pm$ 0.01 & 6.25 $\pm$ 0.04 & 0.159 $\pm$ 0.002 & 0.051 $\pm$ 0.002   \\
228800773 &C10     & 0.372744($\pm$4)   & 3.21 $\pm$ 0.02 & 6.13 $\pm$ 0.05 & 0.187 $\pm$ 0.004 & 0.069 $\pm$ 0.004   \\
228952519 &C10     & 0.40349($\pm$7)    & 3.19 $\pm$ 0.08 & 6.03 $\pm$ 0.10 & 0.195 $\pm$ 0.015 & 0.087 $\pm$ 0.008   \\
248369176 &C10     & 0.424078($\pm$10)  & 3.35 $\pm$ 0.05 & 6.34 $\pm$ 0.11 & 0.205 $\pm$ 0.009 & 0.081 $\pm$ 0.008   \\ [0.00cm] 

225326517 &C11     & 0.4335412($\pm$2)  & 3.64 $\pm$ 0.02 & 6.79 $\pm$ 0.03 & 0.149 $\pm$ 0.002 & 0.077 $\pm$ 0.002   \\
225456697 &C11     & 0.2943527($\pm$1)  & 3.08 $\pm$ 0.05 & 6.70 $\pm$ 0.09 & 0.100 $\pm$ 0.005 & 0.049 $\pm$ 0.005   \\
235631055 &C11     & 0.4327289($\pm$2)  & 3.73 $\pm$ 0.03 & 6.73 $\pm$ 0.07 & 0.199 $\pm$ 0.005 & 0.084 $\pm$ 0.005   \\
235794591 &C11     & 0.4040492($\pm$1)  & 3.45 $\pm$ 0.02 & 6.50 $\pm$ 0.04 & 0.190 $\pm$ 0.003 & 0.077 $\pm$ 0.003   \\
236212613 &C11     & 0.3489577($\pm$2)  & 3.33 $\pm$ 0.03 & 6.40 $\pm$ 0.08 & 0.144 $\pm$ 0.004 & 0.059 $\pm$ 0.004   \\
251248825 &C11     & 0.4041701($\pm$2)  & 3.61 $\pm$ 0.02 & 6.81 $\pm$ 0.05 & 0.226 $\pm$ 0.004 & 0.084 $\pm$ 0.004   \\
251248826 &C11     & 0.3182948($\pm$2)  & 3.85 $\pm$ 0.04 & 6.28 $\pm$ 0.08 & 0.141 $\pm$ 0.006 & 0.071 $\pm$ 0.006   \\
251248827 &C11     & 0.3379169($\pm$1)  & 3.74 $\pm$ 0.03 & 6.01 $\pm$ 0.13 & 0.172 $\pm$ 0.005 & 0.034 $\pm$ 0.004   \\
251248830 &C11     & 0.310017($\pm$8)   & 3.68 $\pm$ 0.12 & 6.37 $\pm$ 0.51 & 0.145 $\pm$ 0.017 & 0.033 $\pm$ 0.016   \\ [0.00cm] 

245974758 &C12     & 0.3533078($\pm$4)  & 3.38 $\pm$ 0.01 & 6.32 $\pm$ 0.01 & 0.161 $\pm$ 0.001 & 0.057 $\pm$ 0.001   \\ 
246058914 &C12     & 0.3361100($\pm$4)  & 3.37 $\pm$ 0.01 & 5.85 $\pm$ 0.04 & 0.171 $\pm$ 0.002 & 0.039 $\pm$ 0.001   \\
251456808 &C12     & 0.348582($\pm$9)   & 3.19 $\pm$ 0.09 & 6.02 $\pm$ 0.21 & 0.175 $\pm$ 0.014 & 0.068 $\pm$ 0.014   \\ [0.00cm] 

247334376 &C13     & 0.403243($\pm$9)   & 3.31 $\pm$ 0.05 & 6.38 $\pm$ 0.10 & 0.193 $\pm$ 0.009 & 0.088 $\pm$ 0.008   \\ [0.00cm] 

201749391 &C14     & 0.3573249($\pm$3)  & 3.31 $\pm$ 0.01 & 6.29 $\pm$ 0.01 & 0.168 $\pm$ 0.001 & 0.063 $\pm$ 0.001   \\
248426222 &C14     & 0.401116($\pm$3)   & 3.32 $\pm$ 0.01 & 6.13 $\pm$ 0.02 & 0.213 $\pm$ 0.002 & 0.077 $\pm$ 0.002   \\
248509474 &C14     & 0.4151037($\pm$4)  & 3.48 $\pm$ 0.01 & 6.27 $\pm$ 0.01 & 0.206 $\pm$ 0.001 & 0.080 $\pm$ 0.001   \\
248514834 &C14     & 0.4344186($\pm$2)  & 3.22 $\pm$ 0.01 & 6.23 $\pm$ 0.01 & 0.250 $\pm$ 0.001 & 0.088 $\pm$ 0.001   \\
248653210 &C14     & 0.350434($\pm$2)   & 3.28 $\pm$ 0.02 & 6.36 $\pm$ 0.04 & 0.143 $\pm$ 0.003 & 0.062 $\pm$ 0.003   \\
248653582 &C14     & 0.4027261($\pm$2)  & 3.34 $\pm$ 0.01 & 6.18 $\pm$ 0.01 & 0.218 $\pm$ 0.001 & 0.081 $\pm$ 0.001   \\
248667792 &C14     & 0.358942($\pm$1)   & 3.26 $\pm$ 0.01 & 6.30 $\pm$ 0.02 & 0.160 $\pm$ 0.002 & 0.064 $\pm$ 0.002   \\
248730795 &C14     & 0.390117($\pm$1)   & 3.26 $\pm$ 0.01 & 6.26 $\pm$ 0.02 & 0.188 $\pm$ 0.001 & 0.076 $\pm$ 0.001   \\
248731983 &C14     & 0.4175000($\pm$1)  & 3.47 $\pm$ 0.01 & 6.39 $\pm$ 0.01 & 0.193 $\pm$ 0.001 & 0.079 $\pm$ 0.001   \\
248827979 &C14     & 0.389714($\pm$8)   & 3.28 $\pm$ 0.02 & 6.00 $\pm$ 0.04 & 0.219 $\pm$ 0.003 & 0.075 $\pm$ 0.003   \\
248845745 &C14     & 0.358376($\pm$2)   & 3.28 $\pm$ 0.01 & 6.51 $\pm$ 0.03 & 0.143 $\pm$ 0.002 & 0.068 $\pm$ 0.002   \\
248871792 &C14     & 0.3763140($\pm$2)  & 3.31 $\pm$ 0.01 & 5.95 $\pm$ 0.01 & 0.213 $\pm$ 0.001 & 0.064 $\pm$ 0.001   \\ [0.00cm] 

249790928 &C15     & 0.4306134($\pm$9)  & 3.80 $\pm$ 0.01 & 7.09 $\pm$ 0.01 & 0.133 $\pm$ 0.001 & 0.072 $\pm$ 0.001   \\
250056977 &C15     & 0.3967475($\pm$2)  & 3.31 $\pm$ 0.01 & 5.93 $\pm$ 0.01 & 0.237 $\pm$ 0.001 & 0.081 $\pm$ 0.001   \\ [0.00cm] 

211665293 &C16     & 0.366065($\pm$1)   & 3.37 $\pm$ 0.01 & 6.59 $\pm$ 0.02 & 0.135 $\pm$ 0.001 & 0.064 $\pm$ 0.001   \\  [0.00cm] 

212467099 &C17     & 0.386354($\pm$3)   & 3.24 $\pm$ 0.01 & 6.22 $\pm$ 0.03 & 0.186 $\pm$ 0.002 & 0.078 $\pm$ 0.002   \\
212498188 &C17     & 0.371357($\pm$2)   & 3.30 $\pm$ 0.01 & 6.00 $\pm$ 0.03 & 0.202 $\pm$ 0.002 & 0.061 $\pm$ 0.002   \\
212615778 &C17     & 0.348889($\pm$1)   & 3.40 $\pm$ 0.01 & 6.20 $\pm$ 0.03 & 0.166 $\pm$ 0.001 & 0.049 $\pm$ 0.001   \\
212819285 &C17     & 0.353446($\pm$4)   & 3.30 $\pm$ 0.03 & 6.36 $\pm$ 0.07 & 0.152 $\pm$ 0.004 & 0.059 $\pm$ 0.004   \\
251521080 &C17     & 0.373014($\pm$5)   & 3.24 $\pm$ 0.03 & 6.21 $\pm$ 0.07 & 0.164 $\pm$ 0.005 & 0.064 $\pm$ 0.005   \\
251629085 &C17     & 0.344183($\pm$4)   & 3.19 $\pm$ 0.04 & 6.77 $\pm$ 0.06 & 0.105 $\pm$ 0.004 & 0.065 $\pm$ 0.004   \\
251809772 &C17     & 0.355035($\pm$3)   & 3.30 $\pm$ 0.02 & 6.13 $\pm$ 0.05 & 0.167 $\pm$ 0.003 & 0.057 $\pm$ 0.003   \\
251809814 &C17     & 0.347284($\pm$3)   & 3.41 $\pm$ 0.02 & 6.28 $\pm$ 0.06 & 0.160 $\pm$ 0.003 & 0.050 $\pm$ 0.003   \\
251809825 &C17     & 0.353727($\pm$3)   & 3.38 $\pm$ 0.02 & 6.39 $\pm$ 0.06 & 0.153 $\pm$ 0.004 & 0.059 $\pm$ 0.004   \\
251809832 &C17     & 0.387565($\pm$5)   & 3.23 $\pm$ 0.03 & 6.25 $\pm$ 0.06 & 0.184 $\pm$ 0.005 & 0.082 $\pm$ 0.005   \\
251809860 &C17     & 0.363480($\pm$8)   & 3.38 $\pm$ 0.05 & 6.48 $\pm$ 0.17 & 0.181 $\pm$ 0.009 & 0.052 $\pm$ 0.009   \\
251809870 &C17     & 0.375881($\pm$14)  & 3.07 $\pm$ 0.09 & 6.06 $\pm$ 0.23 & 0.201 $\pm$ 0.017 & 0.074 $\pm$ 0.017   \\ [0.1cm] 
mean\,$\pm$\,s.e. &&                    & 3.36 $\pm$ 0.02 & 6.33 $\pm$ 0.03 & 0.171 $\pm$ 0.004 & 0.067 $\pm$ 0.001   \\ [0.1cm] %
\hline
\end{tabular}
\end{table*}

\begin{table*}
\fontsize{7}{8.4}\selectfont
\centering

\caption{Fourier parameters for the {\bf radial fundamental mode} pulsations.  (See Table 4 for column descriptions).  }

\label{tab:five}
\begin{tabular}{lclcclll}
\hline
\multicolumn{1}{c}{ EPIC  } & \multicolumn{1}{c}{K2} & \multicolumn{1}{c}{ $P_0$} & \multicolumn{1}{c}{$\phi_{\rm21,0}^s$(Kp)}   & \multicolumn{1}{c}{$\phi_{\rm31,0}^s$(Kp)} & \multicolumn{1}{c} {$R_{\rm 21,0}$(Kp) }  & \multicolumn{1}{c}{$R_{\rm 31,0}$(Kp) }   \\
\multicolumn{1}{c}{ no.   } & \multicolumn{1}{c}{ Campaign } & \multicolumn{1}{c}{[day]}   & \multicolumn{1}{c}{[radians]} & \multicolumn{1}{c} {[radians]}  & \multicolumn{1}{c}{ }   \\	
\multicolumn{1}{c}{(1)} & \multicolumn{1}{c}{(2)} &\multicolumn{1}{c}{(3)}  &\multicolumn{1}{c}{(4)} & \multicolumn{1}{c}{(5)} & \multicolumn{1}{c}{(6)}   & \multicolumn{1}{c}{(7)} \\
\hline  
\\	    

060018653 & E2    & 0.539441($\pm$1)    & 2.35 $\pm$ 0.16 & 4.68 $\pm$ 0.63 & 0.104 $\pm$ 0.014 & 0.022 $\pm$ 0.014 \\
060018662 & E2    & 0.559323($\pm$1)    & 2.23 $\pm$ 0.11 & 4.52 $\pm$ 0.57 & 0.096 $\pm$ 0.011 & 0.017 $\pm$ 0.010 \\ [0.00cm]

201585823 & C1    & 0.4825903($\pm$6)   & 2.32 $\pm$ 0.01 & 5.32 $\pm$ 0.03 & 0.125 $\pm$ 0.001 & 0.015 $\pm$ 0.001 \\ [0.00cm] 

210600482 & C4    & 0.487191($\pm$3)    & 2.39 $\pm$ 0.01 & 5.48 $\pm$ 0.10 & 0.165 $\pm$ 0.002 & 0.022 $\pm$ 0.002 \\
210831816 & C4    & 0.488793($\pm$6)    & 2.45 $\pm$ 0.06 &\multicolumn{1}{c}{5.3 $\pm$ 1.4} & 0.063 $\pm$ 0.004 & 0.003 $\pm$ 0.004 \\
210933539 & C4    & 0.4818008($\pm$5)   & 2.37 $\pm$ 0.01 & 5.33 $\pm$ 0.01 & 0.177 $\pm$ 0.001 & 0.031 $\pm$ 0.001 \\
211072039 & C4    & 0.527041($\pm$11)   & 2.26 $\pm$ 0.08 & 5.78 $\pm$ 0.63 & 0.092 $\pm$ 0.007 & 0.011 $\pm$ 0.007 \\  [0.00cm] 

211694449 & C5,18 & 0.46365024($\pm$6)  & 2.35 $\pm$ 0.01 & 5.27 $\pm$ 0.02 & 0.258 $\pm$ 0.001 & 0.068 $\pm$ 0.001  \\ 
211888680 & C5,16 & 0.4829926($\pm$2)   & 2.37 $\pm$ 0.02 & 5.58 $\pm$ 0.13 & 0.152 $\pm$ 0.003 & 0.023 $\pm$ 0.003 \\ 
211898723 & C5,18 & 0.5058274($\pm$1)   & 2.40 $\pm$ 0.01 & 5.55 $\pm$ 0.06 & 0.176 $\pm$ 0.002 & 0.024 $\pm$ 0.001 \\ [ 0.00cm]

212335848 & C6    & 0.476856($\pm$6)    & 2.32 $\pm$ 0.04 & 5.33 $\pm$ 0.43 & 0.111 $\pm$ 0.004 & 0.010 $\pm$ 0.004 \\
212449019 & C6    & 0.487778($\pm$6)    & 2.38 $\pm$ 0.04 & 5.70 $\pm$ 0.38 & 0.113 $\pm$ 0.004 & 0.011 $\pm$ 0.004 \\
212455160 & C6,17 & 0.469592($\pm$4)    & 2.32 $\pm$ 0.03 & 5.28 $\pm$ 0.20 & 0.168 $\pm$ 0.004 & 0.022 $\pm$ 0.004 \\
212547473 & C6    & 0.545079($\pm$4)    & 2.37 $\pm$ 0.02 & 5.59 $\pm$ 0.12 & 0.125 $\pm$ 0.002 & 0.018 $\pm$ 0.002 \\ [0.00cm]

213514736 & C7    & 0.503583($\pm$9)    & 2.35 $\pm$ 0.03 & 5.37 $\pm$ 0.17 & 0.193 $\pm$ 0.006 & 0.035 $\pm$ 0.006 \\
214147122 & C7    & 0.541040($\pm$14)   & 2.44 $\pm$ 0.19 & 4.48 $\pm$ 0.35 & 0.083 $\pm$ 0.015 & 0.044 $\pm$ 0.015 \\
229228175 & C7    & 0.46954($\pm$5)     & 2.85 $\pm$ 0.30 &\multicolumn{1}{c}{$\dots$}& 0.131 $\pm$ 0.038 &\multicolumn{1}{c}{$\dots$} \\
229228184 & C7    & 0.45943($\pm$4)     & 2.59 $\pm$ 0.11 & 5.53 $\pm$ 0.31 & 0.204 $\pm$ 0.021 & 0.067 $\pm$ 0.020 \\
229228194 & C7    & 0.52263($\pm$10)    & 2.70 $\pm$ 0.11 & 5.7  $\pm$ 1.2  & 0.168 $\pm$ 0.006 & 0.015 $\pm$ 0.006  \\
229228220 & C7    & 0.48774($\pm$3)     & \multicolumn{1}{c}{\dots}  & 4.46 $\pm$ 0.77 & 0.055 $\pm$ 0.029 & 0.037 $\pm$ 0.029 \\ [0.00cm] 

220254937 & C8    & 0.535934($\pm$17)   & 2.42 $\pm$ 0.11 &\multicolumn{1}{c}{6.1 $\pm$ 2.2} & 0.101 $\pm$ 0.011 & 0.005 $\pm$ 0.011 \\
220604574 & C8    & 0.476772($\pm$7)    & 2.44 $\pm$ 0.07 & 5.33 $\pm$ 0.39 & 0.071 $\pm$ 0.005 & 0.012 $\pm$ 0.005 \\
220636134 & C8    & 0.501375($\pm$2)    & 2.27 $\pm$ 0.03 & 5.37 $\pm$ 0.22 & 0.134 $\pm$ 0.004 & 0.018 $\pm$ 0.004 \\
229228811 & C8    & 0.500219($\pm$6)    & 2.30 $\pm$ 0.04 & 5.38 $\pm$ 0.36 & 0.121 $\pm$ 0.004 & 0.012 $\pm$ 0.004 \\ [0.00cm] 

224366356 & C9    & 0.4619359($\pm$6)   & 2.35 $\pm$ 0.06 & 5.01 $\pm$ 0.17 & 0.244 $\pm$ 0.013 & 0.077 $\pm$ 0.013   \\
223051735 & C9    & 0.5677371($\pm$7)   & 2.66 $\pm$ 0.13 &\multicolumn{1}{c}{6.0 $\pm$ 1.0} & 0.252 $\pm$ 0.029 & 0.028 $\pm$ 0.028   \\

201152424 & C10   & 0.481725($\pm$9)   & 2.47 $\pm$ 0.14 & \multicolumn{1}{c}{$\dots$} & 0.052 $\pm$ 0.007 & 0.017 $\pm$ 0.007 \\
201440678 & C10   & 0.505614($\pm$11)  & 2.37 $\pm$ 0.07 & \multicolumn{1}{c}{5.0 $\pm$ 1.1} & 0.083 $\pm$ 0.006 & 0.006 $\pm$ 0.006 \\
201519136 & C10   & 0.463376($\pm$4)   & 2.37 $\pm$ 0.01 & 5.24 $\pm$ 0.07 & 0.190 $\pm$ 0.002 & 0.033 $\pm$ 0.002 \\
228800773 & C10   & 0.49988($\pm$1)    & 2.27 $\pm$ 0.05 & 5.51 $\pm$ 0.34 & 0.149 $\pm$ 0.007 & 0.019 $\pm$ 0.007 \\
228952519 & C10   & 0.5406($\pm$15)    & \multicolumn{1}{c}{\dots} & \multicolumn{1}{c}{\dots} & 0.026 $\pm$ 0.026 & 0.020 $\pm$ 0.026 \\
248369176 & C10   & 0.56828($\pm$3)    & 2.20 $\pm$ 0.14 & 5.97 $\pm$ 0.84 & 0.146 $\pm$ 0.020 & 0.024 $\pm$ 0.020 \\ [0.00cm] 

225326517 & C11   & 0.581816($\pm$1)   & 2.52 $\pm$ 0.16 & 6.70 $\pm$ 0.74 & 0.059 $\pm$ 0.010 & 0.013 $\pm$ 0.010 \\
225456697 & C11   & 0.3998380($\pm$4)  & 2.52 $\pm$ 0.07 & 5.06 $\pm$ 0.53 & 0.136 $\pm$ 0.009 & 0.016 $\pm$ 0.009 \\
235631055 & C11   & 0.5804633($\pm$7)  & 2.51 $\pm$ 0.10 & 5.53 $\pm$ 0.45 & 0.121 $\pm$ 0.012 & 0.027 $\pm$ 0.012 \\
235794591 & C11   & 0.5415829($\pm$5)  & 2.34 $\pm$ 0.10 & 4.71 $\pm$ 0.22 & 0.090 $\pm$ 0.009 & 0.040 $\pm$ 0.009 \\
236212613 & C11   & 0.4692002($\pm$4)  & 2.40 $\pm$ 0.05 & 6.02 $\pm$ 0.33 & 0.143 $\pm$ 0.007 & 0.022 $\pm$ 0.007 \\
251248825 & C11   & 0.5413433($\pm$9)  & 2.11 $\pm$ 0.23 &\multicolumn{1}{c}{6.1 $\pm$ 1.0} & 0.066 $\pm$ 0.015 & 0.015 $\pm$ 0.015 \\
251248826 & C11   & 0.430234($\pm$1)   & 2.32 $\pm$ 0.49 &\multicolumn{1}{c}{$\dots$} & 0.044 $\pm$ 0.021 & 0.009 $\pm$ 0.021 \\
251248827 & C11   & 0.4552683($\pm$2)  & 2.64 $\pm$ 0.02 & 5.63 $\pm$ 0.11 & 0.212 $\pm$ 0.005 & 0.044 $\pm$ 0.005 \\
251248830 & C11   & 0.419592($\pm$14)  & 2.67 $\pm$ 0.08 & 5.16 $\pm$ 0.48 & 0.209 $\pm$ 0.016 & 0.037 $\pm$ 0.016 \\ [0.00cm] 

245974758 & C12   & 0.475291($\pm$5)   & 2.37 $\pm$ 0.01 & 5.42 $\pm$ 0.03 & 0.185 $\pm$ 0.001 & 0.030 $\pm$ 0.001 \\
246058914 & C12   & 0.4529476($\pm$7)  & 2.37 $\pm$ 0.01 & 5.25 $\pm$ 0.03 & 0.240 $\pm$ 0.001 & 0.055 $\pm$ 0.001 \\
251456808 & C12   & 0.46898($\pm$2)    & 2.31 $\pm$ 0.10 & 5.53 $\pm$ 0.74 & 0.188 $\pm$ 0.018 & 0.024 $\pm$ 0.018 \\ [0.00cm] 

247334376 & C13   & 0.53958($\pm$4)    & \multicolumn{1}{c}{\dots} & \multicolumn{1}{c}{\dots} & 0.037 $\pm$ 0.053 & \multicolumn{1}{c}{\dots} \\ [0.00cm] 

201749391 & C14   & 0.479930($\pm$1)   & 2.35 $\pm$ 0.01 & 5.29 $\pm$ 0.02 & 0.162 $\pm$ 0.001 & 0.023 $\pm$ 0.001 \\
248426222 & C14   & 0.537644($\pm$8)   & 2.34 $\pm$ 0.02 & 5.46 $\pm$ 0.12 & 0.143 $\pm$ 0.003 & 0.025 $\pm$ 0.003 \\
248509474 & C14   & 0.557312($\pm$1)   & 2.44 $\pm$ 0.01 & 5.65 $\pm$ 0.02 & 0.179 $\pm$ 0.001 & 0.029 $\pm$ 0.001 \\
248514834 & C14   & 0.581287($\pm$2)   & 2.35 $\pm$ 0.01 & 5.60 $\pm$ 0.11 & 0.064 $\pm$ 0.001 & 0.006 $\pm$ 0.001 \\
248653210 & C14   & 0.471007($\pm$6)   & 2.30 $\pm$ 0.04 & 5.47 $\pm$ 0.23 & 0.143 $\pm$ 0.005 & 0.021 $\pm$ 0.005 \\
248653582 & C14   & 0.5398778($\pm$5)  & 2.36 $\pm$ 0.01 & 5.57 $\pm$ 0.01 & 0.142 $\pm$ 0.001 & 0.023 $\pm$ 0.001 \\
248667792 & C14   & 0.482054($\pm$4)   & 2.32 $\pm$ 0.02 & 5.46 $\pm$ 0.15 & 0.138 $\pm$ 0.003 & 0.018 $\pm$ 0.003 \\
248730795 & C14   & 0.522545($\pm$6)   & 2.29 $\pm$ 0.04 & 5.40 $\pm$ 0.31 & 0.094 $\pm$ 0.004 & 0.012 $\pm$ 0.004 \\
248731983 & C14   & 0.5600838($\pm$5)  & 2.42 $\pm$ 0.01 & 5.63 $\pm$ 0.02 & 0.135 $\pm$ 0.001 & 0.017 $\pm$ 0.001 \\
248827979 & C14   & 0.522807($\pm$10)  & 2.33 $\pm$ 0.03 & 5.36 $\pm$ 0.15 & 0.179 $\pm$ 0.005 & 0.033 $\pm$ 0.005 \\
248845745 & C14   & 0.481123($\pm$8)   & 2.47 $\pm$ 0.07 & 4.55 $\pm$ 0.62 & 0.092 $\pm$ 0.006 & 0.010 $\pm$ 0.006 \\
248871792 & C14   & 0.5054967($\pm$4)  & 2.33 $\pm$ 0.01 & 5.40 $\pm$ 0.01 & 0.211 $\pm$ 0.001 & 0.047 $\pm$ 0.001 \\ [0.00cm] 

249790928 & C15   & 0.578550($\pm$6)   & 2.59 $\pm$ 0.05 & 5.84 $\pm$ 0.86 & 0.066 $\pm$ 0.003 & 0.004 $\pm$ 0.003 \\
250056977 & C15   & 0.532603($\pm$1)   & 2.39 $\pm$ 0.01 & 5.52 $\pm$ 0.01 & 0.203 $\pm$ 0.001 & 0.047 $\pm$ 0.001 \\ [0.00cm] 

211665293 & C16   & 0.491516($\pm$4)   & 2.38 $\pm$ 0.03 & 5.51 $\pm$ 0.48 & 0.093 $\pm$ 0.003 & 0.006 $\pm$ 0.003 \\ [0.00cm] 

212467099 & C17   & 0.517170($\pm$18)  & 2.36 $\pm$ 0.13 & \multicolumn{1}{c}{$\dots$} & 0.073 $\pm$ 0.010 & 0.004 $\pm$ 0.010 \\
212498188 & C17   & 0.498643($\pm$4)   & 2.34 $\pm$ 0.01 & 5.41 $\pm$ 0.05 & 0.207 $\pm$ 0.002 & 0.045 $\pm$ 0.002 \\
212615778 & C17   & 0.469391($\pm$2)   & 2.37 $\pm$ 0.01 & 5.38 $\pm$ 0.04 & 0.196 $\pm$ 0.002 & 0.037 $\pm$ 0.002 \\
212819285 & C17   & 0.474968($\pm$13)  & 2.29 $\pm$ 0.06 & 6.14 $\pm$ 0.51 & 0.131 $\pm$ 0.008 & 0.016 $\pm$ 0.008 \\
251521080 & C17   & 0.50087($\pm$2)    & 2.41 $\pm$ 0.04 & 5.34 $\pm$ 0.26 & 0.172 $\pm$ 0.007 & 0.027 $\pm$ 0.007 \\
251629085 & C17   & 0.46311($\pm$4)    &  \multicolumn{1}{c}{\dots}  &\multicolumn{1}{c}{5.1 $\pm$ 1.1} & 0.043 $\pm$ 0.024 & 0.023 $\pm$ 0.024 \\
251809772 & C17   & 0.477332($\pm$6)   & 2.28 $\pm$ 0.02 & 5.51 $\pm$ 0.14 & 0.186 $\pm$ 0.004 & 0.030 $\pm$ 0.004 \\
251809814 & C17   & 0.467124($\pm$6)   & 2.38 $\pm$ 0.03 & 5.33 $\pm$ 0.12 & 0.170 $\pm$ 0.004 & 0.033 $\pm$ 0.004 \\
251809825 & C17   & 0.475900($\pm$8)   & 2.41 $\pm$ 0.03 & 5.42 $\pm$ 0.20 & 0.171 $\pm$ 0.005 & 0.025 $\pm$ 0.005 \\
251809832 & C17   & 0.51919($\pm$3)    & 2.52 $\pm$ 0.32 & \multicolumn{1}{c}{$\dots$} & 0.067 $\pm$ 0.021 & 0.043 $\pm$ 0.021 \\
251809860 & C17   & 0.488442($\pm$19)  & 2.34 $\pm$ 0.07 & \multicolumn{1}{c}{5.2 $\pm$ 1.8} & 0.181 $\pm$ 0.012 & 0.007 $\pm$ 0.012 \\
251809870 & C17   & 0.50483($\pm$3)    & 2.21 $\pm$ 0.09 & 4.99 $\pm$ 0.39 & 0.235 $\pm$ 0.019 & 0.048 $\pm$ 0.019 \\ [0.1cm] 
mean\,$\pm$\,s.e.&&                    & 2.39 $\pm$ 0.02 & 5.41 $\pm$ 0.05 & 0.141 $\pm$ 0.007 & 0.025 $\pm$ 0.002 \\ [0.1cm] %
\hline
\end{tabular}
\end{table*}

Our best estimates of the pulsation periods and amplitudes  (fundamental and first-overtone radial modes)   for the {\it K2} RRd
stars are given in {\bf
Table\,3}.  Mean magnitudes, useful for distance estimation and for planning
follow-up spectroscopy, are given in columns 2-5.  The $V$ magnitudes
(column 2) are from previously published ground-based photometry.  The  $G$
magnitudes (column 3) are from the {\it {\it Gaia}} mission (from DR3 if
available, otherwise DR2).  For the Campaign\,7 RRd stars in the direction of
the Sagittarius dwarf galaxy (see Appendix~A) the  $V$ magnitudes were
estimated by subtracting 0.30 mag from the mean $B$ magnitudes given by
Cseresnjes (2001).  The {\it Kp} magnitudes given at the \texttt{MAST}
website (column 4) are from Huber {\it et al.} (2016) and were derived using
previous IR and other photometry.  The {\it Kp} magnitudes labelled `{\it
K2}' (column 5) were derived from the present analysis using the {\it K2} light
curves and the flux-to-magnitude calibration given by Lund
{\it et al.} (2015): $ Kp = 25.3 - 2.5\log_{\rm 10}F $, \noindent where $F$ is
the median of the flux time series (units of e$^{-}$/s).  The mean fluxes,
which range from around 50,000 e$^{-}$/s to fewer than 100 e$^{-}$/s, depend on
the size of the aperture and come from either the EAP pipeline, the PDCSAP
estimates calculated by the {\it K2} pipeline, or from our own analyses using
the PyKE software.  In most cases the mean magnitudes in columns 2-5 are
similar, with brightnesses ranging from 14.2 mag (EPIC\,60018653) to 20.4 mag
(EPIC\,248369176).  Outliers are identified with square brackets.  The largest
discrepancies are for stars in the crowded Galactic Bulge fields (Campaigns 9
and 11).  The procedure for calculating the periods and amplitudes in Table~3
was similar to that used in our analysis of peculiar RRd stars (NM21).
For the current paper an additional term was added to the model to
represent non-radial oscillations with  periods $P_{\rm nr}$ near 0.61\,$P_1$
(see Moskalik et al.  2018a,b):
\begin{multline}
m(t) = m_0 + \sum_{i=1}^{N_1} A_{\rm i,1}\sin\,(i\omega_1\,[t-t_0] + \phi_{\rm i,1})\,\,+ \\
      ~~~~~~~\sum_{j=1}^{N_0} A_{\rm j,0}\sin\,(j\omega_0\,[t-t_0] + \phi_{\rm j,0})\,\, + \\
	~~~~~~~\sum_{k=1}^{N_{\rm nr}} A_{\rm k,{\rm nr}}\sin\,(k\omega_{\rm nr}\,[t-t_0] + \phi_{\rm k,{\rm nr}})\,\, + \\
       ~~~~~~~~\sum_{i=1}^{N_1}\sum_{j=1}^{N_0}\,[A^{+}_{\rm i,j}\sin\,((i\omega_1 + j\omega_0)\,[t-t_0] + \phi^{+}_{\rm i,j})\,\,+ \\
                              A^{-}_{\rm i,j}\sin\,((i\omega_1 - j\omega_0)\,[t-t_0] + \phi^{-}_{\rm i,j})],
\end{multline}


\noindent where $m_0$ is the mean magnitude;  $\omega_1$, $\omega_0$ and
$\omega_{\rm nr}$ are the angular frequencies for the first-overtone,
fundamental and non-radial modes; the $A$ and $\phi$ are the amplitudes and
phases of the various terms in the Fourier sums;  and ${N_1}$, $N_0$ and
$N_{\rm nr}$ are, respectively, the number of terms for the (usually dominant)
first-overtone,  the fundamental and the non-radial modes that were included in
the expansion.  Values for ${N_1}$, ${N_0}$ and $N_{\rm nr}$ were adjusted to
include all significant harmonic peaks (typically five harmonics were included) in the amplitude spectra.  The
multi-frequency model (Eqn.\,1) was fitted by non-linear
(Levenberg-Marquardt) least-squares using the \texttt{PROC NONLIN} procedure in
SAS$^\copyright$ version 9.4.

{\bf Figure\,5} shows the component light curves (upper panel) and their sum
(lower panel), for EPIC\,201585823\footnote{EPIC\,201585823 was described by
Kurtz et al. (2016) as a ``rare triple-mode RR Lyrae star''.  We find it to be
a typical intermediate-metallicity classical RRd star (of the type found in
Oosterhoff type I globular clusters) and `rare' only in the sense that RRd
stars in general are relatively rare. The third mode is the very-low-amplitude
non-radial mode that appears to be common in RRd stars.}.  The amplitude
spectrum (shown in Fig.\,3) clearly identifies the main frequencies as  the
radial first-overtone mode (dominant) and the radial fundamental mode
(secondary).  The next most significant contribution comes from the combination
terms involving the two radial modes (red dotted curve in Fig.\,5);  these are
seen to contribute more to the summed light curve than the low amplitude
non-radial component (blue light curve).  When all the terms are added (lower
panel) the standard deviation of the residuals amounts to only $\sigma$=6.6
mmag.

\renewcommand{\thefigure}{5}
\begin{figure} \begin{center}
\begin{overpic}[width=8.4 cm]{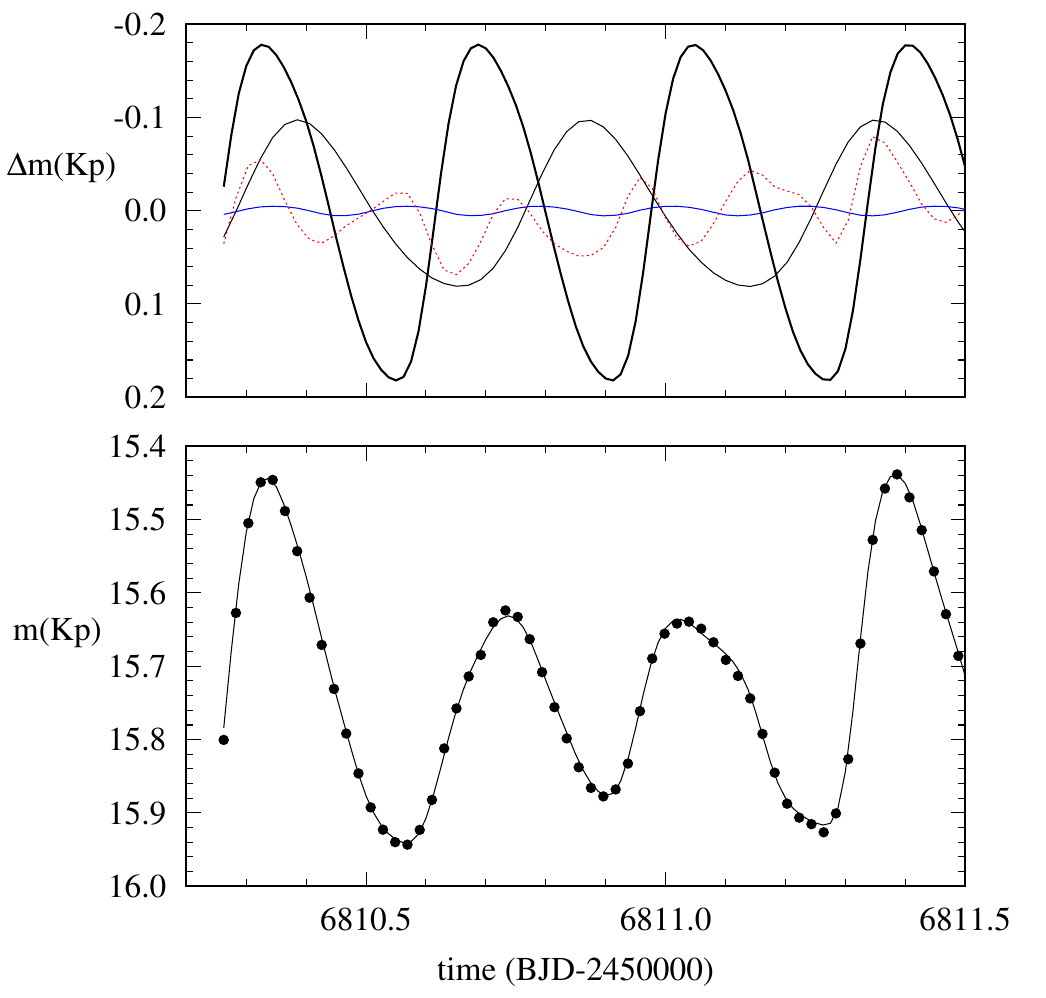}  \put(20,65){(a)}  \put(20,17){(b)}    \end{overpic}    
\end{center}
\vskip-0.4truecm

\caption{Component light curves for the `classical' RRd star EPIC\,201585823.
Non-linear least squares fitting of the EAP photometry was used to estimate the
Fourier parameters, and for clarity only the first 1.2 days of data are
plotted.  Upper panel: the highest amplitude, slightly asymmetric curve with
Fourier 1st-term amplitude $A_1$(Kp)=181.4 mmag is for the first-overtone with
$P_1$=0.3594190\,d; the second largest amplitude curve with  $A_0$(Kp)=88.7
mmag is for the fundamental mode with $P_0$=0.4825903\,d; the lowest amplitude
(blue) curve with $A_{\rm nr}$(Kp)=5.3 mmag is for the non-radial mode with
$P_{\rm nr}$=0.22146\,d; and the dotted (red) curve is the variable amplitude
contribution from the $P_1$ and $P_0$ combination frequencies.  Lower panel:
comparison of the observed {\it K2} photometry (black points) with the
predicted light curve (equal to the sum of the four curves shown in the top
panel).  }  

\label{Fig5} 
\end{figure}

The derived amplitudes and phases were used to calculate two sets of Fourier
decomposition parameters (Simon \& Lee 1981; Simon 1990) for the {\it K2} light
curves: epoch($t_0$)-independent phase differences $\phi_{\rm i1}$ = $\phi_{i}$
-- $i \phi_{1}$ and  amplitude ratios $R_{\rm i1}$ = $A_{i}$/$A_{1}$, where $i$
denotes the $i$th harmonic.  In {\bf Table\,4} these quantities (for i=2,3) are
given for the radial first-overtone component, and in {\bf Table\,5} for the
radial fundamental mode.  In both tables column (3) contains the pulsation
period, either $P_1$ or $P_0$; columns (4-5) contain the corresponding
phase-difference parameters,
$\phi_{\rm21}^s$(=$\phi_{\rm2}^s$$-$2$\phi_{\rm1}^s$) and
$\phi_{\rm31}^s$(=$\phi_{\rm3}^s$$-$3$\phi_{\rm1}^s$), where the `s'
superscripts indicate that the Fourier fits to the {\it Kepler/K2} photometry
are based on sine functions (and not cosine functions, as is the case for the
\texttt{OGLE} survey);  and columns (6-7) contain the amplitude-ratio
parameters, $R_{\rm 21}$(=A$_2$/A$_1$)  and $R_{\rm 31}$(=A$_3$/A$_1$).  Mean
values ($\pm$ standard errors of the mean) are given at the bottom of each
column.  

The precision of the derived Fourier parameters is very high for both pulsation
modes.  For the first-overtone pulsations (see Table\,4) the uncertainties in
$\phi_{\rm 21}^s$ and $\phi_{\rm 31}^s$ typically are $\sim$0.02 and $\sim$0.04
radians, respectively, and for $R_{\rm 21}$ and $R_{\rm 31}$ the uncertainties
are $\sim$0.002.  For the fundamental-mode pulsations (see Table\,5), which are
usually of lower amplitude, the uncertainties are larger, $\sim$0.04 and
$\sim$0.26 radians for $\phi_{\rm 21}^s$ and $\phi_{\rm 31}^s$, respectively,
and $\sim$0.005 for both $R_{\rm 21}$ and $R_{\rm 31}$.  Differences in the
uncertainties for individual stars are due to the non-homogeneous nature of the
sample, which is drawn from Ecliptic Plane and Galactic Bulge fields having
different star densities, and other factors such as the methods used to produce
the detrended and outlier-free photometry.  Owing to the low amplitudes of the
non-radial pulsations their inclusion in the fitted model (Eqn.\,3) was found
to have little effect on the derived Fourier decomposition parameters for the
radial pulsations.

\section{DISCUSSION}

The pulsation properties of RRd stars are determined by their masses,
luminosities, effective temperatures, metal abundances, and other physical
characteristics.  Analysis of correlations among descriptors of the observed
light curve, such as the periods, amplitudes and Fourier parameters, are key to
making inferences about the unknown physical quantities that drive the
oscillations.  In $\S$4.1 within-mode and between-mode correlations among 12
pulsation descriptors  are given for the {\it K2} cRRd stars.  The strongest
correlation is between the fundamental and first overtone periods.  In $\S$4.2
this correlation is discussed within the framework of a simple statistical
model that explains both the observed $P_1$ vs $P_0$  and Petersen diagrams.
Dependencies of the pulsation amplitudes and several Fourier parameters on
period are  discussed in $\S$4.3.  In $\S$4.4 an independent sample of RRd
stars with spectroscopic metal abundances and known periods is used to derive a
period-[Fe/H] calibration equation consistent with the $P_1$-$P_0$ correlation
results in $\S$4.2.   The equation is used to estimate metal abundances for the
cRRd stars observed by {\it K2} and for 2130 cRRd stars observed by the {\it
Gaia} Mission.  The effect of misclassification bias on the estimated [Fe/H]
values is also discussed.

\subsection{Within- and Between-Mode Correlations}

For RRd and other multimode pulsators two kinds of correlations are of
interest: {\it cross-mode} and {\it within-mode}.  Examples of the former
include the correlations between $P_1$ and $P_0$, and between $A_1$ and $A_0$
(see figs.5a,b of NM21), and examples of the latter include the
period-amplitude relations for each of the two radial modes (see Fig.6 below).
For single-mode RRab and RRc stars correlations are necessarily within-mode
correlations.  Pearson correlation coefficients involving 12 descriptors of the
light curves are presented for the 72 {\it K2} cRRd stars in {\bf Table\,6}.
In the top section all pairwise correlations (and {\it p}-values measuring statistical significance)
are given for the $P_1$, $A_1$, $R_{\rm 21}$, $R_{\rm 31}$, $\phi_{\rm 21}$ and
$\phi_{\rm 31}$ parameter estimates for the first-overtone mode; the middle
section gives the corresponding correlations for the fundamental mode; and
cross-mode correlation coefficients are given in the bottom section.

\begin{table}
\fontsize{7}{8.4}\selectfont   %
\centering

\caption{Pearson correlation coefficients ($r$) and their associated {\it
p}-values (in {\it italics}) for the 72 {\it K2} cRRd stars (Tables 4-5). The
correlation coefficient is a measure of the linear association between the
respective row and column variables,  and  $r^2$  is the proportion of
variability in the row (column) variable that is explained by the column (row)
variable, where a positive (negative) value corresponds to a positive
(negative) slope.  Small {\it p}-values ($<$0.05) correspond to statistically
significant  (linear) correlation.   If $p<0.01$ the correlation coefficients are
highlighted in boldface.  Note that Sections (a) and (b)
are symmetric about the diagonal.  }

\label{tab:three}
\begin{tabular}{lrrrrrr}
\hline \\

\multicolumn{7}{c}{(a) First-overtone correlations} \\ [0.1cm] %
\multicolumn{1}{l}{}&\multicolumn{1}{r}{$\phi_{\rm 21,1}$ } & \multicolumn{1}{r}{$\phi_{\rm 31,1}$ } &\multicolumn{1}{r}{$R_{\rm 21,1}$}  & \multicolumn{1}{r}{$R_{\rm 31,1}$} & \multicolumn{1}{r}{$A_1$}& \multicolumn{1}{r}{$P_1$} \\ [0.1cm]
$\phi_{\rm 21,1}$   &       1.00            & {\bf 0.36}             &        --0.05               &        --0.06               &  {\bf 0.48}           &          0.16               \\ %
        	    & \dots                 & {\it 0.002}            &     {\it 0.69}              &     {\it 0.64}              & {\it $<$0.001}        &     {\it 0.19}              \\ [0.05cm] %
$\phi_{\rm 31,1}$   & {\bf 0.36}            &      1.00              &  {\bf --0.54}               &          0.21               &      0.12             &          0.13               \\ %
                    & {\it 0.002}           & \dots                  & {\it $<$0.001}              &     {\it 0.08}              & {\it 0.34}            &     {\it 0.27}              \\ [0.05cm] %
$R_{\rm 21,1}$      &   --0.05              & {\bf --0.54}           &          1.00               &   {\bf 0.45}                &    --0.06              &   {\bf 0.61}                \\ %
		    & {\it 0.69}            & {\it $<$0.001}         & \dots                       & {\it $<$0.001}              & {\it 0.60}            &      $<$0.001               \\ [0.05cm] %
$R_{\rm 31,1}$      &    --0.06             &   0.21                 & {\bf 0.45}                  &          1.00               &    --0.25             &    {\bf 0.80}               \\ %
		    & {\it 0.64}            & {\it 0.08}             & {\it $<$0.001}              & \dots                       & {\it 0.04}            &   $<$0.001                  \\ [0.05cm] %
$A_1$               & {\bf 0.48}            &   0.12                 &    --0.06                   &   --0.25                    &      1.00             &        --0.08               \\ %
		    & {\it 0.001}           & {\it 0.34}             & {\it 0.60}                  &  {\it 0.04}                 & \dots                 &     {\it 0.93}              \\ [0.05cm] %
$P_1$               &    0.16               &  0.13                  & {\bf 0.61}                  & {\bf 0.80}                  &  --0.08               &          1.00               \\ %
		    & {\it 0.19}            & {\it 0.27}             &  {\it $<$0.001}             &  {\it $<$0.001}             & {\it 0.93}            & \dots                       \\ [0.2cm] %
		
\hline \\

\multicolumn{7}{c}{(b) Fundamental-mode correlations} \\ [0.1cm]
\multicolumn{1}{r}{} & \multicolumn{1}{r}{$\phi_{\rm 21,0}$} &\multicolumn{1}{r}{$\phi_{\rm 31,0}$} &\multicolumn{1}{r}{$R_{\rm 21,0}$} &\multicolumn{1}{r}{$R_{\rm 31,0}$} &\multicolumn{1}{r}{$A_0$} & \multicolumn{1}{r}{$P_0$} \\ [0.1cm]
$\phi_{\rm 21,0}$   &       1.00 &      0.08  &      0.05               &          0.13           &      0.06                &       --0.11  \\ %
	            &  \dots     & {\it 0.53} & {\it 0.68}              &     {\it 0.29}          & {\it 0.64}               &    {\it 0.38} \\ [0.05cm] %
$\phi_{\rm 31,0}$   &    0.08    &      1.00  &      0.02               &        --0.26           &    --0.02                &         0.27  \\ %
		    & {\it 0.53} & \dots      & {\it 0.85}              &     {\it 0.04}          & {\it 0.88}               &    {\it 0.03} \\ [0.05cm] %
$R_{\rm 21,0}$      &    0.05    &  0.02      &      1.00               &    {\bf 0.68}           &  {\bf 0.82}              &       --0.28  \\ %
		    & {\it 0.68} & {\it 0.85} & \dots                   & {\it $<$0.001}          & {\it $<$0.001}           &    {\it 0.02} \\ [0.05cm] %
$R_{\rm 31,0}$      &    0.13    &   --0.26   &    {\bf 0.68}           &          1.00           &      {\bf 0.55}          &       --0.22  \\ %
		    & {\it 0.29} & {\it 0.04} & {\it $<$0.001}          & \dots                   & {\it $<$0.001}           &    {\it 0.07} \\ [0.05cm] %
$A_0$               &    0.06    &    --0.02  & {\bf 0.82}              & {\bf 0.55}              &      1.00                &  {\bf --0.43}  \\ %
		    & {\it 0.64} &  {\it 0.88}& {\it $<$0.001}          & {\it $<$0.001}          & \dots                    &   {\it 0.002} \\ [0.05cm] %
$P_0$               &  --0.11    &     0.27   &               --0.28    &               --0.22    &   {\bf --0.43}           &         1.00  \\ %
		    & {\it 0.38} & {\it 0.03} & {\it 0.02}              & {\it 0.07}              & {\it 0.002}              & \dots         \\ [0.2cm] %
	
\hline \\

\multicolumn{7}{c}{(c) Cross correlations: Fund.\,(rows) $\times$ 1st Overtone\,(columns)} \\ [0.1cm] %
\multicolumn{1}{r}{} & \multicolumn{1}{r}{$\phi_{\rm 21,1}$} & \multicolumn{1}{r}{$\phi_{\rm 31,1}$} &\multicolumn{1}{r}{$R_{\rm 21,1}$} &\multicolumn{1}{r}{$R_{\rm 31,1}$} &\multicolumn{1}{r}{$A_1$}& \multicolumn{1}{r}{$P_1$} \\ [0.1cm]
$\phi_{\rm 21,0}$   & {\bf  0.37}    & {\bf 0.32}     &   {\bf --0.39} &        --0.16  &      0.08      &        --0.11  \\ %
                    & {\it 0.002}    &  {\it $<$0.01} & {\it 0.001}    &     {\it 0.20} & {\it 0.50}     &     {\it 0.35} \\ [0.05cm] %
$\phi_{\rm 31,0}$   &  {\bf 0.36}    &      0.15      &          0.11  &          0.14  &      0.18      &          0.27  \\ %
                    & {\it 0.004}    & {\it 0.24}     & {\it 0.38}     &     {\it 0.26} & {\it 0.17}     &     {\it 0.03} \\ [0.05cm] %
$R_{\rm 21,0}$      &  --0.10        & {\bf --0.68}   &          0.29  &   {\bf --0.50} &    --0.03      &        --0.28  \\ %
                    & {\it 0.41}     & {\it $<$0.001} &    {\it 0.02}  & {\it $<$0.001} & {\it 0.81}     &     {\it 0.02} \\ [0.05cm] %
$R_{\rm 31,0}$      &   --0.08       &  {\bf --0.62}  &    0.23        &  {\bf --0.34}  &    --0.22      &        --0.23  \\ %
                    & {\it 0.52}     & {\it $<$0.001} & {\it 0.06}     &    {\it 0.004} & {\it 0.07}     &     {\it 0.06} \\ [0.05cm] %
$A_0$               &    0.19        &  {\bf--0.51}   &    0.12        &   {\bf --0.74} &  {\bf 0.41}    &  {\bf --0.37}  \\ %
                    & {\it 0.10}     & {\it $<$0.001} & {\it 0.33}     & {\it $<$0.001} & {\it $<$0.001} &    {\it 0.001} \\ [0.05cm] %
$P_0$               &   0.16         &   0.14         &  {\bf 0.61}    &    {\bf 0.79}  &    --0.01      & {\bf 0.9999}   \\ %
		    & {\it 0.17}     & {\it 0.26}     & {\it $<$0.001} & {\it $<$0.001} & {\it 0.92}     & {\it $<$0.001} \\ [0.2cm] %
\hline
\hline
\end{tabular}
\end{table}

Not surprisingly, the strongest correlation is the cross correlation between
$P_0$ and $P_1$ (bottom right corner of Table\,6), with $r=0.9999$ and {\it
p}$<$0.001.  There is also evidence that $A_1$ and $A_0$ and all but one of
the four Fourier parameters ($\phi_{\rm 31}$) are cross-correlated for the two
modes ({\it p}$\leq$0.02), where $r>0$ (i.e., the correlation is positive) in all cases except $R_{\rm 31}$
(diagonal entries, Table 6c).  Patterns of within-mode correlation differ for
the first-overtone and fundamental modes.  For instance, the Fourier parameters
$R_{\rm 21}$ and $R_{\rm 31}$ are strongly correlated with period for the
first-overtone (Table 6a), but not for the fundamental mode; whereas, amplitude
is strongly and negatively correlated with period for the fundamental mode
(Table 6b) but shows no significant correlation with period in the
first-overtone case.  The nature of these correlations and their implications
are discussed below.

\subsection{$P_0$, $P_1$ Relationships}

Theoretical pulsation models imply that the strong correlation between $P_0$
and $P_1$ arises because both periods depend on the same unobserved physical
factors: mass, luminosity, temperature, [Fe/H], etc.  Consider the following
statistical model\footnote{Models that express the correlation structure of a
set of observable variables in terms of a system of linear equations involving
a smaller number of unobserved ``common factors'' are known as ``factor
analysis'' models in the statistical literature.  Eqn.\,2 has only two
observable variables ($P_0$ and $P_1$) and one common factor ($X$) but the
model can easily be generalized to include additional observable variables
(e.g., $A_0$ and $A_1$) and more than one common factor.  See Morrison (1976).}
that embodies this idea:

\begin{equation}
	\begin{split}	
  \sqrt{P_0} = a_0 +b_0 X + \epsilon_0  \\
  \sqrt{P_1} = a_1 +b_1 X + \epsilon_1 ,   
	\end{split}
\end{equation}

\noindent where the observed periods are assumed to depend primarily on a
single common (unmeasured or unknown) factor $X$.  (The special case where X=[Fe/H] is discussed in $\S$4.4.2).   The square-root
transformation is applied to the periods to ensure that the functional forms of
the $P_1$-$P_0$ and the $P_1/P_0$ vs $P_0$  relationships are consistent with
the observations - see the discussion that follows.  Assume that $X$ has a normal distribution
	with mean $0$ and variance $1$  (if necessary, replace X with 
$ {X-<X>}\over{\sqrt{{\rm var}X}}  $), and that the measurement errors in $\sqrt{P_0}$
and $\sqrt{P_1}$, $\epsilon_0$ and $\epsilon_1$, are independent normally
distributed random variables with mean $0$ and respective variances
$\sigma_0^2$ and $\sigma_1^2$.

It follows from Eqn.2 that   $(\sqrt{P_0}, \sqrt{P_1} )^T $ has a bivariate
normal distribution with mean $(a_0,a_1)^T,$   variance-covariance matrix 
\[
\left[  {\begin{array}{cc} b_0^2 + \sigma^2_0 & b_0\,b_1 \\ 
	                   b_0\,b_1 & b_1^2+ \sigma^2_1 \\ 
         \end{array}	}  
\right] , 
\] 
\noindent and
correlation coefficient $$ \rho = { {b_0 b_1} \over { \sqrt{(b_0^2 +
\sigma^2_0) \times (b_1^2 + \sigma^2_1) }} }.$$ 

\noindent Notice that if $\sigma_0^2$ and $\sigma_1^2$ are small compared with
$b_0^2$ and $b_1^2$ then $\rho$ is close to 1.

It also follows from the properties of the bivariate normal distribution (see Hogg and Craig 1959) that the
conditional mean and variance of $\sqrt{P_1}$ given $\sqrt{P_0}$ are:

$$  E\Bigl(\sqrt{P_1} \Bigm| \sqrt{P_0}\Bigr) = \Bigl(a_0  -  {b_0 \over b_1} a_1 \Bigr) +  {b_0 \over b_1} \sqrt{P_0}  $$

$$  {\rm Var} \Bigl(\sqrt{P_1} \Bigm| \sqrt{P_0}\Bigr) =   \sigma_0^2  +  \Bigl(  {b_0 \over b_1} \Bigr)^2 \sigma_1^2 ,  $$

\noindent  which imply (by the definition of variance) that


$$
	E \Bigl({P_1} \Bigm| \sqrt{P_0} \Bigr) = {{\rm Var}\Bigl( {\sqrt{P_1} } \Bigm| \sqrt{P_0} \Bigr) } +  \Bigl[ E\Bigl({ \sqrt{P_1} }  \bigm| \sqrt{P_0} \Bigr) \Bigr]^2 .   
$$

\noindent  Since  conditioning on $\sqrt{P_0}$ is equivalent to conditioning on $P_0$ (because $P_0$ is positive), substitution into  the preceding equation gives the conditional mean of $P_1$ given $P_0$:   
\begin{equation}
   E \bigl({P_1} \bigm| P_0 \bigr) =   a + b \sqrt{P_0} + c  P_0  ,  
\end{equation}
\noindent  where
$$  a = {\sigma_0^2} + { \Bigl({b_0 \over b_1}\Bigr)^2 \sigma_1^2 }   + { \Bigl(  a_0 - {b_0 \over b_1} a_1  \Bigr)^2 }     , $$ 
$$  b = 2 \,  {b_0 \over b_1} \, { \Bigl(  a_0 - {b_0 \over b_1} a_1  \Bigr) },   $$  
$$  {\rm and} \hskip0.4truecm   c =   { \Bigl({b_0 \over b_1}\Bigr)^2 }    . $$

\noindent Dividing both sides of Eqn.\,(3) by $P_0$ gives the conditional mean of the ratio   
\begin{equation}
	E \biggl({{P_1}\over{P_0}} \biggm| {P_0} \biggr) = { a \over P_0 } + {b \over \sqrt{P_0}} + c .  
\end{equation}

\noindent Thus the Eqn.\,(2) model implies that the expected $P_1$-$P_0$ and $P_1/P_0$ vs. $P_0$ (Petersen diagram) relationships have 
specific functional forms (given by Eqns.\,3 and 4), which can be compared with observations in order to help validate
the model or rule out competing models.  For example, if $\sqrt{P_0}$ and $\sqrt{P_1}$ in Eqn.\,(2) are replaced with $P_0$ and $P_1$ then
\begin{equation} 
E \biggl({P_1 \over P_0} \biggm| P_0 \biggr) = {a' \over P_0} + b', 
\end{equation}
\noindent which fails to fit the observed Petersen curve for cRRd stars (see the red dotted curve in Fig.10c, and $\S$4.4.2 below).

\renewcommand{\thefigure}{6}
\begin{figure*} 
\begin{center}
\begin{overpic}[width=8.2cm]{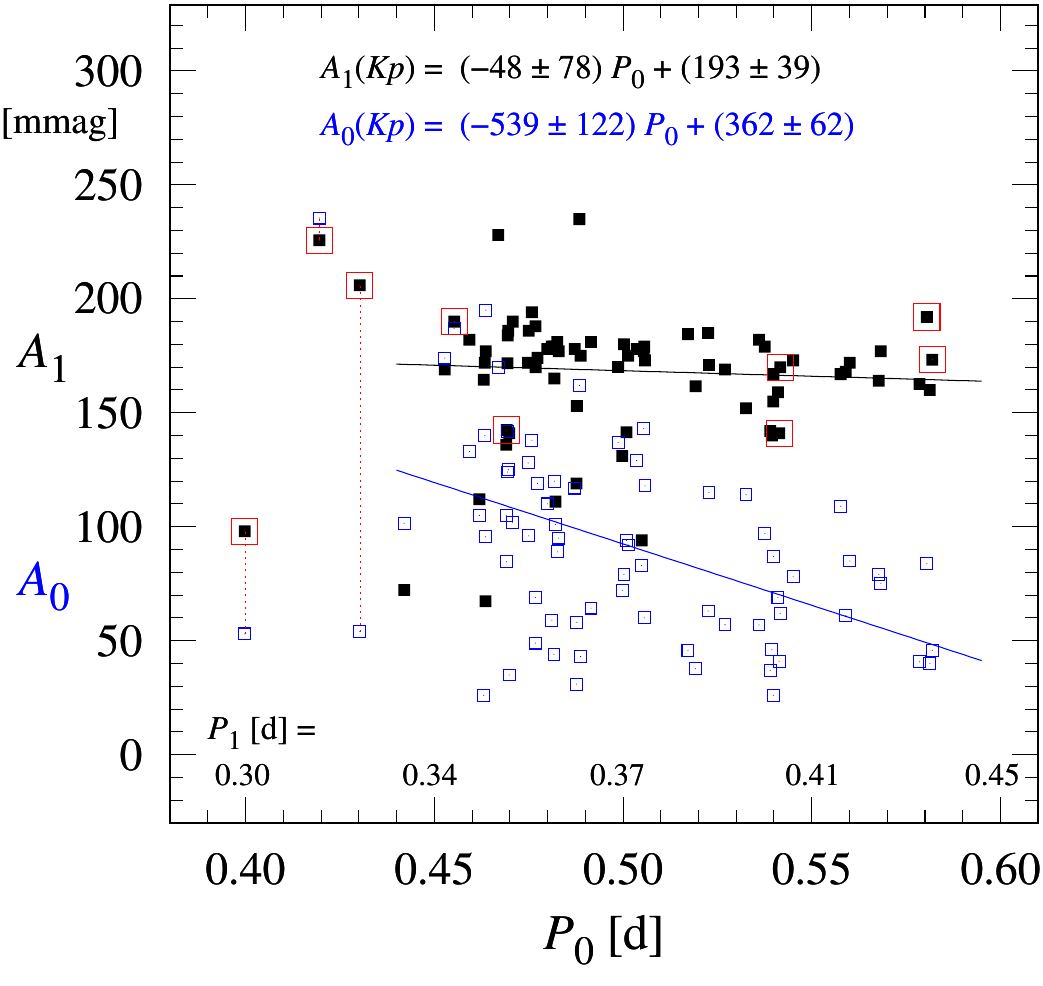}    \put(87,75){(a)} \end{overpic}   
\hskip0.5truecm
\begin{overpic}[width=8.3cm]{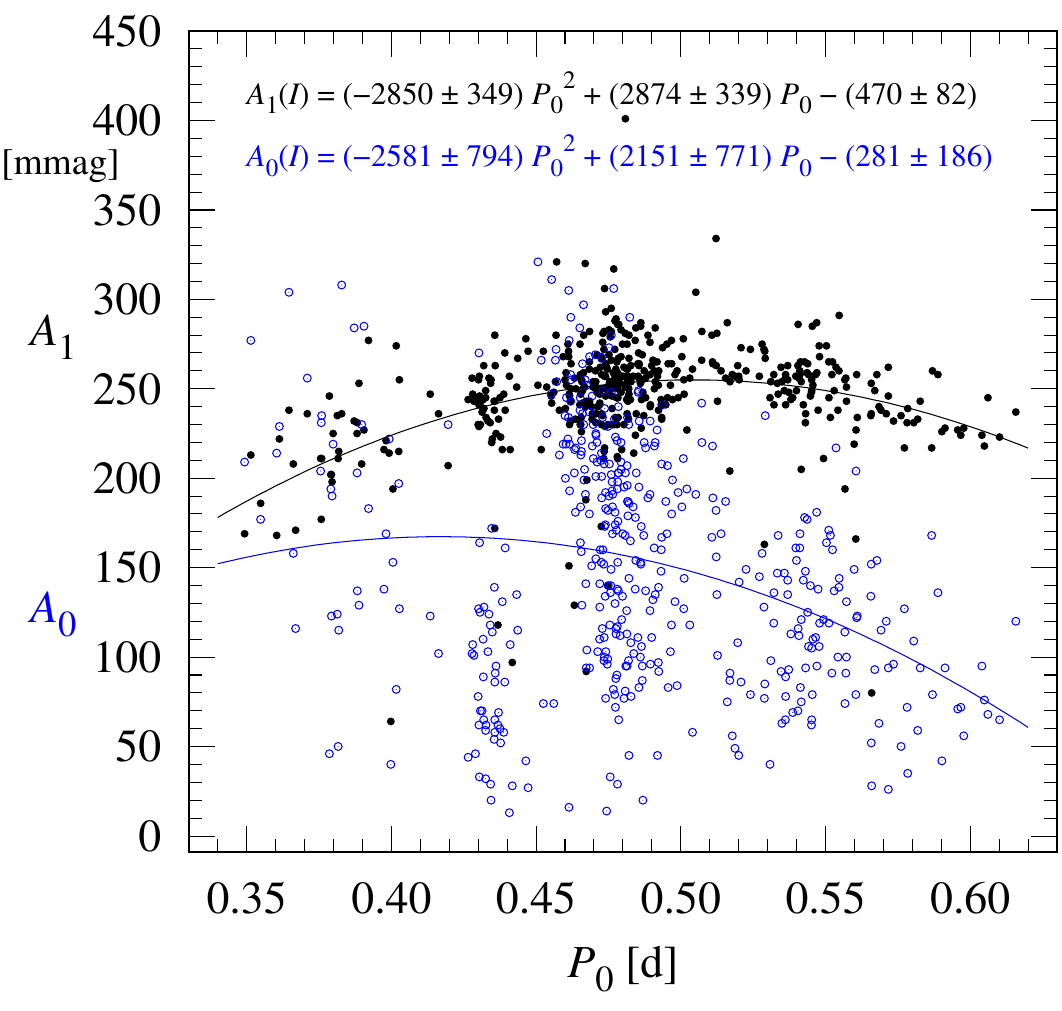} \put(85,70){(b)} \end{overpic}   
\end{center} 

\caption {Period-amplitude ($P$-$A$) diagrams for both radial pulsation modes
of the cRRd stars observed by {\it K2} (left) and by \texttt{OGLE}
(right).   (a) For the {\it K2} stars the first-overtone amplitudes, $A_1$({\it
Kp}) (black solid squares), and the fundamental mode amplitudes $A_0$({\it Kp})
(blue open boxes), are Fourier first-term values.  Red boxes surround the points
representing  the first-overtone components of the Campaign~11 stars.   (b) For
the \texttt{OGLE} stars the $A_1$({\it I}) (black dots) and $A_0$({\it
I}) (blue circles) amplitudes are trough-to-peak (min-to-max) values derived
from $I$-passband photometry. } 	

\label{Fig6}
\end{figure*}

\renewcommand{\thefigure}{7}
\begin{figure*} 
\begin{center}
\begin{overpic}[width=8.5cm]{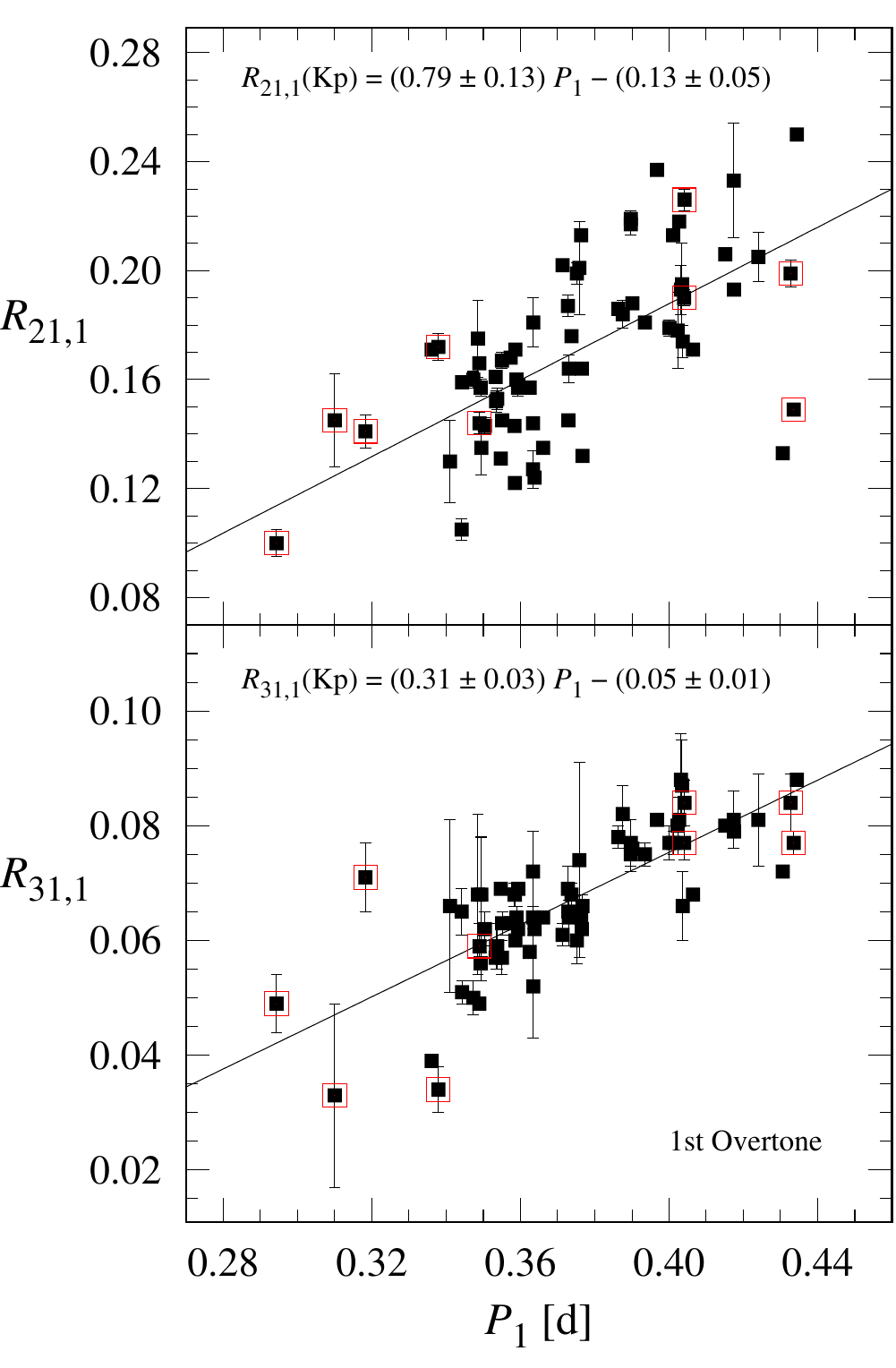}  \put(18,85){(a)}  \put(18,40){(b)}  \end{overpic}  
\hskip0.5truecm
\begin{overpic}[width=8.5cm]{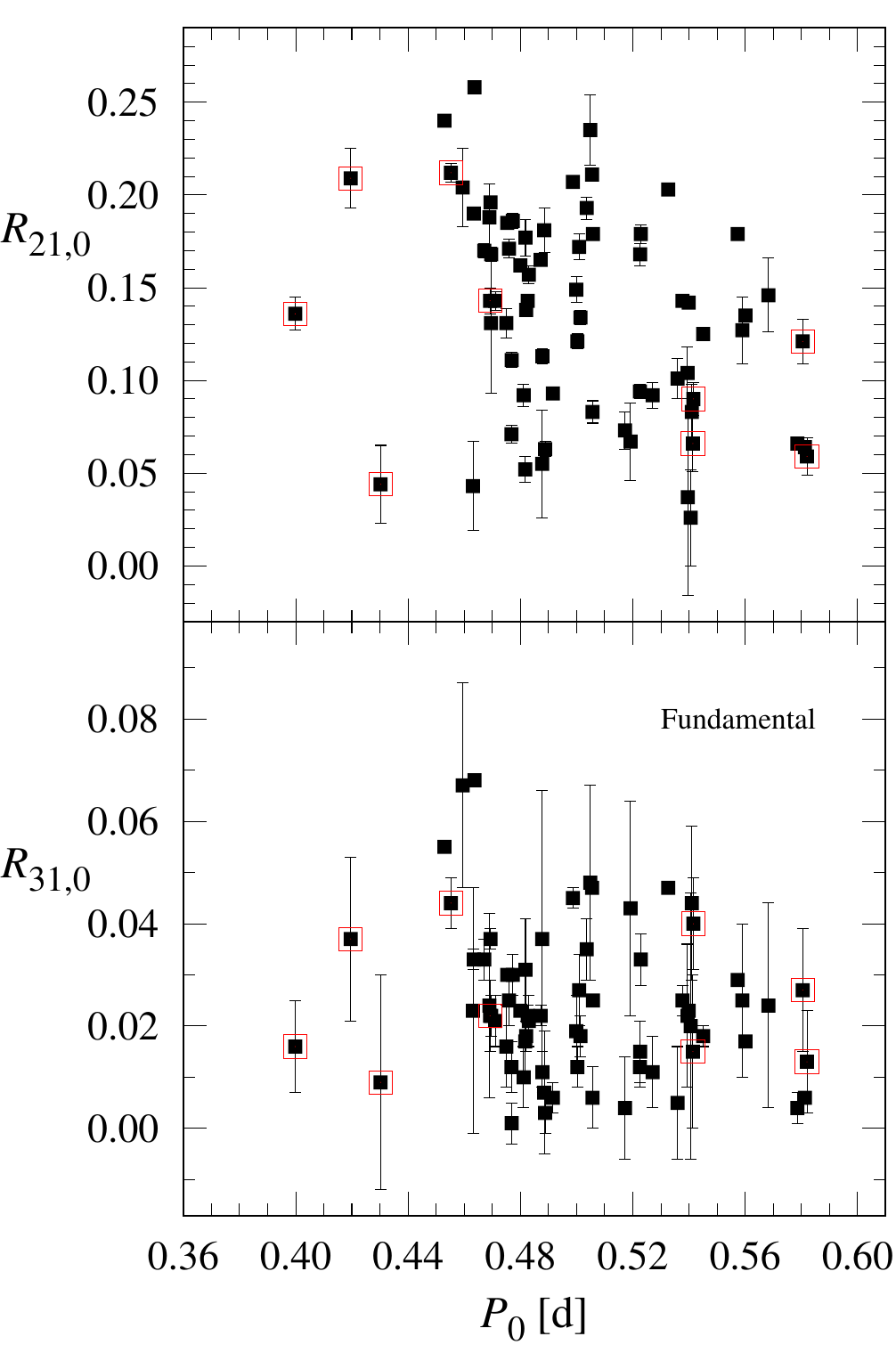}  \put(18,90){(c)}  \put(18,45){(d)}  \end{overpic}  
\end{center} 

\caption {Fourier amplitude-ratio parameters versus period for the 72 Ecliptic
Plane and Galactic Bulge cRRd stars observed by {\it K2}, derived from
decomposition of the {\it Kp}-passband light curves (see Tables 4-5).  The left
panels show the radial first-overtone parameters vs.  $P_1$, and the right
panels show the fundamental-mode parameters vs. $P_0$.   The first-overtone
graphs also show least-squares fitted lines and their equations.
Correlation coefficients for all four panels can be found in Table~6.  The
points enclosed by red boxes identify the nine Campaign~11 (Galactic Bulge)
cRRd stars.  } 	

\label{Fig7} 
\end{figure*}

\renewcommand{\thefigure}{8}
\begin{figure*} 
\begin{center}
\begin{overpic}[width=8.5cm]{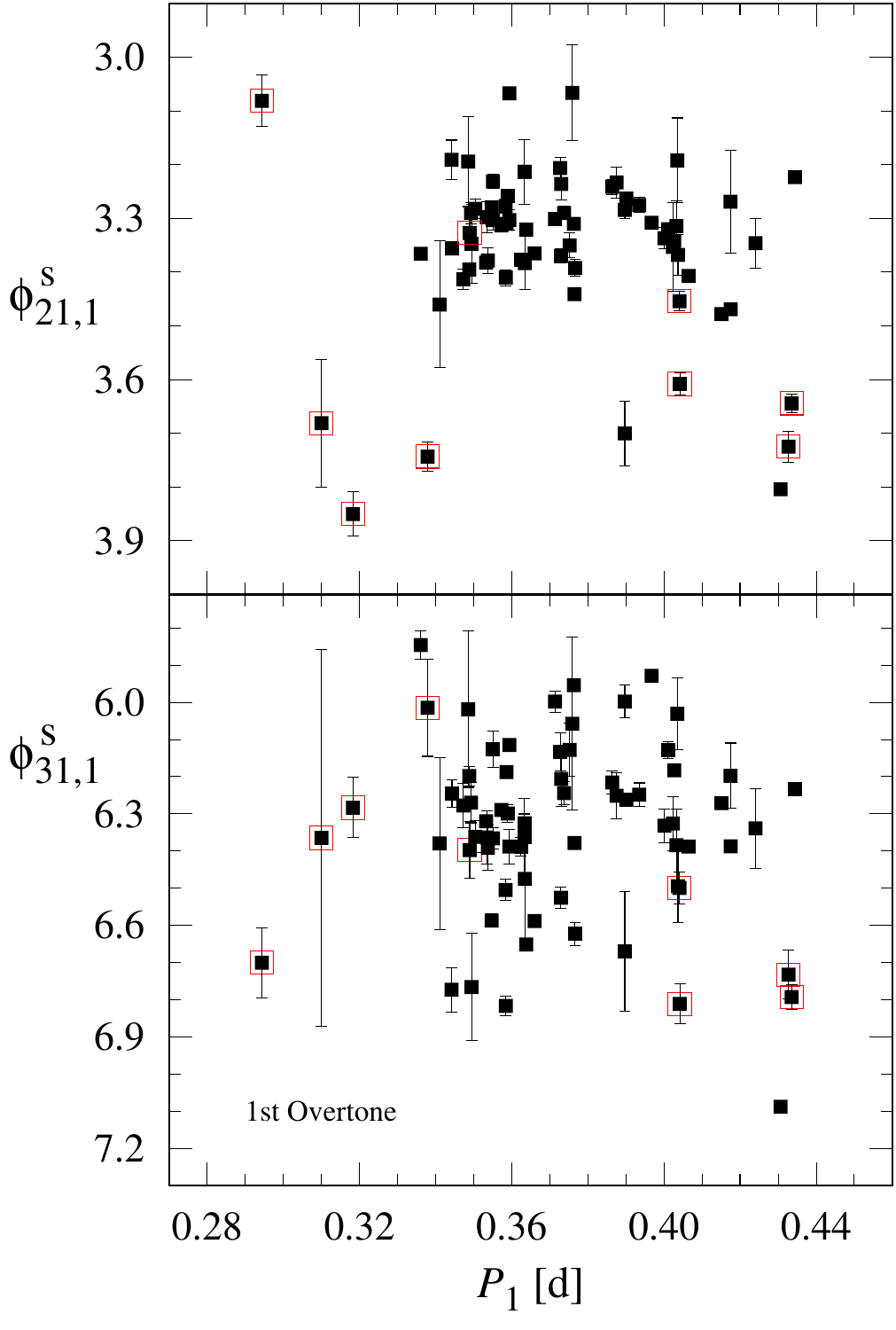}  \put(60,90){(a)}  \put(60,48){(b)}  \end{overpic}  
\hskip0.5truecm
\begin{overpic}[width=8.5cm]{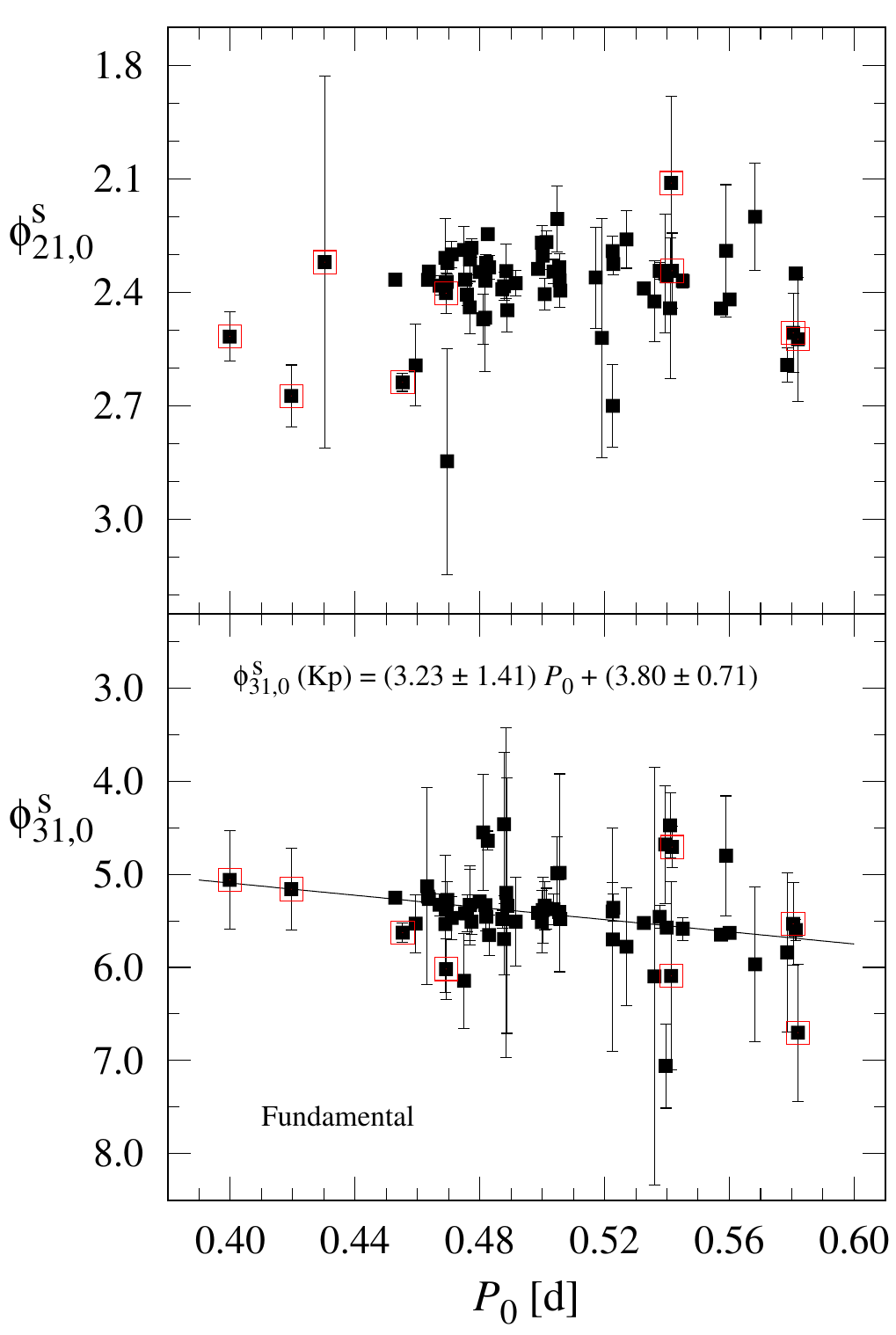}  \put(60,92){(c)}  \put(60,48){(d)}  \end{overpic}  
\end{center} 

\caption{Fourier phase-difference parameters versus period for the
first-overtone (left) and fundamental (right) pulsation modes of the cRRd stars
observed by {\it K2}, where the symbols and panels correspond to the
amplitude-ratio graphs plotted in Fig.17. } 	

\label{Fig8} 
\end{figure*}

\subsection{Dependencies of Amplitudes and Fourier Parameters on Period}

Period-amplitude diagrams have proved useful for distinguishing single-mode
RRab and RRc stars.  The earliest studies revealed that the metal abundances of
RR\,Lyrae stars correlate with period and amplitude (Oosterhoff 1939, Arp 1955,
Preston 1959).  More recently, $P$-$A$ relationships and relationships
involving Fourier decomposition parameters, such as the period-$\phi_{\rm 31}$
diagram, have been used to infer [Fe/H] for single-mode RRab and RRc stars
(e.g., Simon 1990; Kov\'acs \& Jurcsik 1996; Sandage 2004; Morgan et al.  2007;
Nemec et al. 2013; Clementini et al. 2023).  At a given amplitude (or
$\phi_{\rm 31}$) stars of longer period tend to  have higher masses, greater
luminosities and  lower metal abundances (see also the hydrodynamical models
presented in figs.14-15 of Nemec et al.  2011).  The $P$-$A$, $P$-$R_{\rm 21}$,
$P$-$\phi_{\rm 31}$, etc., relationships for the individual modes have not
previously been analyzed in detail.

\subsubsection{Period-Amplitude Diagram for cRRd stars}

\renewcommand{\thefigure}{9}
\begin{figure*} 
\begin{center}
\begin{overpic}[width=8.8cm]{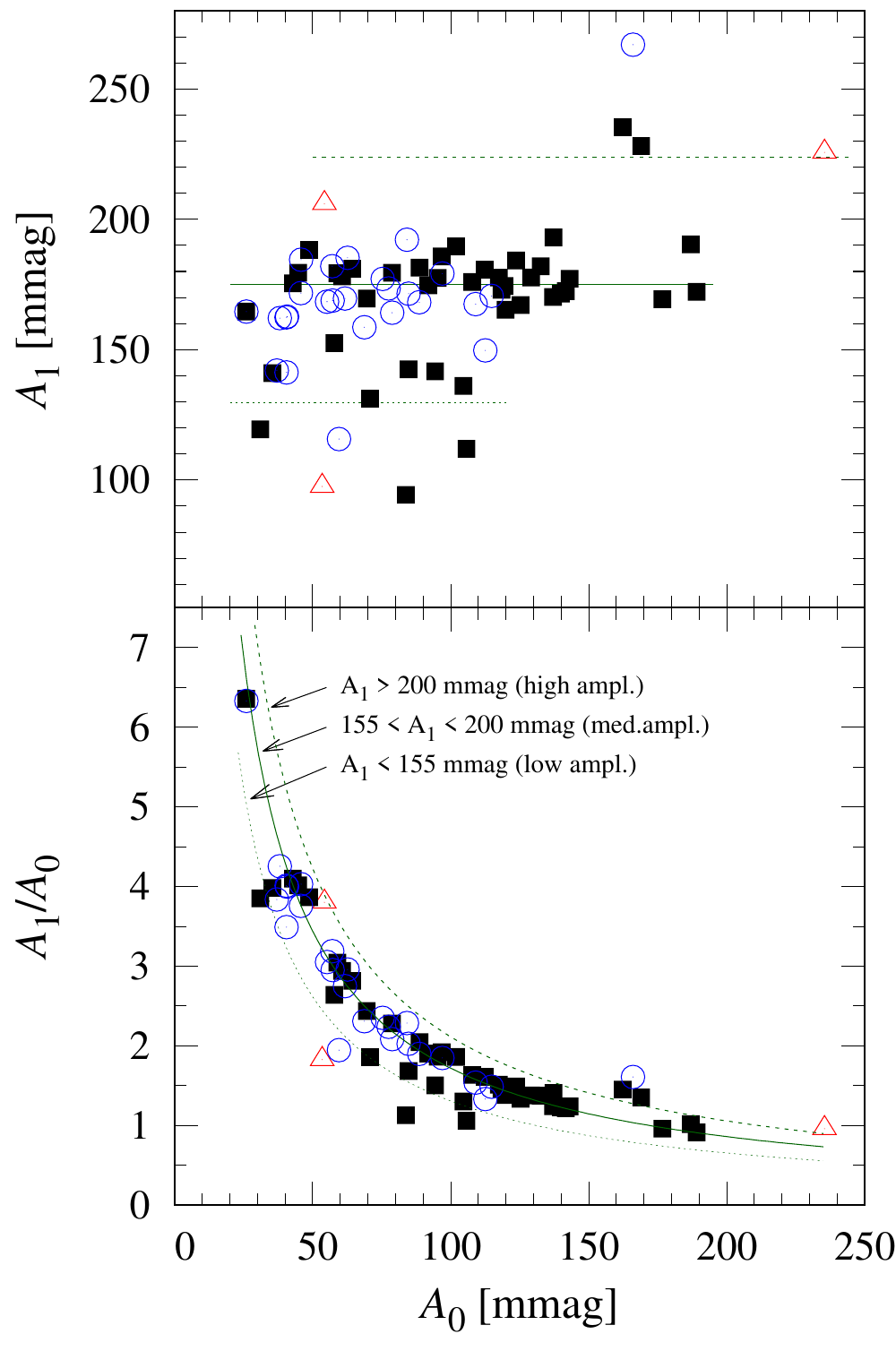}  \put(53,60){(a)}  \put(53,30){(b)}    \end{overpic}   
\hskip0.4truecm
\begin{overpic}[width=8.5cm]{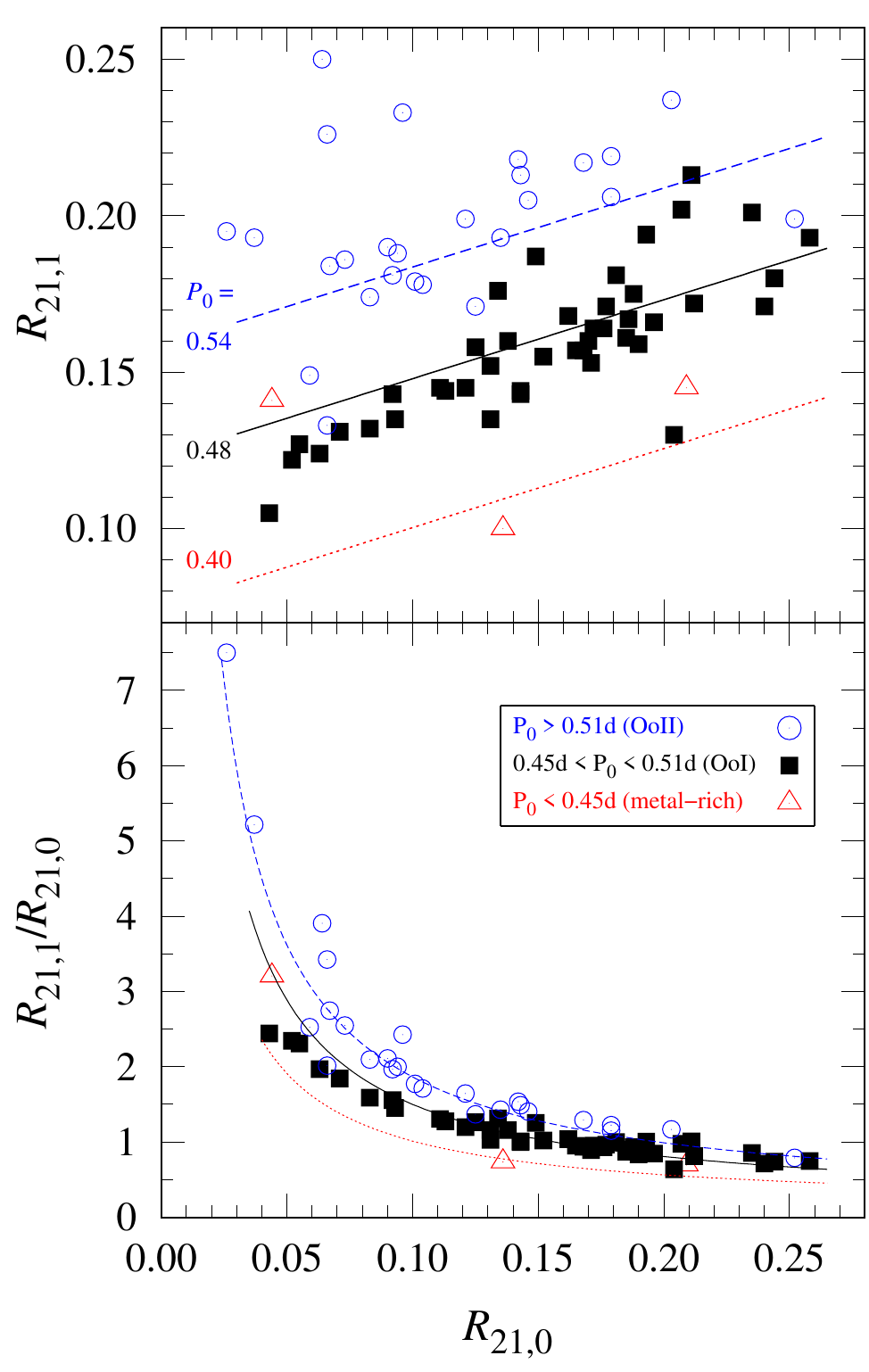}  \put(53,60){(c)}  \put(53,30){(d)}  \end{overpic}   
\end{center} 

\caption{(a) $A_1$ vs $A_0$ diagram for the 72 cRRd stars observed by {\it K2}
through the {\it Kp}-filter, where the stars have been sorted into three period
groups (see legend).  (b) $A_1$/$A_0$ vs $A_0$ diagram for the same stars.
(c) $R_{\rm 21,1}$ vs $R_{\rm 21,0}$ diagram for the same stars, with model
prediction lines given for three periods: $P_0$=0.40\,d (red), 0.48\,d (black)
and 0.54\,d (blue).    (d) $R_{\rm 21,1}$/$R_{\rm 21,0}$ vs $R_{\rm 21,0}$
diagram, again showing a stratification by period (i.e., a family of curves). }
\label{Fig9} 
\end{figure*}

In {\bf Figure\,6} period-amplitude diagrams are plotted for the individual
radial pulsation modes for the {\it K2} cRRd stars (left) and for the 458
\texttt{OGLE-IV} Galactic Disk and Bulge cRRd stars identified by Soszy\'nski
et al. (2019).  For the {\it K2} stars the Fourier first-term amplitudes
(Table\,3) are plotted.  For each mode the equation of the least-squares fitted
line is given at the top of the graph.  Excluded from the fits are the three
short-period Galactic Bulge stars observed during Campaign~11 (the two modes of
which are connected by red vertical dotted lines).  The first-overtone
amplitudes $A_1$ (solid black squares and black line) show a slight, but
statistically insignificant, decrease with period, while the usually-smaller
fundamental mode amplitudes $A_0$ (open blue boxes and blue line) show a
pronounced, statistically significant, decrease with period (see Table\,6a,b).
As a result the amplitude ratios $A_1$/$A_0$ for the cRRd stars with
$P_0>0.44$\,day (see Fig.\,2) increase with increasing period.  It follows that
intermediate-metallicity (Oosterhoff type I) cRRd stars, which have fundamental
mode periods $P_0$ between 0.45 and 0.51\,day, tend to have lower $A_1/A_0$
ratios than more metal-poor (Oosterhoff type II) cRRd stars which have $P_0$
between 0.51 and 0.62\,day.   It is noteworthy that the pronounced downward
trend for the fundamental-mode amplitudes (Fig.\,6), and the much shallower
downward trend for the first-overtone amplitudes, are consistent with the
well-known $P$-$A$ downward trends for single-mode RRab and RRc stars (see, for
example, fig.3 of Soszy\'nski et al.  2009).  

The period-amplitude diagram  for the two radial pulsation modes of the 458
Galactic Disk and Bulge stars observed by \texttt{OGLE-IV} is plotted in
Fig.\,6b.  Unlike the {\it K2} amplitudes, which are Fourier first-term values
derived from {\it Kp} photometry, the \texttt{OGLE} amplitudes are min-to-max
values through the $I$ filter and thus tend to be larger.   Another difference
is the presence in the \texttt{OGLE} sample of many cRRd stars with
fundamental-mode periods shorter than 0.45\,d.  Inclusion of these (presumably
metal-rich) stars suggests that the fundamental and first-overtone
relationships between amplitude and period are non-linear over this broader
range of periods.  Fitted polynomials, the equations of which are given at the
top of the graph, are plotted for the two modes.  The period at which the
fitted first-overtone amplitude reaches a maximum appears to occur at
$P_0$$\sim$0.50\,d.  For the period range where Figs. 6a and 6b overlap (i.e.,
$P_0 > 0.45$\,d)  both samples show approximately-linear downward trends. For
periods shorter than 0.42\,d the fundamental-mode amplitudes exhibit a large
scatter that cannot be explained by period alone. Consequently the large
scatter in both graphs, the confounding effect of  different filters, and the
small range of the period overlap, makes it difficult to establish the precise
functional forms of the relationships.

\subsubsection{Dependence of Fourier Parameters on Period}

In addition to amplitudes, the Fourier amplitude-ratio and phase-difference
parameters (i.e., $R_{\rm 21}$, $R_{\rm 31}$, $\phi^s_{\rm 21}$ and
$\phi^s_{\rm 31}$) have been used to describe the light curves of pulsating
stars and as predictors of the metal abundance of single-mode RR~Lyrae stars.
Thus it is instructive to examine how these four descriptors correlate with
period for cRRd stars.  The four panels of {\bf Figure\,7} show, for each of
the two pulsation components,  the $R_{\rm 21}$ and $R_{\rm 31}$ values for the
{\it K2} cRRd stars plotted against the corresponding pulsation period.  The
first-overtone graphs (left) show fitted least-squares lines, the equations of
which are given in each panel.  Since the {\it K2} survey covers 20 different
fields around the Ecliptic Plane the fitted lines describe only overall trends
for the composite sample.  As expected the $R_{\rm 21}$ values (upper panels)
are larger than the $R_{\rm 31}$ values (lower panels) for both modes.    Note
also that the slopes of the first-overtone lines are positive (Figs.\,7a,b) and
the correlation coefficients for both $R_{\rm 21,1}$ and  $R_{\rm 31,1}$ are
highly significant (Table\,6a), although the $R_{\rm 31,1}$ correlation with
$P_1$ is stronger.  There is no evidence of such a clear positive trend in the
$R_{\rm 21,1}$ vs $\log{P_0}$ graph for the first overtone of 986 RRd stars in
the Large Magellanic Cloud observed by the \texttt{OGLE} survey (fig.2 of
Soszy\'nski et al. 2009).  The dependence of $R_{\rm 21,0}$ and  $R_{\rm 31,0}$
on period is less clear for the fundamental mode (Figs.7c,d).  There is some
evidence that both ratios are negatively correlated with $P_0$ (Table\,6b) but
evidence for a simple linear relationship is lacking. 

Graphs of the Fourier phase-difference parameters  $\phi^s_{\rm 21}$ and
$\phi^s_{\rm 31}$ versus pulsation period are given in {\bf Figure\,8}.  The
layout and symbols match those seen in Fig.\,7.  Only $\phi^s_{\rm 31,0}$ shows
a statistically significant correlation with period (Table\,6b);  the fitted
least squares line and its equation are shown in panel\,(d).  Note
that because the phase-differences are plotted on a reversed scale (see fig.3
of Sandage 2004; and figs.4,12 of Nemec et al. 2013) the line appears to slope
downwards, even though the correlation is positive.

\begin{table*}
{\fontsize{6}{7.2}\selectfont  
\centering

\caption{Physical characteristics for the 16 {\it K2} RRd stars with SDSS spectra, from which [Fe/H] values 
were derived for 14 stars by the \texttt{SEGUE} Stellar Parameter Pipeline (Lee {\it et al.}
2008a,b).  All are classical RRd stars.  The individual
spectra are identified by the plate number, the Modified Julian Date
(MJD-2400000) and the optical fiber number (columns 2-4), and the
signal-to-noise ratios (column\,5) are per pixel $r$-band median values. 
The subheaders for $T_{\rm eff}$, $\log g$ and [Fe/H] (columns 7-12) are the stellar parameter names from the \texttt{SEGUE} `sppParams' table.  }  

\label{tab:rrl}
\begin{tabular}{lrcrrcllcccc}
\hline
\multicolumn{1}{c}{ EPIC } &   \multicolumn{3}{c}{SDSS spectrum}  & \multicolumn{1}{c}{S/N}  &  RV  & \multicolumn{2}{c}{ $T_{\rm eff}$(K) } & \multicolumn{2}{c}{ $\log\,g$ } & \multicolumn{2}{c}{ [Fe/H] } \\ 
	&   plate  &   MJD   &  fiber &   & [km/s]   & \multicolumn{1}{c}{\texttt{TEFFADOP}}   & \multicolumn{1}{c}{\texttt{TEFFSPEC}} & \multicolumn{1}{c}{\texttt{LOGGADOP}} & \multicolumn{1}{c}{\texttt{LOGGSPEC}} & \multicolumn{1}{c}{\texttt{FEHADOP}} & \multicolumn{1}{c}{\texttt{FEHSPEC}} \\
\multicolumn{1}{c}{(1)} & \multicolumn{1}{c}{(2)} &\multicolumn{1}{c}{(3)} &   \multicolumn{1}{c}{(4)}  &  \multicolumn{1}{c}{(5)}  & \multicolumn{1}{c}{ (6)}  & \multicolumn{1}{c}{ (7)} &
	\multicolumn{1}{c}{(8)} &\multicolumn{1}{c}{(9)} &   \multicolumn{1}{c}{(10)}  &  \multicolumn{1}{c}{(11)}  &  \multicolumn{1}{c}{(12)}   
  \\
\hline  
\\
60018662\,(E2)     & 1903 & 53357 &  469 & 100.2 & --289$\pm$1 & 6945$\pm$95 & 7074$\pm$75 & 3.65$\pm$0.27 & 3.54$\pm$0.30 &  --2.06$\pm$0.08  & --2.18$\pm$0.08 \\   
\underline{201585823}\,(C1)    
		   &  514 & 51994 &   47 & 43.0 & --62$\pm$2 & 7156$\pm$95 & 7078$\pm$59 & 3.88$\pm$0.03 & 3.91$\pm$0.02 &  --1.57$\pm$0.03 & --1.57$\pm$0.03 \\    
\underline{211694449}\,(C5,18) & 2274 & 53726 &  435 & 31.2 & +67$\pm$3  & 7108$\pm$75 & 6990$\pm$80 & 3.12$\pm$0.42 & 3.09$\pm$0.34  & --1.23$\pm$0.02 & --1.26$\pm$0.03 \\ 
		   & 3230 & 54860 &  195 & 46.8 & +62$\pm$2  & 6955$\pm$73 & 6842$\pm$86 & 2.84$\pm$0.28 & 2.66$\pm$0.28  & --1.42$\pm$0.09 & --1.37$\pm$0.09 \\ 
211888680\,(C5,16) & 2283 & 53729 &   32 & 11.2 & +53$\pm$10 & 7046$\pm$94 & 7174$\pm$55 & 2.80$\pm$0.40 & 2.74$\pm$0.65  & --1.73$\pm$0.05 & --1.68$\pm$0.04 \\ 
\underline{211898723}\,(C5,18) & 2273 & 53709 &  416 & 34.0 & +343$\pm$3 & 6834$\pm$91 & 6957$\pm$93 & 3.68$\pm$0.27 & 3.74$\pm$0.31  & --1.72$\pm$0.01 & --1.55$\pm$0.11 \\ 
220254937\,(C8)    & 7860 & 57006 & 232 & 33.3   & --135$\pm$3  &  \dots   &  \dots &  \dots &  \dots &  \dots & \dots   \\   
201440678\,(C10)   &  286 & 51999 &  181& 29.9 & +216$\pm$4 & 7232$\pm$107 & 7413$\pm$88 & 3.17$\pm$0.33 & 3.02$\pm$0.08 & --2.02$\pm$0.18 & --1.93$\pm$0.02  \\   
                   & 2892 & 54552 &  144& 32.2 & +240$\pm$4 & 6766$\pm$63  & 6731$\pm$79 & 2.42$\pm$0.28 & 2.58$\pm$0.29 & --1.92$\pm$0.04  & --1.92$\pm$0.04 \\    
201519136\,(C10)   &  288 & 52000 &  577& 33.8 & --63$\pm$3 & 6838$\pm$46  & 6861$\pm$71 & 2.96$\pm$0.16 & 3.07$\pm$0.11  & --1.40$\pm$0.04  & --1.37$\pm$0.06 \\   
228800773\,(C10)   & 2707 & 54144 &  442& 35.0 & +188$\pm$3 & 6781$\pm$52  & 6814$\pm$75 & 2.98$\pm$0.13 & 2.93$\pm$0.14  & --1.91$\pm$0.08  & --1.82$\pm$0.06  \\   
248369176\,(C10)   & 2568 & 54153 & 234 & 7.9  & --17$\pm$18  &  \dots   &  \dots &  \dots &  \dots &  \dots & \dots  \\  
		   & 3847 & 55588 & 480 & 3.7  & --40$\pm$15   &  \dots   &  \dots &  \dots &  \dots &  \dots & \dots \\  
201749391\,(C14)   & 3242 & 54889 & 569 & 56.6  & --170$\pm$2 & 6721$\pm$43 & 6691$\pm$61  & 3.35$\pm$0.15 & 3.40$\pm$0.17 & --1.46$\pm$0.01  & --1.46$\pm$0.01 \\    
248426222\,(C14)   & 275  & 51910 & 382 & 33.8  & +203$\pm$3  & 6939$\pm$51 & 6987$\pm$25  & 3.26$\pm$0.28 & 3.42$\pm$0.27 & --2.20$\pm$0.02  & --1.94$\pm$0.18  \\   
248845745\,(C14)   & 1600 & 53090 & 636 & 20.4  & --11$\pm$5  & 7099$\pm$11 & 7326$\pm$158 & 3.23$\pm$0.33 & 3.21$\pm$0.43 & --1.60$\pm$0.17 & --1.60$\pm$0.01 \\   
248871792\,(C14)   & 1602 & 53117 & 326 & 50.2  & +215$\pm$1  & 7277$\pm$92 & 7416$\pm$87  & 3.43$\pm$0.16 & 3.37$\pm$0.18 & --1.69$\pm$0.07 & --1.68$\pm$0.02  \\  
211665293\,(C16)   & 2435 & 53828 & 249 & 43.0  & --33$\pm$3  & 6979$\pm$83 & 6904$\pm$8   & 4.18$\pm$0.63 & 4.26$\pm$0.08 & --1.58$\pm$0.04 & --1.58$\pm$0.04 \\  

251629085\,(C17)   & 3307 & 54970 & 230 & 64.1  &  +31$\pm$1  & 6989$\pm$75 & 6906$\pm$65  & 3.10$\pm$0.23 & 3.08$\pm$0.30 & --1.47$\pm$0.06 & --1.40$\pm$0.07  \\  
\\
\hline
\end{tabular}
}
\end{table*}

\subsubsection{Amplitudes, Amplitude Ratios and Periods}

In {\bf Figure\,9} Fourier amplitudes and amplitude-ratio
parameters are compared for the two pulsation modes.  The first-overtone
amplitude $A_1$ is plotted against the fundamental-mode amplitude $A_0$ in the
upper left panel (Fig.9a), and the first-overtone amplitude-ratio parameter
$R_{\rm 21,1}$ is plotted against its fundamental-mode counterpart $R_{\rm 21,0}$
in the upper right panel (Fig.9c).  The lower panels show the corresponding
ratios versus $A_0$ (Fig.9b) and versus $R_{\rm 21,0}$ (Fig.9d).  The
amplitudes are the {\it Kp} values given in Table\,3, and the $R_{\rm 21}$
values are given in Tables\,4-5.  Three period groups are plotted with
different symbols: $P_0$$>$0.51\,d (blue open circles); $0.45<P_0<0.51$\,d
(black squares); and $P_0$$<$0.45\,d (red open triangles).

The {\it K2} stars in Fig.\,9a appear to separate into three amplitude groups,
which are also evident in  a histogram (not shown) of the $A_1$ values.
Fifty-three of the 72 stars form a horizontal band with 155$<$$A_1$$<$200 mmag,
where $<$$A_1$$>$=175$\pm$2 mmag and $<$$A_0$$>$=90 mmag.  A second group of 15
stars ($A_1$$<$155 mmag) lies below this band, where $<$$A_1$$>$=130$\pm$3 mmag
and $<$$A_0$$>$=70 mmag, and five stars lie above the band ($A_1$$>$200 mmag)
with $<$$A_1$$>$=224$\pm$6 mmag and $<$$A_0$$>$=160 mmag.  The mean $A_1$
values for the three amplitude groups are indicated  by horizontal lines.  The
positive Pearson correlation coefficient (Table\,6c) reflects the increase in
$<$$A_1$$>$ as $<$$A_0$$>$ increases.  A similar pattern is seen for the 458
Disk and Bulge cRRd stars observed by the \texttt{OGLE} survey (see fig.15b of
NM21; note that the \texttt{OGLE} photometry was through an $I$-filter and the
amplitudes are min-to-max and not Fourier first-term values).  In the
\texttt{OGLE} case the bulk of the cRRd stars have $A_1$($I$)$\sim$250 mmag,
with $<$1$\%$ of the stars having high amplitudes and $\sim$2$\%$ of the stars
having low amplitudes.  Within the {\it K2} and \texttt{OGLE} amplitude groups
$A_1$ does not appear to depend on $A_0$.  Therefore $A_1$/$A_0$  is expected
to decrease inversely with $A_0$, i.e., $E(A_1/A_0) = a/A_0.$  In Fig.9b
$A_1$/$A_0$ is plotted against $A_0$, together with the fitted inverse
relationships for the low, medium and high amplitude groups, where the $a$
values are equal to the $<$$A_1$$>$ values given above.  Agreement between  the
observed and predicted relationships is excellent for both {\it K2} and
\texttt{OGLE}. 

$R_{\rm 21,1}$ vs. $R_{\rm 21,0}$ and  $R_{\rm 21,1}/R_{\rm 21,0}$ vs.  $R_{\rm
21,0}$ diagrams for the {\it K2} cRRd stars are plotted in Figs.\,9c and 9d.
Unlike their amplitude counterparts (Figs.9a and 9b) both diagrams show a clear
(but unexpected) stratification by period. For a given period  $R_{\rm 21,1}$
increases linearly with $R_{\rm 21,0}$. A linear model with a common slope was
fitted to the data (after checking that the slope did not vary significantly
with period): $ R_{\rm 21,1} = (-0.163\pm0.036) + (0.253\pm0.045)\,R_{\rm 21,0}
+ (0.595\pm0.066)\,P_0. $ Three representative lines obtained by substituting
$P_0 = 0.40, 0.48$ and 0.54\,day are shown on the graph. The corresponding
$R_{\rm 21,1}$/$R_{\rm 21,0}$  vs $R_{\rm 21,0}$ curves (i.e., above equation
divided by $R_{\rm 21,0}$) are plotted in Fig.9d.  Analysis of the
\texttt{OGLE} 458 cRRd data  found a similar  stratification by period when
$R_{\rm 21,1}$ was plotted against  $R_{\rm 21,0}$, although in that case the
slope increased with period.  Since period depends on [Fe/H] (see next section)
this period stratification suggests that the $R_{\rm 21}$ (light-curve shape)
parameters for the two components are related to each other via metal
abundance.

\subsection{Metal Abundances and Masses}

It is now well established from observations and theoretical models that
RR~Lyrae stars have metal abundances ranging from less than $1/100^{th}$ solar
({\it i.e.}, [Fe/H]$<$--2 dex) to greater than solar ({\it i.e.}, [Fe/H]$>$0
dex).  In this section, period-[Fe/H] calibration equations are derived and
used to estimate the metal abundances of the {\it K2} cRRd stars.  The
calibration equations are based on the model given in $\S$4.2 (Eqn.\,2) and are
{\it internally consistent} with the observed Petersen diagram for cRRd stars.
Metallicities and approximate masses are also given for 2130 cRRd stars
observed by {\it Gaia}, and the effect of misclassification bias on derived
[Fe/H] values is discussed.

\renewcommand{\thefigure}{10}
\begin{figure*} 
\begin{center}
\begin{overpic}[width=8.3cm]{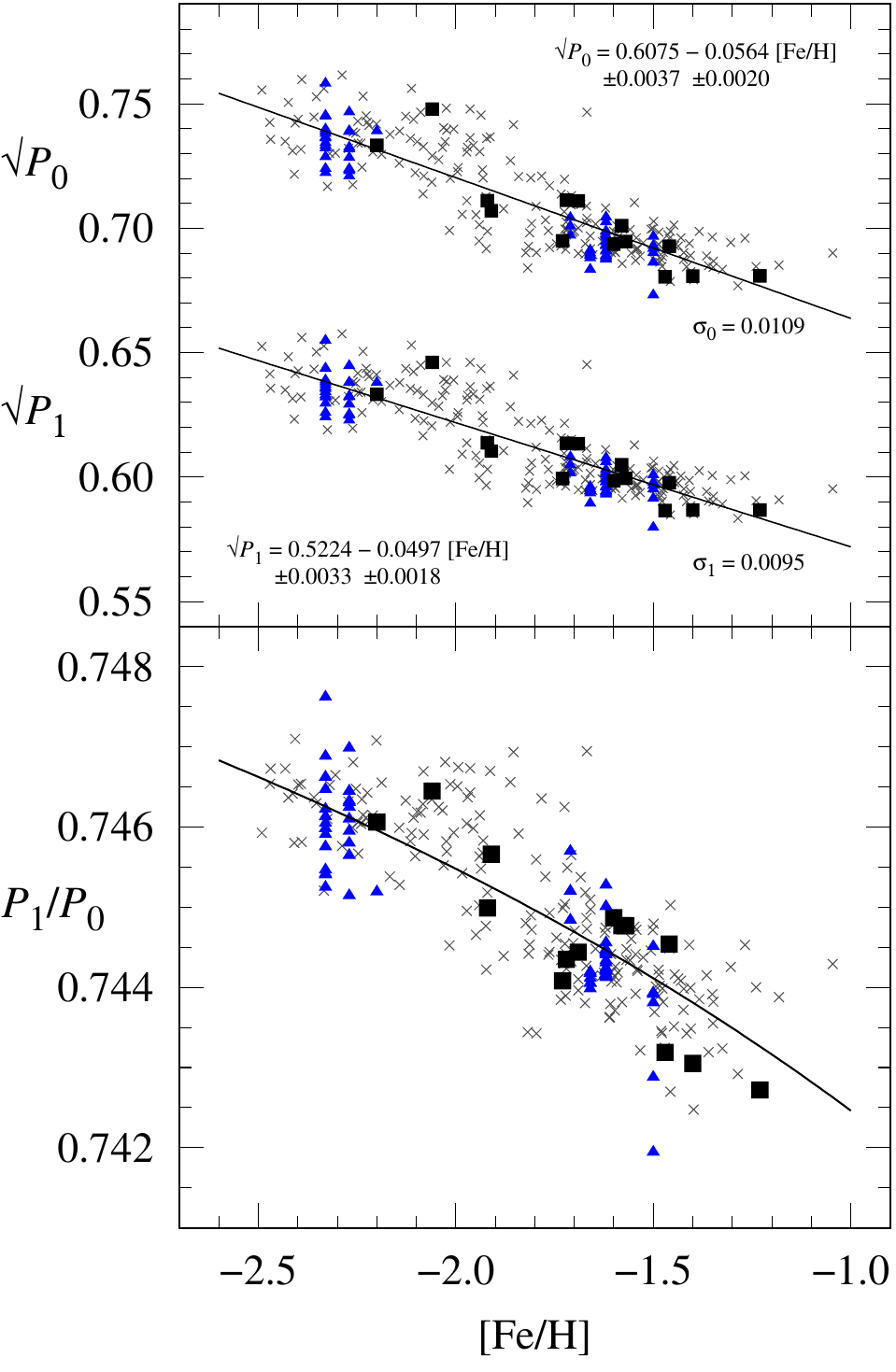} \put(56,85){(a)}  \put(56,47){(b)}  \end{overpic}   %
\hskip0.6truecm
\begin{overpic}[width=8.3cm]{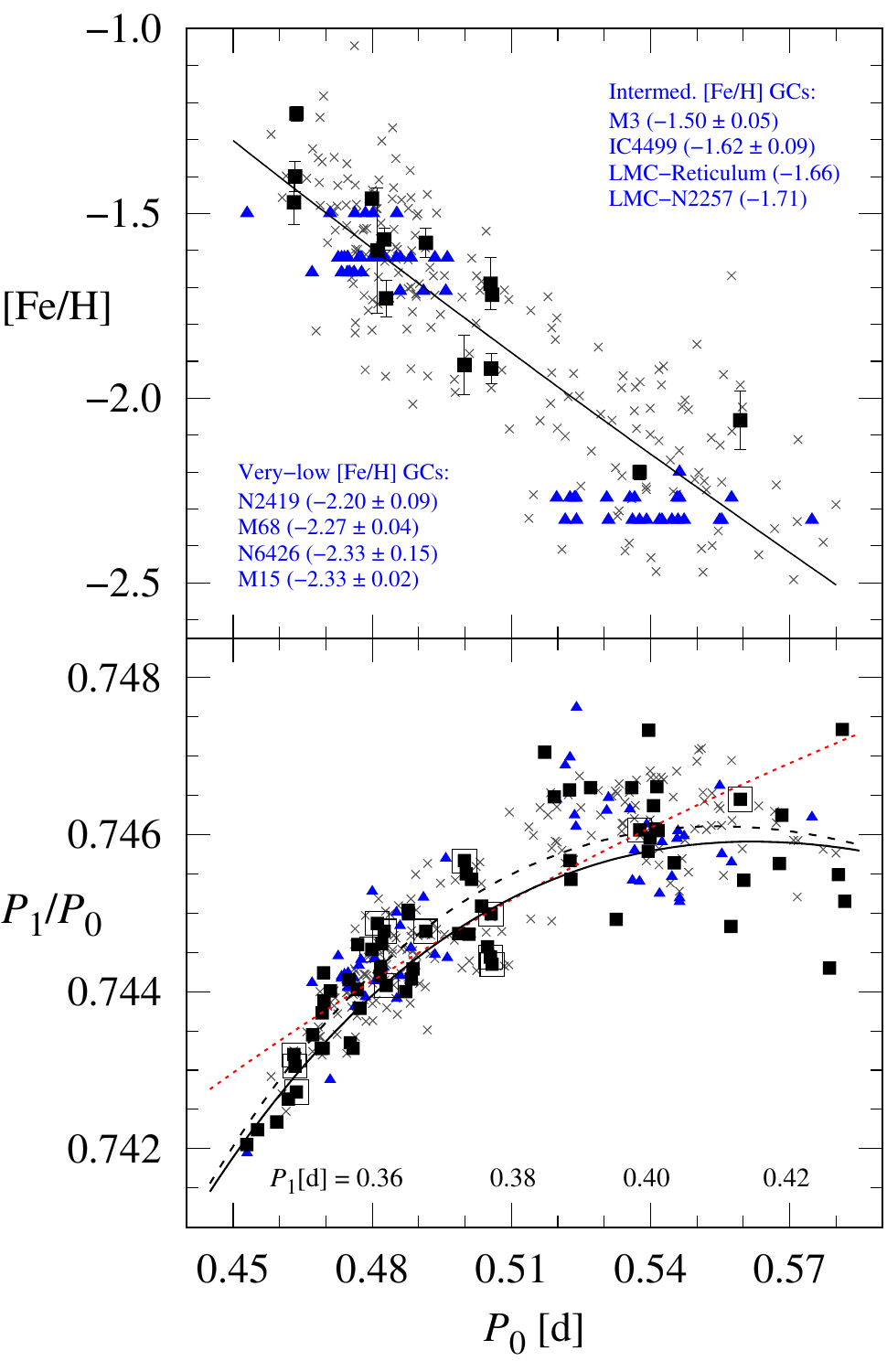}  \put(56,80){(d)} \put(20,47){(c)}   \end{overpic}   
\end{center}

\caption{Period-metallicity relations and Petersen diagram for `classical' RRd
(cRRd) stars.  The symbols are as follows: filled black squares for the 14 {\it
K2} cRRd calibration stars with SDSS/\texttt{SEGUE} metallicities, filled blue
triangles for the 57 cRRd stars in eight globular clusters that have
well-established mean [Fe/H] values, and crosses for the 197 non-{\it K2} Chen
et al. (2023) calibration stars with SDSS/\texttt{SEGUE} metallicities.  The
fitted regression curves are discussed in the text.  (a) Period-metallicity
relations for the fundamental and first-overtone pulsation periods of the
calibration cRRd stars.   (b) Period-ratio vs metallicity diagram for the
calibration stars.    (c) Petersen diagram for the cRRd metallicity calibration
stars and  58 {\it K2} stars not included in the calibration sample (the {\it
K2} calibration stars are identified with boxes around the solid squares).  The
three Galactic Bulge {\it K2} stars with $P_0$$<$0.45\,d and period ratios
$<$0.742 are off-scale (but can be seen in Fig.\,1).  (d) Metallicity vs.
period graph for cRRd stars, where the abscissa is the fundamental period,
$P_0$.  The assumed mean metallicities of the GCs are noted on the graph.    }

\label{Fig10} 
\end{figure*}

\subsubsection{Period-Metallicity Calibration Sample}

Sixteen of the 75 {\it K2} RRd stars were observed spectroscopically by the
Sloan Digital Sky Survey (SDSS).  Although the spectra are of relatively low
resolution (spectrograph resolution 0.2\,nm at 500\,nm), the  S/N ratios are
sufficiently large to give quite accurate [Fe/H] values for 14 of the K2 stars.
Effective temperatures $T_{\rm eff}$, surface gravities $\log g$, and
metallicities [Fe/H]  derived by the \texttt{SEGUE} Stellar Parameter Pipeline
(Lee {\it et al.} 2008a,b) are summarized in {\bf Table\,7}.  Also in the table
are spectrum identifiers (plate, MJD, fiber), S/N ratios and radial velocities.
According to Lee (2008b), when all the systematic offsets are combined the
typical uncertainty in the derived [Fe/H] values is $\sim$0.24 dex, which is
considerably larger than the individual [Fe/H] uncertainties noted at
the SDSS website.   The radial velocities range from --289$\pm$1 km/s to
+343$\pm$3 km/s, consistent with the observed range for Galactic halo RR~Lyrae stars.
Three of the stars were observed twice, presumably at different pulsation
phases. 

The original plan was to use the 14 stars with \texttt{SEGUE} metallicities as
calibrators for deriving [Fe/H] values for the entire sample of {\it K2} cRRd
stars.  However,  none of the 14 stars is more metal-rich than --1.0 dex and
there are only two stars with [Fe/H]$<$--2.0 dex.  To compensate for the lack
of low-metallicity calibration stars 57 cRRd stars in eight globular clusters
(GCs) that have well-determined mean [Fe/H] values were added to the sample,
where the metal abundances are on the `high resolution spectra' scale of
Carretta et al. (2009, hereafter C09).  Since  the cRRd stars within a given GC
are known to show little variation in metal abundance, their metallicities are
assumed to be equal to the cluster mean\footnote{This would not be the case in
most dwarf galaxies, in particular the higher luminosity systems (see Braga et
al. 2022) where the stars are observed to have a range of metallicities.}.
Pulsation periods for the two components of the GC stars were derived from
$B,V,I$ photometry from various sources\footnote{Sources of the photometry (and
preliminary periods) for the GC RRd stars: M68 (Clement et al.  1993; Walker
1994; Brocato et al.  1994; Kains et al. 2015a,b), M15 (Sandage, Katem \&
Sandage 1981; Bingham et al. 1984; Nemec 1985b; Corwin et al 2008), NGC\,2257
(Nemec, Walker \& Jeon 2009), IC\,4499 (Clement et al. 1986; Walker \& Nemec
1996; Kunder et al.  2011);  M3 (Nemec \& Clement 1989); Reticulum (Kuehn et
al. 2013);  NGC\,2419 (Clement \& Nemec 1990; Di Criscienzo et al.  2011) and
NGC\,6426 (Clement \& Nemec 1990; Hatzidimitriou et al.  1999).}.  The periods
were checked using the same methods that were used to analyze the {\it K2} RRd
stars.  Agreement across filters and with previously published values was
excellent.  

Also added to the calibration sample were 207 cRRd stars with
SDSS/\texttt{SEGUE} metallicities and Zwicky Transient Facility (ZTF, DR14) 
photometry (Chen et al. 2023)\footnote{An additional 96 cRRd stars with
\texttt{LAMOST} metallicities (and ZTF photometry) were identified by Chen et
al., five of which are in common with the {\it K2} sample.  However, owing to
apparent systematic differences between the LAMOST and \texttt{SEGUE}
metallicities only the Sloan [Fe/H] values for the Chen stars have been
considered in the present paper.}.  Ten of the 14 {\it K2} calibration stars
were found to be in common with the Chen sample.   The remaining 197 Chen et
al. stars  have an overall distribution that closely matches that of the {\it
K2} stars but includes many more low-metallicity stars.  The Chen sample, like
the {\it K2} sample, does not include short-period (i.e., $P_0$$<$0.45\,d) cRRd
stars.  Chen et al. do not provide amplitudes or Fourier parameters.

\subsubsection{Period-[Fe/H] Calibration}

Panels (a) and (b) of {\bf Figure 10} show, for the combined {\it K2}+GC+Chen
metallicity calibration sample (N=268), the relationships between $P_0$, $P_1$,
$P_1$/$P_0$ and [Fe/H].  The fitted curves in the four panels are based on the
model given by Eqn.\,2, where $X$=[Fe/H] is assumed to be a common factor
linking the periods.  In {\bf Fig.\,10a} the fitted lines  relating
$\sqrt{P_0}$ and $\sqrt{P_1}$ to [Fe/H] are plotted.  The estimated slopes,
$b_0$=$-0.0564\pm0.0020$ and $b_1$=$-0.0497\pm0.0018$, differ significantly
($p$$<$0.0001),  i.e., the lines are not parallel.  Notice that Eqn.\,2 implies
(by squaring both sides and calculating the mean) that the mean period for a
given [Fe/H] is non-linear in [Fe/H] for both pulsation modes: 

\begin{equation}
\begin{split}	
E(P_0) = \bigl(a_0 + b_0 {\rm [Fe/H]}\bigr)^2 + \sigma_0^2  \\ 
E(P_1) = \bigl(a_1 + b_1 {\rm [Fe/H]}\bigr)^2 + \sigma_1^2.  
\end{split} 
\end{equation} 

\noindent Since $b_0 \neq b_1$  the difference between the mean periods depends on
[Fe/H], which contradicts the Braga et al. (2022) conclusion that ``their
difference is constant over a broad range in pulsation periods and in metal
abundance.''

In {\bf Fig.\,10b} the ratio $P_1$/$P_0$ is plotted against [Fe/H].   The ratio of
the mean periods (Eqn.\,6), which, using a first-order Taylor expansion, is
approximately equal to the mean of $P_1$/$P_0$, is also plotted.  A similar
diagram was plotted by Braga et al. (2022), who fitted a line to their data
(see their eqn.\,2 and fig.9).  Owing to the relatively large scatter and
limited [Fe/H] range of both samples ($-1.0$ to $-2.5$ dex) it is difficult to
determine from the data which form provides a better fit.  However, a linear
relationship between $P_1$/$P_0$ and [Fe/H] can be ruled out because it is
inconsistent with the assumed quadratic relationships between the individual
periods and [Fe/H] (Fig.10a).  Notice also that there would be a similar lack
of consistency if the $P_0$-[Fe/H] and $P_1$-[Fe/H]  relationships were assumed
to be linear.  

To validate the functional form of the Eqn.\,(2) model, Eqn.\,(4) was fitted to
the combined sample of {\it K2}+\texttt{OGLE} cRRd stars shown in Fig.\,1.
This sample includes the 72 {\it K2} cRRd stars and the 458 \texttt{OGLE}
Galactic Disk and Bulge cRRd stars from Soszy\'nski et al. (2019), and spans the
entire period range of known cRRd stars, $0.35$$<$$P_0$$<$$0.62$\,day.  The
fitted curve is given by Eqn.\,(4) with $a=-0.1634\pm0.0032$,
$b=0.4359\pm0.0093$, and $c=0.4552\pm0.0068$, where the rms-error is 0.0007,
and is plotted with a solid black line in Figs.\,1 and 10(c).  Eqn.\,(4) was
also fitted to the cRRd stars shown in the Petersen diagram plotted in {\bf
Fig.10c}.  In this case the sample of cRRd stars consists of the 268
metallicity calibration stars ({\it K2}+GC+Chen) plus the 58 {\it K2} stars not
included in the metallicity calibration (i.e., those stars with unknown
spectroscopic [Fe/H]).  The fit is plotted as a black dashed curve in
Fig.10(c), where the coefficients are $a=-0.1870\pm0.0115$,
$b=0.5023\pm0.0323$, $c=0.4088\pm0.0227$, with rms-error 0.0005.   The red  
dotted curve in Fig.10(c) was obtained by fitting Eqn.\,(5) to the same data,
where now the coefficients are $a' = -0.0085\pm0.0003$  and $b' =
0.7617\pm0.0006$, with rms-error 0.0007.  Comparison of the three fitted curves
shows that the Eqn.\,4 curves (black) are consistent with the two samples,
while the Eqn.\,5 curve (red dotted) is not (see $\S$4.2).

Figs.\,10a-c demonstrate that the Eqn.\,2 model provides a solid framework
relating $P_0$, $P_1$ and [Fe/H].  The model fits the observations and is
internally consistent.  Inverting and fitting Eqn.\,2 gives the following
period-metallicity calibration curves:  
\begin{equation}
\begin{split}	
	{\rm [Fe/H]} = (7.59\pm0.34) - (13.25\pm0.47) \sqrt{P_0} \\  
	{\rm [Fe/H]} = (7.42\pm0.33) - (15.08\pm0.53) \sqrt{P_1},
\end{split}
\end{equation}
\noindent where in both cases the root-mean-square error is $\pm0.17$\,dex,
and the standard errors of the mean typically are $\pm$0.01\,dex (rising to $\pm$0.04 dex at the
extremes of the period ranges).
Either equation  can be used to estimate metal abundance.  The calibration
curve plotted in {\bf Fig.\,10d} is the $P_0$ version, which is used below for deriving
metallicities for cRRd stars with $P_0$ in the range 0.45-0.59\,d.

\subsubsection{Metallicities and Masses for the K2 cRRd stars}

Metal abundances for the 72 {\it K2} cRRd stars obtained by applying Eqn.\,7
are given in {\bf Table\,8}, where the stars are ordered by increasing $P_0$.
Also in the table are {\it Gaia} Identification numbers for the 53 stars in
common with ESA's {\it Gaia} Mission (col.\,3; see $\S$4.4.4), the {\it Gaia}
DR2 and DR3 RR\,Lyrae classifications (cols.\,4-5), fundamental-mode periods
and period ratios (cols.\,6-7).   Approximate masses for the {\it K2} cRRd
stars were estimated by substituting $P_1$/$P_0$ and $Z$ into the following
formula derived specifically for RRd stars by Marconi {\it et al.} (2015,
eqn.\,5) from hydrodynamical models: 

\begin{equation} 
\begin{split} \log M/M_{\odot} = -0.85(\pm0.05) - 2.8(\pm0.3)\log(P_1/P_0)  \\ - 0.097(\pm0.003)\log Z, \end{split}
\end{equation}

\noindent where  $Z$ represents the fraction by mass of elements heavier than
hydrogen and helium, and is related to [Fe/H] according to [Fe/H] = $\log
Z/Z_{\odot}$ (assuming $X = X_{\odot}$).  In this paper the values adopted for
the Sun are $Z_{\odot}$=0.0139$\pm$0.0006 and $X_{\odot}$=0.7438$\pm$0.0054
(Asplund et al. 2021).  Column 9 of Table 8 contains the mass based on scaled
solar abundances and assuming no enhancement with respect to iron of the
$\alpha$-elements (O, Ne, Mg, Si, S and C), i.e., [$\alpha$/Fe]=0, in which
case $\log Z$=${\rm [Fe/H]}-1.857$.  Column 10 contains the mass derived
assuming an $\alpha$-element enhancement [$\alpha$/Fe]=+0.30 dex (i.e., the
average $\alpha$-element abundance is twice the scaled solar value), in which
case $\log Z = {\rm [Fe/H]} - 1.635$, where the constant follows from the
VandenBerg et al. (2000, tables 1-2; 2006 table 1) metallicities after
adjusting to correct for $Z_{\odot}$=0.0188 assumed by VandenBerg.    The
resulting mass estimates range from $\sim$$0.57\,M_\odot$ for the most
metal-rich stars in the sample to $0.81\,M_\odot$ for the most metal-poor
stars, with the $\alpha$-enhanced masses typically $\sim$$0.03\,M_\odot$
smaller than the [$\alpha$/Fe]=0 case.  The derived mass range for the RRd
stars is consistent with the range established from horizontal-branch evolution
models (see VandenBerg and Denissenkov 2018), from hydrodynamical models
(Moln\'ar et al. 2015) and from asteroseismology (Netzel, Moln\'ar \& Joyce
2023).

\begin{table*}
\fontsize{7}{8.4}\selectfont   %
\centering

\caption{Metal abundances and masses for the 72 cRRd stars observed during
NASA's {\it K2} Mission.  The values for the three shortest period stars are
enclosed in parentheses because they are outside the period range of the [Fe/H]
calibration curve.   {\it Gaia} IDs and classifications are also given
for the 53 cRRd stars in common with either DR2 or DR3. The
masses were derived assuming no enhancement of $\alpha$ elements with respect
to iron (i.e., [$\alpha$/Fe]=0.0 dex), and an enhancement [$\alpha$/Fe]=0.3 dex.}

\vskip0.1cm
\label{tab:three}
\begin{tabular}{llrllllccc}
\hline 
\multicolumn{1}{c}{ EPIC } & \multicolumn{1}{c}{K2} & \multicolumn{1}{c}{{\it Gaia}} &\multicolumn{2}{c}{{\it Gaia} classif.}  &  \multicolumn{1}{c}{$P_0$}& \multicolumn{1}{c}{$P_1/P_0$}  & \multicolumn{1}{c}{[Fe/H] }   & \multicolumn{2}{c}{$M/M_{\sun}$} \\
\multicolumn{1}{c}{No.} & \multicolumn{1}{c}{Campaign } & \multicolumn{1}{c}{Identification no. } &  \multicolumn{1}{c}{DR2} &  \multicolumn{1}{c}{DR3} &   \multicolumn{1}{c}{[day]} &  \multicolumn{1}{c}{ } &  \multicolumn{1}{c}{$\pm$0.17\,dex }  & \multicolumn{1}{c}{[$\alpha$/Fe]=0.0 }  & \multicolumn{1}{c}{[$\alpha$/Fe]=0.3}  \\
\multicolumn{1}{c}{(1)} & \multicolumn{1}{c}{(2)}  &\multicolumn{1}{c}{(3)} & \multicolumn{1}{c}{(4)} & \multicolumn{1}{c}{(5)} & \multicolumn{1}{c}{(6)} & \multicolumn{1}{c}{(7)} & \multicolumn{1}{c}{(8)} & \multicolumn{1}{c}{(9)} &  \multicolumn{1}{c}{(10)} \\
\hline 
\\
225456697  &     C11     & \multicolumn{1}{c}{\dots} &   \dots     &  \dots      &   0.399838  &   0.73618   &  (--0.79)   & (0.57) & (0.57)  \\ %
251248830  &     C11     & \multicolumn{1}{c}{\dots} &   \dots     &  \dots      &   0.419592  &   0.73885   &  (--1.00)   & (0.62) & (0.59)  \\ %
251248826  &     C11     &    4059683185121550336    &   \dots     &     RRc     &   0.430234  &   0.73982   &  (--1.11)   & (0.64) & (0.61)  \\ %
246058914  &     C12     &    2438582821787698176    &     RRab    &  \dots      &   0.452948  &   0.74205   &   --1.33    & 0.66 & 0.63  \\ %
251248827  &     C11     & \multicolumn{1}{c}{\dots} &    \dots    &  \dots      &   0.455268  &   0.74224   &   --1.35    & 0.67 & 0.63  \\ %
229228184  &      C7     &    4071397068375975296    &     RRc     &     RRc     &   0.459430  &   0.74234   &   --1.40    & 0.67 & 0.64  \\ %
224366356  &      C9     & \multicolumn{1}{c}{\dots} &   \dots     &  \dots      &   0.461936  &   0.74263   &   --1.42    & 0.68 & 0.64  \\ %
251629085  &     C17     &    3686704514188694144    &     RRc     &     RRc     &   0.463110  &   0.74320   &   --1.43    & 0.68 & 0.64  \\ %
201519136  &     C10     &    3699831549153899648    &     RRd     &     RRd     &   0.463376  &   0.74305   &   --1.43    & 0.68 & 0.64 \\ %
211694449  &    C5,18    & \multicolumn{1}{c}{\dots} &   \dots     &  \dots      &   0.463650  &   0.74272   &   --1.44    & 0.68 & 0.65  \\ %
251809814  &     C17     & \multicolumn{1}{c}{\dots} &   \dots     &  \dots      &   0.467124  &   0.74345   &   --1.47    & 0.68 & 0.65  \\ %
251456808  &     C12     & \multicolumn{1}{c}{\dots} &   \dots     &  \dots      &   0.468980  &   0.74328   &   --1.49    & 0.68 & 0.65  \\ %
236212613  &     C11     &    4107786711365354496    &     RRc     &     RRd     &   0.469200  &   0.74373   &   --1.49    & 0.68 & 0.65  \\ %
212615778  &     C17     &    3624326573845415552    &     RRab    &     RRc     &   0.469391  &   0.74328   &   --1.49    & 0.68 & 0.65  \\ %
212455160  &    C6,17    & \multicolumn{1}{c}{\dots} &   \dots     &   \dots     &   0.469592  &   0.74388   &   --1.49    & 0.68 & 0.65  \\ %
229228175  &      C7     &    4071509081124919040    &     RRc     &     RRc     &   0.469540  &   0.74424   &   --1.49    & 0.68 & 0.65  \\ %
248653210  &     C14     &    3862737081709822208    &   \dots     &     RRc     &   0.471007  &   0.74401   &   --1.51    & 0.69 & 0.65  \\ %
245974758  &     C12     & \multicolumn{1}{c}{\dots} &  \dots      &   \dots     &   0.475291  &   0.74335   &   --1.55    & 0.69 & 0.66  \\ %
212819285  &     C17     & \multicolumn{1}{c}{\dots} &  \dots      &   \dots     &   0.474968  &   0.74415   &   --1.55    & 0.69 & 0.66  \\ %
251809825  &     C17     &    3630514930231792128    &  \dots      &     RRd     &   0.475900  &   0.74328   &   --1.56    & 0.69 & 0.66  \\ %
220604574  &      C8     &    2579544339932337152    &  \dots      &     RRc     &   0.476772  &   0.74403   &   --1.56    & 0.69 & 0.66  \\ %
212335848  &      C6     &    3604456989982044800    &  \dots      &     RRc     &   0.476856  &   0.74460   &   --1.56    & 0.69 & 0.66   \\ %
251809772  &     C17     & \multicolumn{1}{c}{\dots} &  \dots      &    \dots    &   0.477332  &   0.74379   &   --1.57    & 0.70 & 0.66  \\ %
201749391  &     C14     & \multicolumn{1}{c}{\dots} &  \dots      &    \dots    &   0.479930  &   0.74454   &   --1.59    & 0.70 & 0.66  \\ %
248845745  &     C14     &    3870825497264938752    &  \dots      &     RRc     &   0.481123  &   0.74487   &   --1.61    & 0.70 & 0.66  \\ %
201152424  &     C10     &    3596646712214016768    &  \dots      &     RRc     &   0.481725  &   0.74420   &   --1.61    & 0.70 & 0.67 \\ %
210933539  &      C4     &    61543999430570496      &     RRd     &    \dots    &   0.481801  &   0.74432   &   --1.61    & 0.70 & 0.67  \\ %
248667792  &     C14     &    3863597548342360448    &  \dots      &     RRc     &   0.482054  &   0.74461   &   --1.61    & 0.70 & 0.67  \\ %
211888680  &    C5,16    &    612194609624700928     &     RRc     &     RRc     &   0.482993  &   0.74408   &   --1.62    & 0.70 & 0.67  \\ %
201585823  &      C1     &    3796490612783265152    &     RRc     &     RRc     &   0.482590  &   0.74477   &   --1.62    & 0.70 & 0.67  \\ %
210600482  &      C4     &    44250085978293504      &   \dots     &     RRc     &   0.487191  &   0.74400   &   --1.66    & 0.71 & 0.68  \\ %
229228220  &      C7     &    4073132888018172160    &   \dots     &     RRc     &   0.487740  &   0.74504   &   --1.67    & 0.71 & 0.67  \\ %
212449019  &      C6     &    3620942277055055488    &     RRc     &     RRc     &   0.487778  &   0.74499   &   --1.67    & 0.71 & 0.67   \\ %
251809860  &     C17     & \multicolumn{1}{c}{\dots} &   \dots     &   \dots     &   0.488442  &   0.74416   &   --1.67    & 0.71 & 0.68   \\ %
210831816  &      C4     &    51156844364167552      &     RRd     &     RRc     &   0.488793  &   0.74429   &   --1.68    & 0.71 & 0.68   \\ %
211665293  &     C16     &    610414019262262912     &     RRc     &     RRd     &   0.491516  &   0.74477   &   --1.70    & 0.71 & 0.68   \\ %
212498188  &     C17     & \multicolumn{1}{c}{\dots} &   \dots     &   \dots     &   0.498643  &   0.74474   &   --1.77    & 0.72 & 0.69   \\ %
228800773  &     C10     & \multicolumn{1}{c}{\dots} &   \dots     &   \dots     &   0.499880  &   0.74567   &   --1.78    & 0.72 & 0.69   \\ %
229228811  &      C8     &    2576293393286532224    &     RRc     &     RRc     &   0.500219  &   0.74550   &   --1.79    & 0.73 & 0.69   \\ %
251521080  &     C17     &    3684381207464081792    &   \dots     &     RRab    &   0.500870  &   0.74473   &   --1.79    & 0.73 & 0.69   \\ %
220636134  &      C8     &    2580012972403894528    &     RRc     &     RRc     &   0.501375  &   0.74543   &   --1.80    & 0.73 & 0.69   \\ %
213514736  &      C7     & \multicolumn{1}{c}{\dots} &   \dots     &    \dots    &   0.503583  &   0.74509   &   --1.82    & 0.73 & 0.70   \\ %
251809870  &     C17     & \multicolumn{1}{c}{\dots} &   \dots     &    \dots    &   0.504830  &   0.74457   &   --1.83    & 0.73 & 0.70   \\ %
248871792  &     C14     &    3872607878628627840    &     RRc     &     RRab    &   0.505497  &   0.74444   &   --1.83    & 0.74 & 0.70   \\ %
201440678  &     C10     &    3698706061563300608    &     RRd     &     RRc     &   0.505614  &   0.74499   &   --1.84    & 0.73 & 0.70   \\ %
211898723  &    C5,18    &    662527846763392000     &   \dots     &     RRc     &   0.505827  &   0.74435   &   --1.84    & 0.74 & 0.70   \\ %
212467099  &     C17     &    3609194923025131648    &   \dots     &     RRc     &   0.517170  &   0.74705   &   --1.94    & 0.75 & 0.71   \\ %
251809832  &     C17     &    3606923499505619968    &   \dots     &     RRc     &   0.519190  &   0.74648   &   --1.96    & 0.75 & 0.72   \\ %
248730795  &     C14     &    3869100230377688448    &   \dots     &     RRc     &   0.522545  &   0.74657   &   --1.99    & 0.76 & 0.72   \\ %
229228194  &      C7     &    4071405658308934144    &     RRc     &    \dots    &   0.522630  &   0.74567   &   --1.99    & 0.76 & 0.72   \\ %
248827979  &     C14     & \multicolumn{1}{c}{\dots} &   \dots     &    \dots    &   0.522807  &   0.74543   &   --1.99    & 0.76 & 0.72   \\ %
211072039  &      C4     &    54010936031517440      &   \dots     &     RRc     &   0.527041  &   0.74660   &   --2.03    & 0.76 & 0.73   \\ %
250056977  &     C15     &    6265195826928713600    &   \dots     &     RRc     &   0.532603  &   0.74492   &   --2.08    & 0.78 & 0.74   \\ %
220254937  &      C8     &    2558508311670941824    &   \dots     &     RRc     &   0.535934  &   0.74660   &   --2.11    & 0.78 & 0.74   \\ %
248426222  &     C14     &    3807285480505638784    &   \dots     &     RRc     &   0.537644  &   0.74606   &   --2.13    & 0.78 & 0.74   \\ %
060018653  &      E2     &    2739784862463040512    &     RRd     &     RRc     &   0.539441  &   0.74579   &   --2.15    & 0.79 & 0.75   \\ %
247334376  &     C13     &    3414155402936376832    &    \dots    &     RRc     &   0.539580  &   0.74733   &   --2.15    & 0.78 & 0.74   \\ %
248653582  &     C14     &    3865590520542557184    &    \dots    &     RRab    &   0.539878  &   0.74596   &   --2.15    & 0.79 & 0.75   \\ %
228952519  &     C10     &    3682596906250805632    &     RRc     &     RRc     &   0.540600  &   0.74637   &   --2.16    & 0.79 & 0.75   \\ %
214147122  &      C7     &    4072051140361909632    &     RRc     &     RRc     &   0.541040  &   0.74607   &   --2.16    & 0.79 & 0.75   \\ %
251248825  &     C11     &    4059675454173637632    &   \dots     &     RRc     &   0.541343  &   0.74661   &   --2.16    & 0.79 & 0.75   \\ %
235794591  &     C11     &    4059476025896752384    &     RRc     &    \dots    &   0.541583  &   0.74605   &   --2.17    & 0.79 & 0.75   \\ %
212547473  &    C6,17    &    3610631916003219328    &     RRc     &     RRc     &   0.545079  &   0.74564   &   --2.20    & 0.79 & 0.76   \\ %
248509474  &     C14     &    3856963644936004352    &   \dots     &     RRc     &   0.557312  &   0.74483   &   --2.31    & 0.82 & 0.78   \\ %
060018662  &      E2     &    2642992895363833088    &   \dots     &     RRc     &   0.559323  &   0.74645   &   --2.32    & 0.81 & 0.78   \\ %
248731983  &     C14     &    3869066244300744064    &   \dots     &     RRc     &   0.560084  &   0.74542   &   --2.33    & 0.82 & 0.78   \\ %
223051735  &      C9     & \multicolumn{1}{c}{\dots} &   \dots     &   \dots     &   0.567737  &   0.74563   &   --2.40    & 0.83 & 0.79   \\ %
248369176  &     C10     &    3698207256945765760    &     RRc     &     RRd     &   0.568280  &   0.74625   &   --2.40    & 0.83 & 0.79    \\ %
249790928  &     C15     &    6255192229621483136    &   \dots     &     RRc     &   0.578550  &   0.74430   &   --2.49    & 0.85 & 0.81  \\ %
235631055  &     C11     &    4059259044129556352    &   \dots     &     RRc     &   0.580463  &   0.74549   &   --2.51    & 0.85 & 0.81  \\ %
248514834  &     C14     &    3857004812197962880    &   \dots     &     RRc     &   0.581287  &   0.74734   &   --2.52    & 0.85 & 0.81   \\ %
225326517  &     C11     &    4116711825239025152    &   \dots     &     RRc     &   0.581816  &   0.74515   &   --2.52    & 0.86 & 0.81   \\ [0.1cm] %

\hline
\end{tabular}
\end{table*}

\subsubsection{{\it Gaia} Observations of the {\it K2} RRd Stars}

The 75 RRd stars observed by {\it K2} were matched using RA and DEC coordinates
to the {\it Gaia} DR2 and DR3 RR\,Lyrae catalogues (Clementini et al.  2019,
2023).  Fifty-four stars were found in one or both of the catalogues: the 53
cRRd stars identified in col.2 of Table\,8, and one of the three aRRd stars
(EPIC\,205209951).  The RR~Lyrae types given by {\it Gaia} were ``obtained
using the period-amplitude diagram in the G-band, the plots of the Fourier
parameters $R_{\rm 21}$ and $\phi_{\rm 21}$ versus period, and the Petersen
diagram" (see fig.7 of Clementini et al. 2023).  Of the 49 {\it K2} cRRd stars
listed in DR3 (see col.5 of Table\,8) only five were correctly classified `RRd'
(10$\%$), 41 were misclassified `RRc' (84$\%$) and three were misclassified
`RRab' (6$\%$).  Five of the 24 {\it K2} cRRd stars in DR2 (see col.4 of
Table\,8) were correctly classified `RRd' (21$\%$), 17 were misclassified `RRc'
(71$\%$) and two were misclassified `RRab' (8$\%$).  Only one of the 20 stars
found in both {\it Gaia} catalogues was correctly classified in both.   

Comparison of the periods given in the  DR3 catalogue with those in Table\,3
reveals that only  the first-overtone period was detected in all the cRRd stars
misclassified `RRc' (and in three of the five stars misclassified `RRab'),
i.e.,  only the shorter-period higher-amplitude component was detected.   This
suggests that a significant number of `RRc' stars (and possibly some `RRab'
stars) in the {\it Gaia} catalogues may actually be cRRd stars.   Moreover, for
two of the nine {\it K2} stars correctly classified `RRd' in DR2 or DR3,  the
{\it Gaia} fundamental period and period-ratio differ from the high-precision
{\it K2} values given in Table\,3 by amounts that place the stars significantly
above or below the Petersen curve for cRRd stars.

\subsubsection{Metallicities and Masses for 2130 {\it Gaia} cRRd Stars}

The latest {\it Gaia} Survey, DR3 (Clementini et al. 2023), gives photometric
[Fe/H] estimates for 113202  (65$\%$)  of the 175350 stars classified `RRab'
and for 20375 (22$\%$) of the 94422 stars classified `RRc'.  No metal
abundances are reported for the 2378 stars classified `RRd' in DR2 or for the
2007 stars classified `RRd' in DR3.   To fill in this gap Eqn.\,6 was applied
to those stars classified `RRd' in one or both {\it Gaia} catalogues, and which
have periods and period-ratios within the ranges of the [Fe/H] calibration
data, i.e., 0.45$<$$P_0$$<$0.59\,day and 0.7418$<$$P_1/P_0$$<$0.7477 (see
Fig.\,10), and which lie on or close to the Petersen curve.  A total of 2253 of
the 3714 stars classified `RRd' in either DR2 or DR3 have periods and
period-ratios within the [Fe/H] calibration range.  Of the 2253 stars, 123 have
locations more than 2$\sigma$  above or below the `{\it K2}+\texttt{OGLE}'
Petersen curve shown in Fig.\,1  (i.e., with $|\Delta$($P_1/P_0$)$|$$>$0.0014)
and were eliminated, leaving 2130 stars.  A Petersen diagram for the 2130
`presumably {\it bona fide}' cRRd stars is shown in {\bf Figure\,11}, together
with the  `K2+\texttt{OGLE}' Petersen curve (Fig.\,1).   The distribution of
the 623 stars classified `RRd' in both catalogues (plotted with black filled
dots) is similar to that of the 1507 stars classified `RRd' only in one or the
other but not both (blue crosses).

Metallicities and masses for the 2130 {\it Gaia} cRRd stars are given in {\bf
Table\,9}.  Also in the table are {\it Gaia} identification numbers (usually
from DR3)\footnote{ {\it Gaia} identification numbers are usually the same in
DR2 and DR3, but not always. An example of where this is not the case is
RR~Lyrae itself: in DR2 it is 2125982599341232896, and in DR3 it is
2125982599343482624, the difference occuring in the last seven digits.   Of the
stars classified `RRd' in either DR2 or DR3, 93 stars were found to have
similar coordinates (RA,DEC) but different identification numbers (again, the
differences tending to occur in the last $\sim$8 digits of the 19 digit
number).  A match by RA and DEC revealed that 93 of the 3714 stars classified
`RRd' in either DR2 or DR3 were found to have different {\it Gaia}
identification numbers in DR2 and DR3.  The 93 stars are identified in {\bf
Table 10}.  Also given in the table are the  periods and period-ratios given in
DR2 and DR3;  close agreement of the DR2 and DR3 periods and period-ratios
increases the confidence that it is the same star.  The last column indicates
whether or not the star is a `classical' RRd star within 2$\sigma$ of the {\it
K2}+\texttt{OGLE} Petersen curve.}, coordinates of the stars (RA,DEC), the
RR\,Lyr classifications given in DR2 and DR3, pulsation periods and period
ratios, and masses.  If the star is listed in DR3 then the periods from that
catalogue were used to calculate [Fe/H], otherwise the periods from DR2 were
used.    The resulting [Fe/H] values range from --1.32 to --2.58 dex, with
uncertainties of $\pm$0.17 dex, and the masses range from 0.63 to
0.86\,$M_{\odot}$.  Until more is known about their physical characteristics
Eqn.\,7 should not be applied to aRRd or pRRd stars or any other stars that
deviate significantly from the Petersen curve.  The masses were calculated
using the same Marconi et al.  formula used to derive masses for the K2 cRRd
stars, again both without and with $\alpha$-element abundance enhancements.

\renewcommand{\thefigure}{11}
\begin{figure} \begin{center}
\begin{overpic}[width=8.2cm]{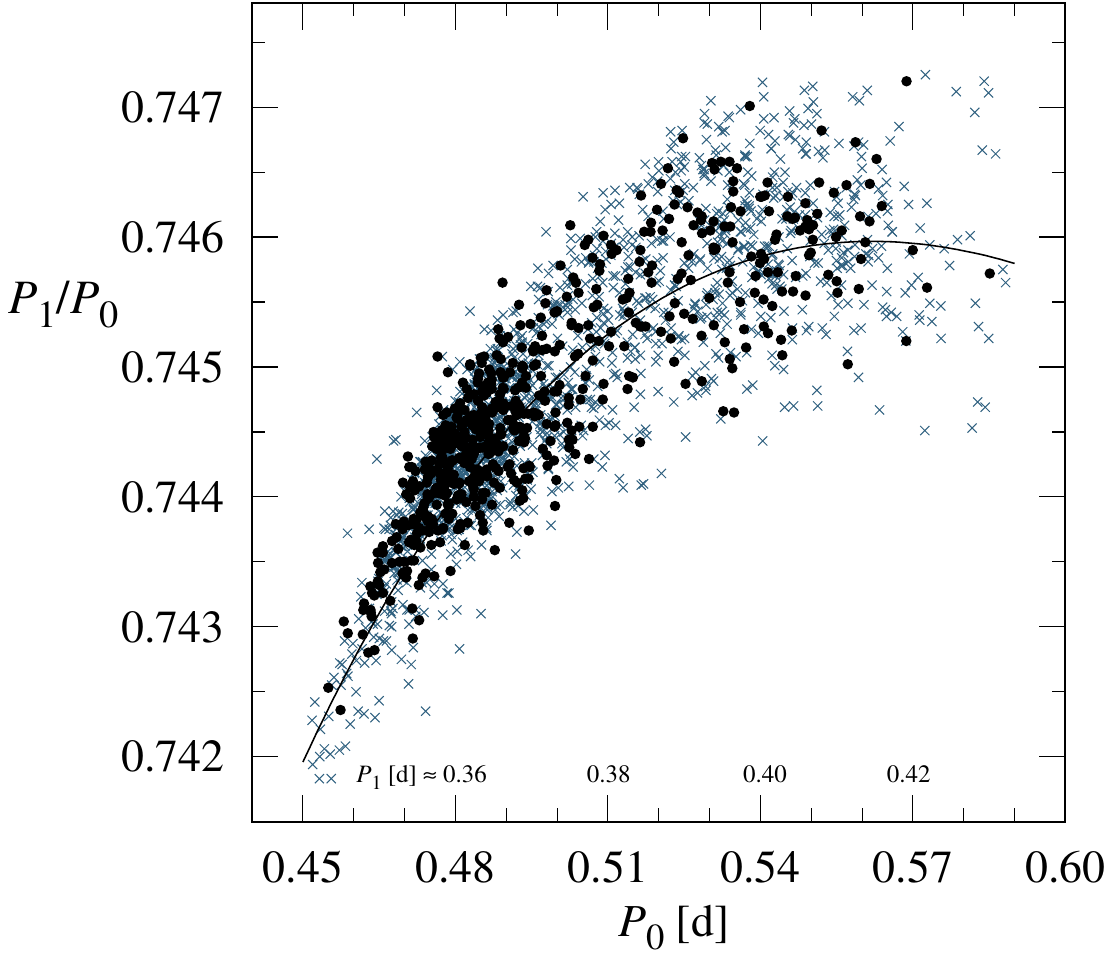}  \end{overpic} \end{center}  

\caption{Petersen diagram for 2130 {\it Gaia} cRRd stars.  The 1507 (blue) crosses
represent the stars classified `RRd' in either DR2 or DR3 but not both, and the
623 (black) dots represent the stars classified `RRd' in both catalogues (see
Table\,9).   Also shown is
the long-period portion of the `{\it K2}+\texttt{OGLE}' Petersen curve (black
curves in Figs.\,1 and 10c).  } 

\label{Fig11} 
\end{figure}

\renewcommand{\thetable}{9}
\begin{table*}
\fontsize{7}{8.4}\selectfont
\centering

\caption{Metallicities and masses for the 2130 {\it Gaia} stars classified
`RRd' by either DR2 or DR3 and that are on the cRRd curve in the Petersen
diagram (see Fig.\,11).  The [Fe/H] values (col.\,9) were estimated using the $P_0$-[Fe/H]
calibration formula (Eqn.\,7).  The masses estimates were made assuming no
$\alpha$-element enhancements (col.10) and assuming $\alpha$ element abundances
enhanced by a factor two (col.11).  The complete table is given online in the
Supporting Information.  }

\label{tab:three}
\begin{tabular}{rlllllccccc}
\hline
\multicolumn{1}{c}{ {\it Gaia} Id  } &   \multicolumn{1}{c}{RA} & \multicolumn{1}{c}{DEC} &  \multicolumn{2}{c}{{\it Gaia} Classif.} & \multicolumn{1}{c}{ $P_1$}   & \multicolumn{1}{c}{ $P_0$ } & \multicolumn{1}{c} { $P_1/P_0$ }  & \multicolumn{1}{c}{[Fe/H] }  & \multicolumn{2}{c}{$M/M_{\sun}$} \\
\multicolumn{1}{c}{  } &  \multicolumn{2}{c}{(2016.0)}  &DR2  &DR3    &\multicolumn{1}{c}{[day]}  &\multicolumn{1}{c}{[day]} & & \multicolumn{1}{c}{ $\pm$0.17 }  & \multicolumn{1}{c}{[$\alpha$/Fe]=0.0} & \multicolumn{1}{c}{[$\alpha$/Fe]=0.3 }   \\
\multicolumn{1}{c}{(1)} & \multicolumn{1}{c}{(2)} &\multicolumn{1}{c}{(3)}  &\multicolumn{1}{c}{(4)} & \multicolumn{1}{c}{(5)} & \multicolumn{1}{c}{(6)}   & \multicolumn{1}{c}{(7)}  & \multicolumn{1}{c}{(8)}   & \multicolumn{1}{c}{(9)} &\multicolumn{1}{c}{(10)} &\multicolumn{1}{c}{(11)}   \\
\hline  
\\	    
4109411961354217856  & 258.5835748  & $-25.70825502$ & RRab & RRd & 0.335361 & 0.451800 & 0.74228 &--1.32&0.66&0.63 \\
1797412030619582848  & 323.6745277  & $+23.89800184$ & RRd  & RRc & 0.335297 & 0.451917 & 0.74194 &--1.32&0.66&0.63\\
6028840233297277440  & 258.3737969  & $-29.79608735$ & RRd  & RRc & 0.335828 & 0.452341 & 0.74242 &--1.33&0.66&0.63\\
1494770600375550336  & 218.7012557  & $+46.44626505$ &\dots & RRd & 0.336212 & 0.453218 & 0.74183 &--1.33&0.66&0.63\\
6786048671278168448  & 322.7495052  & $-29.83430933$ &\dots & RRd & 0.336362 & 0.453317 & 0.74200 &--1.34&0.66&0.63\\
4541576557133679360  & 259.6822537  & $+13.59407288$ &\dots & RRd & 0.336546 & 0.453436 & 0.74221 &--1.34&0.66&0.63\\
6582029310878687104  & 320.4805041  & $-41.500339  $ &\dots & RRd & 0.337054 & 0.454214 & 0.74206 &--1.34&0.67&0.63\\
4686535719166593408  & 22.66688193  & $-72.73111643$ & RRd  & RRd & 0.337847 & 0.454993 & 0.74253 &--1.35&0.67&0.63\\
3464599248369119104  & 176.9940688  & $-35.49997468$ & RRd  & RRc & 0.337824 & 0.455095 & 0.74231 &--1.35&0.67&0.63\\
5341057708232415104  & 165.5449429  & $-55.53852362$ & RRab & RRd & 0.337885 & 0.455360 & 0.74202 &--1.36&0.67&0.64\\
\multicolumn{1}{c}{\dots} \\ [0.1cm]
\hline
\end{tabular}
\end{table*}

\renewcommand{\thetable}{10}
\begin{table*}
\fontsize{6}{7.2}\selectfont  
\centering

\caption{Ninety-three {\it Gaia} stars classified `RRd' in  DR2 or DR3,  with
similar RA,DEC coordinates but different identification numbers in DR2 and DR3.
The complete table is given online in the Supporting Information.}

\label{tab:three}
\begin{tabular}{llccclllcccc}
\hline \\
\multicolumn{5}{c}{{\it Gaia} DR2} &\multicolumn{1}{c}{} & \multicolumn{5}{c}{{\it Gaia} DR3} \\ %
\cline{1-5} \cline{7-11} \\ %
\multicolumn{1}{c}{DR2  } & \multicolumn{1}{c}{RR}   & \multicolumn{1}{c}{ $P_1$ }    & \multicolumn{1}{c} { $P_0$ }  &  \multicolumn{1}{c}{$P_1/P_0$} & & \multicolumn{1}{c}{DR3}  
	& \multicolumn{1}{c}{RR} &  \multicolumn{1}{c}{$P_1$ } & \multicolumn{1}{c}{ $P_0$}   & \multicolumn{1}{c}{ $P_1/P_0$ }   & \multicolumn{1}{c}{cRRd?}   \\
\multicolumn{1}{c}{Identification No. }   & \multicolumn{1}{c}{class} & \multicolumn{1}{c}{[day]} &\multicolumn{1}{c}{[day]}  & & &  \multicolumn{1}{c}{Identification No. } &  \multicolumn{1}{c}{class} 
	& \multicolumn{1}{c}{[day]} & \multicolumn{1}{c}{[day]} & \multicolumn{1}{c}{ }   \\
\multicolumn{1}{c}{(1)} & \multicolumn{1}{c}{(2)} &\multicolumn{1}{c}{(3)}  &\multicolumn{1}{c}{(4)} & \multicolumn{1}{c}{(5)} && \multicolumn{1}{c}{(6)}  & \multicolumn{1}{c}{(7)}   & \multicolumn{1}{c}{(8)}
	& \multicolumn{1}{c}{(9)} & \multicolumn{1}{c}{(10)} &  \multicolumn{1}{c}{(11)} \\
\hline  
\\	    
1442424496748916480   &  RRd   &  0.404211   &  0.541729   &  0.74615   &&  1442424501044402432   &  RRd   &  0.404205   &  0.541687   &  0.74620 & yes\\
1470192632844893568   &  RRd   &  0.382444   &  0.513261   &  0.74513   &&  1470192632845260288   &  RRd   &  0.382446   &  0.513238   &  0.74516 & yes\\
1554867810007895296   &  RRd   &  0.350923   &  0.471790   &  0.74381   &&  1554867810004952192   &  RRd   &  0.350912   &  0.471762   &  0.74383 & yes\\
1639360120343239808   &  RRc   &  0.361888   &  \dots      &   \dots    &&  1639360124638557440   &  RRd   &  0.361899   &  0.486449   &  0.74396 & yes\\
1745948362385628416   &  RRab  &   \dots     &  0.567080   &   \dots    &&  1745948362391096832   &  RRd   &  0.413354   &  0.554544   &  0.74539 & yes\\
2510409037347211264   &  RRab  &   \dots     &  0.353731   &   \dots    &&  2510409041642733568   &  RRd   &  0.353735   &  0.475571   &  0.74381 & yes\\
2862259978075970816   &  RRd   &  0.397965   &  0.534996   &  0.74387   &&  2862259978077052160   &  RRab  &   \dots     &  0.534997   &   \dots  & no\\
2972392044878569984   &  RRd   &  0.385688   &  0.516862   &  0.74621   &&  2972392044879268096   &  RRc   &  0.385673   &  \dots      &  \dots   & yes\\
3059619325966429824   &  RRd   &  0.396827   &  0.542776   &  0.73111   &&  3059619325975593600   &  RRab  &   \dots     &  0.542772   &  \dots   & no\\
3746580820769812096   &  RRc   &  0.394827   &  \dots      &  \dots     &&  3746580825061729536   &  RRd   &  0.394837   &  0.528923   &  0.74649 & yes\\
\multicolumn{1}{c}{\dots} &    &             &             &            &&                        &        &             &             &          &    \\
\hline
\end{tabular}
\end{table*}

\renewcommand{\thefigure}{12}
\begin{figure*} 
\begin{center}
\begin{overpic}[width=8.7cm]{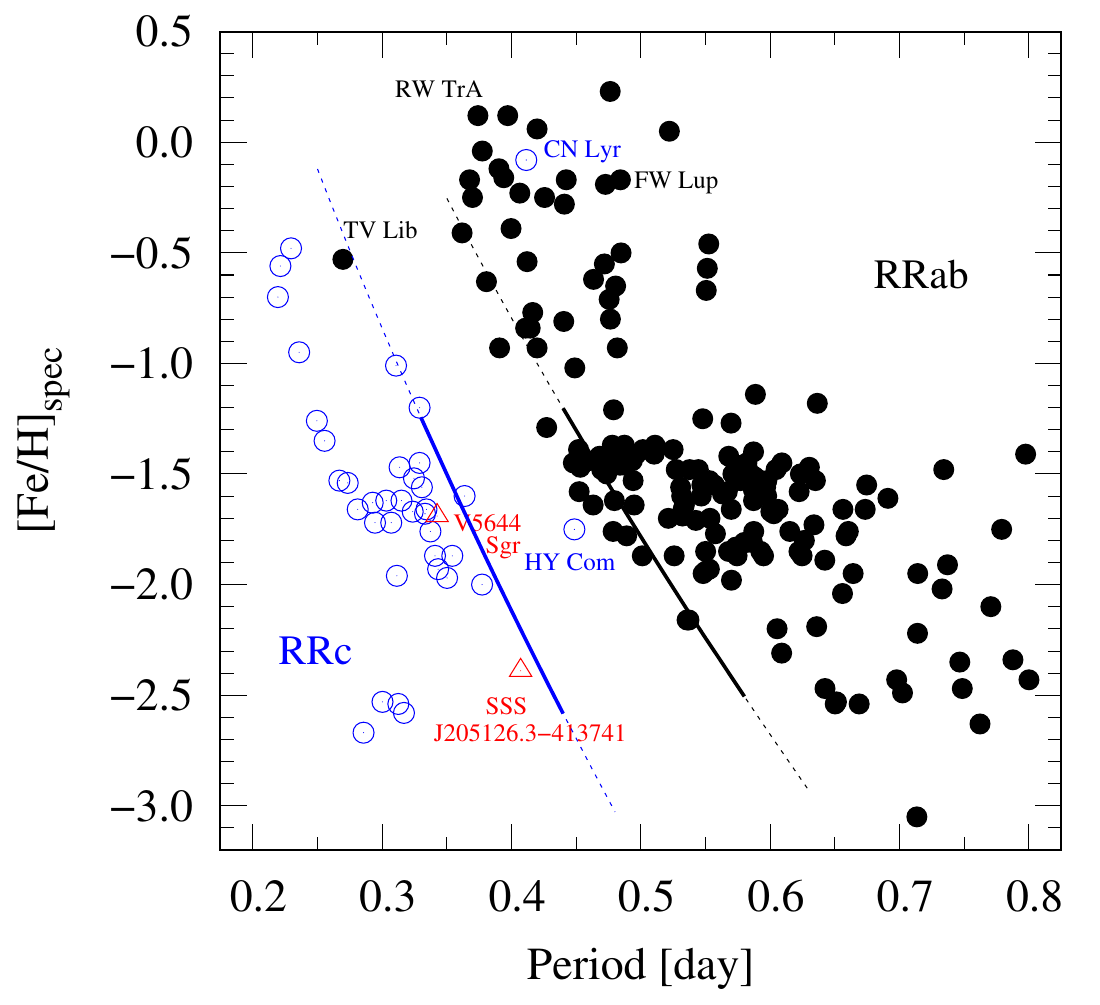}  \put(84,78){(a)}  \end{overpic}  
\hskip0.3truecm
\begin{overpic}[width=8.5cm]{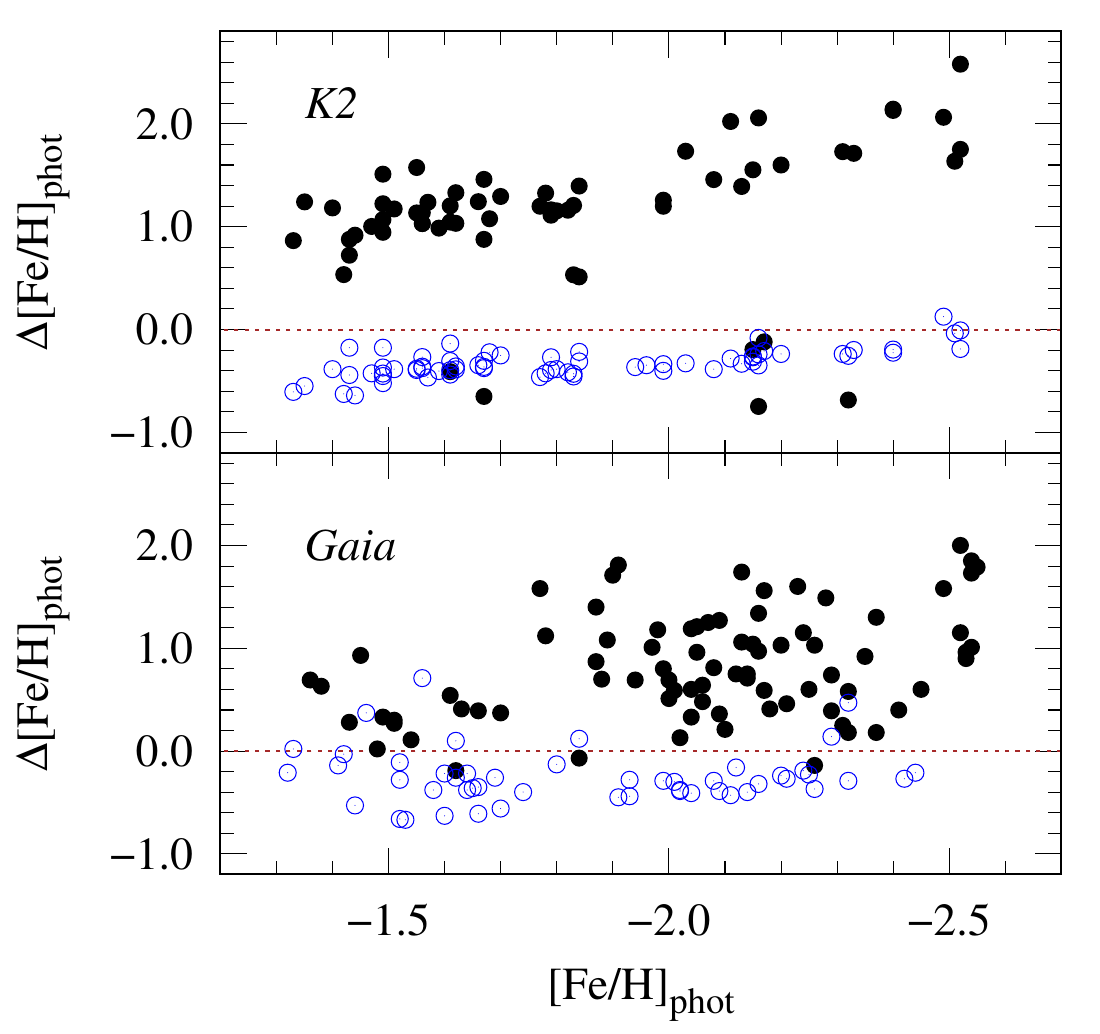}  \put(84,72){(b)} \end{overpic} 
\end{center}  %

\caption{(a) Period-metallicity diagram for 206 single-mode RRab or RRc stars
(and two possible RRd stars) with [Fe/H]$_{\rm spec}$ values derived from
high-resolution spectra (Crestani et al. 2021a,b).  The diagonal `lines' are
the first-overtone and fundamental mode metallicity regressions for cRRd stars,
and the labelled stars are discussed in the text.
(b) Misclassification error, $\Delta$[Fe/H]$_{\rm phot}$, as a function of
metal abundance for the 72 {\it K2} cRRd stars (top panel), and for 131 cRRd
stars misclassified in the {\it Gaia} DR2 or DR3 catalogues (bottom panel), where 
the black dots correspond to the bias if misclassified `RRab', the blue circles
to the bias if misclassified `RRc'.  } 

\label{Fig12} \end{figure*}

\subsubsection{Misclassification Bias}

In {\bf Figure\,12} the period-metallicity relations for single-mode RRab and
RRc stars are compared with the [Fe/H] calibration curves for cRRd stars
(Eqn.\,6).  The figure also illustrates the effect on the derived metal
abundance of misclassifying a cRRd star as `RRab' or `RRc'.   The period-[Fe/H]
diagram plotted in {\bf Fig.\,12a} (left panel) includes 208 RR~Lyr stars
having spectroscopic metal abundances derived from high-resolution spectra by
Crestani et al.  (2021b; tables 2,6).    The thicker portions of the cRRd
calibration curves (Eqn.\,7) correspond to the period ranges of the calibration
stars (see Fig.\,10) and the dashed portions represent extrapolations beyond
these ranges.  The graph shows that the cRRd relationship for the
first-overtone (left) coincides with the long-period edge for the RRc stars,
and the fundamental-mode curve (right) coincides with the short-period edge for
the RRab stars, with a clear gap separating the two types of stars.  As
previously observed, both RR\,Lyr types show a tendency to increase in
metallicity with decreasing pulsation period (see figs.\,12 and 13 of Nemec et
al. 2013; and fig.7 of Sneden et al. 2018).

Seven stars labelled in Fig.\,12a have unusual locations or questionable
RR\,Lyr types.  The pulsation period for TV\,Lib (0.270\,d) is extremely short
for an `RRab star, and the period for HY\,Com (0.449\,d) is long
for an `RRc' star.  V5644\,Sgr and SSS\,J205126.3-413741 may be RRd stars but
the evidence is inconclusive: V5644\,Sgr is classified `RRd' by Crestani et al.
but `RRab' by {\it Gaia} DR3;  and SSS\,J205126.3-413741 is classified `RRc' by
Crestani et al. but `RRd' in DR2 and `RRab' in DR3.   The RR\,Lyr types given
by Crestani et al. for CN\,Lyr (`RRc') and RW\,TrA (`RRab') are also
inconsistent with those given in the {\it Gaia} catalogues (`RRab' and `RRc',
respectively).  Finally, the low pulsation amplitude ($A_V$$\sim$0.4 mag) for
the metal-rich star FW\,Lup, classified `RRab' by Crestani et al., locates it
among the RRc stars in the period-amplitude diagram.  More comments on these
seven stars are given in Appendix\,A. 

In the absence of high-precision photometry, or where the number and spacing of
the photometry is inadequate, cRRd stars are likely to be misclassified as
`RRc', or sometimes `RRab'.  For example, only 10$\%$ of the {\it K2} cRRd
stars in the {\it Gaia} DR3 catalogue were correctly classified as `RRd', while
84$\%$ were misclassified `RRc' and 6$\%$ were misclassified `RRab' (see column
5 of Table\,8).  Misclassification affects the photometric estimation of
metallicity, [Fe/H]$_{\rm phot}$, since the calibration  equations are
different for RRab, RRc and RRd stars.  To investigate the effect of
misclassification  [Fe/H]$_{\rm phot}$ was calculated for the {\it K2} cRRd
stars assuming first an `RRc' misclassification and then an `RRab'
misclassification.  In the RRc case, first-overtone period and $\phi_{\rm 31}$
(Table\,4) were substituted into eqn.\,3 of Nemec et al. (2013), and in the
RRab case, the fundamental mode values (Table\,5) were substituted into eqn.\,4
of the same paper.  The resulting difference between [Fe/H]$_{\rm phot}$ based
on the wrong type (RRc or RRab) and the correct [Fe/H]$_{\rm phot}$ value
(column\,9 of Table\,8) is plotted in the top panel of {\bf Fig.\,12b}
(right panel).  A similar graph is plotted in the lower panel for the {\it
Gaia} stars in Table\,9 that were misclassified in DR2 or DR3 and for which
[Fe/H]$_{\rm phot}$ (calculated using the same Nemec et al. calibrations - see
$\S$5.1 of Clementini et al. 2023) is given in the respective {\it Gaia}
catalogue.  The error due to misclassification, $\Delta$[Fe/H]$_{\rm phot}$, is
the difference between the {\it Gaia} estimate and that given in Table\,9.  In
general, misclassifying a cRRd star as `RRab' leads to systematic
overestimation of [Fe/H] while misclassification as `RRc' leads to systematic
underestimation.  Fig.\,12b also shows that the bias is more serious in the
case of RRab stars than in the case of RRc stars, and tends to increase with
decreasing metal abundance (increasing period).

\section{SUMMARY}

Seventy-five double-mode RR~Lyrae (RRd) stars  observed  by the {\it Kepler}
space telescope during NASA's {\it K2} Mission have been identified and
studied.  Seventy-two of the stars are `classical' RRd (cRRd) stars with period
ratios $P_1$/$P_0$$\sim$0.745, and, in most cases, amplitude ratios
$A_1/A_0$$>$1; none of the cRRd stars shows evidence  of Blazhko amplitude or
phase modulations.  The other three stars are  `anomalous' RRd (aRRd) stars.

High precision periods, amplitudes, and Fourier parameters were derived for the
72 cRRd stars.  Within- and between-mode correlations among the periods,
amplitudes, and four low-order Fourier parameters ($R_{\rm 21}$, $R_{\rm 31}$,
$\phi_{\rm 21}$ and $\phi_{\rm 31}$) were analyzed.    The results show that
the within-mode period-amplitude relationships differ significantly for the two
pulsation modes (see Fig.6).  The first-overtone {\it Kp}-amplitude tends to be
around 175 mmag and decreases slightly with period, while the fundamental-mode
amplitude, which is almost always lower, decreases more rapidly.  Comparison
with \texttt{OGLE} data found a similar result for the period range where the
two samples overlap.   These findings are consistent with the observed increase
in $A_1/A_0$ with increasing period (seen in Fig.2).  The within-mode
dependencies of the Fourier parameters on period were also investigated and
compared for the two pulsation modes.  Both  $R_{\rm 21}$ and $R_{\rm 31}$ show
a significant positive correlation with period for the first overtone but not
for the fundamental mode (see Fig.7).  Neither $\phi_{\rm 21}$ nor $\phi_{\rm
31}$ shows a clear dependency on period for either pulsation mode (see Fig.8).  

Three cross-mode correlations and their relationships to [Fe/H] are of
particular interest:   $P_1$ vs $P_0$, $A_1$ vs $A_0$, and $R_{\rm 21,1}$ vs
$R_{\rm 21,0}$.  The $P_1$-$P_0$ diagram (see fig.15a of NM21) and 
Petersen diagram (see Fig.10c) were modelled by relating the two periods to
[Fe/H].  An [Fe/H] calibration equation was derived from the same model.  In
the $A_1$-$A_0$ plane (see Fig.9a) most of the {\it K2} cRRd stars have
$A_1$$\sim$175 mmag regardless of the $A_0$ value, with a smaller fraction of
the stars having lower $A_1$ amplitudes, and a few with higher amplitudes.  The
same pattern was found in  \texttt{OGLE} data (see fig.5b of NM21).
The $R_{\rm 21,1}$ and $R_{\rm 21,0}$ parameters show a different pattern of
correlation: when the stars are sorted into three period classes a stratified
linear relationship emerges (see Fig.\,9c).  For all three period classes
$R_{\rm 21,1}$ increases linearly with $R_{\rm 21,0}$, with approximately
constant slopes and offsets that depend on period.  Since [Fe/H] depends on
period the $R_{\rm 21,1}$-$R_{\rm 21,0}$  diagram might be used in conjunction
with period for metal abundance estimation of cRRd stars.  No separation of the
three period classes was evident in the $A_1$-$A_0$ plane.

A  sample of 268 cRRd stars with known spectroscopic metal abundances (see
Fig.\,10) was used to derive $P_0$-[Fe/H] and  $P_1$-[Fe/H] calibration
equations for cRRd stars (Eqn.\,6).  The $P_0$ version of these was used to
estimate metallicities for the full sample of 72 {\it} {\it K2} cRRd stars
(Table\,8) and for 2130 cRRd stars in the {\it Gaia} DR2 and DR3 catalogues
(Fig.\,11 and Table\,9).  Forty-nine of the 72 {\it K2} cRRd stars are in the
Gaia DR3 catalogue.  Of these, 84$\%$ are misclassified `RRc', 6$\%$ are
misclassified `RRab', and only 10$\%$ are correctly classified `RRd'.  The
resulting metallicity bias when the wrong calibration curve is used (see
Fig.\,12b) was found to be more serious when cRRd stars were misclassified
`RRab' than when they were misclassified `RRc', with the error tending to
increase with decreasing metal abundance (i.e, increasing period).

\section*{Acknowledgements} Funding for the {\it Kepler}/{\it K2} Mission was
provided by the NASA Science Mission directorate.   JMN thanks International
Statistics \& Research Corporation and the Camosun College Faculty Association
for supporting his travel to various {\it Kepler} conferences.  He acknowledges
interesting discussions with Rados\l{}aw Poleski, Johanna Jurcsik and Geza
Kov\'acs.  The research was also supported by the `SeismoLab' KKP-137523 \'Elvonal
grant of the Hungarian Research, Development and Innovation Office (NKFIH).  \\
\\
\\

\noindent \textsc{\bf DATA AVAILABILITY} \\
The data underlying this article are available from the MAST website,
at https://archive.stsci.edu/k2/, and all the datasets were derived from
sources in the public domain. \\

\vfill \eject

\vskip 0.5truecm

\noindent {\bf SUPPORTING INFORMATION} \\ 
Supplementary files available at {\it MNRAS} online:  \\ [0.2cm]
\noindent {\bf Appendix A} - Notes on Individual Stars   \\  [0.1cm]
\noindent {\bf Appendix B} - Fitted light curves for the RRd stars \\ observed during {\it K2} Campaigns 7-18 (Figs. B1-B10);  \\ [0.1cm]
\noindent {\bf EPIC205209951\_animation.gif} - Animation showing the light variations of both modes of the aRRd star EPIC\,205209951 (see Figs.\,A1-A3 in Appendix A); \\ [0.1cm]
\noindent {\bf Table 9} - Metallicities and masses for 2130 {\it Gaia} stars. \\ [0.1cm]
\noindent {\bf Table\,10} - Ninety-three {\it Gaia} `RRd stars' with different DR2 and DR3 identification numbers.
\clearpage


\vfill \eject 

\appendix \section{Notes on Individual Stars} 

The following notes are ordered by Ecliptic Plane Input Catalog (EPIC) number.
{\it K2} Campaign numbers are given in parentheses, with underlined campaign numbers
indicating that short cadence as well as long cadence {\it K2} observations were
available for analysis.  The non-radial pulsations that are present in
all of the well-studied stars (see Moskalik et al. 2018a,b) will be discussed elsewhere.    

\vskip0.5truecm

%

%



\noindent {\bf EPIC\,60018653} (E2) - The periods and amplitudes given in
Table~3 (top row) are from the detailed analysis performed by Moln\'ar {\it et
al.} (2015) of data from the 8.9-day {\it K2} Two-Wheel-Concept Engineering
Run.  The more precise periods given in the bottom row were derived by
combining the K2-E2 and CSS photometry.  The {\it K2} and CSS mean magnitudes
were expected to be similar, but the mean Kp magnitude was found to be
$\sim$0.5 mag brighter than the mean $V$ magnitude, 13.765 vs. 14.210 mag (this
was also the case for the other RRd star observed during the Engineering Run,
EPIC\,60018662), discussed next). 
  \\

\noindent {\bf EPIC\,60018662} (E2) - The long periods are consistent
with the low spectroscopic metallicity derived by the \texttt{SDSS/SEGUE }pipeline from a
single SDSS spectrum: [Fe/H] = --2.18$\pm$0.08 dex (see Table~7).  The mean Kp magnitude
derived using the Lund {\it et al.} formula is brighter by $\sim$0.5 mag than
the mean $V$ magnitude derived using the CSS $V$ photometry, 14.22 versus 14.76
mag.  Unfortunately Kp magnitudes are not given in the EPIC catalog (MAST
website) for the stars observed during the Engineering Run.  
 \\

\noindent {\bf EPIC\,201152424}\,(C10) - After prewhitening with the
first-overtone frequency, $f_1$, there is considerable unremoved power left in
the Fourier amplitude spectrum.   \\

\noindent {\bf EPIC\,201440678}\,(C10) = V369\,Vir = Zinn {\it et al.} (2014)
star 436 = Vivas {\it et al.} (2004) star 177.    The period given in the GCVS
(Samus 2017), 0.604369\,d, is not correct, and in other papers the star is not
recognized as an RRd star.  It is not among the RRd stars found by Poleski
(2014) in his systematic search for RRd stars using \texttt{LINEAR} data.  
The {\it K2} photometry clearly shows it to be a `classical' RRd star.  See Table\,7 for \texttt{SDSS/SEGUE} physical
characteristics.  \\

\noindent {\bf EPIC\,201519136}\,(C10) = \texttt{LINEAR}\,3384231  - The RRd nature of this star was
discovered by Poleski (2014) during his reanalysis of \texttt{LINEAR}
photometry.  The mean magnitudes derived from the {\it K2} fluxes
(Kp=17.03\,mag derived here and 17.019\,mag derived by Moln\'ar {\it et al.}
2018) agree with  G=17.024 mag (range 16.2-17.6) from the {\it {\it Gaia}}\,(DR2)
survey;  the mean magnitude given in the EPIC catalog, Kp(MAST)=17.947\,mag, is
considerably fainter.   The periods and period ratio derived by Poleski are
confirmed by the {\it K2} data.  Table~7 contains physical
characteristics from the \texttt{SDSS/SEGUE} pipeline.   \\

\noindent {\bf EPIC\,201585823}\,(C\underline{1})\,=\,\texttt{LINEAR}\,2122319 - The RRd nature was
discovered by Poleski (2014) using 519 \texttt{LINEAR} brightness measurements.
Reanalysis of the \texttt{LINEAR} data revealed 14 frequencies more significant
than \texttt{sig}$>$5, mostly combination frequencies.  The  mean Kp magnitudes
derived from the {\it K2} fluxes, 15.73 mag derived here and 15.774 mag
derived by Moln\'ar {\it et al.} (2018), are consistent with the EPIC catalog
value, 15.83 mag, and all are similar to the {\it {\it Gaia}}\,(DR2) mean
magnitude, G=15.839 mag (range 15.2-16.3).  See Table\,7 for \texttt{SDSS/SEGUE}
physical characteristics.   A detailed analysis of the  {\it K2} short-cadence
photometry was performed by Kurtz {\it et al.} (2016). \\

\noindent {\bf EPIC\,201749391}\,(C\underline{14}) = \texttt{LINEAR}\,23675270 - Similar periods and period
ratios are derived from the {\it K2}, CSS and \texttt{LINEAR} photometry.   The
amplitudes derived from the long-cadence photometry (\texttt{EAP} pipeline) are
slightly larger than those derived from the short-cadence photometry
(\texttt{PyKE} pipeline), with little effect on the period and amplitude
ratios.   \texttt{SDSS/SEGUE} physical characteristics are given in Table~7.  \\



\begin{figure}  
\begin{center}
	\begin{overpic}[width=7.8cm]{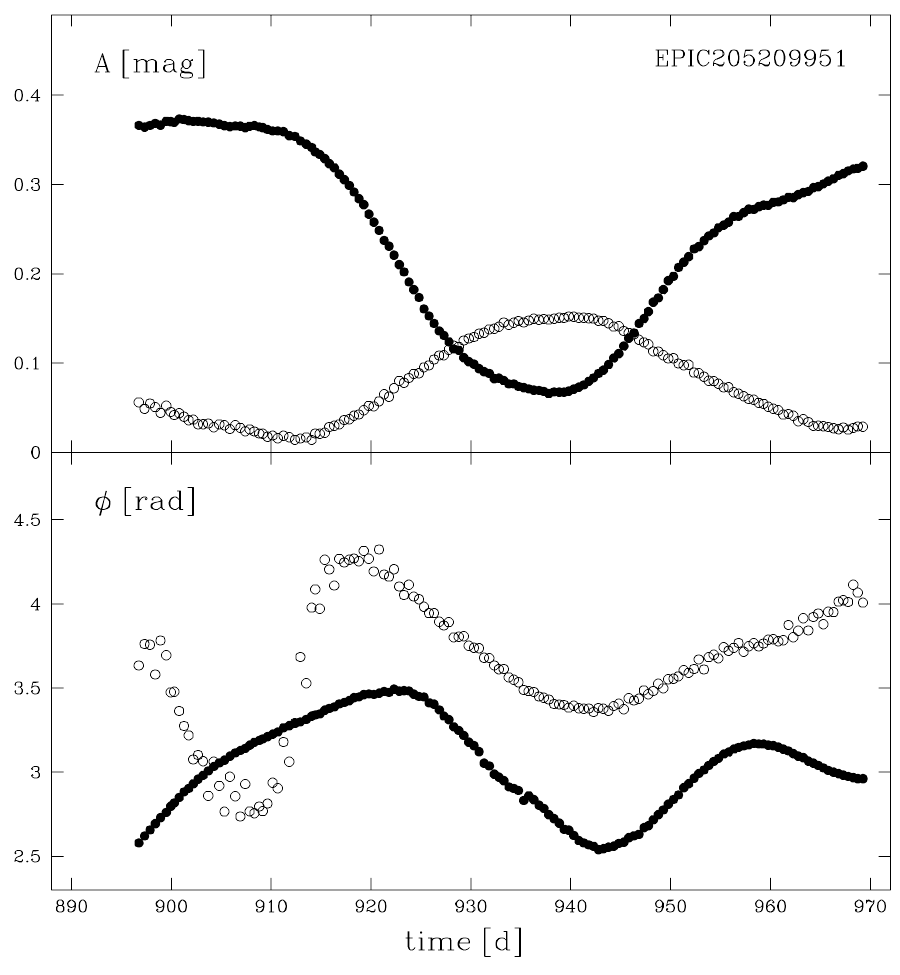}  \put(-4,76){Kp}  \put(80,85){$A_0$}   \put(80,62){$A_1$}  \put(80,42){$\phi_1$} \put(80,16){$\phi_0$}    \end{overpic}
\end{center} 

\caption{Amplitude and phase variations for the anomalous RRd star
	EPIC\,205209951, showing changes as a function of time for the
	fundamental and first-overtone radial modes (similar to fig.\,3 of
	Plachy et al. 2017, but with higher resolution).  Because the Blazhko
	modulation period is somewhat longer than the 73-day
	length of the observing run it was necessary to use time-dependent
	prewhitening to construct the graph.  The times are BJD minus 2456000.}	

\label{Fig_A1}  \end{figure}

\noindent {\bf EPIC\,205209951}\,(C\underline{2}, aRRd) - In the Petersen
diagram this `anomalous' RRd star lies below the curve defined by the classical
RRd stars, its light curve is dominated by the fundamental mode, and it
exhibits Blazhko variations of the amplitudes and phases (see Plachy et al.
2017b).   The mean {\it Kp} magnitude is $\sim$14.70, which is brighter than the
mean magnitude given in the EPIC catalog, {\it Kp}(MAST)=14.91\,mag.  In the
{\it Gaia} DR2 and DR3 catalogues its identification number is
6248239227324924416.  The amplitude and phase variations were investigated by
Plachy {\it et al.} (2017b).   {\bf Figure A1} is a higher resolution version of
the figure given there, showing that the amplitudes of the two modes are
anticorrelated: the overtone is high when the fundamental mode is low, and vice
versa.  {\bf Figure A2} extends the Plachy et al.  analysis to variations of
the $R_{\rm 21}$ and $\phi_{\rm 21}$ Fourier parameters for each mode.  For the
first-overtone the $R_{\rm 21}$ and $\phi_{\rm 21}$ values do not change very
much;  however, for the fundamental mode the variation is very pronounced
showing that when the amplitude goes through a minimum the $\phi_{\rm 21}$
increases by $2 \pi$.  At the same time the amplitude ratio, $R_{\rm 21}$ also
shows a very pronounced variability.  The variability of $A$, $R_{\rm 21}$ and
$\phi_{\rm 21}$ look like going through the resonance center, which is very
curious.  These amplitude and phase variations are readily seen in the 
animated gif file given in the Supporting Information.  Ten frames from the
animation are shown in {\bf Figure A3}.  One sees a bump showing up first on
the descending branch of the fundamental-mode light curve, growing, then trading places with the
main maximum.  The original maximum then becomes a bump on the ascending branch
of the lightcurve and finally disappears.  This sequence resembles the
Hertzsprung progression seen for classical Cepheids (especially for their
radial velocity curves).  This behaviour again suggests a crossing of the
resonance centre.  This is the first time that the modulation of an aRRd star
has been seen in such great detail.  \\

\begin{figure}  
\begin{center}
\begin{overpic}[width=8.5cm]{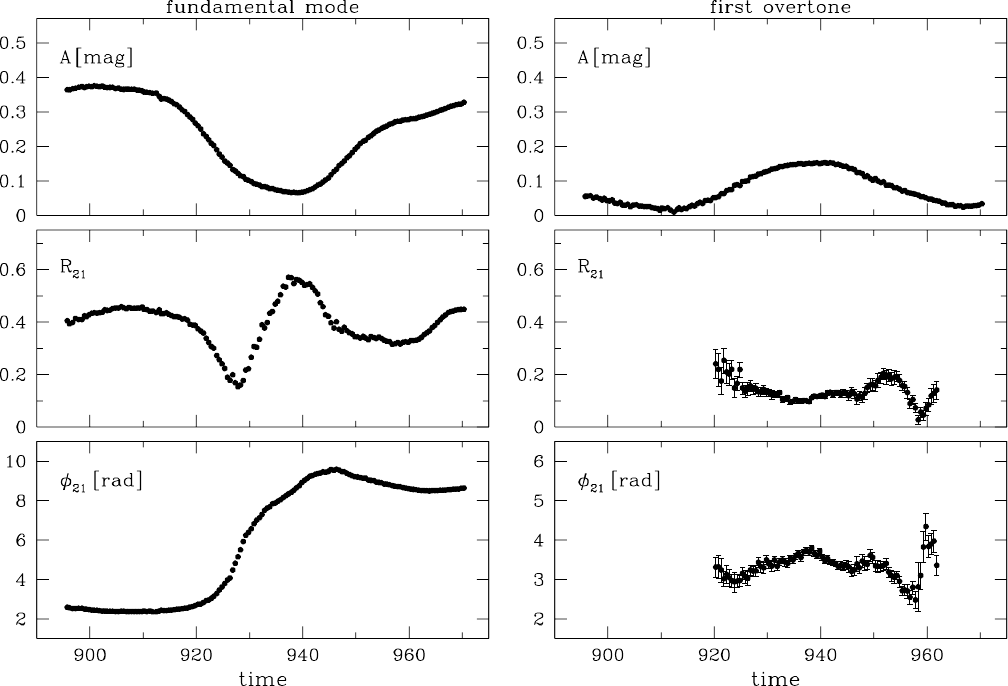}  \end{overpic}
\end{center} 
\caption{EPIC\,205209951\,(C2, aRRd) - Graphs showing the changes as a function of time of the {\it Kp} light curves and Fourier parameters
for the fundamental mode (left) and first-overtone mode (right).  The times are BJD minus 2456000.  } 	
\label{Fig_A1}  \end{figure}

\begin{figure*}  
\begin{center}
\begin{overpic}[width=16.0cm]{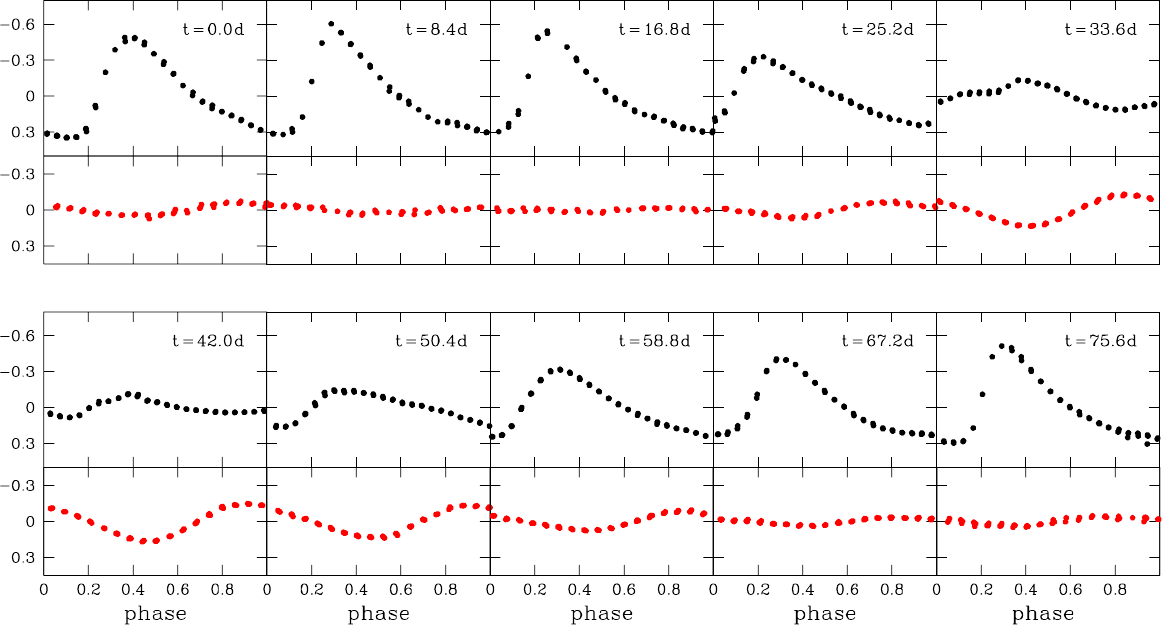} \put(-3,14){Kp}  \put(-3,40){Kp}  \end{overpic}
\end{center} 

\caption{Light curve variations for the aRRd star EPIC\,205209951, where the
	ten different frames were extracted from the `animated gif' file
	included in the Supporting Information.   For each frame the top panel
	shows the fundamental mode ($P_0$=0.470741\,d) light curve (black) and
	the bottom panel shows the first-overtone mode ($P_1$=0.348780\,d)
	light curve (red).  Note how the light curve shapes change, in
	particular, the anticorrelation of the amplitudes of the two modes.   }

\label{Fig_A2}  \end{figure*}

\vskip0.2truecm

\noindent {\bf EPIC\,210600482}\,(C4) -  The best nonlinear least-squares fit
to the K2SC-EAP photometry found residuals that are not normally distributed.
In Fig.\,1 of Plachy {\it et al.} (2017a)  the EAP photometry is compared with
photometry from other pipelines.   \\

\noindent {\bf EPIC\,210831816}\,(C4) - The mean magnitude derived from the
{\it K2} fluxes, Kp=15.56 mag, and that  derived by Moln\'ar {\it et al.}
(2018), Kp=15.541,  agree with the {\it {\it Gaia}}\,(DR2) mean magnitude, G=15.504
mag (range 14.8-16.2), while the mean magnitude given in the EPIC catalog is
considerably fainter, Kp(MAST)=15.85 mag.   A time variability analysis found
constancy of the radial modes.  \\

\noindent {\bf EPIC\,211665293}\,(C16) = \texttt{LINEAR}\,5974119 -  See Table\,7 for \texttt{SDSS/SEGUE} physical
characteristics.   The \texttt{PyKE}-detrended {\it K2} photometry is excellent.    \\

\noindent {\bf EPIC\,211694449}\,(C5, C\underline{18}) - The {\it K2} and CSS
photometry both show that this star lies on the `classical' RRd curve (Petersen
diagram) among the stars with intermediate periods, consistent with the
\texttt{SDSS/SEGUE} spectroscopic metallicity, [Fe/H]$\sim$--1.3 dex.  However, the dominant
radial pulsation is the fundamental mode, which is unusual for a `classical'
RRd star but not for an `anomalous' RRd star.  Since no Blazhko modulations are
seen and since the fundamental mode amplitude is only 10$\%$ larger than the
first-overtone amplitude we tentatively conclude that it as a `classical'
RRd star.    See Table\,7 for \texttt{SDSS/SEGUE} physical
characteristics.  \\

\noindent {\bf EPIC\,211888680}\,(C5,16) - The {\it {\it Gaia}}\,(DR2) mean
magnitude, G=18.786 mag (range 18.0-19.6 mag), compares favourably with that
derived from the {\it K2} fluxes, Kp=18.91 mag.  Owing to its faintness the
secondary fundamental-mode pulsation was not detectable in a reanalysis of the
Catalina Sky Survey data.  The \texttt{SDSS/SEGUE} pipeline (see Table\,7) measured a spectroscopic
[Fe/H]=--1.68$\pm$0.04 dex. An `extended aperture photometry' light curve is
shown in fig.11 of Bodi et al. (2022).   \\

\noindent {\bf EPIC\,211898723}\,(C\underline{5},\underline{18}) - The radial
fundamental mode (which was not detected in the CSS data) is clearly resolved
by the {\it K2} photometry.
See Table\,7 for \texttt{SDSS/SEGUE} physical
characteristics;  the radial velocity derived from a single SDSS spectrum
was found to be very high, $+343\pm3$\,km/s (Lee {\it et al.} 2008a,b). \\

\noindent {\bf EPIC\,212449019}\,(C6) -
The {\it {\it Gaia}}\,(DR2) mean magnitude, G=16.517 mag (range
15.9-17.0 mag), agrees with that derived from the {\it K2} fluxes, Kp=16.52
mag.  \\

\noindent {\bf EPIC\,212455160}\,(C6,\underline{17}) - Unfortunately many of
the C17 images of this star were located on the edge of `CCD module 20' hence
much of that photometry is questionable.  The most reliable C17 observations
were made between BJD\,2458199 and 2458239, but even
these measurements have $\sim$30$\%$ smaller amplitudes and much larger residuals
than the C6 data.  For these reasons our results are based on the C6
observations only.  
  \\

%



\noindent {\bf EPIC\,212467099}\,(C17) -  From a reanalysis of the Catalina Sky
Survey photometry only the dominant first-overtone frequency could be
recovered.  The \texttt{PyKE} detrending of the {\it K2} photometry was found
to be superior to the \texttt{PDCsap} data.  In the Petersen diagram the star
is located slightly above the classical RRd curve. Neither dataset showed
evidence for Blazhko modulations, and since the first-overtone dominates over
the fundamental mode we conclude that this star probably is a `classical'
(rather than `anomalous') RRd star.  
  \\

\noindent {\bf EPIC\,212547473}\,(C\underline{6},\underline{17}) -  The long
pulsation periods suggest that the metallicity is very low.   The merging of
the C6 and C17 short-cadence photometry successfully resulted in very precise
periods and amplitudes (see Table 3).  
The {\it {\it Gaia}}\,(DR2) mean magnitude, G=15.674 mag (range 15.0-16.2 mag),
compares favourably with that derived by Moln\'ar {\it et al.} (2018) from the
{\it K2} fluxes, Kp=15.619 mag.   \\

\noindent {\bf EPIC\,212615778}\,(C17) -  There is almost a 0.9 mag difference
between the MAST {\it Kp} magnitude and that derived from our Pyke:LC extraction. \\

\noindent {\bf EPIC\,212819285}\,(C17) - After prewhitening with $f_1$ =
2.8293\,d$^{-1}$ significant excess power is seen at $f$=2.8354 c/d.   \\

\noindent {\bf EPIC\,213514736}\,(C7)\,=\,vs11f158\,(Cseresnjes\,2001)\,- The
double-mode nature of the six RRd stars observed during Campaign\,7 were
discovered by Cseresjnes.  The brightness  of this star strongly suggests that
it lies in the foreground of the Sagittarius dwarf galaxy.  The mean magnitude
derived here, $<${\it Kp}$>$=17.46 mag,  and $<${\it Kp}$>$=17.208 derived by
Moln\'ar {\it et al.} (2018), are both similar to {\it Kp}(MAST)=17.295 mag.
Only the 46.5 days of {\it K2} photometry before BJD~2457348 were used in our analysis owing
to significantly larger photometric variance after this time.  The period
ratios derived by Cseresjnes and from the {\it K2} photometry are similar.  \\

\noindent {\bf EPIC\,214147122}\,(C7)\,=\,vs1f148\,(Cseresnjes 2001) - This
star is the brightest of the six cRRd stars observed during Campaign~7.  The
mean magnitude derived from the {\it K2} fluxes, {\it Kp}=15.90 mag, agrees
well with that from {\it {\it Gaia}}\,(DR2), {\it G}=15.832 mag (range
15.0-16.4). This brightness strongly suggests that it lies in the foreground of
the Sagittarius dwarf galaxy.  The periods, period ratios and amplitude ratios
derived from the {\it K2} data are in accord with the values given by
Cseresnjes. \\

\noindent {\bf EPIC\,220254937}\,(C8) - A single spectrum was taken by the \texttt{SDSS}
(see Table\,7), with radial velocity --135$\pm$3 Km/s, but no physical
characteristics were given by the \texttt{SEGUE} pipeline.   \\

\noindent {\bf EPIC\,220636134}\,(C\underline{8}) - The mean magnitudes derived
from the {\it K2} fluxes (Kp=17.36 mag derived here,  Kp=17.390 derived by
Moln\'ar {\it et al.} 2018),  Kp(MAST)=17.299 from the EPIC catalog, and G=17.320
mag (range 16.6-17.8) from the {\it {\it Gaia}}\,(DR2) survey, are all in agreement.
Slightly brighter is Kp=17.16 mag based on the \texttt{Everest} (Luger et al. 2016)  short cadence photometry
(mean flux 1803 e$^{-1}$/s).   
  \\

\noindent {\bf EPIC\,223051735}\,(C9)\,=\,\texttt{OGLE-BLG-RRLYR-12804} - The
{\it K2} photometry of this low-[FeH] Galactic Bulge cRRd star suffers from
significant contamination by a close red star to the East (see the
\texttt{OGLE-IV} finding chart) which is almost certainly responsible for the
excessively bright mean magnitude given in the EPIC catalog, Kp=13.47 (Table
3).  With a mean PDCsap flux level of 3376 e$^{-1}$/s the Kp magnitude derived
using the Lund {\it et al.} formula is very similar to the mean $V$ magnitude
from the \texttt{OGLE} photometry (16.48 vs. 16.50 mag).  The precise pulsation
periods given throughout the present paper were determined from the combined
{\it K2} (2016) and \texttt{OGLE-III,IV} (2001-09, 2010-17) data sets shown in
{\bf Figure\,A4}.  The \texttt{OGLE-IV} photometry appears to suffer a
faintward photometric drift over the last several years of observations, a
trend shared with EPIC\,225045562.  \\

\begin{figure} 
\begin{center}

\begin{overpic}[width=8.1cm]{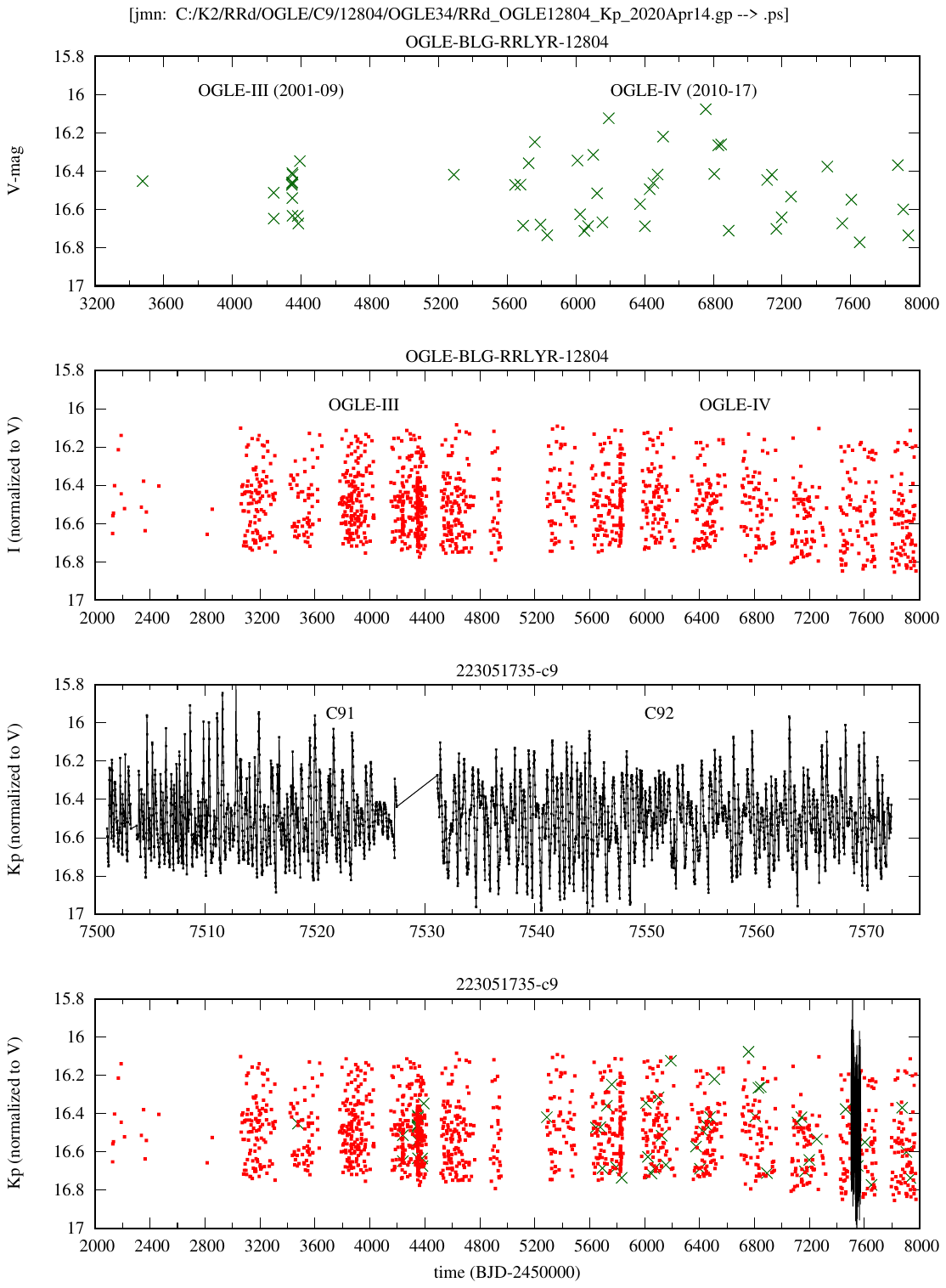}   \end{overpic}
\end{center} \caption{EPIC\,223051735(C9) - Comparing the  \texttt{OGLE-III,IV}
photometry (green crosses for the $V$-photometry, red dots for the
$I$-photometry transformed to the $V$-passband) and the {\it K2} photometry
(transformed to the $V$-passband). } 	

\label{fig:A1} 
\end{figure}

\noindent {\bf EPIC\,224366356}\,(C9)\,=\,\texttt{OGLE-BLG-RRLYR-36501} -  The
$A_1$ and $A_0$ {\it Kp}-amplitudes are similar.  The \texttt{OGLE-IV} $V$- and
$I$-photometry (2010-2017) are compared with the {\it K2} photometry in {\bf
Figure\,A5}.  \\

\begin{figure} 
\begin{center}
\begin{overpic}[width=8.1cm]{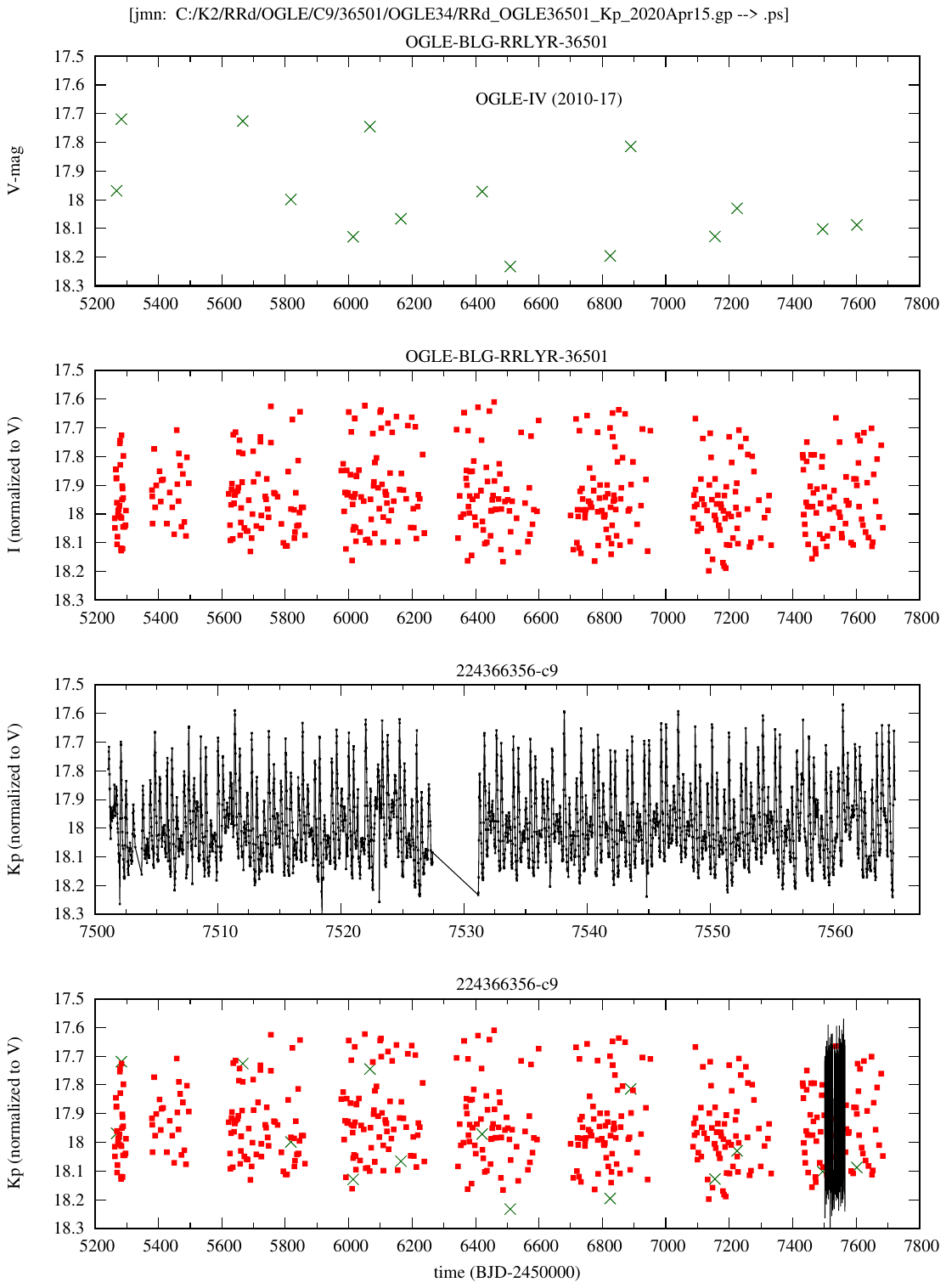}   \end{overpic}
\end{center} 

\caption{EPIC\,224366356 (C9) - Comparing the \texttt{OGLE-IV} $V$- and
$I$-photometry with the {\it K2} photometry.  To make the comparison the $I$-
and {\it Kp}-photometry were transformed to the $V$-passband.    } 	

\label{Fig_A2} 
\end{figure}





\noindent {\bf EPIC\,225045562}\,(C9, aRRd)\,=\,\texttt{OGLE-BLG-RRLYR-02530} -
This star was identified as an aRRd star by the \texttt{OGLE-III} survey.  The
Petersen diagram clearly shows its period ratio, $P_1/P_0$=0.735, to be much
lower than that expected for a cRRd star with $P_1$=0.34\,d.  Smolec {\it et
al.} (2015a) analyzed the 2010-2013 \texttt{OGLE-IV} photometry and found the
fundamental mode dominating over the first-overtone mode
($A_0$/$A_1$=95.6/67.3=1.42) and long-period Blazhko modulations
($P_B\sim469$\,d), both characteristics being typical for an aRRd star.  The
more recent \texttt{OGLE-IV} data (through 2017) and our {\it K2} photometry
both confirm the low period ratio and support the Blazhko conclusion (with the
\texttt{OGLE-IV} $I$-photometry from 2010 to 2017 showing the variance
increasing).  An additional faintward drift with the mean $I$ magnitude
increasing from 14.82 to 14.90 is also seen.   However, by the time of the
Campaign 9 observations in 2016 the first-overtone mode appears to have become
dominant with $A_1$/$A_0$=1.5.  We conclude that either the dominant pulsation
mode changed between 2010-2013 and 2016, or one or both of the data sets is
problematic.  Given that the finding charts reveal very close and bright
neighbouring stars, probably also causing the \texttt{EPIC} $<$Kp$>$ magnitude
to be almost five magnitudes brighter than that inferred from the
\texttt{OGLE-IV} photometry (13.69 versus 18.07 mag!) the latter conclusion
seems more likely.  A comparison of the OGLE\,II-IV $V$ and $I$ photometry and
the {\it Kepler/K2} Kp-photometry, all transformed to the $V$-passband, is
shown in {\bf Figure A6}.  \\

\begin{figure}  \begin{center}
\begin{overpic}[width=8.1cm]{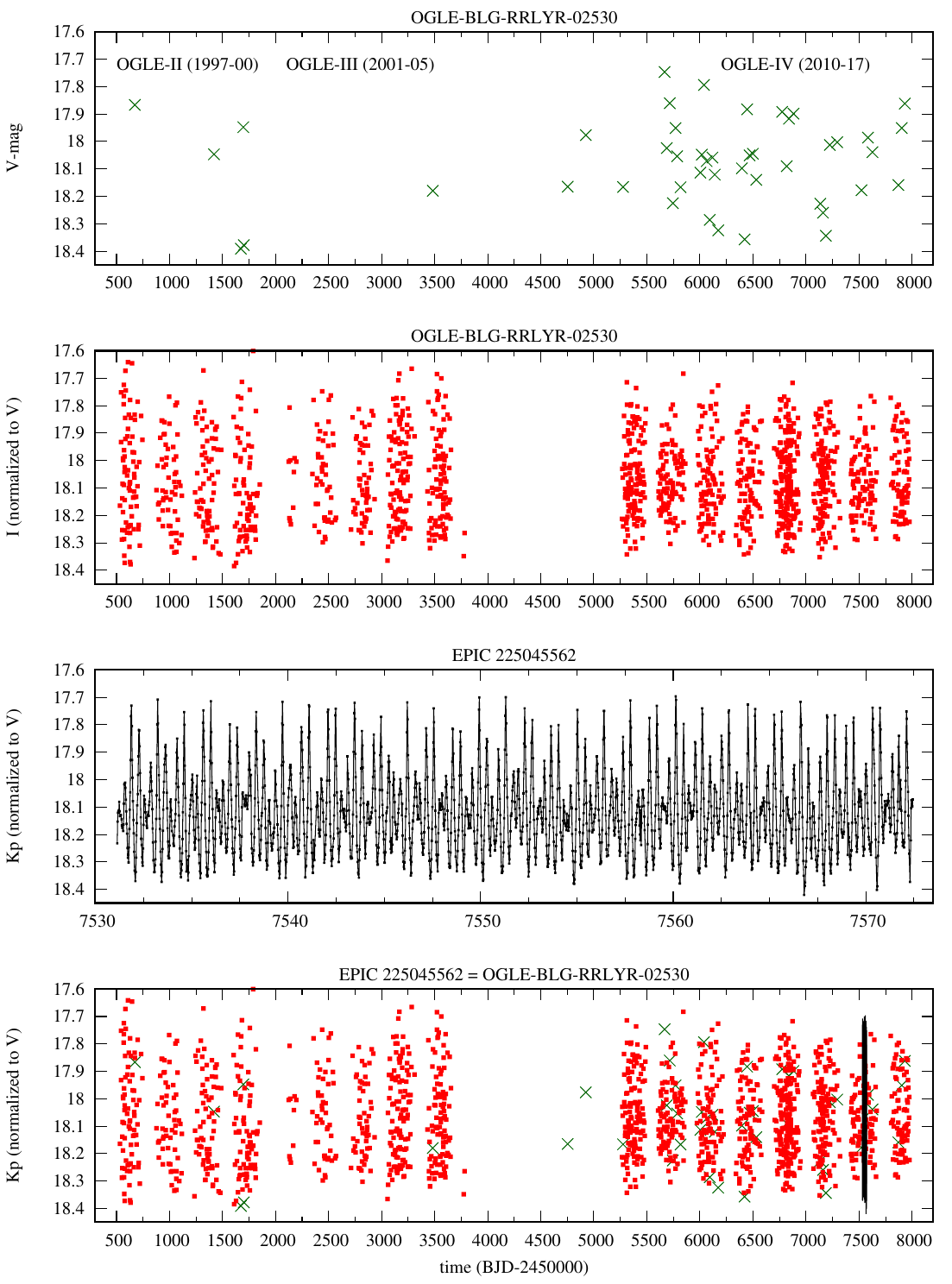} \end{overpic}
\end{center} 

\caption{EPIC\,225045562\,(C9) -  Comparison of the 1997-2017 \texttt{OGLE-II,III,IV} photometry
(green crosses for $V$ photometry, red dots for $I$ photometry transformed to
$V$-passband) and the {\it K2} photometry (transformed to the $V$-passband). } 	

\label{Fig_A3}  \end{figure}

\noindent {\bf EPIC\,225326517}\,(C11)= \texttt{OGLE-BLG-RRLYR-00978} - This
cRRd star has the third longest (first-overtone) period in the {\it K2} sample and is very metal poor.  The optimum data set
was the combined the \texttt{OGLE} and {\it K2} photometry. \\

\noindent {\bf EPIC\,225456697}\,(C11) =  \texttt{OGLE-BLG-RRLYR-21574} - This
Galactic Bulge star is the shortest-period RRd star in the {\it K2} sample, with [Fe/H]$_{\rm phot}$$\sim$0.8 dex.   
The amplitudes of both radial modes are low, almost
certainly due to instrumental problems.  A very weak
signature of modulation of $P_1$ was seen, with modulation period 11.9\,d. \\

\noindent {\bf EPIC\,228800773}\,(C10) -  The {\it K2} photometry of this
18.3-mag star is contaminated by the light of an 18.6 mag star to the west (see
\texttt{Everest} image plots).  The light curves derived using the Catalina survey photometry
(266\,CSS, 89\,MLS and 67\,SSS observations) are consistent with the {\it K2}
first-overtone period, $P_1$=0.372750\,day.  See Table\,7 for \texttt{SDSS/SEGUE} physical
characteristics. \\

\noindent {\bf EPIC\,228952519}\,(C10) - The {\it K2} data collection ended
prematurely when Module~4 failed, resulting in only 7.2\,d of long cadence
photometry (C102).  Thus the periods and amplitudes are uncertain.  Also
uncertain are the mean magnitudes, which show a considerable range:  from the
{\it K2} fluxes, Kp=17.90 mag (here) and 18.500 (Moln\'ar {\it et al.} 2018);
from the EPIC catalog Kp(MAST)=18.193;  and from the {\it {\it Gaia}}\,(DR2) survey,
G=18.053 (range 17.2-18.6 mag).   These photometric differences probably arise from
contamination by a close neighbour $\sim$5 arcsec to the southeast of this
star.  Combining the C102 and C101 data resulted in 650 data points acquired
over 13.44 days.  The limited {\it K2} data are in accord with the CSS period
of $P_1$=0.40349\,d.   \\

\noindent {\bf EPIC\,229228175}\,(C7)\,=\,vs2f266 (Cseresnjes 2001) is in the Sagittarius dwarf galaxy.  The {\it
K2} photometry suffers from crowding by a nearby relatively bright star located
to the East of this star (see \texttt{Everest} image panels), which probably explains
why the mean magnitude from {\it {\it Gaia}}\,(DR2), {\it G}=18.531 mag (range
17.6-19.2), is significantly brighter than that derived from the {\it K2}
fluxes, {\it Kp}=18.86 here, and {\it Kp}=19.18 from Moln\'ar {\it et al.}
(2018).  The {\it K2} data acquired after BJD 2457353, and between
2457323.5 and 2457324.3, were found to be unreliable and were not used.
Also, the {\it Kp} magnitude range is unusually low.  Despite these problems
the period and amplitude ratios derived from the {\it K2} data agree well with
Cseresjnes' values derived from photographic plates.  \\

\noindent {\bf EPIC\,229228184}\,(C7)\,=\,vs1f243 (Cseresnjes 2001) is in the Sagittarius dwarf
galaxy.  The periods and amplitudes derived from the {\it K2}
photometry are in accord with (and more accurate than) the values derived by
Cseresjnes from photographic plates.   Two close bright stars appear to have
caused the detrending to be poor,  the considerable amplitude variation seen in
the PDCsap data (which probably is instrumental), and may explain why the {\it
{\it Gaia}}\,(DR2) mean magnitude, {\it G}=18.162 mag (range 17.4-18.8 mag), is
considerably brighter than that derived from the {\it K2} fluxes, {\it
Kp}=18.79 mag, but is consistent with the MAST value, {\it Kp}(MAST)=18.10 mag.
Similar problems exist for EPIC\,229228194.  \\

\noindent {\bf EPIC\,229228194}\,(C7)\,=\,vs1f242 in the Sagittarius dwarf
galaxy (Cseresnjes 2001) - The RRd nature of this star was discovered by
Cseresjnes.  The periods derived from the {\it K2} photometry (PDCsap data) are
in accord with those given by Cseresjnes from $B$ photometry.  Owing to the
star's faintness, $<${\it Kp}$>$=19.3 mag, and probable contamination by
neighbouring stars, the $A_1$ and $A_0$ amplitudes given in Table\,3 probably
are too large.  \\

\noindent {\bf EPIC\,229228220}\,(C7)\,=\,vs3f170 in the Sagittarius dwarf
galaxy (Cseresnjes 2001) - The mean {\it Kp} magnitude is uncertain: Huber {\it
et al.} (2016) give 19.20 mag, whereas the \texttt{PDCsap} fluxes transformed
with {\it Kp}$=25.3-2.5\log(F)$ give 19.97 mag.  Thus the \texttt{PDCsap}
amplitudes are uncertain.  The \texttt{Pyke} data were normalized to 19.20 mag,
the mean magnitude given in the EPIC catalog, and the \texttt{Pyke} residuals
were found to be significantly lower than the \texttt{PDCsap} residuals, 32
mmag {\it vs.} 89 mmag.  \\

\noindent {\bf EPIC\,229228811}\,(C8) - This star previously was observed by
the Catalina Sky Survey (302\,CSS and 172 MLS epochs), where the mean $V$
magnitudes from the CSS and MLS data sets differ by 0.17 mag.  The mean
magnitudes derived from the {\it K2} fluxes, Kp=18.14 mag derived here and
Kp=18.651 derived by Moln\'ar {\it et al.} (2018), are both fainter than
Kp(MAST)=17.700 mag;  these man magnitudes are to be compared with that from
the {\it {\it Gaia}}\,(DR2) survey, G=18.050 mag (range 17.4-18.6).  The dominant
periods derived from the CSS and {\it K2} photometry are in good agreement,
0.372905\,d (CSS) versus 0.372910\,d ({\it K2}). \\

\noindent {\bf EPIC\,235631055}\,(C11) = \texttt{OGLE-BLG-RRLYR-17146} -  With
$P_1$=0.43278\,d this is one of the longest period `classical' RRd stars known,
hence [Fe/H] probably is very low.  \\

%



\noindent {\bf EPIC\,235794591}\,(C11) = \texttt{OGLE-BLG-RRLYR-00727} -
Long-period `classical' RRd star, probably quite metal poor.  \\

\noindent {\bf EPIC\,236212613}\,(C11) = \texttt{OGLE-BLG-RRLYR-16881} - The radial periods were derived by analyzing the 
combined \texttt{OGLE-IV} and {\it K2} photometry.  \\

\noindent {\bf EPIC\,245974758}\,(C\underline{12}, C\underline{19}) - For all
three C12 RRd stars the \texttt{EAP} data corrected with the \texttt{K2SC}
algorithm were found to be superior to basic \texttt{EAP} photometry.  The mean
magnitudes derived here from the {\it K2} fluxes, Kp=17.13\,mag, and that derived by Moln\'ar {\it et al.} (2018), 17.281\,mag, are in agreement. 
A cross-match of the EPIC coordinates with those in the  {\it Gaia}\, DR2 and DR3 catalogues 
did not result in a match (contrary to the match with the DR2 Id 2413839863087928064 by Moln\'ar et al. 2018).  
Considerably fainter is the mean magnitude given by the EPIC catalog,
Kp(MAST)=17.681\,mag. \\

\noindent {\bf EPIC\,246058914}\,(C\underline{12}, C\underline{19}) - The mean
magnitudes derived from the {\it K2} fluxes (Kp=17.02\,mag derived here and
17.105\,mag derived by Moln\'ar {\it et al.} 2018), by the EPIC catalog,
Kp(MAST)=16.891\,mag, and  from the {\it {\it Gaia}}\,(DR2) survey, G=17.044\,mag
(range 16.2-17.6), are all in agreement.  In the Petersen diagram this star
lies along the smooth curve defined by the `classical' RRd stars.  However,
both the {\it K2} and Catalina Sky Survey data both show the fundamental mode
to be more powerful than the first-overtone mode: $A_0/A_1$=1.04 ({\it K2}
data), which is unusual for a classical RRd star.    Many significant
combination frequencies were identified in a \texttt{SigSpec}/\texttt{Combine}
analysis of the short-cadence data (\texttt{PyKE} pipeline).   
  \\

\noindent {\bf EPIC\,247334376}\,(C13) - This long-period, probable
low-metallicity RRd stars  stands out for having the second largest amplitude
ratio, $A_1$/$A_0$=5.7 (caused primarily by the low value of $A_0=26.8\pm1.4$
mmag), resulting in the light curve showing little amplitude variation.   \\

\noindent {\bf EPIC\,248369176}\,(C10) - This star is one of the faintest in
the {\it K2} RRd sample.  The mean magnitudes derived from the {\it K2} fluxes,
Kp=20.22 mag derived here and Kp=20.952 derived by Moln\'ar {\it et al.} (2018),
are both faint and roughly consistent with G=20.317 mag (range 19.0-22.5)  from
the {\it {\it Gaia}}\,(DR2) survey.  The period $P_1$ derived from the {\it K2}
photometry is in agreement with the (sole) period derived from the Catalina Sky
Survey (CSS) data.  The long periods suggest that the metallicity is very low.
Unfortunately the two available SDSS spectra have low S/N ratios and cannot
confirm the suspected low [Fe/H] (see Table\,7 for \texttt{SDSS/SEGUE} physical
characteristics).   A time series analysis found very little variability in the
Fourier $A_1$ and $A_0$ amplitudes. 
\\

\noindent {\bf EPIC\,248426222}\,(C14) - 
See Table\,7 for \texttt{SDSS/SEGUE} physical characteristics.  \\

\noindent {\bf EPIC\,248509474}\,(C\underline{14}) =\,\texttt{LINEAR}\,22657637 - In the Petersen diagram this long-period RRd star has a relatively small $P_1$/$P_0$ ratio, but not so
low as to claim that it is an aRRd star.  \\

\noindent {\bf EPIC\,248514834}\,(C\underline{14}) = \texttt{LINEAR}\,22316675 - This is the longest period
RRd star in the {\it K2} sample and therefore probably is very metal weak. It was
first identified as an RRd star by Poleski (2014) in his analysis of the
\texttt{LINEAR} data.  
   \\

\noindent {\bf EPIC\,{248653582}}\,(\underline{C14}) = \texttt{LINEAR}\,23135759 - 
A time variability analysis suggests that there is no evidence for amplitude or
phase variability.  \\

\noindent {\bf EPIC\,248667792}\,(C14) - The {\it K2} photometry confirms (and
improves upon) the dominant period derived from the CSS data.  Also identified
is the fundamental mode period.  \\

\noindent {\bf EPIC\,248845745}\,(C14) -  
SDSS spectroscopy (see Table\,7) via the
SEGUE pipeline shows [Fe/H]=--1.60$\pm$0.01 dex.   
\\

\noindent {\bf EPIC\,248871792}\,(C\underline{14}) = \texttt{LINEAR}\,23184879
- No evidence is seen for amplitude or phase modulation of the radial modes.
With \texttt{SigSpec} a total of 497 frequencies more significant than
\texttt{sig}$=$5 were identified in the SC data, almost all of which are
combination frequencies.   See Table\,7 for physical characteristics derived by
the \texttt{SDSS/SEGUE} pipeline.   \\

\noindent {\bf EPIC\,249790928}\,(C15) - This {\it Kp}=14.58 mag star is one of the
brighter RRd stars in the {\it K2} sample.  The pulsation periods are long,
suggesting low metallicity.   The (dominant) first-overtone mode exhibits weak
evidence of non-stationary modulation with period $P_{\rm mod}\sim$21
days.  Sidepeaks with similar separations are
seen, corresponding to $P_{\rm mod}$ ranging from 20\,d to 25\,d. 
   \\

\noindent {\bf EPIC\,251248825}\,(C11) = \texttt{OGLE-BLG-RRLYR-17931}   
  \\

\noindent {\bf EPIC\,251248826}\,(C11) =  \texttt{OGLE-BLG-RRLYR-18439} -   In the
Petersen diagram this cRRd star lies among the ``OGLE-clump'' of short-period
stars identified by that survey (Soszynski {\it et al.} 2014b).
\\

\noindent {\bf EPIC\,251248827}\,(C11) = \texttt{OGLE-BLG-RRLYR-18912}  - This
star sits on the classical RRd curve, with approximately equal power in the
fundamental and first-overtone modes.    \\

\noindent {\bf EPIC\,251248828}\,(C11, aRRd) = \texttt{OGLE-BLG-RRLYR-24137}.  This
`anomalous' RRd star was studied in detail by Smolec {\it et al.} (2015b) and 
characterized there as an ``intriguing triple-mode RR Lyrae star with period
doubling".  In the Petersen diagram it lies significantly below the `classical'
RRd curve but not so low as to be among the `peculiar' RRd stars.  The fundamental
mode dominates over the first-overtone mode, as is typical for aRRd stars.
  \\

%



\noindent {\bf EPIC\,251248830}\,(C11) - This faint (20$^{\rm th}$ magnitude)
star was not observed by the \texttt{OGLE} survey thus the periods and
amplitudes are based on the {\it K2} photometry only.  Its RRd nature is well
established.  The periods are short and typical for a metal-rich cRRd star.
The amplitudes of the two modes are similar with the fundamental mode slightly
larger than that of the first-overtone.  \\

%
%


\noindent {\bf EPIC\,251456808}\,(C12) - At Kp=20.31 this star is one of the
faintest in the {\it K2} RRd-star sample.  The CSS photometry (144
measurements) established that $<$$V$$>$=20.08$\pm$0.04 mag but are too
uncertain for reliable period determinations. 
 \\

\noindent {\bf EPIC\,251521080}\,(C17) - The {\it K2} photometry appears to be
contaminated by the light from a reddish star of comparable brightness
($\sim$18th mag) located $\sim$5 arcsec to the north.  Similar contamination is
seen in the Catalina CSS data but does not seem to have much affected the
Catalina MLS measurements.  \\

\noindent {\bf EPIC\,251629085}\,(C17) -  In the Petersen diagram this star
lies on the curve occupied by `classical' RRd stars, on the short period side,
with $P_1$=0.344182\,d.  The very large amplitude ratio,
$A_1$/$A_0$=6.36$\pm$0.18, is caused primarily by the low $A_0=25.9\pm0.6$, thus the light
curve shows little amplitude variation.   See Table\,7 for \texttt{SDSS/SEGUE} physical
characteristics, where the measured spectroscopic
[Fe/H]=--1.40$\pm$0.07 agrees extremely well with the metal abundance derived
from the {\it K2} photometry (Table\,8). \\

\noindent {\bf EPIC\,251809825}\,(C17) - Both radial pulsation modes are non-stationary.   \\

%
%
%


\noindent {\bf EPIC\,251809832}\,(C17) - This star is faint, with Kp=19.5 mag.   An additional
significant signal, possibly due to light contamination, is seen at $P$=0.07665\,d
(amplitude 4.6 mmag, period ratio relative to $P_1$ of 0.1977).  \\

%
%


\noindent {\bf EPIC\,251809860}\,(C17) -  The {\it K2} photometry of this faint star 
is sufficient to identify it as a cRRd star.   \\





\noindent {\bf EPIC\,251809870}\,(C17) - Very faint, with Kp = 20.45
mag.     \\

\noindent {\bf TV\,Libr{\ae}} = {\it Gaia} 6321161342439508480,
with period 0.2696\,d and [Fe/H]=--0.53\,dex, is classified `RRab' in DR3 and
by Crestani et al. (2021b), but in Fig.\,12a it appears to be located among the
metal-rich RRc stars.  The RRab classification is consistent with its
asymmetric light curve and with the high amplitudes given by Crestani et al.
($A_V$=1.241 mag) and by {\it Gaia} ($A_G$=0.85 mag), both of which are
considerably higher than the 0.55 mag upper limit that one usually sees for RRc
stars.  Its extremely short period appears to be unique among RRab stars, and
its spectrum is unusual, showing overabundant [$\alpha$/Fe] ratios (Liu et al.
2013).   TV\,Lib almost certainly is an unusual short-period RRab star.    \\

\noindent {\bf HY\,Com{\ae} Berenices} ={\it Gaia}
3946316423735761536, with period 0.4486\,d and [Fe/H] = $-1.75\pm0.02$\,dex
(Crestani et al. 2021a,b) was classified `RRc' by Crestani, DR2 and DR3.
Consistent with this are the sinusoidal light curves seen in Fig.4 of Soper et
al. (2022), and its location among the RRc stars in the period-$A_V$ diagram
constructed from the Crestani data.  However, in Fig.\,12a it is located among
the RRab stars.  For an RRc star its period is unusually long.  \\

\noindent {\bf SSS\,J205126.3-413741} = {\it Gaia}
6677541412082758144, has period 0.406994\,d, [Fe/H] = $-2.39\pm0.11$ dex, and
was classified `RRc' by Crestani et al. (2021a,b).  However, in the {\it Gaia}
DR2 catalogue it was classified `RRd' (with $P_1=0.40699$\,day,
$P_0=0.54652$\,day and $P_1/P_0$=0.7447).  In the DR3 catalogue it was
classified `RRab' (with $P_0=0.40699$\,day) but its low amplitude ($A_V$=0.36
mag) is not typical for short-period RRab stars.  We conclude that this probably
is an RRd star, in accord with the DR2 classification.   \\

\noindent {\bf RW\,Trianguli\,Australis} = {\it Gaia}
5815008831122635520, with $P_0$=0.37404\,d,  [Fe/H]$=+0.12\pm0.04$\,dex, and
relatively high amplitude ($A_V$=0.72 mag), was classified `RRab' by Crestani
et al. (2021a,b) and by the \texttt{GEOS} RR\,Lyr database (Le Borgne et al. 2007,2012).
However the {\it Gaia} DR2 and DR3 catalogues, despite finding the same period,
classify it as `RRc'.  Given its relatively high amplitude and location among
the short-period metal-rich RRab stars in Fig.\,12a, the `RRab' classification
is favoured.   \\

\noindent {\bf V5644\,Sagittarii} = {\it Gaia}
6736741901293501056 (DR3) = ASAS\,183952-3200.9, is, according to Crestani et
al. (2021a,b),  an `RRd' star with metal abundance  [Fe/H] =
$-1.69\pm0.01$\,dex and period 0.3424736\,d (only one of the two periods is
given).  It is not in the {\it Gaia} DR2 catalogue (and not in the \texttt{GEOS} RR\,Lyr
database) but the DR3 and ASAS catalogues both give $P_0$$\sim$0.46125\,d and
type `RRab'.  If both detected periods are present then V5644\,Sgr may be
doubly-periodic with $P_1/P_0$=0.7425. In Fig.12a its location is close to the
expected location for an RRd star. \\

\noindent {\bf CN\,Lyr\ae}\,=\,{\it Gaia}\,4539434124372063744,
with period $0.41138$\,d, amplitude $A_V=0.50$\,mag,  [Fe/H]$_{\rm
spec}$=$-0.08\pm0.05$\,dex, and location in the period-$A_V$ diagram amongst
the long-period RRc stars, is classified `RRc' by Crestani et al. (2021a,b).
However, in the {\it Gaia} DR2 and DR3 catalogues, and in the \texttt{GEOS} RR\,Lyr
database, CN\,Lyr is classified `RRab', which is consistent with its location
among the metal-rich RRab stars in Fig.\,12a.  \\

\noindent {\bf FW\,Lupi} = {\it Gaia} 6005656897473385600, with
period $0.48417$\,d, amplitude $A_V$=0.39\,mag, and [Fe/H]$_{\rm spec}$ =
$-0.17\pm0.02$\,dex, is classified `RRab' by Crestani et al. (2021a,b) and by
both {\it Gaia} catalogues.  Such a classification is also consistent with its
location among the metal-rich RRab stars in Fig.\,12a.  However, in the
period-amplitude diagram its $A_V$ is very low for a (non-Blazhko) short-period
RRab star, which brings this classification into question.

 \clearpage  
\vfill \eject 

\section{Fitted Light Curves for the RRd stars observed during Campaigns 7-18}

The 10 figures in this Appendix ({\bf Figures B1-B10}) show the observed {\it Kp}-photometry and the fitted light curves
for the RRd stars observed by {\it K2} during Campaigns 7 to 18.

\renewcommand{\thefigure}{B1}
\begin{figure*} \begin{center}
\begin{overpic}[width=7.9cm]{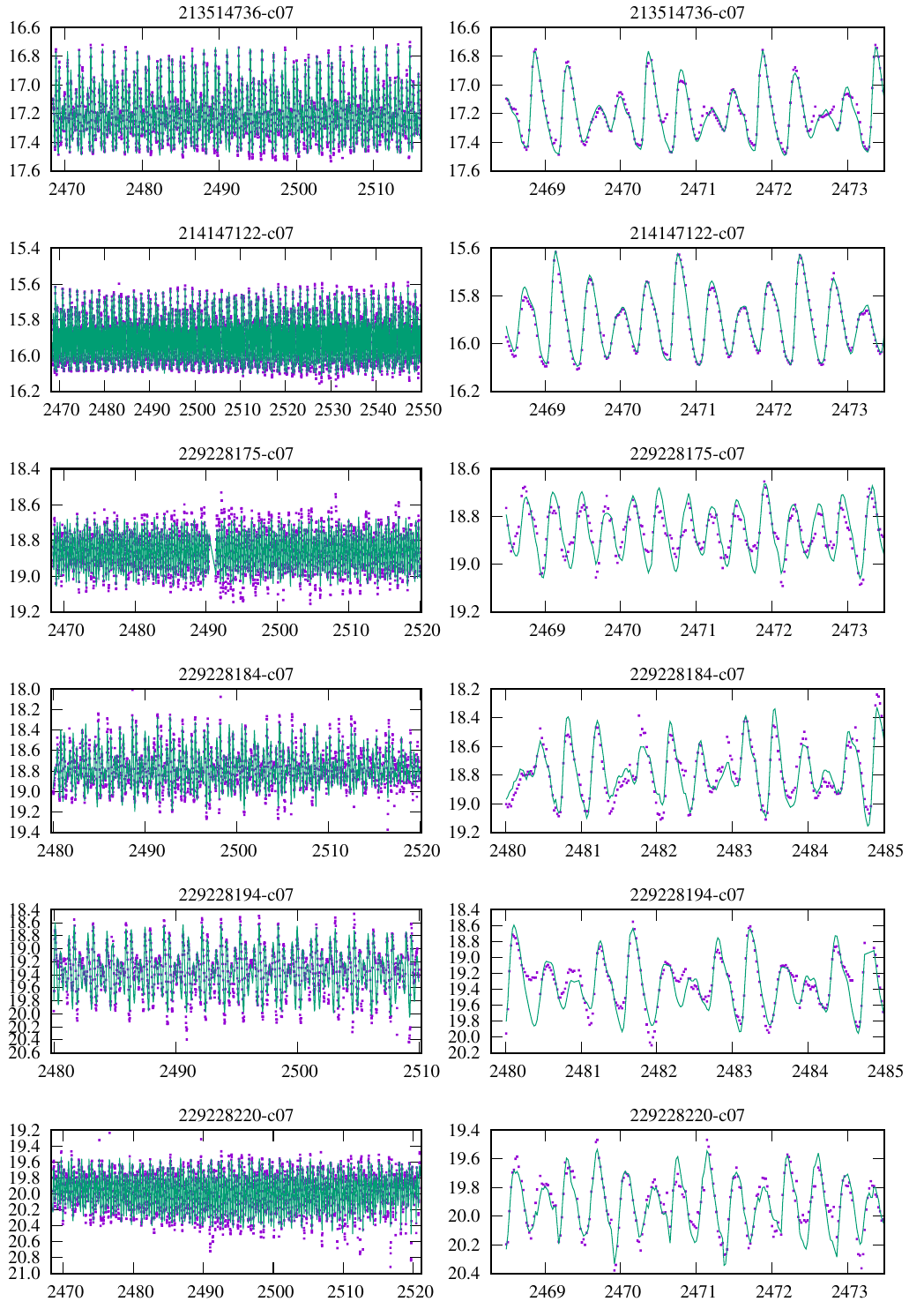} \put(25,-3.0){BJD - 2454833}    \end{overpic}
\end{center}

\caption{Observed Kp photometry (long cadence \texttt{PDCsap} data) and fitted
(\texttt{Period04}) light curves for the six `Galactic Bulge' cRRd stars
observed during {\it K2} \underline{Campaign\,7}.    Left panels: all measured
{\it K2} photometry.  Right panels: first five days of observations. } 

\label{FigB1} 
\end{figure*}

\renewcommand{\thefigure}{B2}
\begin{figure*} \begin{center}
\begin{overpic}[width=8.1cm]{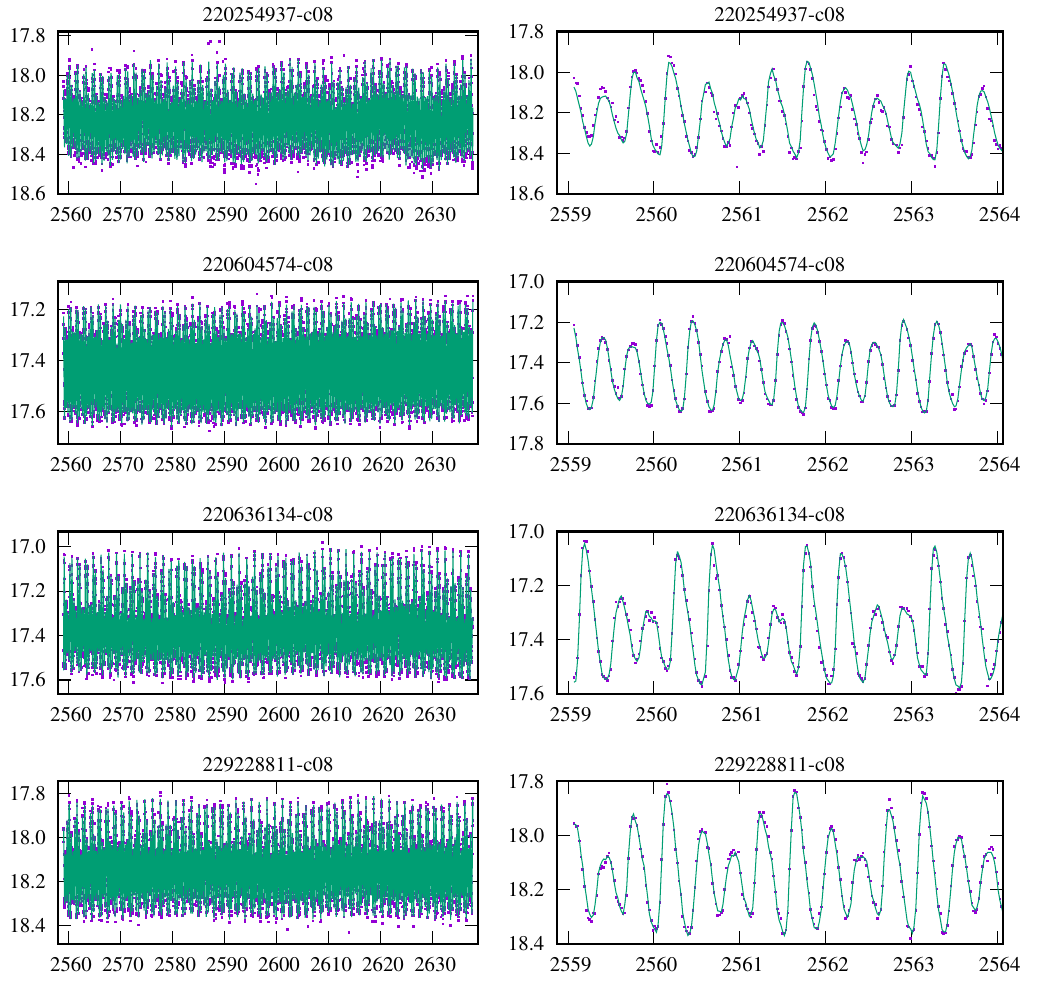} \put(35,-3.0){BJD - 2454833}    \end{overpic}
\end{center} 

\caption{Observed Kp photometry (long cadence \texttt{EAP} reduction) and fitted light curves (\texttt{Period04}) for the four `Equatorial Plane' RRd stars
observed during {\it K2} \underline{Campaign~8}.   Left panels: all 78.7\,d of photometry.
Right panels: first five days of data.  EPIC\,220636134 also was observed at
short-cadence.   } 

\label{FigB2} 
\end{figure*}

\renewcommand{\thefigure}{B3}
\begin{figure*} \begin{center}
\begin{overpic}[width=8.3cm]{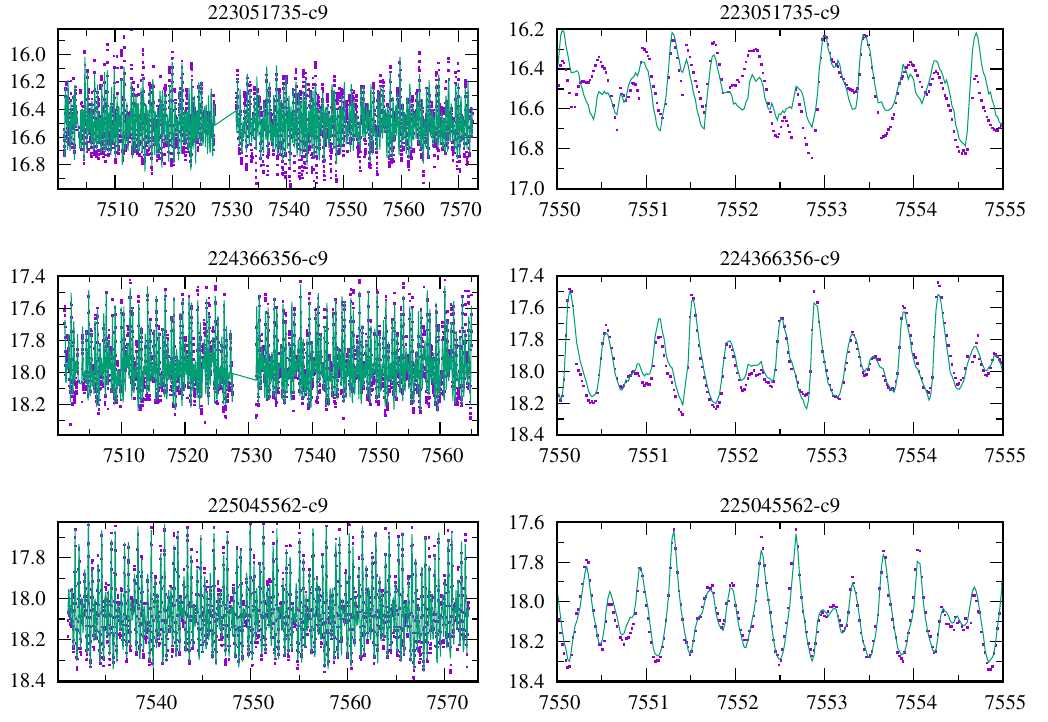} \put(35,-3.0){BJD - 2450000}    \end{overpic}
\end{center}
 
\caption{Observed Kp photometry (long cadence \texttt{PyKE} analysis) and
fitted light curves (\texttt{Period04}) for the three `Galactic Bulge' RRd
stars observed during {\it K2} \underline{Campaign~9}.  The data were
calibrated using \texttt{OGLE} photometry, and the photometry for
EPIC\,223051735 suffers from contamination by a nearby bright star.
EPIC\,225045562 is an `anomalous' RRd star.  Left panels: all available {\it
K2} photometry.  Right panels: five days of observations.      } 

\label{FigB3} 
\end{figure*}

\renewcommand{\thefigure}{B4}
\begin{figure*} \begin{center}
\begin{overpic}[width=7cm]{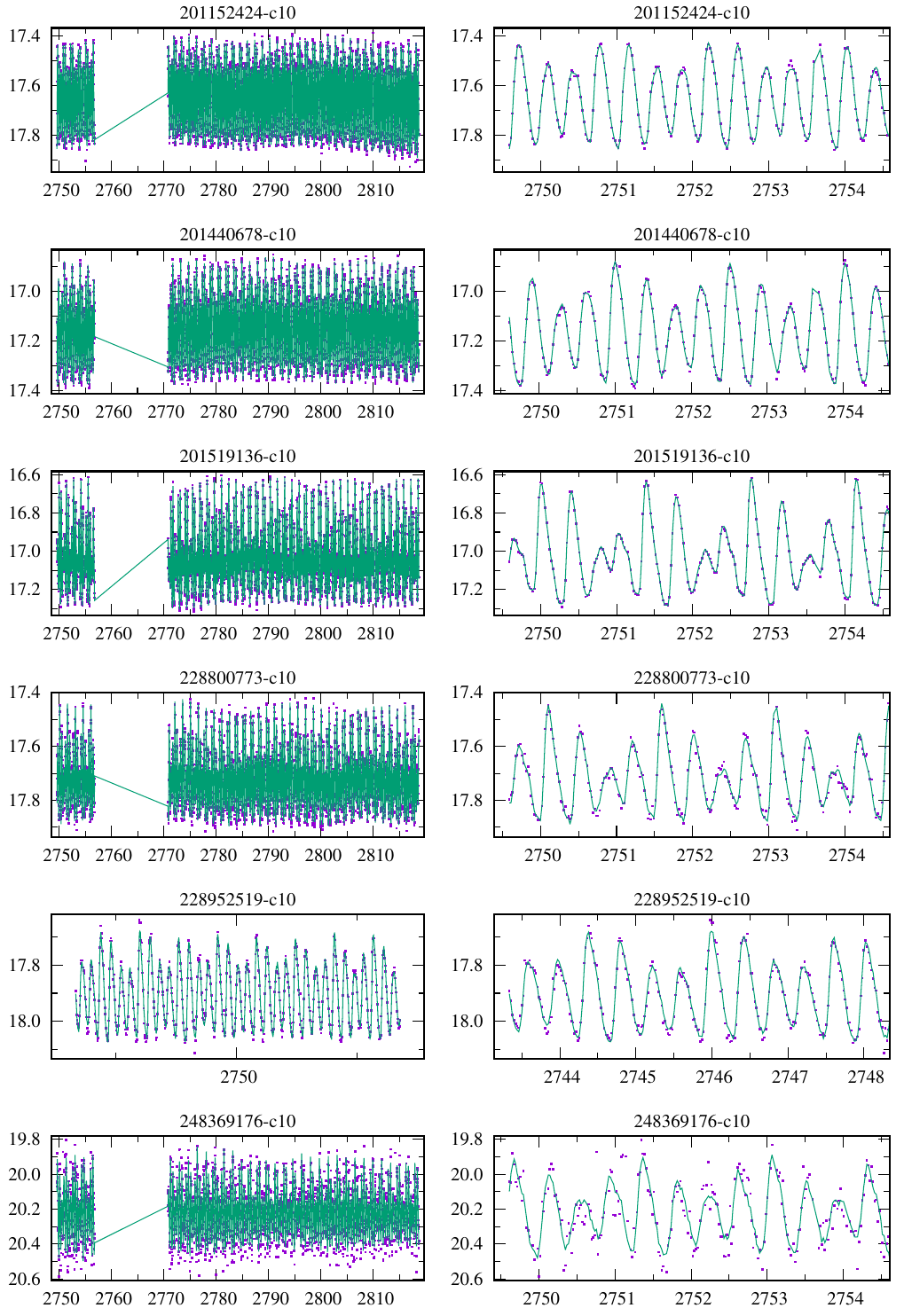} \put(25,-3.0){BJD - 2454833}   \end{overpic}
\end{center} 

\caption{Observed Kp photometry (\texttt{EAP} pipeline with systematics
correction) and fitted light curves (\texttt{Period04}) for the six `North
Galactic Cap' RRd stars observed during {\it K2} \underline{Campaign~10}.  Left
panels: all the measured long-cadence Kp photometry.     Right panels:  first
five days of data.  The observations for EPIC\,228952519 ended prematurely when
{\it K2} module\,4 failed. } 

\label{FigB4} 
\end{figure*}

\renewcommand{\thefigure}{B5}
\begin{figure*}  
\begin{center}
%
\begin{overpic}[width=7.1cm]{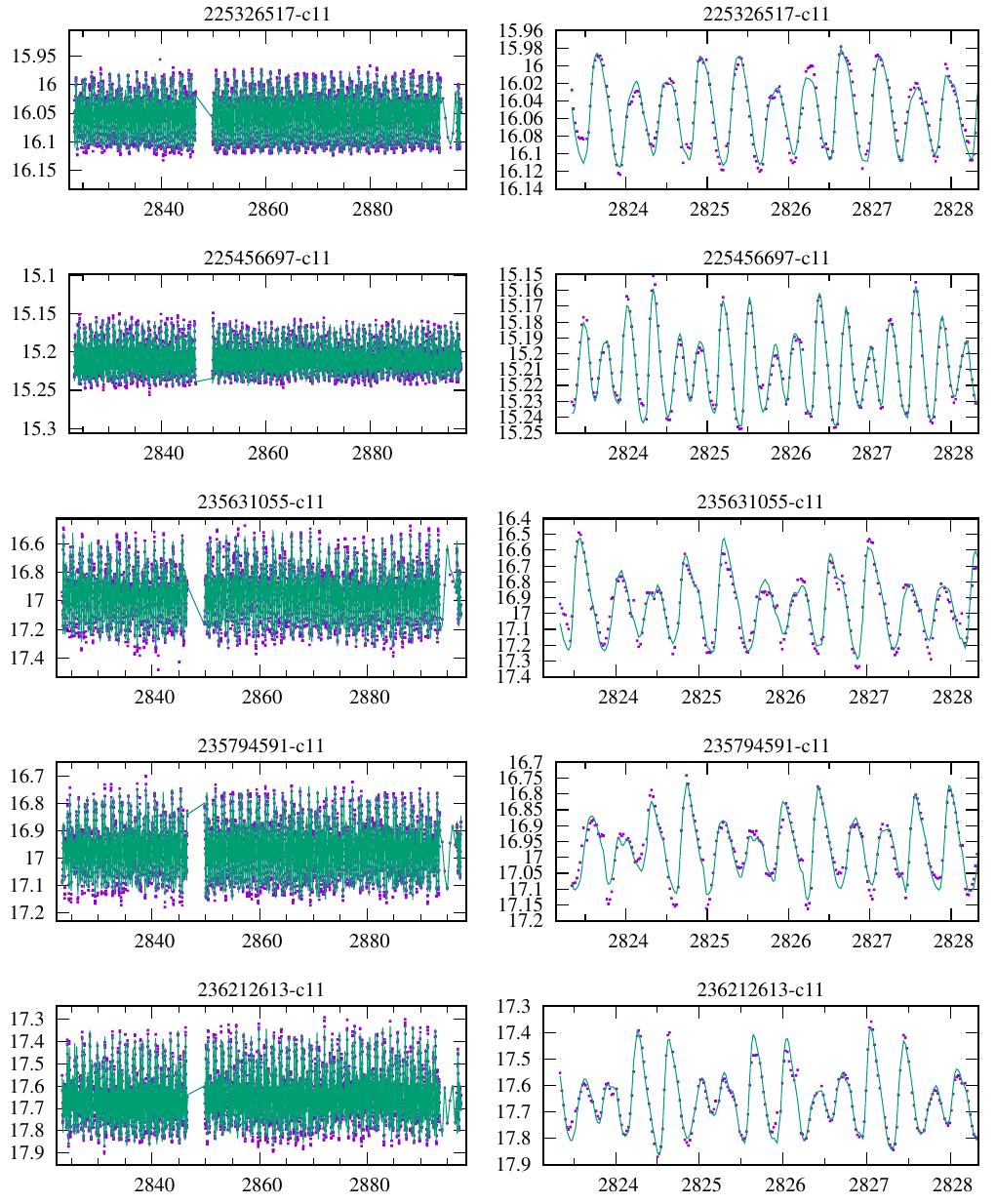} \put(25,-3.0){BJD - 2454833}   \end{overpic}  \hskip0.5truecm    
\begin{overpic}[width=7.1cm]{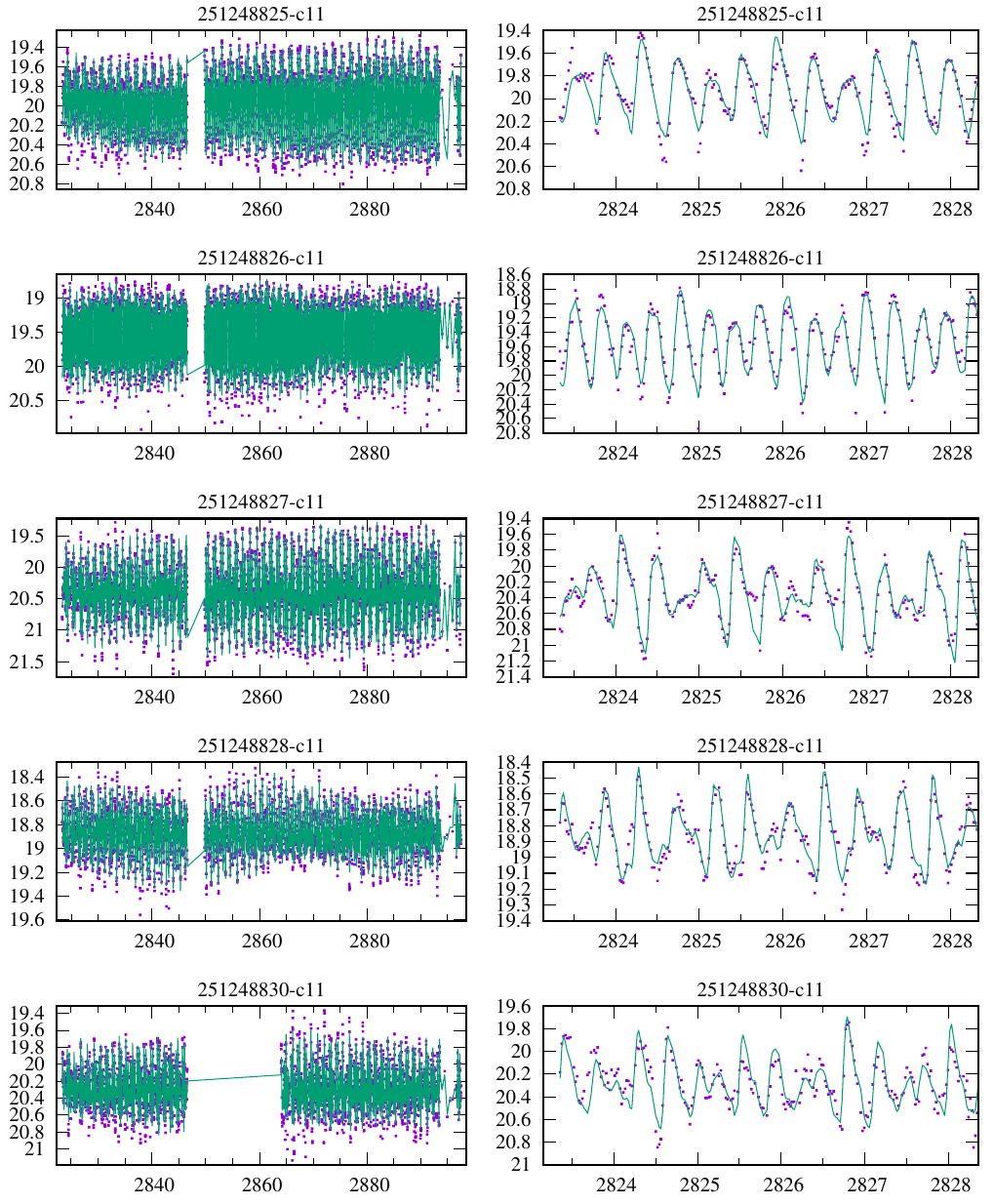} \put(25,-3.0){BJD - 2454833}   \end{overpic}   \end{center} 

\caption{Observed Kp photometry (long cadence, \texttt{PDCsap} pipeline) and fitted light curves (\texttt{Period04}) for the ten `Galactic Bulge' RRd
stars observed during {\it K2} \underline{Campaign\,11}. All are cRRd stars except
EPIC\,251248828 which is an aRRd star.   The gap in the data at 2847 ({\it
i.e.}, at BJD\,2457680) separates the C111 data from the C112 data.  The mean
magnitudes were either set equal to the mean values given at the \texttt{MAST} website
or were calculated assuming the transformation Kp=25.3--2.5 $\log_{10}$(flux).
Left panels: full sample of long cadence photometry.   Right panels: the first five
days of photometry (illustrating cycle-to-cycle variations).  } 	

\label{FigB5}  
\end{figure*}

\renewcommand{\thefigure}{B6}
\begin{figure*} \begin{center}
\begin{overpic}[width=8.1cm]{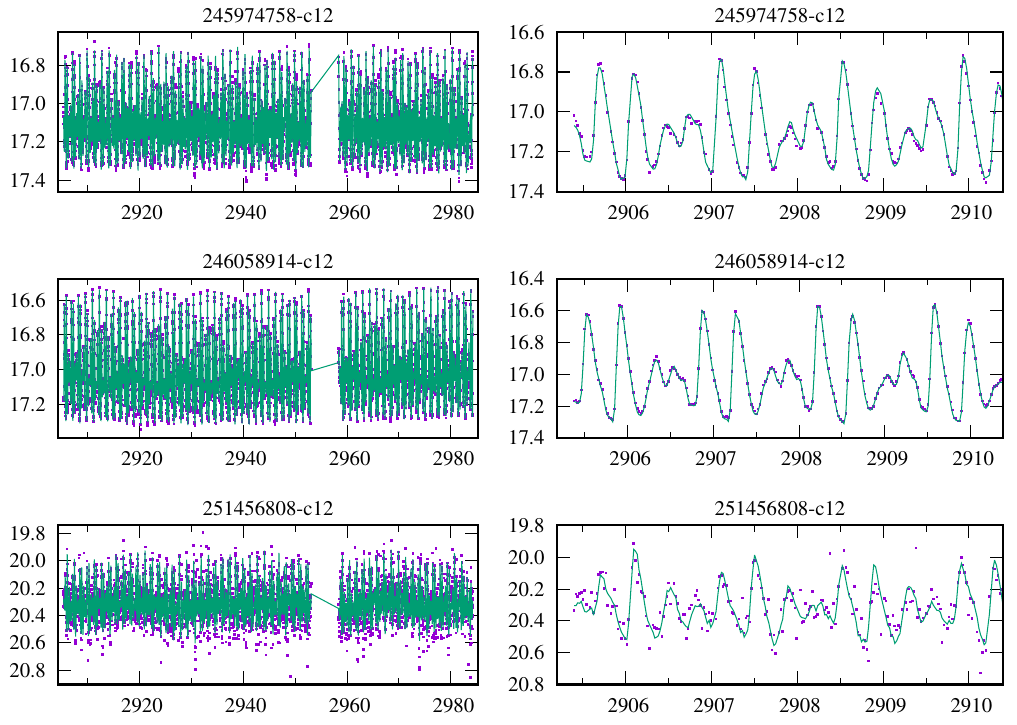}    \end{overpic}   %
\end{center}
\begin{center}
\begin{overpic}[width=8.1cm]{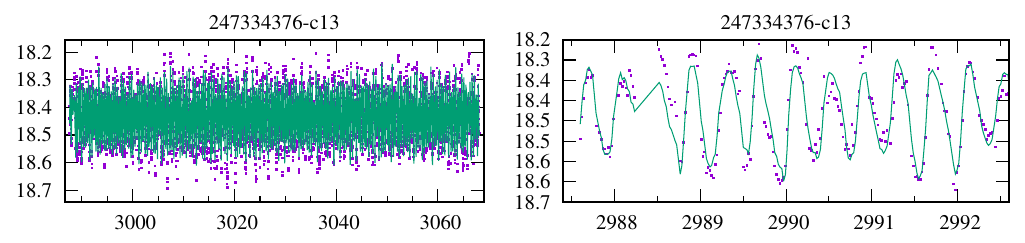} \put(38,-4){BJD - 2454833}   \end{overpic}
\end{center}

\caption{Observed photometry (\texttt{EAP} pipeline) and fitted light curves (\texttt{Period04}
analysis) for the three high Galactic latitude
($b$$\sim$60$^\circ$) RRd stars observed during {\it K2}
\underline{Campaign~12}  and for the lone `Galactic anti-centre' RRd star
observed during  \underline{Campaign~13} (which includes the Hyades and Taurus
star clusters).  Left panels: all available long-cadence Kp photometry.  Right panels: first five days of data.  For EPIC\,246058914 the fundamental and
first-overtone modes are comparable in strength.  EPIC\,251456808 and
EPIC\,247334376 are fainter than the other three stars and therefore their
light curves are noisy.  } 

\label{FigB6} 
\end{figure*}

\renewcommand{\thefigure}{B7}
\begin{figure*} 
\begin{center}
\begin{overpic}[width=7.1cm]{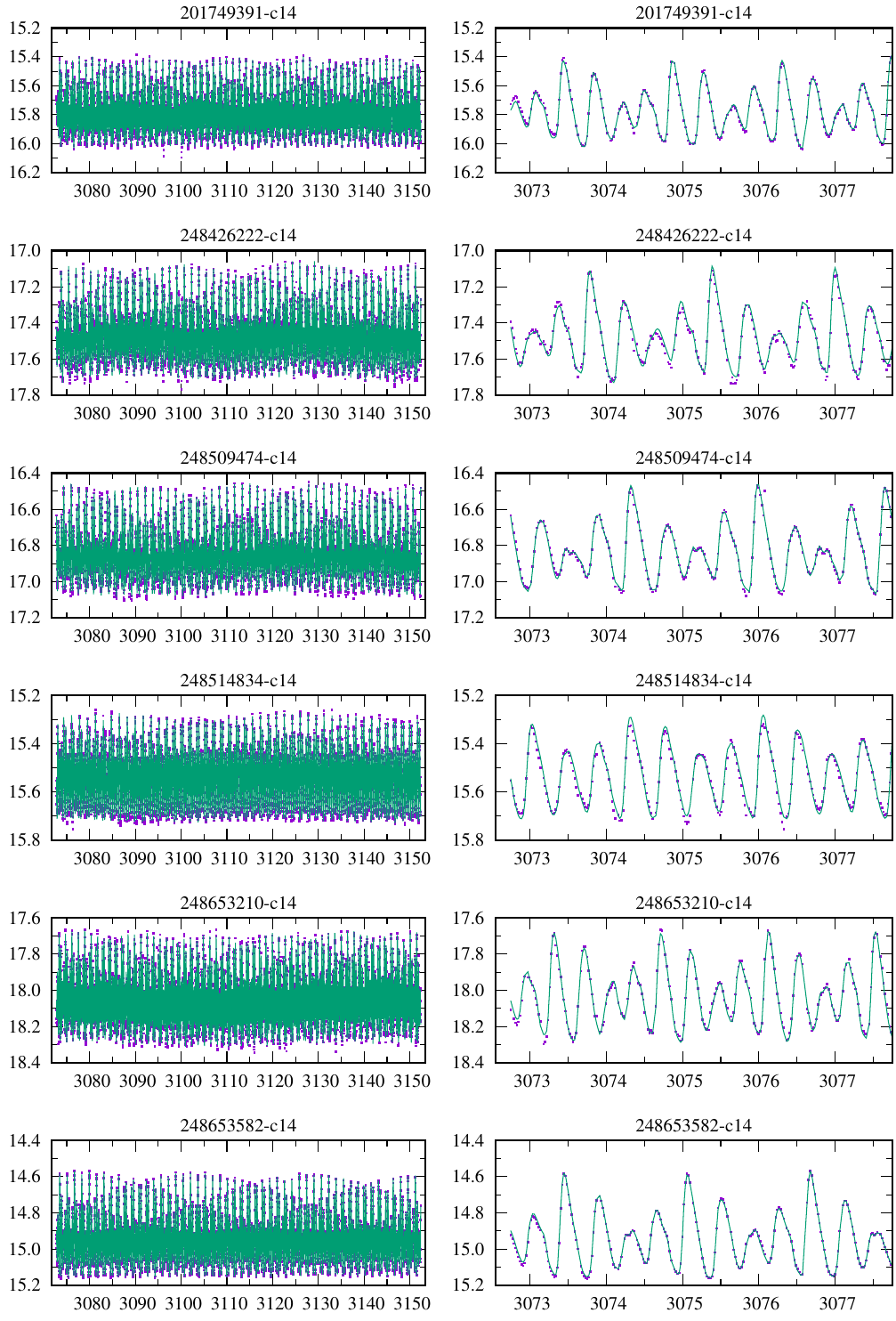} \put(27,-2.5){BJD - 2454833} \end{overpic} \hskip0.5truecm
\begin{overpic}[width=7.1cm]{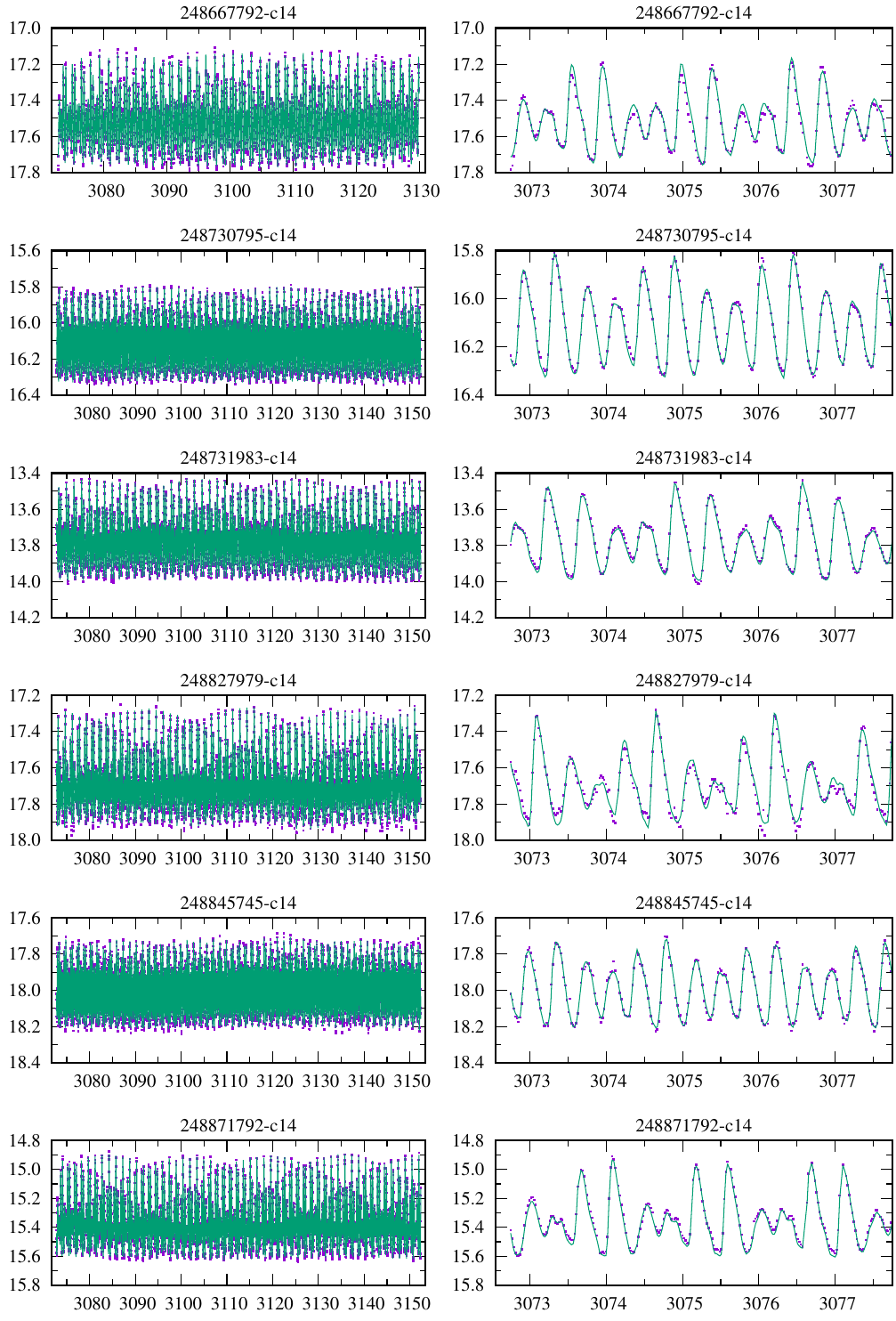} \put(27,-2.5){BJD - 2454833} \end{overpic}
\end{center} 

\caption{Observed photometry (\texttt{EAP} pipeline) and fitted light curves (\texttt{Period04} analysis) for the 12 RRd stars observed
during {\it K2} \underline{Campaign\,14}.   Left panels: full sample of 81 days of
long-cadence Kp photometry (\texttt{EAP} pipeline, \texttt{Period04} analysis).
Right panels: first five days of photometry. } 

\label{FigB7} 
\end{figure*}

\renewcommand{\thefigure}{B8}
\begin{figure*} 
\begin{center}
\begin{overpic}[width=8.1cm]{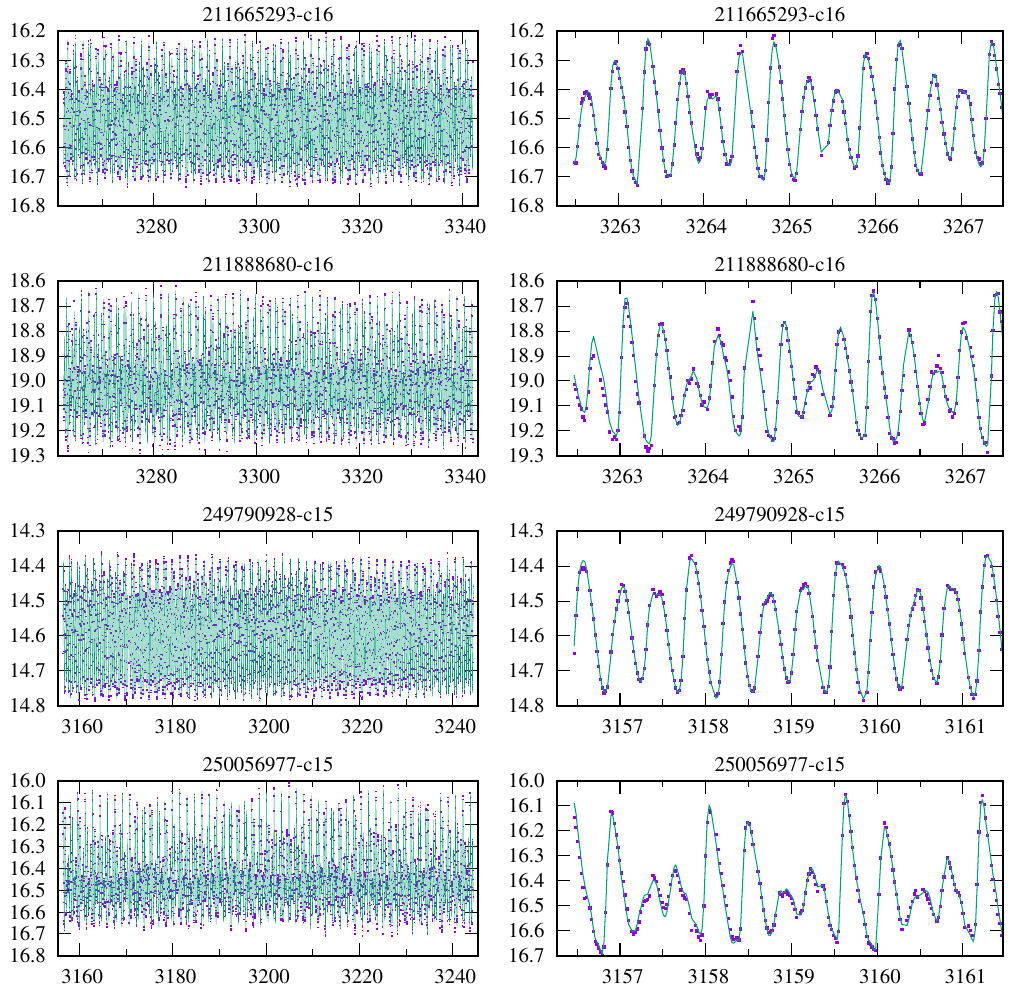} \put(32,-4){BJD - 2454833}   \end{overpic}
\end{center} 

\caption{Observed photometry (\texttt{PyKE} extractions) and fitted light
curves (\texttt{Period04} analysis) for the four RRd stars observed during {\it
K2} \underline{Campaigns\,15-16}.   Left panels: all the measured long-cadence
Kp photometry.   Right panels:  first five days of data.  EPIC\,211888680 was
observed also in Campaign\,5;  also, short-cadence observations are available
for EPIC\,250056977. } 

\label{FigB8} 
\end{figure*}

\renewcommand{\thefigure}{B9}
\begin{figure*} 
\begin{center}
\begin{overpic}[width=8.2cm]{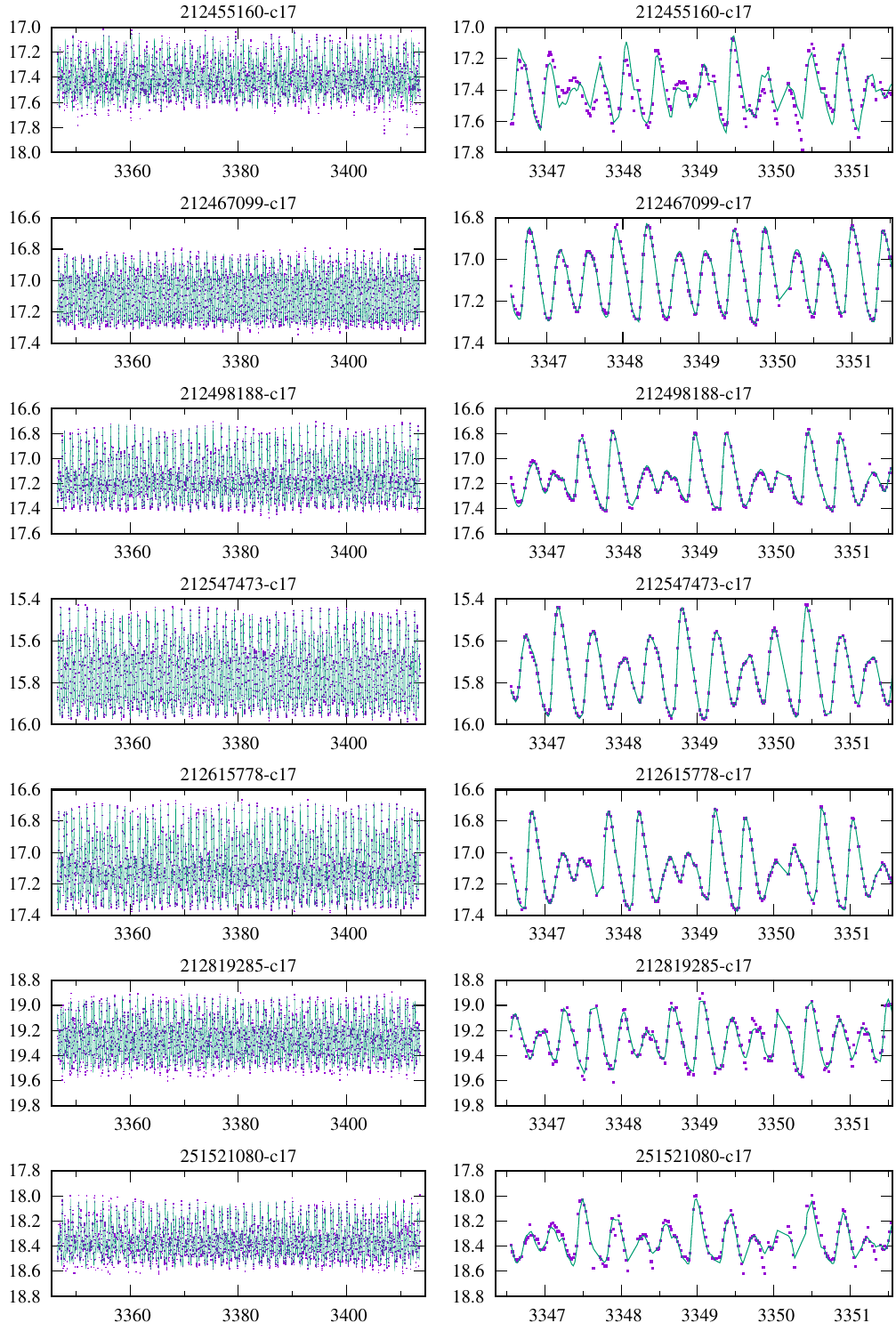} \put(27,-2){BJD - 2454833}   \end{overpic}  \hskip0.6cm
\begin{overpic}[width=8.2cm]{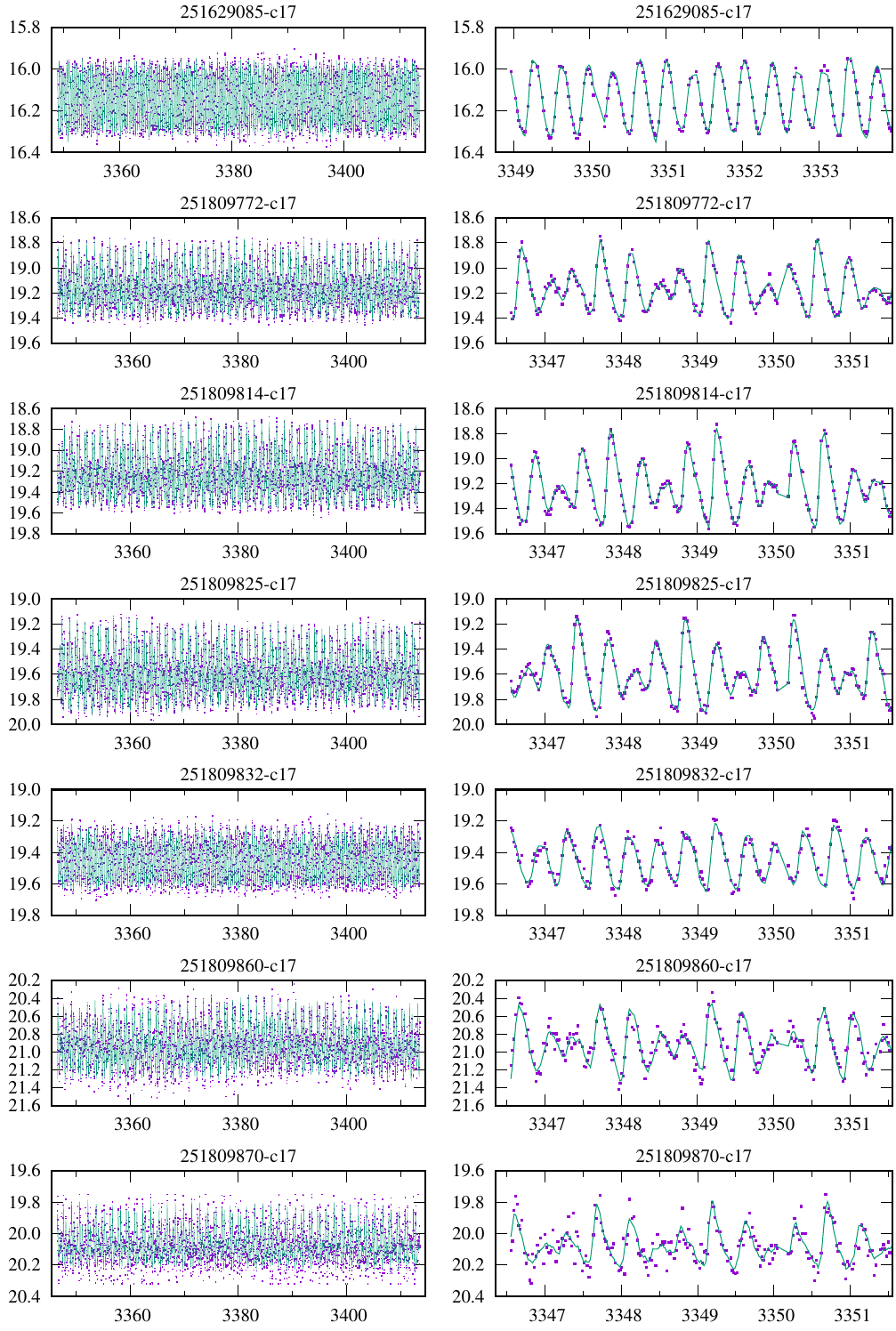} \put(27,-2){BJD - 2454833}   \end{overpic}
\end{center} 
\vskip0.1truecm

\caption{Observed  and fitted light curves (\texttt{Period04} analysis) for the
14 RRd stars observed during {\it K2} \underline{Campaign\,17} (ordered by EPIC
number).  Most of the photometry (Kp-passband) is from \texttt{PyKE}
extractions (but some is \texttt{PDCsap}).  Left panels: all 67.1 days of measured
photometry.   Right panels: first five days of data.  EPIC\,212455160 and 212547473 were
also observed during Campaign\,6.  } 

\label{FigB9} 
\end{figure*}

\renewcommand{\thefigure}{B10}
\begin{figure*} 
\begin{center}
\begin{overpic}[width=7.2cm]{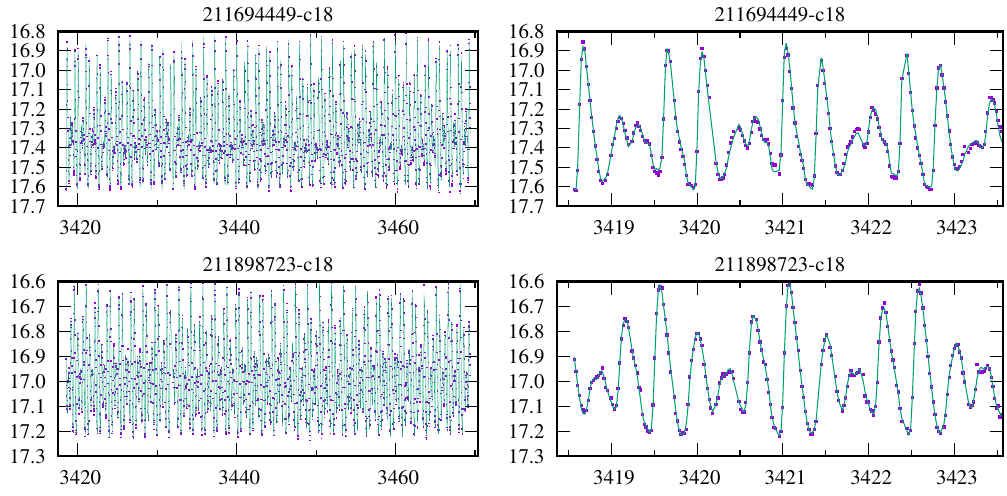}    \end{overpic}
\end{center}
\begin{center}
\begin{overpic}[width=7.2cm]{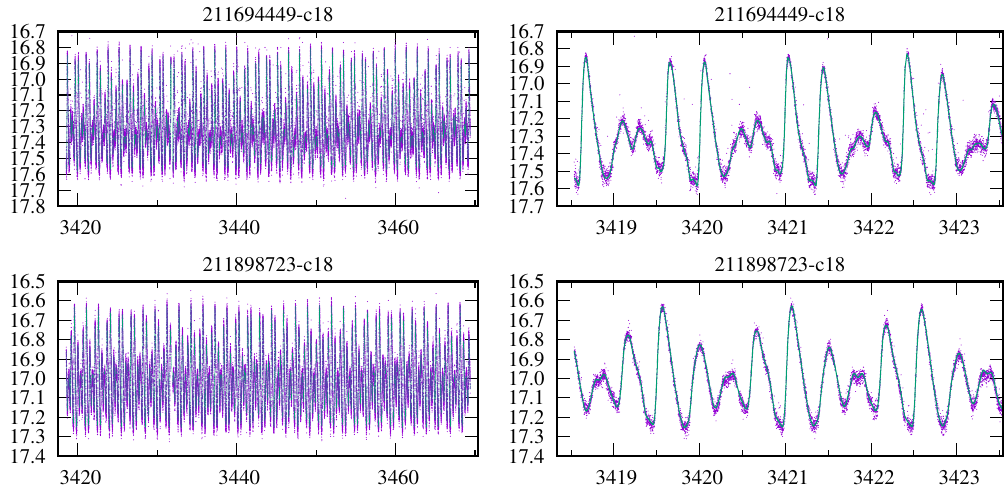} \put(39,-3){BJD - 2454833}   \end{overpic}
\end{center} 
\vskip0.1truecm

\caption{Observed and fitted light curves for the two RRd stars observed during
{\it K2} \underline{Campaign\,18}.  Left panels: all 50.8 days of measured Kp photometry
(\texttt{Period04} analysis).   Right panels: the first five days.  The top
two panels show long cadence data while the bottom two show short
cadence data.  Both stars also were observed during Campaign\,5 (see Fig.2).   } 
\label{FigB10}

\end{figure*}

 \clearpage
\vfill \eject 
\renewcommand{\thetable}{9}
\begin{table*}
\fontsize{7}{8.4}\selectfont
\centering
\caption{Metallicities and masses for the 2130 {\it Gaia} stars classified
`RRd' by either DR2 or DR3 and that are on the cRRd curve in the Petersen
diagram.  The [Fe/H] values (col.\,9) were estimated using the $P_0$-[Fe/H]
calibration formula (Eqn.\,6).  The masses estimates were made assuming no
$\alpha$-element enhancements (col.10) and assuming $\alpha$ element abundances
enhanced by a factor two (col.11).  The complete table is given here and in the online
Supporting Information.}
\begin{tabular}{lllllcccccc}
\hline
\multicolumn{1}{c}{ {\it Gaia} Id  } &   \multicolumn{1}{c}{RA} & \multicolumn{1}{c}{DEC} &  \multicolumn{2}{c}{{\it Gaia} Classif.} & \multicolumn{1}{c}{ $P_1$}   & \multicolumn{1}{c}{ $P_0$ } & \multicolumn{1}{c} { $P_1/P_0$ }  & \multicolumn{1}{c}{[Fe/H] }  & \multicolumn{2}{c}{$M/M_{\sun}$} \\
\multicolumn{1}{c}{  } &  \multicolumn{2}{c}{(2016.0)}  &DR2  &DR3    &\multicolumn{1}{c}{[day]}  &\multicolumn{1}{c}{[day]} & & \multicolumn{1}{c}{ $\pm$0.17 }  & \multicolumn{1}{c}{[$\alpha$/Fe]=0.0} & \multicolumn{1}{c}{[$\alpha$/Fe]=0.3 }   \\
\multicolumn{1}{c}{(1)} & \multicolumn{1}{c}{(2)} &\multicolumn{1}{c}{(3)}  &\multicolumn{1}{c}{(4)} & \multicolumn{1}{c}{(5)} & \multicolumn{1}{c}{(6)}   & \multicolumn{1}{c}{(7)}  & \multicolumn{1}{c}{(8)}   & \multicolumn{1}{c}{(9)} &\multicolumn{1}{c}{(10)} &\multicolumn{1}{c}{(11)}   \\
\hline  
\\	    
    4109411961354217856 & 258.5835748 & -25.70825502 & RRab  & RRd   & 0.335361 & 0.451800 & 0.74228 & -1.32 & 0.66  & 0.63 \\
    1797412030619582848 & 323.6745277 & 23.89800184 & RRd   & RRc   & 0.335297 & 0.451917 & 0.74194 & -1.32 & 0.66  & 0.63 \\
    6028840233297277440 & 258.3737969 & -29.79608735 & RRd   & RRc   & 0.335828 & 0.452341 & 0.74242 & -1.33 & 0.66  & 0.63 \\
    1494770600375550336 & 218.7012557 & 46.44626505 &       & RRd   & 0.336212 & 0.453218 & 0.74183 & -1.33 & 0.66  & 0.63 \\
\hline
\end{tabular}
\end{table*}

\begin{table*}
\fontsize{7}{8.4}\selectfont
\centering
\contcaption{  }
\begin{tabular}{lllllcccccc}
\hline
	\multicolumn{1}{c}{ {\it Gaia} Id  } &   \multicolumn{1}{c}{RA} & \multicolumn{1}{c}{DEC} &  \multicolumn{2}{c}{{\it Gaia} Classif.} & \multicolumn{1}{c}{ $P_1$}   & \multicolumn{1}{c}{ $P_0$ } & \multicolumn{1}{c} { $P_1/P_0$ }  & \multicolumn{1}{c}{[Fe/H] }  & \multicolumn{2}{c}{$M/M_{\sun}$} \\
\multicolumn{1}{c}{  } &  \multicolumn{2}{c}{(2016.0)}  &DR2  &DR3    &\multicolumn{1}{c}{[day]}  &\multicolumn{1}{c}{[day]} & & \multicolumn{1}{c}{ $\pm$0.17 }  & \multicolumn{1}{c}{[$\alpha$/Fe]=0.0} & \multicolumn{1}{c}{[$\alpha$/Fe]=0.3 }   \\
\multicolumn{1}{c}{(1)} & \multicolumn{1}{c}{(2)} &\multicolumn{1}{c}{(3)}  &\multicolumn{1}{c}{(4)} & \multicolumn{1}{c}{(5)} & \multicolumn{1}{c}{(6)}   & \multicolumn{1}{c}{(7)}  & \multicolumn{1}{c}{(8)}   & \multicolumn{1}{c}{(9)} &\multicolumn{1}{c}{(10)} &\multicolumn{1}{c}{(11)}   \\
\hline  
\\	    
    4121912863155228288 & 262.9776813 & -18.38227059 &       & RRd   & 0.345764 & 0.465180 & 0.74329 & -1.45 & 0.68  & 0.65 \\
    6637876560175534080 & 283.4532433 & -56.13410487 & RRd   & RRd   & 0.345837 & 0.465195 & 0.74342 & -1.45 & 0.68  & 0.65 \\
    4660461458086001280 & 79.91329159 & -66.57467274 & RRd   & RRd   & 0.346285 & 0.465709 & 0.74357 & -1.46 & 0.68  & 0.65 \\
    6776831052985351168 & 316.8281313 & -36.5397445 & RRd   & RRd   & 0.346172 & 0.465747 & 0.74326 & -1.46 & 0.68  & 0.65 \\
\hline
\end{tabular}
\end{table*}

\begin{table*}
\fontsize{7}{8.4}\selectfont
\centering
\contcaption{  }
\begin{tabular}{llllllccccc}
\hline
\multicolumn{1}{c}{ {\it Gaia} Id  } &   \multicolumn{1}{c}{RA} & \multicolumn{1}{c}{DEC} &  \multicolumn{2}{c}{{\it Gaia} Classif.} & \multicolumn{1}{c}{ $P_1$}   & \multicolumn{1}{c}{ $P_0$ } & \multicolumn{1}{c} { $P_1/P_0$ }  & \multicolumn{1}{c}{[Fe/H] }  & \multicolumn{2}{c}{$M/M_{\sun}$} \\
\multicolumn{1}{c}{  } &  \multicolumn{2}{c}{(2016.0)}  &DR2  &DR3    &\multicolumn{1}{c}{[day]}  &\multicolumn{1}{c}{[day]} & & \multicolumn{1}{c}{ $\pm$0.17 }  & \multicolumn{1}{c}{[$\alpha$/Fe]=0.0} & \multicolumn{1}{c}{[$\alpha$/Fe]=0.3 }   \\
\multicolumn{1}{c}{(1)} & \multicolumn{1}{c}{(2)} &\multicolumn{1}{c}{(3)}  &\multicolumn{1}{c}{(4)} & \multicolumn{1}{c}{(5)} & \multicolumn{1}{c}{(6)}   & \multicolumn{1}{c}{(7)}  & \multicolumn{1}{c}{(8)}   & \multicolumn{1}{c}{(9)} &\multicolumn{1}{c}{(10)} &\multicolumn{1}{c}{(11)}   \\
\hline  
\\	    

    5838828234405256832 & 195.1738961 & -73.01939675 & RRd   & RRc   & 0.349041 & 0.469663 & 0.74317 & -1.49 & 0.69  & 0.65 \\
    4352301059250803200 & 247.6054993 & -6.143988165 & RRd   & RRd   & 0.349489 & 0.469674 & 0.74411 & -1.50 & 0.68  & 0.65 \\
    2884374867763847296 & 89.06696969 & -36.92478291 &       & RRd   & 0.349180 & 0.469729 & 0.74336 & -1.50 & 0.69  & 0.65 \\
    4108720957020811392 & 256.7169659 & -27.38380235 &       & RRd   & 0.349368 & 0.469729 & 0.74376 & -1.50 & 0.68  & 0.65 \\
\hline
\end{tabular}
\end{table*}

\begin{table*}
\fontsize{7}{8.4}\selectfont
\centering
\contcaption{  }
\begin{tabular}{lllllcccccc}
\hline
\multicolumn{1}{c}{ {\it Gaia} Id  } &   \multicolumn{1}{c}{RA} & \multicolumn{1}{c}{DEC} &  \multicolumn{2}{c}{{\it Gaia} Classif.} & \multicolumn{1}{c}{ $P_1$}   & \multicolumn{1}{c}{ $P_0$ } & \multicolumn{1}{c} { $P_1/P_0$ }  & \multicolumn{1}{c}{[Fe/H] }  & \multicolumn{2}{c}{$M/M_{\sun}$} \\
\multicolumn{1}{c}{  } &  \multicolumn{2}{c}{(2016.0)}  &DR2  &DR3    &\multicolumn{1}{c}{[day]}  &\multicolumn{1}{c}{[day]} & & \multicolumn{1}{c}{ $\pm$0.17 }  & \multicolumn{1}{c}{[$\alpha$/Fe]=0.0} & \multicolumn{1}{c}{[$\alpha$/Fe]=0.3 }   \\
\multicolumn{1}{c}{(1)} & \multicolumn{1}{c}{(2)} &\multicolumn{1}{c}{(3)}  &\multicolumn{1}{c}{(4)} & \multicolumn{1}{c}{(5)} & \multicolumn{1}{c}{(6)}   & \multicolumn{1}{c}{(7)}  & \multicolumn{1}{c}{(8)}   & \multicolumn{1}{c}{(9)} &\multicolumn{1}{c}{(10)} &\multicolumn{1}{c}{(11)}   \\
\hline  
\\	    

    4662478439122071168 & 70.39830315 & -66.79943741 & RRd   & RRd   & 0.351664 & 0.472821 & 0.74376 & -1.53 & 0.69  & 0.66 \\
    5003198778816655360 & 14.97444688 & -33.78111075 & RRd   & RRd   & 0.351376 & 0.472884 & 0.74305 & -1.53 & 0.69  & 0.66 \\
    4676021948462223104 & 63.70590377 & -63.91995223 & RRd   & RRd   & 0.351872 & 0.472895 & 0.74408 & -1.53 & 0.69  & 0.65 \\
    1079573775408821120 & 175.3097673 & 75.70619479 &       & RRd   & 0.351758 & 0.473000 & 0.74367 & -1.53 & 0.69  & 0.66 \\
\hline
\end{tabular}
\end{table*}

\begin{table*}
\fontsize{7}{8.4}\selectfont
\centering
\contcaption{  }
\begin{tabular}{lllllcccccc}
\hline
\multicolumn{1}{c}{ {\it Gaia} Id  } &   \multicolumn{1}{c}{RA} & \multicolumn{1}{c}{DEC} &  \multicolumn{2}{c}{{\it Gaia} Classif.} & \multicolumn{1}{c}{ $P_1$}   & \multicolumn{1}{c}{ $P_0$ } & \multicolumn{1}{c} { $P_1/P_0$ }  & \multicolumn{1}{c}{[Fe/H] }  & \multicolumn{2}{c}{$M/M_{\sun}$} \\
\multicolumn{1}{c}{  } &  \multicolumn{2}{c}{(2016.0)}  &DR2  &DR3    &\multicolumn{1}{c}{[day]}  &\multicolumn{1}{c}{[day]} & & \multicolumn{1}{c}{ $\pm$0.17 }  & \multicolumn{1}{c}{[$\alpha$/Fe]=0.0} & \multicolumn{1}{c}{[$\alpha$/Fe]=0.3 }   \\
\multicolumn{1}{c}{(1)} & \multicolumn{1}{c}{(2)} &\multicolumn{1}{c}{(3)}  &\multicolumn{1}{c}{(4)} & \multicolumn{1}{c}{(5)} & \multicolumn{1}{c}{(6)}   & \multicolumn{1}{c}{(7)}  & \multicolumn{1}{c}{(8)}   & \multicolumn{1}{c}{(9)} &\multicolumn{1}{c}{(10)} &\multicolumn{1}{c}{(11)}   \\
\hline  
\\	    
  
    4639336021447251840 & 51.16012682 & -75.86617301 & RRc   & RRd   & 0.353659 & 0.475237 & 0.74418 & -1.55 & 0.69  & 0.66 \\
    3478707494382792960 & 174.8330345 & -31.28140957 & RRd   & RRd   & 0.353399 & 0.475239 & 0.74363 & -1.55 & 0.69  & 0.66 \\
    4726052033924416384 & 40.27218104 & -59.58580803 &       & RRd   & 0.353647 & 0.475255 & 0.74412 & -1.55 & 0.69  & 0.66 \\
    5003196407994372992 & 14.9184054 & -33.88434728 & RRd   & RRd   & 0.353736 & 0.475271 & 0.74428 & -1.55 & 0.69  & 0.66 \\
\hline
\end{tabular}
\end{table*}

\begin{table*}
\fontsize{7}{8.4}\selectfont
\centering
\contcaption{  }
\begin{tabular}{lllllcccccc}
\hline
\multicolumn{1}{c}{ {\it Gaia} Id  } &   \multicolumn{1}{c}{RA} & \multicolumn{1}{c}{DEC} &  \multicolumn{2}{c}{{\it Gaia} Classif.} & \multicolumn{1}{c}{ $P_1$}   & \multicolumn{1}{c}{ $P_0$ } & \multicolumn{1}{c} { $P_1/P_0$ }  & \multicolumn{1}{c}{[Fe/H] }  & \multicolumn{2}{c}{$M/M_{\sun}$} \\
\multicolumn{1}{c}{  } &  \multicolumn{2}{c}{(2016.0)}  &DR2  &DR3    &\multicolumn{1}{c}{[day]}  &\multicolumn{1}{c}{[day]} & & \multicolumn{1}{c}{ $\pm$0.17 }  & \multicolumn{1}{c}{[$\alpha$/Fe]=0.0} & \multicolumn{1}{c}{[$\alpha$/Fe]=0.3 }   \\
\multicolumn{1}{c}{(1)} & \multicolumn{1}{c}{(2)} &\multicolumn{1}{c}{(3)}  &\multicolumn{1}{c}{(4)} & \multicolumn{1}{c}{(5)} & \multicolumn{1}{c}{(6)}   & \multicolumn{1}{c}{(7)}  & \multicolumn{1}{c}{(8)}   & \multicolumn{1}{c}{(9)} &\multicolumn{1}{c}{(10)} &\multicolumn{1}{c}{(11)}   \\
\hline  
\\	    

    6648425412177481856 & 277.363665 & -56.20874787 & RRd   & RRd   & 0.355125 & 0.477125 & 0.74430 & -1.57 & 0.69  & 0.66 \\
    5275124187801866624 & 113.3386579 & -67.07900786 & RRd   & RRd   & 0.355203 & 0.477130 & 0.74446 & -1.57 & 0.69  & 0.66 \\
    5288584340431063552 & 112.901974 & -63.42018134 & RRd   & RRd   & 0.355052 & 0.477137 & 0.74413 & -1.57 & 0.69  & 0.66 \\
    6848271894018143488 & 306.4226729 & -24.98503121 & RRd   & RRc   & 0.354942 & 0.477154 & 0.74387 & -1.57 & 0.69  & 0.66 \\
\hline
\end{tabular}
\end{table*}

\begin{table*}
\fontsize{7}{8.4}\selectfont
\centering
\contcaption{  }
\label{tab:three}
\begin{tabular}{lllllcccccc}
\hline
\multicolumn{1}{c}{ {\it Gaia} Id  } &   \multicolumn{1}{c}{RA} & \multicolumn{1}{c}{DEC} &  \multicolumn{2}{c}{{\it Gaia} Classif.} & \multicolumn{1}{c}{ $P_1$}   & \multicolumn{1}{c}{ $P_0$ } & \multicolumn{1}{c} { $P_1/P_0$ }  & \multicolumn{1}{c}{[Fe/H] }  & \multicolumn{2}{c}{$M/M_{\sun}$} \\
\multicolumn{1}{c}{  } &  \multicolumn{2}{c}{(2016.0)}  &DR2  &DR3    &\multicolumn{1}{c}{[day]}  &\multicolumn{1}{c}{[day]} & & \multicolumn{1}{c}{ $\pm$0.17 }  & \multicolumn{1}{c}{[$\alpha$/Fe]=0.0} & \multicolumn{1}{c}{[$\alpha$/Fe]=0.3 }   \\
\multicolumn{1}{c}{(1)} & \multicolumn{1}{c}{(2)} &\multicolumn{1}{c}{(3)}  &\multicolumn{1}{c}{(4)} & \multicolumn{1}{c}{(5)} & \multicolumn{1}{c}{(6)}   & \multicolumn{1}{c}{(7)}  & \multicolumn{1}{c}{(8)}   & \multicolumn{1}{c}{(9)} &\multicolumn{1}{c}{(10)} &\multicolumn{1}{c}{(11)}   \\
\hline  
\\	    

    4469229615442034304 & 268.5799448 & 3.137606294 &       & RRd   & 0.356296 & 0.478570 & 0.74450 & -1.58 & 0.70  & 0.66 \\
    4892460644887994368 & 66.14179656 & -27.13345652 & RRd   & RRd   & 0.356058 & 0.478603 & 0.74395 & -1.58 & 0.70  & 0.66 \\
    4663993398637189376 & 74.64199998 & -64.20664981 & RRd   & RRd   & 0.356308 & 0.478621 & 0.74445 & -1.58 & 0.70  & 0.66 \\
    4412961150114707840 & 240.4161644 & 3.433022557 & RRd   & RRc   & 0.356385 & 0.478642 & 0.74458 & -1.58 & 0.70  & 0.66 \\
\hline
\end{tabular}
\end{table*}

\begin{table*}
\fontsize{7}{8.4}\selectfont
\centering
\contcaption{  }
\begin{tabular}{lllllcccccc}
\hline
	\multicolumn{1}{c}{ {\it Gaia} Id  } &   \multicolumn{1}{c}{RA} & \multicolumn{1}{c}{DEC} &  \multicolumn{2}{c}{{\it Gaia} Classif.} & \multicolumn{1}{c}{ $P_1$}   & \multicolumn{1}{c}{ $P_0$ } & \multicolumn{1}{c} { $P_1/P_0$ }  & \multicolumn{1}{c}{[Fe/H] }  & \multicolumn{2}{c}{$M/M_{\sun}$} \\
\multicolumn{1}{c}{  } &  \multicolumn{2}{c}{(2016.0)}  &DR2  &DR3    &\multicolumn{1}{c}{[day]}  &\multicolumn{1}{c}{[day]} & & \multicolumn{1}{c}{ $\pm$0.17 }  & \multicolumn{1}{c}{[$\alpha$/Fe]=0.0} & \multicolumn{1}{c}{[$\alpha$/Fe]=0.3 }   \\
\multicolumn{1}{c}{(1)} & \multicolumn{1}{c}{(2)} &\multicolumn{1}{c}{(3)}  &\multicolumn{1}{c}{(4)} & \multicolumn{1}{c}{(5)} & \multicolumn{1}{c}{(6)}   & \multicolumn{1}{c}{(7)}  & \multicolumn{1}{c}{(8)}   & \multicolumn{1}{c}{(9)} &\multicolumn{1}{c}{(10)} &\multicolumn{1}{c}{(11)}   \\
\hline  
\\	    

    5771812382215480960 & 225.1985335 & -82.22500625 & RRc   & RRd   & 0.357280 & 0.479977 & 0.74437 & -1.59 & 0.70  & 0.66 \\
    5895265616510752000 & 213.2466796 & -56.72671066 &       & RRd   & 0.357460 & 0.479977 & 0.74474 & -1.59 & 0.70  & 0.66 \\
    1200935498061703552 & 246.5322066 & 19.13923185 & RRd   &       & 0.357207 & 0.480035 & 0.74413 & -1.59 & 0.70  & 0.66 \\
    2797778362750981504 & 4.353840646 & 20.0948773 &       & RRd   & 0.357129 & 0.480044 & 0.74395 & -1.59 & 0.70  & 0.67 \\
\hline
\end{tabular}
\end{table*}

\begin{table*}
\fontsize{7}{8.4}\selectfont
\centering
\contcaption{  }
\begin{tabular}{lllllcccccc}
\hline
\multicolumn{1}{c}{ {\it Gaia} Id  } &   \multicolumn{1}{c}{RA} & \multicolumn{1}{c}{DEC} &  \multicolumn{2}{c}{{\it Gaia} Classif.} & \multicolumn{1}{c}{ $P_1$}   & \multicolumn{1}{c}{ $P_0$ } & \multicolumn{1}{c} { $P_1/P_0$ }  & \multicolumn{1}{c}{[Fe/H] }  & \multicolumn{2}{c}{$M/M_{\sun}$} \\
\multicolumn{1}{c}{  } &  \multicolumn{2}{c}{(2016.0)}  &DR2  &DR3    &\multicolumn{1}{c}{[day]}  &\multicolumn{1}{c}{[day]} & & \multicolumn{1}{c}{ $\pm$0.17 }  & \multicolumn{1}{c}{[$\alpha$/Fe]=0.0} & \multicolumn{1}{c}{[$\alpha$/Fe]=0.3 }   \\
\multicolumn{1}{c}{(1)} & \multicolumn{1}{c}{(2)} &\multicolumn{1}{c}{(3)}  &\multicolumn{1}{c}{(4)} & \multicolumn{1}{c}{(5)} & \multicolumn{1}{c}{(6)}   & \multicolumn{1}{c}{(7)}  & \multicolumn{1}{c}{(8)}   & \multicolumn{1}{c}{(9)} &\multicolumn{1}{c}{(10)} &\multicolumn{1}{c}{(11)}   \\
\hline  
\\	    

    4121067716689661184 & 262.1912195 & -20.35583951 &       & RRd   & 0.358478 & 0.481507 & 0.74449 & -1.61 & 0.70  & 0.67 \\
    4664087578672513792 & 79.59050354 & -64.17141431 & RRc   & RRd   & 0.358283 & 0.481525 & 0.74406 & -1.61 & 0.70  & 0.67 \\
    5359861285517982336 & 165.3120292 & -52.38792753 &       & RRd   & 0.358374 & 0.481553 & 0.74420 & -1.61 & 0.70  & 0.67 \\
    4412246639356459648 & 242.5216573 & 2.705817084 & RRc   & RRd   & 0.358374 & 0.481583 & 0.74416 & -1.61 & 0.70  & 0.67 \\
\hline
\end{tabular}
\end{table*}

\begin{table*}
\fontsize{7}{8.4}\selectfont
\centering
\contcaption{  }
\begin{tabular}{lllllcccccc}
\hline
\multicolumn{1}{c}{ {\it Gaia} Id  } &   \multicolumn{1}{c}{RA} & \multicolumn{1}{c}{DEC} &  \multicolumn{2}{c}{{\it Gaia} Classif.} & \multicolumn{1}{c}{ $P_1$}   & \multicolumn{1}{c}{ $P_0$ } & \multicolumn{1}{c} { $P_1/P_0$ }  & \multicolumn{1}{c}{[Fe/H] }  & \multicolumn{2}{c}{$M/M_{\sun}$} \\
\multicolumn{1}{c}{  } &  \multicolumn{2}{c}{(2016.0)}  &DR2  &DR3    &\multicolumn{1}{c}{[day]}  &\multicolumn{1}{c}{[day]} & & \multicolumn{1}{c}{ $\pm$0.17 }  & \multicolumn{1}{c}{[$\alpha$/Fe]=0.0} & \multicolumn{1}{c}{[$\alpha$/Fe]=0.3 }   \\
\multicolumn{1}{c}{(1)} & \multicolumn{1}{c}{(2)} &\multicolumn{1}{c}{(3)}  &\multicolumn{1}{c}{(4)} & \multicolumn{1}{c}{(5)} & \multicolumn{1}{c}{(6)}   & \multicolumn{1}{c}{(7)}  & \multicolumn{1}{c}{(8)}   & \multicolumn{1}{c}{(9)} &\multicolumn{1}{c}{(10)} &\multicolumn{1}{c}{(11)}   \\
\hline  
\\	    

    3717777361329045504 & 202.7748211 & 5.385535618 &       & RRd   & 0.359330 & 0.483112 & 0.74378 & -1.62 & 0.70  & 0.67 \\
    4684326250549633792 & 10.3284766 & -76.46602962 & RRd   & RRab  & 0.359716 & 0.483112 & 0.74458 & -1.62 & 0.70  & 0.67 \\
    6248564854564620416 & 237.9857875 & -18.0553581 &       & RRd   & 0.359795 & 0.483123 & 0.74473 & -1.62 & 0.70  & 0.67 \\
    2026242498011872768 & 293.7373372 & 28.54287461 & RRc   & RRd   & 0.359841 & 0.483147 & 0.74479 & -1.62 & 0.70  & 0.67 \\
\hline
\end{tabular}
\end{table*}

\begin{table*}
\fontsize{7}{8.4}\selectfont
\centering
\contcaption{  }
\begin{tabular}{lllllcccccc}
\hline
	\multicolumn{1}{c}{ {\it Gaia} Id  } &   \multicolumn{1}{c}{RA} & \multicolumn{1}{c}{DEC} &  \multicolumn{2}{c}{{\it Gaia} Classif.} & \multicolumn{1}{c}{ $P_1$}   & \multicolumn{1}{c}{ $P_0$ } & \multicolumn{1}{c} { $P_1/P_0$ }  & \multicolumn{1}{c}{[Fe/H] }  & \multicolumn{2}{c}{$M/M_{\sun}$} \\
\multicolumn{1}{c}{  } &  \multicolumn{2}{c}{(2016.0)}  &DR2  &DR3    &\multicolumn{1}{c}{[day]}  &\multicolumn{1}{c}{[day]} & & \multicolumn{1}{c}{ $\pm$0.17 }  & \multicolumn{1}{c}{[$\alpha$/Fe]=0.0} & \multicolumn{1}{c}{[$\alpha$/Fe]=0.3 }   \\
\multicolumn{1}{c}{(1)} & \multicolumn{1}{c}{(2)} &\multicolumn{1}{c}{(3)}  &\multicolumn{1}{c}{(4)} & \multicolumn{1}{c}{(5)} & \multicolumn{1}{c}{(6)}   & \multicolumn{1}{c}{(7)}  & \multicolumn{1}{c}{(8)}   & \multicolumn{1}{c}{(9)} &\multicolumn{1}{c}{(10)} &\multicolumn{1}{c}{(11)}   \\
\hline  
\\	    

    5834504752877789440 & 241.8700217 & -59.68328735 & RRd   & RRd   & 0.360787 & 0.484446 & 0.74474 & -1.64 & 0.70  & 0.67 \\
    4636037142966203776 & 16.60512086 & -77.35329175 & RRd   & RRd   & 0.360876 & 0.484453 & 0.74491 & -1.64 & 0.70  & 0.67 \\
    4688475532570193024 & 6.023324779 & -73.72816699 & RRd   & RRd   & 0.360519 & 0.484457 & 0.74417 & -1.64 & 0.71  & 0.67 \\
    4685675626227398912 & 13.37272215 & -73.67654114 &       & RRd   & 0.360797 & 0.484464 & 0.74473 & -1.64 & 0.70  & 0.67 \\
\hline
\end{tabular}
\end{table*}

\begin{table*}
\fontsize{7}{8.4}\selectfont
\centering
\contcaption{  }
\begin{tabular}{lllllcccccc}
\hline
	\multicolumn{1}{c}{ {\it Gaia} Id  } &   \multicolumn{1}{c}{RA} & \multicolumn{1}{c}{DEC} &  \multicolumn{2}{c}{{\it Gaia} Classif.} & \multicolumn{1}{c}{ $P_1$}   & \multicolumn{1}{c}{ $P_0$ } & \multicolumn{1}{c} { $P_1/P_0$ }  & \multicolumn{1}{c}{[Fe/H] }  & \multicolumn{2}{c}{$M/M_{\sun}$} \\
\multicolumn{1}{c}{  } &  \multicolumn{2}{c}{(2016.0)}  &DR2  &DR3    &\multicolumn{1}{c}{[day]}  &\multicolumn{1}{c}{[day]} & & \multicolumn{1}{c}{ $\pm$0.17 }  & \multicolumn{1}{c}{[$\alpha$/Fe]=0.0} & \multicolumn{1}{c}{[$\alpha$/Fe]=0.3 }   \\
\multicolumn{1}{c}{(1)} & \multicolumn{1}{c}{(2)} &\multicolumn{1}{c}{(3)}  &\multicolumn{1}{c}{(4)} & \multicolumn{1}{c}{(5)} & \multicolumn{1}{c}{(6)}   & \multicolumn{1}{c}{(7)}  & \multicolumn{1}{c}{(8)}   & \multicolumn{1}{c}{(9)} &\multicolumn{1}{c}{(10)} &\multicolumn{1}{c}{(11)}   \\
\hline  
\\	    

    2006010315803593472 & 335.0039322 & 55.81377753 &       & RRd   & 0.361665 & 0.485882 & 0.74435 & -1.65 & 0.71  & 0.67 \\
    4689057659585426048 & 13.1656348 & -72.30013289 &       & RRd   & 0.361829 & 0.485911 & 0.74464 & -1.65 & 0.71  & 0.67 \\
    5283002467803149312 & 95.6593484 & -67.30386067 & RRd   & RRd   & 0.361774 & 0.485929 & 0.74450 & -1.65 & 0.71  & 0.67 \\
    4638092164560734592 & 21.69082853 & -74.75030484 &       & RRd   & 0.361821 & 0.485931 & 0.74459 & -1.65 & 0.71  & 0.67 \\
\hline
\end{tabular}
\end{table*}

\begin{table*}
\fontsize{7}{8.4}\selectfont
\centering
\contcaption{  }
\label{tab:three}
\begin{tabular}{lllllcccccc}
\hline
	\multicolumn{1}{c}{ {\it Gaia} Id  } &   \multicolumn{1}{c}{RA} & \multicolumn{1}{c}{DEC} &  \multicolumn{2}{c}{{\it Gaia} Classif.} & \multicolumn{1}{c}{ $P_1$}   & \multicolumn{1}{c}{ $P_0$ } & \multicolumn{1}{c} { $P_1/P_0$ }  & \multicolumn{1}{c}{[Fe/H] }  & \multicolumn{2}{c}{$M/M_{\sun}$} \\
\multicolumn{1}{c}{  } &  \multicolumn{2}{c}{(2016.0)}  &DR2  &DR3    &\multicolumn{1}{c}{[day]}  &\multicolumn{1}{c}{[day]} & & \multicolumn{1}{c}{ $\pm$0.17 }  & \multicolumn{1}{c}{[$\alpha$/Fe]=0.0} & \multicolumn{1}{c}{[$\alpha$/Fe]=0.3 }   \\
\multicolumn{1}{c}{(1)} & \multicolumn{1}{c}{(2)} &\multicolumn{1}{c}{(3)}  &\multicolumn{1}{c}{(4)} & \multicolumn{1}{c}{(5)} & \multicolumn{1}{c}{(6)}   & \multicolumn{1}{c}{(7)}  & \multicolumn{1}{c}{(8)}   & \multicolumn{1}{c}{(9)} &\multicolumn{1}{c}{(10)} &\multicolumn{1}{c}{(11)}   \\
\hline  
\\	    

    4658956742049654144 & 79.08037347 & -66.92698841 & RRd   & RRd   & 0.362824 & 0.487243 & 0.74465 & -1.66 & 0.71  & 0.67 \\
    5842807573141418240 & 184.3659495 & -71.66939404 & RRc   & RRd   & 0.362632 & 0.487312 & 0.74415 & -1.66 & 0.71  & 0.68 \\
    4658912143096142336 & 79.16400077 & -67.21651038 & RRd   &       & 0.362561 & 0.487316 & 0.74400 & -1.66 & 0.71  & 0.68 \\
    5217685739027372672 & 144.567228 & -73.5686423 & RRc   & RRd   & 0.363083 & 0.487345 & 0.74502 & -1.66 & 0.71  & 0.67 \\
\hline
\end{tabular}
\end{table*}

\begin{table*}
\fontsize{7}{8.4}\selectfont
\centering
\contcaption{  }
\label{tab:three}
\begin{tabular}{lllllcccccc}
\hline
	\multicolumn{1}{c}{ {\it Gaia} Id  } &   \multicolumn{1}{c}{RA} & \multicolumn{1}{c}{DEC} &  \multicolumn{2}{c}{{\it Gaia} Classif.} & \multicolumn{1}{c}{ $P_1$}   & \multicolumn{1}{c}{ $P_0$ } & \multicolumn{1}{c} { $P_1/P_0$ }  & \multicolumn{1}{c}{[Fe/H] }  & \multicolumn{2}{c}{$M/M_{\sun}$} \\
\multicolumn{1}{c}{  } &  \multicolumn{2}{c}{(2016.0)}  &DR2  &DR3    &\multicolumn{1}{c}{[day]}  &\multicolumn{1}{c}{[day]} & & \multicolumn{1}{c}{ $\pm$0.17 }  & \multicolumn{1}{c}{[$\alpha$/Fe]=0.0} & \multicolumn{1}{c}{[$\alpha$/Fe]=0.3 }   \\
\multicolumn{1}{c}{(1)} & \multicolumn{1}{c}{(2)} &\multicolumn{1}{c}{(3)}  &\multicolumn{1}{c}{(4)} & \multicolumn{1}{c}{(5)} & \multicolumn{1}{c}{(6)}   & \multicolumn{1}{c}{(7)}  & \multicolumn{1}{c}{(8)}   & \multicolumn{1}{c}{(9)} &\multicolumn{1}{c}{(10)} &\multicolumn{1}{c}{(11)}   \\
\hline  
\\	    

    5432535666915585664 & 145.1980519 & -39.18283553 &       & RRd   & 0.364320 & 0.488898 & 0.74519 & -1.68 & 0.71  & 0.67 \\
    4636147953122913664 & 18.51447312 & -76.62069969 & RRd   & RRd   & 0.364283 & 0.488929 & 0.74506 & -1.68 & 0.71  & 0.68 \\
    1968430790761100032 & 316.6322335 & 40.04908209 &       & RRd   & 0.363841 & 0.488953 & 0.74412 & -1.68 & 0.71  & 0.68 \\
    4605813187524865920 & 274.0028173 & 35.50114911 & RRd   & RRd   & 0.364062 & 0.489013 & 0.74448 & -1.68 & 0.71  & 0.68 \\
\hline
\end{tabular}
\end{table*}

\begin{table*}
\fontsize{7}{8.4}\selectfont
\centering
\contcaption{  }
\label{tab:three}
\begin{tabular}{lllllcccccc}
\hline
	\multicolumn{1}{c}{ {\it Gaia} Id  } &   \multicolumn{1}{c}{RA} & \multicolumn{1}{c}{DEC} &  \multicolumn{2}{c}{{\it Gaia} Classif.} & \multicolumn{1}{c}{ $P_1$}   & \multicolumn{1}{c}{ $P_0$ } & \multicolumn{1}{c} { $P_1/P_0$ }  & \multicolumn{1}{c}{[Fe/H] }  & \multicolumn{2}{c}{$M/M_{\sun}$} \\
\multicolumn{1}{c}{  } &  \multicolumn{2}{c}{(2016.0)}  &DR2  &DR3    &\multicolumn{1}{c}{[day]}  &\multicolumn{1}{c}{[day]} & & \multicolumn{1}{c}{ $\pm$0.17 }  & \multicolumn{1}{c}{[$\alpha$/Fe]=0.0} & \multicolumn{1}{c}{[$\alpha$/Fe]=0.3 }   \\
\multicolumn{1}{c}{(1)} & \multicolumn{1}{c}{(2)} &\multicolumn{1}{c}{(3)}  &\multicolumn{1}{c}{(4)} & \multicolumn{1}{c}{(5)} & \multicolumn{1}{c}{(6)}   & \multicolumn{1}{c}{(7)}  & \multicolumn{1}{c}{(8)}   & \multicolumn{1}{c}{(9)} &\multicolumn{1}{c}{(10)} &\multicolumn{1}{c}{(11)}   \\
\hline  
\\	    

    4684376901098145280 & 13.81736293 & -75.71753628 &       & RRd   & 0.365778 & 0.491018 & 0.74494 & -1.70 & 0.71  & 0.68 \\
    5750875520452673408 & 131.6724985 & -8.311029795 &       & RRd   & 0.365771 & 0.491083 & 0.74483 & -1.70 & 0.71  & 0.68 \\
    535845548689447936 & 19.73513032 & 74.60316383 & RRd   & RRd   & 0.365720 & 0.491095 & 0.74470 & -1.70 & 0.71  & 0.68 \\
    5926599503169493248 & 258.2952657 & -49.49153805 &       & RRd   & 0.365831 & 0.491102 & 0.74492 & -1.70 & 0.71  & 0.68 \\
\hline
\end{tabular}
\end{table*}

\begin{table*}
\fontsize{7}{8.4}\selectfont
\centering
\contcaption{  }
\label{tab:three}
\begin{tabular}{lllllcccccc}
\hline
	\multicolumn{1}{c}{ {\it Gaia} Id  } &   \multicolumn{1}{c}{RA} & \multicolumn{1}{c}{DEC} &  \multicolumn{2}{c}{{\it Gaia} Classif.} & \multicolumn{1}{c}{ $P_1$}   & \multicolumn{1}{c}{ $P_0$ } & \multicolumn{1}{c} { $P_1/P_0$ }  & \multicolumn{1}{c}{[Fe/H] }  & \multicolumn{2}{c}{$M/M_{\sun}$} \\
\multicolumn{1}{c}{  } &  \multicolumn{2}{c}{(2016.0)}  &DR2  &DR3    &\multicolumn{1}{c}{[day]}  &\multicolumn{1}{c}{[day]} & & \multicolumn{1}{c}{ $\pm$0.17 }  & \multicolumn{1}{c}{[$\alpha$/Fe]=0.0} & \multicolumn{1}{c}{[$\alpha$/Fe]=0.3 }   \\
\multicolumn{1}{c}{(1)} & \multicolumn{1}{c}{(2)} &\multicolumn{1}{c}{(3)}  &\multicolumn{1}{c}{(4)} & \multicolumn{1}{c}{(5)} & \multicolumn{1}{c}{(6)}   & \multicolumn{1}{c}{(7)}  & \multicolumn{1}{c}{(8)}   & \multicolumn{1}{c}{(9)} &\multicolumn{1}{c}{(10)} &\multicolumn{1}{c}{(11)}   \\
\hline  
\\	    

    1732577334622610816 & 318.17565 & 4.915774743 & RRc   & RRd   & 0.367458 & 0.493105 & 0.74519 & -1.72 & 0.72  & 0.68 \\
    4686400895853896064 & 18.43925096 & -73.58505002 & RRd   & RRd   & 0.367313 & 0.493117 & 0.74488 & -1.72 & 0.72  & 0.68 \\
    4661762730090966528 & 74.18333758 & -66.94198558 &       & RRd   & 0.367214 & 0.493125 & 0.74467 & -1.72 & 0.72  & 0.68 \\
    4037592553382324480 & 272.9595902 & -37.6592069 & RRab  & RRd   & 0.367501 & 0.493132 & 0.74524 & -1.72 & 0.72  & 0.68 \\
\hline
\end{tabular}
\end{table*}

\begin{table*}
\fontsize{7}{8.4}\selectfont
\centering
\contcaption{  }
\label{tab:three}
\begin{tabular}{lllllcccccc}
\hline
	\multicolumn{1}{c}{ {\it Gaia} Id  } &   \multicolumn{1}{c}{RA} & \multicolumn{1}{c}{DEC} &  \multicolumn{2}{c}{{\it Gaia} Classif.} & \multicolumn{1}{c}{ $P_1$}   & \multicolumn{1}{c}{ $P_0$ } & \multicolumn{1}{c} { $P_1/P_0$ }  & \multicolumn{1}{c}{[Fe/H] }  & \multicolumn{2}{c}{$M/M_{\sun}$} \\
\multicolumn{1}{c}{  } &  \multicolumn{2}{c}{(2016.0)}  &DR2  &DR3    &\multicolumn{1}{c}{[day]}  &\multicolumn{1}{c}{[day]} & & \multicolumn{1}{c}{ $\pm$0.17 }  & \multicolumn{1}{c}{[$\alpha$/Fe]=0.0} & \multicolumn{1}{c}{[$\alpha$/Fe]=0.3 }   \\
\multicolumn{1}{c}{(1)} & \multicolumn{1}{c}{(2)} &\multicolumn{1}{c}{(3)}  &\multicolumn{1}{c}{(4)} & \multicolumn{1}{c}{(5)} & \multicolumn{1}{c}{(6)}   & \multicolumn{1}{c}{(7)}  & \multicolumn{1}{c}{(8)}   & \multicolumn{1}{c}{(9)} &\multicolumn{1}{c}{(10)} &\multicolumn{1}{c}{(11)}   \\
\hline  
\\	    

    6661057941781000832 & 285.1828801 & -49.67974334 &       & RRd   & 0.369601 & 0.495937 & 0.74526 & -1.75 & 0.72  & 0.68 \\
    4685857698462831744 & 12.95996698 & -73.69081813 & RRab  & RRd   & 0.369354 & 0.495985 & 0.74469 & -1.75 & 0.72  & 0.69 \\
    3482402295830820992 & 173.0457003 & -28.83238017 &       & RRd   & 0.369598 & 0.496090 & 0.74502 & -1.75 & 0.72  & 0.69 \\
    4663630972124543360 & 79.09675097 & -65.28023906 & RRd   & RRc   & 0.369934 & 0.496145 & 0.74562 & -1.75 & 0.72  & 0.68 \\
\hline
\end{tabular}
\end{table*}

\begin{table*}
\fontsize{7}{8.4}\selectfont
\centering
\contcaption{  }
\label{tab:three}
\begin{tabular}{lllllcccccc}
\hline
	\multicolumn{1}{c}{ {\it Gaia} Id  } &   \multicolumn{1}{c}{RA} & \multicolumn{1}{c}{DEC} &  \multicolumn{2}{c}{{\it Gaia} Classif.} & \multicolumn{1}{c}{ $P_1$}   & \multicolumn{1}{c}{ $P_0$ } & \multicolumn{1}{c} { $P_1/P_0$ }  & \multicolumn{1}{c}{[Fe/H] }  & \multicolumn{2}{c}{$M/M_{\sun}$} \\
\multicolumn{1}{c}{  } &  \multicolumn{2}{c}{(2016.0)}  &DR2  &DR3    &\multicolumn{1}{c}{[day]}  &\multicolumn{1}{c}{[day]} & & \multicolumn{1}{c}{ $\pm$0.17 }  & \multicolumn{1}{c}{[$\alpha$/Fe]=0.0} & \multicolumn{1}{c}{[$\alpha$/Fe]=0.3 }   \\
\multicolumn{1}{c}{(1)} & \multicolumn{1}{c}{(2)} &\multicolumn{1}{c}{(3)}  &\multicolumn{1}{c}{(4)} & \multicolumn{1}{c}{(5)} & \multicolumn{1}{c}{(6)}   & \multicolumn{1}{c}{(7)}  & \multicolumn{1}{c}{(8)}   & \multicolumn{1}{c}{(9)} &\multicolumn{1}{c}{(10)} &\multicolumn{1}{c}{(11)}   \\
\hline  
\\	    

    4669720475525190400 & 62.38589329 & -65.67355867 & RRd   & RRd   & 0.371905 & 0.499509 & 0.74454 & -1.78 & 0.73  & 0.69 \\
    4630639198004086656 & 15.43042971 & -81.17356928 & RRd   & RRd   & 0.372229 & 0.499554 & 0.74512 & -1.78 & 0.73  & 0.69 \\
    4685435589084082048 & 9.820128153 & -75.1505751 & RRc   & RRd   & 0.371997 & 0.499579 & 0.74462 & -1.78 & 0.73  & 0.69 \\
    4663020395271859328 & 72.15936081 & -65.63690644 & RRd   & RRd   & 0.371667 & 0.499597 & 0.74393 & -1.78 & 0.73  & 0.69 \\
\hline
\end{tabular}
\end{table*}

\begin{table*}
\fontsize{7}{8.4}\selectfont
\centering
\contcaption{  }
\label{tab:three}
\begin{tabular}{lllllcccccc}
\hline
	\multicolumn{1}{c}{ {\it Gaia} Id  } &   \multicolumn{1}{c}{RA} & \multicolumn{1}{c}{DEC} &  \multicolumn{2}{c}{{\it Gaia} Classif.} & \multicolumn{1}{c}{ $P_1$}   & \multicolumn{1}{c}{ $P_0$ } & \multicolumn{1}{c} { $P_1/P_0$ }  & \multicolumn{1}{c}{[Fe/H] }  & \multicolumn{2}{c}{$M/M_{\sun}$} \\
\multicolumn{1}{c}{  } &  \multicolumn{2}{c}{(2016.0)}  &DR2  &DR3    &\multicolumn{1}{c}{[day]}  &\multicolumn{1}{c}{[day]} & & \multicolumn{1}{c}{ $\pm$0.17 }  & \multicolumn{1}{c}{[$\alpha$/Fe]=0.0} & \multicolumn{1}{c}{[$\alpha$/Fe]=0.3 }   \\
\multicolumn{1}{c}{(1)} & \multicolumn{1}{c}{(2)} &\multicolumn{1}{c}{(3)}  &\multicolumn{1}{c}{(4)} & \multicolumn{1}{c}{(5)} & \multicolumn{1}{c}{(6)}   & \multicolumn{1}{c}{(7)}  & \multicolumn{1}{c}{(8)}   & \multicolumn{1}{c}{(9)} &\multicolumn{1}{c}{(10)} &\multicolumn{1}{c}{(11)}   \\
\hline  
\\	    

    4703555128372152576 & 8.248214411 & -68.50224579 & RRc   & RRd   & 0.375530 & 0.503503 & 0.74583 & -1.82 & 0.73  & 0.69 \\
    6906523672162974336 & 310.1291591 & -8.165373388 & RRd   & RRd   & 0.375162 & 0.503521 & 0.74508 & -1.82 & 0.73  & 0.70 \\
    6076848419952483072 & 180.555574 & -54.31000784 &       & RRd   & 0.375386 & 0.503578 & 0.74544 & -1.82 & 0.73  & 0.70 \\
    4658949629598632320 & 79.94077382 & -66.89912354 & RRd   & RRd   & 0.374834 & 0.503584 & 0.74433 & -1.82 & 0.73  & 0.70 \\
\hline
\end{tabular}
\end{table*}

\begin{table*}
\fontsize{7}{8.4}\selectfont
\centering
\contcaption{  }
\label{tab:three}
\begin{tabular}{lllllcccccc}
\hline
	\multicolumn{1}{c}{ {\it Gaia} Id  } &   \multicolumn{1}{c}{RA} & \multicolumn{1}{c}{DEC} &  \multicolumn{2}{c}{{\it Gaia} Classif.} & \multicolumn{1}{c}{ $P_1$}   & \multicolumn{1}{c}{ $P_0$ } & \multicolumn{1}{c} { $P_1/P_0$ }  & \multicolumn{1}{c}{[Fe/H] }  & \multicolumn{2}{c}{$M/M_{\sun}$} \\
\multicolumn{1}{c}{  } &  \multicolumn{2}{c}{(2016.0)}  &DR2  &DR3    &\multicolumn{1}{c}{[day]}  &\multicolumn{1}{c}{[day]} & & \multicolumn{1}{c}{ $\pm$0.17 }  & \multicolumn{1}{c}{[$\alpha$/Fe]=0.0} & \multicolumn{1}{c}{[$\alpha$/Fe]=0.3 }   \\
\multicolumn{1}{c}{(1)} & \multicolumn{1}{c}{(2)} &\multicolumn{1}{c}{(3)}  &\multicolumn{1}{c}{(4)} & \multicolumn{1}{c}{(5)} & \multicolumn{1}{c}{(6)}   & \multicolumn{1}{c}{(7)}  & \multicolumn{1}{c}{(8)}   & \multicolumn{1}{c}{(9)} &\multicolumn{1}{c}{(10)} &\multicolumn{1}{c}{(11)}   \\
\hline  
\\	    

    4649571237752810496 & 80.48119231 & -74.16518458 & RRc   & RRd   & 0.379465 & 0.508803 & 0.74580 & -1.87 & 0.74  & 0.70 \\
    4112050759261291264 & 256.4672773 & -25.72525978 & RRab  & RRd   & 0.378748 & 0.508910 & 0.74423 & -1.87 & 0.74  & 0.71 \\
    4636407025550252800 & 15.21349112 & -76.04285724 & RRc   & RRd   & 0.379519 & 0.508975 & 0.74565 & -1.87 & 0.74  & 0.70 \\
    6652688871587133184 & 274.2934273 & -54.62689962 & RRd   & RRd   & 0.379070 & 0.508987 & 0.74475 & -1.87 & 0.74  & 0.70 \\
\hline
\end{tabular}
\end{table*}

\begin{table*}
\fontsize{7}{8.4}\selectfont
\centering
\contcaption{  }
\label{tab:three}
\begin{tabular}{lllllcccccc}
\hline
	\multicolumn{1}{c}{ {\it Gaia} Id  } &   \multicolumn{1}{c}{RA} & \multicolumn{1}{c}{DEC} &  \multicolumn{2}{c}{{\it Gaia} Classif.} & \multicolumn{1}{c}{ $P_1$}   & \multicolumn{1}{c}{ $P_0$ } & \multicolumn{1}{c} { $P_1/P_0$ }  & \multicolumn{1}{c}{[Fe/H] }  & \multicolumn{2}{c}{$M/M_{\sun}$} \\
\multicolumn{1}{c}{  } &  \multicolumn{2}{c}{(2016.0)}  &DR2  &DR3    &\multicolumn{1}{c}{[day]}  &\multicolumn{1}{c}{[day]} & & \multicolumn{1}{c}{ $\pm$0.17 }  & \multicolumn{1}{c}{[$\alpha$/Fe]=0.0} & \multicolumn{1}{c}{[$\alpha$/Fe]=0.3 }   \\
\multicolumn{1}{c}{(1)} & \multicolumn{1}{c}{(2)} &\multicolumn{1}{c}{(3)}  &\multicolumn{1}{c}{(4)} & \multicolumn{1}{c}{(5)} & \multicolumn{1}{c}{(6)}   & \multicolumn{1}{c}{(7)}  & \multicolumn{1}{c}{(8)}   & \multicolumn{1}{c}{(9)} &\multicolumn{1}{c}{(10)} &\multicolumn{1}{c}{(11)}   \\
\hline  
\\	    

    4701796459523626624 & 2.693289457 & -71.80053646 &       & RRd   & 0.383341 & 0.514465 & 0.74512 & -1.92 & 0.75  & 0.71 \\
    6593615753957003392 & 337.2871682 & -40.04563191 & RRc   & RRd   & 0.383965 & 0.514738 & 0.74594 & -1.92 & 0.75  & 0.71 \\
    825734538130102400 & 143.0859754 & 48.63424123 & RRc   & RRd   & 0.383996 & 0.514738 & 0.74600 & -1.92 & 0.75  & 0.71 \\
    5014730632863412864 & 19.72044177 & -33.22518401 &       & RRd   & 0.383748 & 0.514810 & 0.74542 & -1.92 & 0.75  & 0.71 \\
\hline
\end{tabular}
\end{table*}

\begin{table*}
\fontsize{7}{8.4}\selectfont
\centering
\contcaption{  }
\label{tab:three}
\begin{tabular}{lllllcccccc}
\hline
	\multicolumn{1}{c}{ {\it Gaia} Id  } &   \multicolumn{1}{c}{RA} & \multicolumn{1}{c}{DEC} &  \multicolumn{2}{c}{{\it Gaia} Classif.} & \multicolumn{1}{c}{ $P_1$}   & \multicolumn{1}{c}{ $P_0$ } & \multicolumn{1}{c} { $P_1/P_0$ }  & \multicolumn{1}{c}{[Fe/H] }  & \multicolumn{2}{c}{$M/M_{\sun}$} \\
\multicolumn{1}{c}{  } &  \multicolumn{2}{c}{(2016.0)}  &DR2  &DR3    &\multicolumn{1}{c}{[day]}  &\multicolumn{1}{c}{[day]} & & \multicolumn{1}{c}{ $\pm$0.17 }  & \multicolumn{1}{c}{[$\alpha$/Fe]=0.0} & \multicolumn{1}{c}{[$\alpha$/Fe]=0.3 }   \\
\multicolumn{1}{c}{(1)} & \multicolumn{1}{c}{(2)} &\multicolumn{1}{c}{(3)}  &\multicolumn{1}{c}{(4)} & \multicolumn{1}{c}{(5)} & \multicolumn{1}{c}{(6)}   & \multicolumn{1}{c}{(7)}  & \multicolumn{1}{c}{(8)}   & \multicolumn{1}{c}{(9)} &\multicolumn{1}{c}{(10)} &\multicolumn{1}{c}{(11)}   \\
\hline  
\\	    

    4685741768759398016 & 9.409288786 & -74.1339489 & RRd   & RRd   & 0.388460 & 0.520436 & 0.74641 & -1.97 & 0.75  & 0.72 \\
    4687698899388301824 & 22.98435732 & -72.01195409 & RRd   & RRd   & 0.387889 & 0.520471 & 0.74527 & -1.97 & 0.76  & 0.72 \\
    5699103744158077824 & 120.1308499 & -23.3960038 & RRd   & RRab  & 0.387365 & 0.520524 & 0.74418 & -1.97 & 0.76  & 0.72 \\
    6412942461388570112 & 329.103664 & -55.7730763 & RRab  & RRd   & 0.388363 & 0.520612 & 0.74597 & -1.97 & 0.76  & 0.72 \\
\hline
\end{tabular}
\end{table*}

\begin{table*}
\fontsize{7}{8.4}\selectfont
\centering
\contcaption{  }
\label{tab:three}
\begin{tabular}{lllllcccccc}
\hline
	\multicolumn{1}{c}{ {\it Gaia} Id  } &   \multicolumn{1}{c}{RA} & \multicolumn{1}{c}{DEC} &  \multicolumn{2}{c}{{\it Gaia} Classif.} & \multicolumn{1}{c}{ $P_1$}   & \multicolumn{1}{c}{ $P_0$ } & \multicolumn{1}{c} { $P_1/P_0$ }  & \multicolumn{1}{c}{[Fe/H] }  & \multicolumn{2}{c}{$M/M_{\sun}$} \\
\multicolumn{1}{c}{  } &  \multicolumn{2}{c}{(2016.0)}  &DR2  &DR3    &\multicolumn{1}{c}{[day]}  &\multicolumn{1}{c}{[day]} & & \multicolumn{1}{c}{ $\pm$0.17 }  & \multicolumn{1}{c}{[$\alpha$/Fe]=0.0} & \multicolumn{1}{c}{[$\alpha$/Fe]=0.3 }   \\
\multicolumn{1}{c}{(1)} & \multicolumn{1}{c}{(2)} &\multicolumn{1}{c}{(3)}  &\multicolumn{1}{c}{(4)} & \multicolumn{1}{c}{(5)} & \multicolumn{1}{c}{(6)}   & \multicolumn{1}{c}{(7)}  & \multicolumn{1}{c}{(8)}   & \multicolumn{1}{c}{(9)} &\multicolumn{1}{c}{(10)} &\multicolumn{1}{c}{(11)}   \\
\hline  
\\	    
	
   5263730807797824768 & 109.8285031 & -72.85165929 & RRd   & RRc   & 0.392311 & 0.526639 & 0.74493 & -2.03 & 0.77  & 0.73 \\
    4651999066835116672 & 78.74773229 & -70.84965769 & RRd   & RRd   & 0.392584 & 0.526700 & 0.74537 & -2.03 & 0.77  & 0.73 \\ 
    1147806271573934336 & 144.1590575 & 84.93838156 & RRd   & RRab  & 0.393189 & 0.526794 & 0.74638 & -2.03 & 0.76  & 0.73 \\
    6799569297045200768 & 308.7381038 & -27.00311618 & RRd   & RRd   & 0.393065 & 0.526833 & 0.74609 & -2.03 & 0.76  & 0.73 \\
\hline
\end{tabular}
\end{table*}

\begin{table*}
\fontsize{7}{8.4}\selectfont
\centering
\contcaption{  }
\label{tab:three}
\begin{tabular}{lllllcccccc}
\hline
\multicolumn{1}{c}{ {\it Gaia} Id  } &   \multicolumn{1}{c}{RA} & \multicolumn{1}{c}{DEC} &  \multicolumn{2}{c}{{\it Gaia} Classif.} & \multicolumn{1}{c}{ $P_1$}   & \multicolumn{1}{c}{ $P_0$ } & \multicolumn{1}{c} { $P_1/P_0$ }  & \multicolumn{1}{c}{[Fe/H] }  & \multicolumn{2}{c}{$M/M_{\sun}$} \\
\multicolumn{1}{c}{  } &  \multicolumn{2}{c}{(2016.0)}  &DR2  &DR3    &\multicolumn{1}{c}{[day]}  &\multicolumn{1}{c}{[day]} & & \multicolumn{1}{c}{ $\pm$0.17 }  & \multicolumn{1}{c}{[$\alpha$/Fe]=0.0} & \multicolumn{1}{c}{[$\alpha$/Fe]=0.3 }   \\
\multicolumn{1}{c}{(1)} & \multicolumn{1}{c}{(2)} &\multicolumn{1}{c}{(3)}  &\multicolumn{1}{c}{(4)} & \multicolumn{1}{c}{(5)} & \multicolumn{1}{c}{(6)}   & \multicolumn{1}{c}{(7)}  & \multicolumn{1}{c}{(8)}   & \multicolumn{1}{c}{(9)} &\multicolumn{1}{c}{(10)} &\multicolumn{1}{c}{(11)}   \\
\hline  
\\	    

    3705768362187755776 & 192.8904775 & 5.010929455 & RRd   & RRc   & 0.397186 & 0.531819 & 0.74685 & -2.08 & 0.77  & 0.73 \\
    4690525713725289344 & 15.6723101 & -71.97241721 & RRc   & RRd   & 0.396569 & 0.531905 & 0.74556 & -2.08 & 0.77  & 0.74 \\
    4704968829741128320 & 18.93756354 & -66.19425346 & RRab  & RRd   & 0.396727 & 0.531929 & 0.74583 & -2.08 & 0.77  & 0.74 \\
    5623698240288344448 & 138.5888073 & -35.46908878 & RRc   & RRd   & 0.396859 & 0.531967 & 0.74602 & -2.08 & 0.77  & 0.74 \\
\hline
\end{tabular}
\end{table*}

\begin{table*}
\fontsize{7}{8.4}\selectfont
\centering
\contcaption{  }
\label{tab:three}
\begin{tabular}{lllllcccccc}
\hline
	\multicolumn{1}{c}{ {\it Gaia} Id  } &   \multicolumn{1}{c}{RA} & \multicolumn{1}{c}{DEC} &  \multicolumn{2}{c}{{\it Gaia} Classif.} & \multicolumn{1}{c}{ $P_1$}   & \multicolumn{1}{c}{ $P_0$ } & \multicolumn{1}{c} { $P_1/P_0$ }  & \multicolumn{1}{c}{[Fe/H] }  & \multicolumn{2}{c}{$M/M_{\sun}$} \\
\multicolumn{1}{c}{  } &  \multicolumn{2}{c}{(2016.0)}  &DR2  &DR3    &\multicolumn{1}{c}{[day]}  &\multicolumn{1}{c}{[day]} & & \multicolumn{1}{c}{ $\pm$0.17 }  & \multicolumn{1}{c}{[$\alpha$/Fe]=0.0} & \multicolumn{1}{c}{[$\alpha$/Fe]=0.3 }   \\
\multicolumn{1}{c}{(1)} & \multicolumn{1}{c}{(2)} &\multicolumn{1}{c}{(3)}  &\multicolumn{1}{c}{(4)} & \multicolumn{1}{c}{(5)} & \multicolumn{1}{c}{(6)}   & \multicolumn{1}{c}{(7)}  & \multicolumn{1}{c}{(8)}   & \multicolumn{1}{c}{(9)} &\multicolumn{1}{c}{(10)} &\multicolumn{1}{c}{(11)}   \\
\hline  
\\	    
    4661424935258381824 & 76.55035964 & -67.93546241 & RRc   & RRd   & 0.400390 & 0.536773 & 0.74592 & -2.12 & 0.78  & 0.74 \\
    1707689824503315968 & 246.3125828 & 78.75120502 & RRd   & RRc   & 0.400757 & 0.536917 & 0.74640 & -2.12 & 0.78  & 0.74 \\
    1705038592731948416 & 255.7504274 & 77.01459294 & RRd   & RRc   & 0.400908 & 0.536955 & 0.74663 & -2.12 & 0.78  & 0.74 \\
    4639486280879048832 & 50.46522103 & -75.57661992 & RRd   & RRc   & 0.400907 & 0.536962 & 0.74662 & -2.12 & 0.78  & 0.74 \\
\hline
\end{tabular}
\end{table*}

\begin{table*}
\fontsize{7}{8.4}\selectfont
\centering
\contcaption{  }
\label{tab:three}
\begin{tabular}{lllllcccccc}
\hline
	\multicolumn{1}{c}{ {\it Gaia} Id  } &   \multicolumn{1}{c}{RA} & \multicolumn{1}{c}{DEC} &  \multicolumn{2}{c}{{\it Gaia} Classif.} & \multicolumn{1}{c}{ $P_1$}   & \multicolumn{1}{c}{ $P_0$ } & \multicolumn{1}{c} { $P_1/P_0$ }  & \multicolumn{1}{c}{[Fe/H] }  & \multicolumn{2}{c}{$M/M_{\sun}$} \\
\multicolumn{1}{c}{  } &  \multicolumn{2}{c}{(2016.0)}  &DR2  &DR3    &\multicolumn{1}{c}{[day]}  &\multicolumn{1}{c}{[day]} & & \multicolumn{1}{c}{ $\pm$0.17 }  & \multicolumn{1}{c}{[$\alpha$/Fe]=0.0} & \multicolumn{1}{c}{[$\alpha$/Fe]=0.3 }   \\
\multicolumn{1}{c}{(1)} & \multicolumn{1}{c}{(2)} &\multicolumn{1}{c}{(3)}  &\multicolumn{1}{c}{(4)} & \multicolumn{1}{c}{(5)} & \multicolumn{1}{c}{(6)}   & \multicolumn{1}{c}{(7)}  & \multicolumn{1}{c}{(8)}   & \multicolumn{1}{c}{(9)} &\multicolumn{1}{c}{(10)} &\multicolumn{1}{c}{(11)}   \\
\hline  
\\	    

    4731414867529564416 & 55.13003151 & -55.70404523 &       & RRd   & 0.404895 & 0.542972 & 0.74570 & -2.18 & 0.79  & 0.75 \\
    5658173289934917504 & 144.5573516 & -26.36745352 &       & RRd   & 0.404909 & 0.543040 & 0.74563 & -2.18 & 0.79  & 0.75 \\
    2478808592129008000 & 22.44704479 & -6.046846805 &       & RRd   & 0.404971 & 0.543062 & 0.74572 & -2.18 & 0.79  & 0.75 \\
    4662238543739598464 & 75.21237554 & -65.982835 & RRd   & RRd   & 0.405198 & 0.543143 & 0.74602 & -2.18 & 0.79  & 0.75 \\
\hline
\end{tabular}
\end{table*}

\begin{table*}
\fontsize{7}{8.4}\selectfont
\centering
\contcaption{  }
\label{tab:three}
\begin{tabular}{lllllcccccc}
\hline
	\multicolumn{1}{c}{ {\it Gaia} Id  } &   \multicolumn{1}{c}{RA} & \multicolumn{1}{c}{DEC} &  \multicolumn{2}{c}{{\it Gaia} Classif.} & \multicolumn{1}{c}{ $P_1$}   & \multicolumn{1}{c}{ $P_0$ } & \multicolumn{1}{c} { $P_1/P_0$ }  & \multicolumn{1}{c}{[Fe/H] }  & \multicolumn{2}{c}{$M/M_{\sun}$} \\
\multicolumn{1}{c}{  } &  \multicolumn{2}{c}{(2016.0)}  &DR2  &DR3    &\multicolumn{1}{c}{[day]}  &\multicolumn{1}{c}{[day]} & & \multicolumn{1}{c}{ $\pm$0.17 }  & \multicolumn{1}{c}{[$\alpha$/Fe]=0.0} & \multicolumn{1}{c}{[$\alpha$/Fe]=0.3 }   \\
\multicolumn{1}{c}{(1)} & \multicolumn{1}{c}{(2)} &\multicolumn{1}{c}{(3)}  &\multicolumn{1}{c}{(4)} & \multicolumn{1}{c}{(5)} & \multicolumn{1}{c}{(6)}   & \multicolumn{1}{c}{(7)}  & \multicolumn{1}{c}{(8)}   & \multicolumn{1}{c}{(9)} &\multicolumn{1}{c}{(10)} &\multicolumn{1}{c}{(11)}   \\
\hline  
\\	    
 
    4660419985878997376 & 81.20849528 & -66.63646889 & RRd   & RRd   & 0.409786 & 0.549214 & 0.74613 & -2.23 & 0.80  & 0.76 \\
    3676707135993695872 & 191.1037536 & -6.615507514 & RRd   & RRab  & 0.410376 & 0.549370 & 0.74699 & -2.24 & 0.80  & 0.76 \\
    4566681671773510272 & 258.6992565 & 19.95027836 & RRd   & RRab  & 0.409568 & 0.549525 & 0.74531 & -2.24 & 0.80  & 0.76 \\
    1618372955925759104 & 218.000715 & 61.03200569 & RRd   & RRd   & 0.410004 & 0.549531 & 0.74610 & -2.24 & 0.80  & 0.76 \\
\hline
\end{tabular}
\end{table*}

\begin{table*}
\fontsize{7}{8.4}\selectfont
\centering
\contcaption{  }
\label{tab:three}
\begin{tabular}{lllllcccccc}
\hline
	\multicolumn{1}{c}{ {\it Gaia} Id  } &   \multicolumn{1}{c}{RA} & \multicolumn{1}{c}{DEC} &  \multicolumn{2}{c}{{\it Gaia} Classif.} & \multicolumn{1}{c}{ $P_1$}   & \multicolumn{1}{c}{ $P_0$ } & \multicolumn{1}{c} { $P_1/P_0$ }  & \multicolumn{1}{c}{[Fe/H] }  & \multicolumn{2}{c}{$M/M_{\sun}$} \\
\multicolumn{1}{c}{  } &  \multicolumn{2}{c}{(2016.0)}  &DR2  &DR3    &\multicolumn{1}{c}{[day]}  &\multicolumn{1}{c}{[day]} & & \multicolumn{1}{c}{ $\pm$0.17 }  & \multicolumn{1}{c}{[$\alpha$/Fe]=0.0} & \multicolumn{1}{c}{[$\alpha$/Fe]=0.3 }   \\
\multicolumn{1}{c}{(1)} & \multicolumn{1}{c}{(2)} &\multicolumn{1}{c}{(3)}  &\multicolumn{1}{c}{(4)} & \multicolumn{1}{c}{(5)} & \multicolumn{1}{c}{(6)}   & \multicolumn{1}{c}{(7)}  & \multicolumn{1}{c}{(8)}   & \multicolumn{1}{c}{(9)} &\multicolumn{1}{c}{(10)} &\multicolumn{1}{c}{(11)}   \\
\hline  
\\	    

    6867352165972138240 & 299.8063234 & -18.70102033 &       & RRd   & 0.415954 & 0.557759 & 0.74576 & -2.31 & 0.81  & 0.78 \\
    4687462165108740480 & 17.34971879 & -72.37296948 & RRd   & RRab  & 0.416327 & 0.557910 & 0.74623 & -2.31 & 0.81  & 0.77 \\
    1900642359851788032 & 337.3599522 & 30.90351848 & RRd   & RRab  & 0.416971 & 0.558137 & 0.74708 & -2.31 & 0.81  & 0.77 \\
    3173553022940488704 & 72.80581235 & -14.37000551 & RRd   & RRc   & 0.416708 & 0.558375 & 0.74629 & -2.32 & 0.81  & 0.77 \\
\hline
\end{tabular}
\end{table*}

\begin{table*}
\fontsize{7}{8.4}\selectfont
\centering
\contcaption{  }
\label{tab:three}
\begin{tabular}{lllllcccccc}
\hline
\multicolumn{1}{c}{ {\it Gaia} Id  } &   \multicolumn{1}{c}{RA} & \multicolumn{1}{c}{DEC} &  \multicolumn{2}{c}{{\it Gaia} Classif.} & \multicolumn{1}{c}{ $P_1$}   & \multicolumn{1}{c}{ $P_0$ } & \multicolumn{1}{c} { $P_1/P_0$ }  & \multicolumn{1}{c}{[Fe/H] }  & \multicolumn{2}{c}{$M/M_{\sun}$} \\
\multicolumn{1}{c}{  } &  \multicolumn{2}{c}{(2016.0)}  &DR2  &DR3    &\multicolumn{1}{c}{[day]}  &\multicolumn{1}{c}{[day]} & & \multicolumn{1}{c}{ $\pm$0.17 }  & \multicolumn{1}{c}{[$\alpha$/Fe]=0.0} & \multicolumn{1}{c}{[$\alpha$/Fe]=0.3 }   \\
\multicolumn{1}{c}{(1)} & \multicolumn{1}{c}{(2)} &\multicolumn{1}{c}{(3)}  &\multicolumn{1}{c}{(4)} & \multicolumn{1}{c}{(5)} & \multicolumn{1}{c}{(6)}   & \multicolumn{1}{c}{(7)}  & \multicolumn{1}{c}{(8)}   & \multicolumn{1}{c}{(9)} &\multicolumn{1}{c}{(10)} &\multicolumn{1}{c}{(11)}   \\
\hline  
\\	    

    1300083989160033792 & 251.9718735 & 24.63474227 &       & RRd   & 0.427406 & 0.572733 & 0.74626 & -2.44 & 0.84  & 0.80 \\
    4927618285099847168 & 19.20151865 & -51.62454311 & RRd   & RRd   & 0.427085 & 0.572804 & 0.74561 & -2.44 & 0.84  & 0.80 \\
    4440234707843633408 & 248.1910362 & 8.524400043 & RRd   & RRc   & 0.427657 & 0.573314 & 0.74594 & -2.45 & 0.84  & 0.80 \\
    4685983970589274496 & 13.70699748 & -72.71655924 & RRc   & RRd   & 0.427403 & 0.573317 & 0.74549 & -2.45 & 0.84  & 0.80 \\
\hline
\end{tabular}
\end{table*}    \clearpage 
\vfill \eject \renewcommand{\thetable}{10}
\begin{table*}
\fontsize{6}{7.2}\selectfont  
\centering

\caption{Ninety-three {\it Gaia} stars classified `RRd' in  DR2 or DR3,  with
similar RA,DEC coordinates but different identification numbers in DR2 and DR3.
The complete table is given here and online in the Supporting Information. }

\label{tab:ten}
\begin{tabular}{llccclllcccc}
\hline \\
\multicolumn{5}{c}{{\it Gaia} DR2} &\multicolumn{1}{c}{} & \multicolumn{5}{c}{{\it Gaia} DR3} \\ %
\cline{1-5} \cline{7-11} \\ %
\multicolumn{1}{c}{DR2  } & \multicolumn{1}{c}{RR}   & \multicolumn{1}{c}{ $P_1$ }    & \multicolumn{1}{c} { $P_0$ }  &  \multicolumn{1}{c}{$P_1/P_0$} & & \multicolumn{1}{c}{DR3}  
	& \multicolumn{1}{c}{RR} &  \multicolumn{1}{c}{$P_1$ } & \multicolumn{1}{c}{ $P_0$}   & \multicolumn{1}{c}{ $P_1/P_0$ }   & \multicolumn{1}{c}{cRRd?}   \\
\multicolumn{1}{c}{Identification No. }   & \multicolumn{1}{c}{class} & \multicolumn{1}{c}{[day]} &\multicolumn{1}{c}{[day]}  & & &  \multicolumn{1}{c}{Identification No. } &  \multicolumn{1}{c}{class} 
	& \multicolumn{1}{c}{[day]} & \multicolumn{1}{c}{[day]} & \multicolumn{1}{c}{ }   \\
\multicolumn{1}{c}{(1)} & \multicolumn{1}{c}{(2)} &\multicolumn{1}{c}{(3)}  &\multicolumn{1}{c}{(4)} & \multicolumn{1}{c}{(5)} && \multicolumn{1}{c}{(6)}  & \multicolumn{1}{c}{(7)}   & \multicolumn{1}{c}{(8)}
	& \multicolumn{1}{c}{(9)} & \multicolumn{1}{c}{(10)} &  \multicolumn{1}{c}{(11)} \\
\hline  
\\	    
1442424496748916480   &  RRd   &  0.404211   &  0.541729   &  0.74615   &&  1442424501044402432   &  RRd   &  0.404205   &  0.541687   &  0.74620 & yes\\
1470192632844893568   &  RRd   &  0.382444   &  0.513261   &  0.74513   &&  1470192632845260288   &  RRd   &  0.382446   &  0.513238   &  0.74516 & yes\\
1554867810007895296   &  RRd   &  0.350923   &  0.471790   &  0.74381   &&  1554867810004952192   &  RRd   &  0.350912   &  0.471762   &  0.74383 & yes\\
1639360120343239808   &  RRc   &  0.361888   &  \dots      &   \dots    &&  1639360124638557440   &  RRd   &  0.361899   &  0.486449   &  0.74396 & yes\\
1745948362385628416   &  RRab  &   \dots     &  0.567080   &   \dots    &&  1745948362391096832   &  RRd   &  0.413354   &  0.554544   &  0.74539 & yes\\
2510409037347211264   &  RRab  &   \dots     &  0.353731   &   \dots    &&  2510409041642733568   &  RRd   &  0.353735   &  0.475571   &  0.74381 & yes\\
2862259978075970816   &  RRd   &  0.397965   &  0.534996   &  0.74387   &&  2862259978077052160   &  RRab  &   \dots     &  0.534997   &   \dots  & no\\
2972392044878569984   &  RRd   &  0.385688   &  0.516862   &  0.74621   &&  2972392044879268096   &  RRc   &  0.385673   &  \dots      &  \dots   & yes\\
3059619325966429824   &  RRd   &  0.396827   &  0.542776   &  0.73111   &&  3059619325975593600   &  RRab  &   \dots     &  0.542772   &  \dots   & no\\
3746580820769812096   &  RRc   &  0.394827   &  \dots      &  \dots     &&  3746580825061729536   &  RRd   &  0.394837   &  0.528923   &  0.74649 & yes\\
4110892114557448320   &  RRd   &  0.308733   &  0.424991   &  0.72644   &&  4110892114567467648   &  RRc   &  0.308732   &   \dots     &   \dots  & no\\
4235196473826059776   &  RRab  &   \dots     &  0.502739   &   \dots    &&  4235196473831446016   &  RRd   &  0.376671   &  0.502762   &  0.74920 & no\\
4324043335540793856   &  RRd   &  0.316984   &  0.428706   &  0.73940   &&  4324043335542461568   &  RRc   &  0.316983   &  \dots      &   \dots  & no\\
4633516241977845888   &  RRd   &  0.398702   &  0.534480   &  0.74596   &&  4633516241978319616   &  RRd   &  0.398715   &  0.534498   &  0.74596 & yes\\
4636274160737689600   &  RRc   &  0.405238   &   \dots     &   \dots    &&  4636274160738387200   &  RRd   &  0.405276   &  0.405272   &  1.00001 & no\\
4648222210001502464   &  RRc   &  0.373544   &   \dots     &   \dots    &&  4648222210002456832   &  RRd   &  0.369157   &  0.495738   &  0.74466 & yes\\
4651282013425522432   &  RRd   &  0.400835   &  0.537268   &  0.74606   &&  4651282013450091776   &  RRab  &   \dots     &  0.537268   &   \dots  & yes\\
4658683032358041600   &  RRd   &  0.376682   &  0.516473   &  0.72934   &&  4658683032375958272   &  RRab  &   \dots     &  0.516449   &   \dots  & no\\
4658878779783703680   &  RRd   &  0.360987   &  0.484868   &  0.74451   &&  4658878779816813824   &  RRd   &  0.360978   &  0.484887   &  0.74446 & yes\\
4658922382303351424   &  RRd   &  0.350045   &  0.471305   &  0.74271   &&  4658922382326774400   &  RRab  &   \dots     &  0.471294   &   \dots  & yes\\ 
4658949625262354176   &  RRd   &  0.374826   &  0.503561   &  0.74435   &&  4658949629598632320   &  RRd   &  0.374834   &  0.503584   &  0.74433 & yes\\
4659528934779761152   &  RRd   &  0.344433   &  0.463521   &  0.74308   &&  4659528934791553280   &  RRd   &  0.344428   &  0.463511   &  0.74308 & yes\\
4659637580246932480   &  RRd   &  0.447465   &  0.614569   &  0.72810   &&  4659637580261331200   &  RRab  &   \dots     &  0.614569   &  \dots   & no\\
4659654931927007616   &  RRd   &  0.367872   &  0.494357   &  0.74414   &&  4659654931928897024   &  RRd   &  0.367880   &  0.494368   &  0.74414 & yes\\
4659671763888265472   &  RRd   &  0.380903   &  0.511526   &  0.74464   &&  4659671768199981824   &  RRd   &  0.380875   &  0.510599   &  0.74594 & yes\\ 
4659675513403593344   &  RRd   &  0.350825   &  0.471448   &  0.74414   &&  4659675509114516224   &  RRd   &  0.350835   &  0.471862   &  0.74351 & yes\\
4659692826408274176   &  RRd   &  0.370706   &  0.498111   &  0.74422   &&  4659692830708745472   &  RRd   &  0.370707   &  0.498099   &  0.74424 & yes\\
4659700183683419520   &  RRd   &  0.360174   &  0.483727   &  0.74458   &&  4659700183698734976   &  RRd   &  0.360174   &  0.483738   &  0.74456 & yes\\
4659826077766598912   &  RRd   &  0.382680   &  0.512992   &  0.74598   &&  4659826077803805696   &  RRc   &  0.382678   &  \dots      &  \dots   & yes\\
4659834908223144704   &  RRd   &  0.397647   &  0.536067   &  0.74179   &&  4659834908254445440   &  RRab  &  \dots      &  0.536064   &   \dots  & no \\
4659908197564758528   &  RRd   &  0.353227   &  0.474748   &  0.74403   &&  4659908197576707968   &  RRd   &  0.353225   &  0.474770   &  0.74399 & yes\\
4660156236217361024   &  RRd   &  0.354026   &  0.483068   &  0.73287   &&  4660156240543558016   &  RRd   &  0.354031   &  0.475831   &  0.74403 & yes\\ 
4660200667680115840   &  RRd   &  0.366549   &  0.492696   &  0.74397   &&  4660200667688016896   &  RRd   &  0.366532   &  0.492668   &  0.74397 & yes\\
4660233893536343296   &  RRd   &  0.350483   &  0.470950   &  0.74420   &&  4660233893551211648   &  RRd   &  0.350482   &  0.470935   &  0.74423 & yes\\
4660303609425096960   &  RRd   &  0.404299   &  0.542342   &  0.74547   &&  4660303609463662592   &  RRd   &  0.404285   &  0.542324   &  0.74547 & yes\\
4660310928052039296   &  RRd   &  0.364227   &  0.489297   &  0.74439   &&  4660310928085981184   &  RRc   &  0.364223   &   \dots     &  \dots   & yes\\
4660388748544243840   &  RRd   &  0.356839   &  0.479427   &  0.74430   &&  4660388752886260992   &  RRd   &  0.356837   &  0.479440   &  0.74428 & yes\\
4660457884649832320   &  RRc   &  0.381989   &   \dots     &   \dots    &&  4660457884679688832   &  RRd   &  0.381947   &  0.512286   &  0.74557 & yes\\
4660494099824933504   &  RRd   &  0.362307   &  0.486757   &  0.74433   &&  4660494099842102528   &  RRd   &  0.362300   &  0.486740   &  0.74434 & yes\\
4661704279871117056   &  RRd   &  0.405208   &  0.553905   &  0.73155   &&  4661704284176412928   &  RRab  &  \dots      &  0.553881   &   \dots  & no \\
4661711946372879104   &  RRc   &  0.354868   &   \dots     &   \dots    &&  4661711946399141760   &  RRd   &  0.354845   &  0.477160   &  0.74366 & yes\\
4661827562621191680   &  RRd   &  0.398174   &  0.534498   &  0.74495   &&  4661827566918655616   &  RRd   &  0.398185   &  0.534484   &  0.74499 & yes\\
4661898618539942016   &  RRd   &  0.355655   &  0.478020   &  0.74402   &&  4661898622837169280   &  RRd   &  0.355654   &  0.478019   &  0.74402 & yes\\
4661935417833462528   &  RRd   &  0.384496   &  0.527400   &  0.72904   &&  4661935422148092928   &  RRab  &   \dots     &  0.527396   &   \dots  & no \\
4662096530637361920   &  RRd   &  0.353748   &  0.475625   &  0.74375   &&  4662096534955211008   &  RRd   &  0.353741   &  0.475592   &  0.74379 & yes\\
4662153915714378624   &  RRc   &  0.355059   &   \dots     &   \dots    &&  4662153915714378752   &  RRd   &  0.361888   &  0.486373   &  0.74405 & yes\\
4662164150585065728   &  RRd   &  0.361402   &  0.485739   &  0.74403   &&  4662164154918918912   &  RRc   &  0.361397   &  \dots      &  \dots   & yes\\
4662190165209306368   &  RRc   &  0.316507   &   \dots     &  \dots     &&  4662190165231759616   &  RRd   &  0.316515   &  0.436459   &  0.72519 & no\\
4662494725632593280   &  RRd   &  0.351306   &  0.472274   &  0.74386   &&  4662494729932836736   &  RRd   &  0.351305   &  0.472298   &  0.74382 & yes\\
4662541012490316928   &  RRd   &  0.405468   &  0.544130   &  0.74517   &&  4662541012500986752   &  RRd   &  0.405450   &  0.544074   &  0.74521 & yes\\
4662548258109049216   &  RRd   &  0.357558   &  0.488019   &  0.73267   &&  4662548262402840320   &  RRd   &  0.357544   &  0.480576   &  0.74399 & yes\\
4662759261253713920   &  RRd   &  0.366206   &  0.492147   &  0.74410   &&  4662759265550901120   &  RRd   &  0.366213   &  0.492147   &  0.74411 & yes\\
4663021048101937920   &  RRd   &  0.456682   &  0.619004   &  0.73777   &&  4663021052404858112   &  RRab  &   \dots     &  0.619009   &   \dots  & no \\
4663539192970237312   &  RRd   &  0.363657   &  0.486239   &  0.74790   &&  4663539197292422400   &  RRd   &  0.363649   &  0.488726   &  0.74407 & yes\\
4664649463478345344   &  RRc   &  0.358344   &  \dots      &  \dots     &&  4664649463479394304   &  RRd   &  0.358351   &  0.481367   &  0.74444 & yes\\
4664687396624659968   &  RRc   &  0.365079   &  \dots      &  \dots     &&  4664687396630451968   &  RRd   &  0.365090   &  0.490241   &  0.74472 & yes\\
4672266978814687232   &  RRab  &   \dots     &  0.386582   &  \dots     &&  4672266983111107584   &  RRd   &  0.386557   &  0.386552   &  1.00001 & no\\
4684212695910169472   &  RRc   &   \dots     &  0.358098   &  \dots     &&  4684212695910903552   &  RRd   &  0.358100   &  0.358110   &  0.99997 & no \\
4685971463639445760   &  RRab  &   \dots     &  0.368998   &  \dots     &&  4685971467924582144   &  RRd   &  0.369089   &  0.495622   &  0.74470 & yes\\
4687708485755389696   &  RRab  &   \dots     &  0.529218   &  \dots     &&  4687708490052845824   &  RRd   &  0.388596   &  0.529171   &  0.73435 & no \\
4691099693149222144   &  RRc   &  0.376480   &   \dots     &  \dots     &&  4691099693150235520   &  RRd   &  0.376476   &  0.504787   &  0.74581 & yes\\
4696179432573411968   &  RRab  &   \dots     &  0.552794   &  \dots     &&  4696179436869474560   &  RRd   &  0.409570   &  0.552797   &  0.74090 & no\\
4702317632329902720   &  RRd   &  0.377764   &  0.507086   &  0.74497   &&  4702317632330480000   &  RRd   &  0.377758   &  0.507025   &  0.74505 & yes\\
4702523103565603456   &  RRc   &  0.393048   &  \dots      &  \dots     &&  4702523103566168576   &  RRd   &  0.385375   &  0.517648   &  0.74447 & yes\\
4761538629975910144   &  RRc   &  0.363155   &   \dots     &   \dots    &&  4761538634272724992   &  RRd   &  0.363138   &  0.487908   &  0.74428 & yes\\
4989634657916320384   &  RRc   &  0.311268   &   \dots     &   \dots    &&  4989634657916646784   &  RRd   &  0.450773   &  0.450781   &  0.99998 & no\\
5027218538038604416   &  RRab  &   \dots     &  0.356637   &   \dots    &&  5027215587395447936   &  RRd   &  0.356610   &  0.479222   &  0.74414 & yes\\
5027218572398373632   &  RRd   &  0.355602   &  0.477989   &  0.74395   &&  5027218576695545984   &  RRd   &  0.355613   &  0.477969   &  0.74401 & yes\\
5207713821393001088   &  RRc   &  0.380464   &   \dots     &   \dots    &&  5207713821395190272   &  RRd   &  0.380554   &  0.510787   &  0.74503 & yes\\
5217685739026607744   &  RRc   &  0.363085   &  \dots      &   \dots    &&  5217685739027372672   &  RRd   &  0.363083   &  0.487345   &  0.74502 & yes\\
5280076869210120320   &  RRd   &  0.334732   &  0.461674   &  0.72504   &&  5280076873508611840   &  RRc   &  0.334733   &  \dots      &   \dots  & no \\
5280966928168741888   &  RRd   &  0.406993   &  0.544409   &  0.74759   &&  5280966928168742016   &  RRab  &   \dots     &  0.544438   &  \dots   & no \\
5283274425126682112   &  RRd   &  0.358503   &  0.481713   &  0.74423   &&  5283274425132263552   &  RRc   &  0.358504   &  \dots      &   \dots  & yes\\
5283378191529964032   &  RRc   &  0.387615   &   \dots     &   \dots    &&  5283378191535050624   &  RRd   &  0.387618   &  0.519719   &  0.74582 & yes\\
5283597436723545088   &  RRd   &  0.362140   &  0.486305   &  0.74468   &&  5283597441025607680   &  RRd   &  0.362137   &  0.486319   &  0.74465 & yes\\
5283961826060223488   &  RRd   &  0.361554   &  0.485932   &  0.74404   &&  5283961826061998976   &  RRd   &  0.361553   &  0.485941   &  0.74403 & yes\\
5284028793186044544   &  RRd   &  0.389680   &  0.523056   &  0.74501   &&  5284028788878285056   &  RRd   &  0.389684   &  0.523041   &  0.74504 & yes\\
5284228831277638016   &  RRd   &  0.397327   &  0.532285   &  0.74646   &&  5284228835579972224   &  RRd   &  0.397327   &  0.532194   &  0.74658 & yes\\
5291980560050240384   &  RRd   &  0.358912   &  0.489331   &  0.73347   &&  5291980564347504128   &  RRc   &  0.358927   &   \dots     &   \dots  & no \\
5374090129908676864   &  RRab  &  \dots      &  0.549998   &   \dots    &&  5374090134210225024   &  RRd   &  0.398523   &  0.549971   &  0.72463 & no\\
5386160091817778560   &  RRd   &  0.394330   &  0.528531   &  0.74609   &&  5386160091818906752   &  RRd   &  0.394341   &  0.528495   &  0.74616 & yes\\
5388635603592569856   &  RRd   &  0.449834   &  0.612815   &  0.73404   &&  5388635607890342016   &  RRab  &   \dots     &  0.612822   &   \dots  & no\\
5806470058176552320   &  RRd   &  0.316214   &  0.427444   &  0.73978   &&  5806470058179357056   &  RRc   &  0.427444   &   \dots     &   \dots  & no \\
\hline
\end{tabular}
\end{table*}

\begin{table*}
\fontsize{6}{7.2}\selectfont  
\centering
\contcaption{ }

\label{tab:nine}
\begin{tabular}{llccclllcccc}
\hline \\
\multicolumn{5}{c}{{\it Gaia} DR2} &\multicolumn{1}{c}{} & \multicolumn{5}{c}{{\it Gaia} DR3} \\ %
\cline{1-5} \cline{7-11} \\ %
\multicolumn{1}{c}{DR2  } & \multicolumn{1}{c}{RR}   & \multicolumn{1}{c}{ $P_1$ }    & \multicolumn{1}{c} { $P_0$ }  &  \multicolumn{1}{c}{$P_1/P_0$} & & \multicolumn{1}{c}{DR3}  
	& \multicolumn{1}{c}{RR} &  \multicolumn{1}{c}{$P_1$ } & \multicolumn{1}{c}{ $P_0$}   & \multicolumn{1}{c}{ $P_1/P_0$ }   & \multicolumn{1}{c}{cRRd?}   \\
\multicolumn{1}{c}{Identification No. }   & \multicolumn{1}{c}{class} & \multicolumn{1}{c}{[day]} &\multicolumn{1}{c}{[day]}  & & &  \multicolumn{1}{c}{Identification No. } &  \multicolumn{1}{c}{class} 
	& \multicolumn{1}{c}{[day]} & \multicolumn{1}{c}{[day]} & \multicolumn{1}{c}{ }   \\
\multicolumn{1}{c}{(1)} & \multicolumn{1}{c}{(2)} &\multicolumn{1}{c}{(3)}  &\multicolumn{1}{c}{(4)} & \multicolumn{1}{c}{(5)} && \multicolumn{1}{c}{(6)}  & \multicolumn{1}{c}{(7)}   & \multicolumn{1}{c}{(8)}
	& \multicolumn{1}{c}{(9)} & \multicolumn{1}{c}{(10)} &  \multicolumn{1}{c}{(11)} \\
\hline  
\\	    
5890460273979029120   &  RRab  &   \dots     &  0.454348   &   \dots    &&  5890460273985358080   &  RRd   &  0.330941   &  0.454332   &  0.72841 & no\\
5917634669503367040   &  RRd   &  0.361348   &  0.484975   &  0.74508   &&  5917634669509111424   &  RRd   &  0.361345   &  0.484981   &  0.74507 & yes\\
5949519677744321152   &  RRd   &  0.334097   &  0.446238   &  0.74870   &&  5949519682079629952   &  RRc   &  0.334094   &   \dots     &  \dots   & no \\
5987053431312452352   &  RRd   &  0.425895   &  0.583084   &  0.73042   &&  5987053435637092736   &  RRc   &  0.425893   &   \dots     &  \dots   & no \\
6001562144383429504   &  RRd   &  0.357057   &  0.483966   &  0.73777   &&  6001562144385811712   &  RRab  &   \dots     &  0.483966   &  \dots   & no \\
6073397121654113024   &  RRd   &  0.351762   &  0.483078   &  0.72817   &&  6073397121676369664   &  RRab  &  \dots      &  0.483080   &  \dots   & no \\
6205639538360433536   &  RRd   &  0.421985   &  0.569154   &  0.74142   &&  6205639542660016512   &  RRab  &  \dots      &  0.569154   &   \dots  & no \\
6428241341059336832   &  RRd   &  0.383928   &  0.520263   &  0.73795   &&  6428241341060660352   &  RRab  &   \dots     &  0.520274   &   \dots  & no \\
6599225878894338560   &  RRab  &   \dots     &  0.364276   &   \dots    &&  6599225878894766720   &  RRd   &  0.364276   &  0.489257   &  0.74455 & yes\\
6635412039221031424   &  RRd   &  0.454742   &  0.614328   &  0.74023   &&  6635412043523100928   &  RRab  &   \dots     &  0.614334   &   \dots  & no \\
\hline
\end{tabular}
\end{table*}

   \clearpage

\end{document}